\documentclass{Thesis_Ari}

\ifpdf
    \pdfcatalog { /PageMode (/UseOutlines)
               /OpenAction (fitbh) }
\fi

\title{Self-consistent \\ Green's functions \\ with three-body forces}

\ifpdf
	\author{Arianna Carbone}
  	\collegeordept{Departament d'Estructura i Constituents de la Mat\`eria}
  	\university{Universitat de Barcelona}
  	\faculty{Facultat de F\'isica}
  	\crest{\includegraphics[width=60mm]{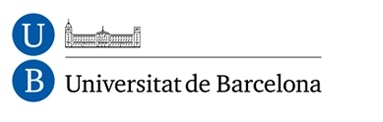}}
\else
  	\author{Arianna Carbone}
  	\collegeordept{Departament d'Estructura i Constituents de la Mat\`eria}
  	\university{Universitat de Barcelona}
  	\faculty{Facultat de F\'isica}
 	\crest{\centering\includegraphics[width=60mm]{logo_home_nou.eps}}
\fi

\degree{Doctor per la Universitat de Barcelona}
\degreedate{11 d'abril de 2014}
\supervisors{Dr. Artur Polls Mart\'i i Dr. Arnau Rios Huguet}
\tutors{Dr. Artur Polls Mart\'i}
\prog{F\'{i}sica}

\newcommand{\beq}{\begin{equation}}
\newcommand{\enq}{\end{equation}}
\newcommand{\beqn}{\begin{eqnarray}}
\newcommand{\enqn}{\end{eqnarray}}
\newcommand{\al}{\alpha}
\newcommand{\be}{\beta}
\newcommand{\ga}{\gamma}
\newcommand{\de}{\delta}
\newcommand{\ep}{\epsilon}
\newcommand{\lm}{\lambda}
\newcommand{\om}{\omega}
\newcommand{\vep}{\varepsilon}
\newcommand{\ta}{\theta}
\newcommand{\sig}{\sigma}
\def\t{\tau}
\newcommand{\gz}{G^{(0)}}
\newcommand{\f}[2]{\frac{#1}{#2}}
\renewcommand{\d}{\mathrm d}
\newcommand{\lphizero}{\langle\Phi_0^N|}
\newcommand{\rphizero}{|\Phi_0^N\rangle}
\newcommand{\lPsizero}{\langle\Psi_0^N|}
\newcommand{\rPsizero}{|\Psi_0^N\rangle}
\newcommand{\h}[1]{\hat{#1}}
\newcommand{\dg}{\dagger}
\newcommand{\T}{{\cal T}}
\newcommand{\nn}{\nonumber}
\newcommand{\dev}{\frac{\partial}{\partial t}}
\renewcommand{\a}[2]{a_{#1#2}}
\newcommand{\ad}[1]{a_{#1}^\dagger}
\newcommand{\ii}{\textrm{i}}
\def\ran{\rangle}
\def\lan{\langle}
\newcommand{\bd}{\boldsymbol}

\newcommand{\eq}[1]{Eq.~(\ref{#1})}

\begin{document}

%\newpage
%\null\thispagestyle{empty}

%\newpage
%\null\thispagestyle{empty}
%\clearpage{\pagestyle{empty}}
%\input{Dedication/copertina}

\maketitle
\newpage
\null\thispagestyle{empty}

\frontmatter % book mode only

%\input{Dedication/dedication}
%\newpage
%\null\thispagestyle{empty}

%\input{Dedication/phrase}
%\newpage
%\null\thispagestyle{empty}

%\input{Acknowledgement/acknowledgements}
%\newpage
%\null\thispagestyle{empty}
%\clearpage{\pagestyle{empty}\cleardoublepage}

%\input{Acknowledgement/ringraziamenti}
%\newpage
%\null\thispagestyle{empty}
%\clearpage{\pagestyle{empty}\cleardoublepage}

\pagestyle{toc}
\renewcommand{\contentsname}{Table of contents}
\tableofcontents
\clearpage{\pagestyle{empty}\cleardoublepage}

%\pagestyle{simpleresum}
%\input{Summary/summary}
%\clearpage{\pagestyle{empty}\cleardoublepage}

\mainmatter % book mode only

\pagestyle{mainmatter}

%%%%%%%%%%%%%%%%%%%% INTRODUCTION %%%%%%%%%%%%%%%%%%%%%%%

\chapter{Introduction}
\label{chapter:introduction}

The quantum many-body problem is an everlasting challenge for theoretical physics. The aim sought is to analyze, at the quantum level, the observables arising from a group of $N$ interacting particles. If the $N$ particles in the many-body system are non-interacting, the system is ideal and can be solved as the sum of $N$ one-body (1B) problems. If the system is interacting, the only analytically solvable system is the two-body (2B) one. This can be recast in the form of two 1B problems, one for the center of mass and the other for the relative coordinate of the two particles. The three-body (3B) problem is already not solvable analytically, but an exact solution could be obtained by solving the Faddeev equations. As the number of particles increases, one is forced to resort to approximate solutions. The objective of many-body physics has constantly been that of dealing with systems formed from few to many interacting particles, from finite systems, such as nuclei, atoms or molecules, to the infinite ones, such as the electron gas or nuclear matter present in the interior of stars. The quest is to provide the best description of the system via the application of consistent approximations, with the aim of achieving results to be contrasted with the experimental ones.

In the specific case of nuclear physics, the study and understanding of interacting nuclear systems has spanned from the finite nuclei to the infinite nuclear matter case. In order to characterize such systems, the interactions which govern the behavior of nucleons in their interior need to be sorted out. The nucleon-nucleon (NN) interactions devised to do so can be divided into two categories: \emph{microscopic} and \emph{phenomenological}. The microscopic potentials are models constructed to fit experimental data of NN scattering and properties of the deuteron. All microscopic potentials have a common one-pion exchange (OPE) long-range part and usually differ in the treatment of the intermediate and especially short-range part. Due to the strong repulsive short-range part, and the presence of tensor components, these potentials cannot be used directly in perturbative calculations and need to be renormalized through the use of many-body methods. Phenomenological potentials are simpler, and are constructed fitting a number of coupling constants to empirical saturation properties of nuclear matter and observables of ground states of nuclei. For this reason their validity far from these conditions can be questionable. These potentials can be of a relativistic or nonrelativistic kind and are used directly in mean-field calculations, such as relativistic mean field theory or the Hartree-Fock (HF) approximation.

In the present thesis we will focus our study on infinite nuclear matter, analyzing the properties of the two systems which stand at the extremes in terms of isospin asymmetry: symmetric nuclear matter (SNM), with an equal number of neutrons and protons, and pure neutron matter (PNM), with only neutrons. We will base our approach on the use of realistic microscopic potentials, hence we will need to choose a many-body technique in order to treat the potential. We choose to perform our calculations in the framework of the self-consistent Green's function's (SCGF) theory.  This approach is based on a diagrammatic expansion of the single-particle (SP) Green's function, or \emph{propagator}. This method was devised to treat the correlated, i.e. beyond mean-field, behavior of strongly interacting systems, such as nuclear matter. 

Unfortunately, it is a well established fact that whatever microscopic two-nucleon forces (2NFs) are used in the many-body calculation, empirical saturation properties of nuclear matter fail to be reproduced. Saturation densities appear at high values, with energies which are too attractive, overbinding nuclear matter. This discrepancy with empirical values bears some similarities to the case of light nuclei where, on the contrary, the 2NF based theory underbinds experimental data. The inclusion of three-nucleon forces (3NFs) has been the indispensable strategy to cure this deficiency. Hence, microscopic 3NFs have been mostly devised to provide attraction in light finite systems, small densities, and repulsion in infinite systems, high densities. However, phenomenological ingredients have often been chosen to model these 3NFs. To bypass the need to adjust the potential with \emph{ad hoc} contributions, an alternative has been provided by chiral effective effective field theory ($\chi$EFT). 

Featuring the consistent Hamiltonian at the 2B and 3B level provided by $\chi$EFT, in this thesis we will present calculations including both 2NFs and 3NFs. This will be dealt within an extended SCGF formalism, specifically formulated to include consistently 3B forces (3BFs).

In the next section we present an overview of the different many-body approaches which have been used to solve the infinite nuclear matter many-body problem. These approaches can treat the particles either relativistically or non-relativistically. The non-relativistic assumption seems reasonable, given that the energies involved in nuclear matter provide Fermi velocities which are less than one third the speed of light. All these approaches try to deal with the strong short-range force which acts in between nucleons. We will then focus, in the second section of this introduction, on the variety of microscopic potentials which, through the decades, have been devised to describe this kind of interactions. Finally, we will outline the general program of the thesis.

%%%%%%%%%%%%%%%%%%% The nuclear many-body problem %%%%%%%%%%%%%%%

\section{The nuclear many-body problem}
\label{section:mb_problem}

In theoretical nuclear physics, the goal to describe from an \emph{ab initio} point of view the basic properties of nuclear systems in terms of realistic interactions, has pushed the development of a variety of many-body methods. The quest to solve the nuclear many-body problem from first principles, i.e. starting from the Schr\"odinger equation, is appealing and very challenging. In the specific case of infinite nuclear matter, this aim is confronted with the lack of experimental results needed to constrain the construction of the theory from its building blocks, i.e. the Hamiltonian. However, the infinite system of nucleons interacting via the strong force, in spite of being ideal, is characterized by well-known empirical properties. These properties can be obtained extrapolating experimental results from finite nuclei. In this sense, the semi-empirical mass formula of Bethe and Weizs\"acker \citep{Weizs1935,Bethe1936}, provides a way to quantify these empirical properties:
\beq
B(N,Z)=a_VA-a_SA^{2/3}-a_C\frac{Z^2}{A^{1/3}}-a_A\frac{(N-Z)^2}{A}+\delta(A)\,.
\label{mass_formula}
\enq
The formula models the nucleus as an incompressible quantum drop \citep{Baym1971} formed by a number $N$ of neutrons and $Z$ of protons, with mass number $A=N+Z$. Each term in Eq.~(\ref{mass_formula}) captures a specific physical property of the nucleus. The first two contributions, the volume and surface terms, define respectively the binding energy in the interior and the correction due to the existence of the surface of the nucleus. The third contribution accounts for the Coulomb repulsion in between protons. The next term is known as symmetry energy, and shows how the nuclei which are isospin symmetric, i.e. $N=Z$, are energetically favored. The last term in Eq.~(\ref{mass_formula}) takes into account pairing effects and depends on wether the nuclei are even or odd in number of nucleons. All coefficients in Eq.~(\ref{mass_formula}), as well as $\delta(A)$, are fitted to reproduce the properties of nuclei along the nuclear chart.

Via experiments of electron scattering on nuclei, it has been tested that the charge distribution in the interior of heavy nuclei reaches a constant value. This charge density is proportional to the density inside the nucleus. The extrapolated value for very large mass number $A$ defines the \emph{empirical saturation density} of nuclear matter, $\rho_0=0.16$ fm$^{-3}$. In this limit, $A\rightarrow\infty$, the binding energy per nucleon of a symmetric nucleus, $N=Z$, can be obtained from the semi-empirical mass formula, Eq.~(\ref{mass_formula}), providing the value $a_V=16$ MeV. The empirical saturation properties of nuclear matter can then be defined as \citep{Mackie1977}:
\beq
\frac{E}{A}(\rho_0)=-16 \mathrm{MeV}\,, \quad \rho_0=0.16\,\mathrm{fm}^{-3}\,.
\enq
Any \emph{ab initio} many-body theory aims at reproducing these values in the energy curve of SNM. Furthermore, from Eq.~(\ref{mass_formula}), the value for the symmetry energy of nuclear matter can be obtained. In the case where the number of nucleons is taken to infinity, but the density $\rho$ is kept at a fixed value, we can rewrite the semi-empirical mass formula as
\beq
\frac{B(N,Z)}{A}=a_V-\alpha^2a_A\,,
\label{mass_form_2}
\enq
where we have neglected the Coulomb term which would drive the drop model unstable. In Eq.~(\ref{mass_form_2}), the isospin asymmetry parameter is identified with $\alpha=(N-Z)/A$. The symmetry energy accounts for the loss in binding energy for the system going from the symmetric to the asymmetric isospin case. The value of the symmetry energy in infinite matter cannot be directly inferred from the parameter $a_A$ \citep{Ste2005}. However, when the energy per nucleon of the system is known for different isospin asymmetries, the symmetry energy can be calculated as (note that $B(N,Z)=-E(\rho,\alpha))$:
\beq
\frac{S}{A}(\rho,\alpha=0)=\frac{1}{2}\frac{\partial ^2}{\partial\alpha^2}\frac{E(\rho,\alpha)}{A}\bigg|_{\alpha=0}\,.
\enq
The accepted value for this quantity at saturation density $\rho_0$ is $\sim32$ MeV \citep{Tsang2012}. This value is fairly well constrained from both experimental results on nuclei and theoretical models which try to predict its value. This quantity impacts various aspects in nuclear physics, from phenomena in nuclear astrophysics, all the way to nuclear structure and nuclear reactions. However, even though the value at saturation density can be considered as reliable, the density dependence of the symmetry energy remains up to today not completely known \citep{Horowitz2014}.  

It is customary to characterize the density dependence behavior of the energy in symmetric nuclear matter with a Taylor expansion \citep{Vid2009,Piek2009}:
\beq
E(\rho,\alpha=0)= E(\rho_0,\alpha=0)+\frac{1}{2}K_0\left(\frac{\rho-\rho_0}{3\rho_0}\right)^2+{\cal O}(3)\,.
\enq
The quantity $K_0$ is known as the compressibility modulus of nuclear matter and can be considered as a further empirical value to constrain the many-body theory under hand. Its formal expression is:
\beq
K=9\rho_0^2\frac{\partial ^2}{\partial\rho^2} \frac{E(\rho,\alpha)}{A}\bigg|_{\rho=\rho_0}\,;
\enq
the compressibility characterizes in some sense the behavior of the energy of SNM around saturation density. Its value can be estimated from the study of giant monopole resonances, and it falls between a narrow band of $K=240\pm10$ MeV \citep{Piek2009}.
\\

A reliable many-body formalism is such if it can reproduce these empirical properties for an infinite nuclear system. 
%The first step for the construction of an \emph{ab initio} method is the definition of the Hamiltonian which describes the way nucleons interact. Once the Hamiltonian is chosen, the many-body problem for the nuclear system has to be solved. 

The simplest approach to the nuclear many-body problem is the mean-field or HF approximation \citep{Hartree1928,Fock1930,Slater1930}. The idea underlying this kind of approximation is that the interacting system of fermions may be described in terms of a single Slater determinant. In other words, each particle is restated as a \emph{quasi}particle feeling a mean-field potential which accounts for its interaction with all the other particles of the system. This procedure can yield good results if used together with the phenomenological potentials, such as Skyrme or Gogny interactions, which are constructed as to fit empirical saturation properties of nuclear matter or finite nuclei observables  \citep{Ventura1992,Rikov2003}. The strength of these mean-field theories lies in the fact that they can be recast in terms of functionals of the energy. Nuclear Density Functional Theory (DFT) is built on theorems showing the existence of universal energy functionals for many-body systems, which include, in principle, the energy effect of all many-body correlations \citep{lalazis2004}.

However, in the case of using realistic potentials, the HF method drastically fails leading to unbound nuclei and nuclear matter. The strong short-range part of the potential, to which particles are exposed in the mean-field approximation, induces correlations in between the particles which cannot be solved within this kind of approach. Over the last 50 years, many-body formalisms capable to deal with the solution of this problem have been devised. 

Several approaches have been based on the variational principle to obtain the energy of the many-body ground state. To evaluate the Hamiltonian expectation value, a trial wave function, typically of a Jastrow kind \citep{Jastrow1955}, is built to incorporate the effects of correlations. This is done by the definition of correlation operators on top of the uncorrelated Slater determinant wave function. These Jastrow factors suppress the short-range and tensor components present in the wave function of two-close particles. If the correlated trial basis is close enough to the exact one, one can perform perturbation theory in the frame of the so called \emph{Correlated Basis Function} (CBF) theory \citep{Clark1966,Fantoni1998}. Alternatively, further variational techniques have been devised to compute the energy from the trial wave function. The energy evaluation using the correlated basis can be performed by solving the \emph{Fermi-Hyper-Netted-Chain} (FHNC) equations \citep{Pand1976}, which are obtained via a diagrammatic cluster expansion \citep{Fantoni1974,Fantoni1975}. In solving the FHNC equations, the operatorial structure of the Hamiltonian can lead to additional approximations to be considered, such as the \emph{Single Operator Chain} (SOC) one \citep{Pand1979}. This approximation is implemented to deal with the non-commutativity of the correlation functions, when solving the set of integral equations. If the variational method is exactly performed, results should provide an upper bound to the total energy per particle. 

\emph{Quantum Monte Carlo} (QMC) approaches have also been widely used to provide solution to the nuclear many-body problem. On a first step, one can compute the expectation value of the ground state by means of the \emph{Variational Monte Carlo} (VMC) calculation. A VMC calculation uses Monte Carlo integration to minimize the expectation value of the Hamiltonian using a variational trial wave function \citep{McMillan1965}. Subsequently, the output configurations of this calculation can be used as an input to more extensive approaches. According to the different samplings of the spin/isospin configuration space performed, these approaches exist in different versions. By treating the Schr\"odinger equation as a diffusion equation, these methods work by projecting out the exact lowest energy eigenstate from a trial wave function which is stochastically evolved. Whereas the \emph{Green's Functions Monte Carlo} (GFMC) method \citep{Ceperley1979} sums over all possible spin/isospin states of the system to build the trial wave function, the \emph{Auxiliary Field Diffusion Monte Carlo} (AFDMC) method \citep{Schmidt1999} relies on the Hubbard-Stratonovich transformation which reduces the number of states to be sampled. This allows the AFDMC method to be performed in systems with a higher number of nucleons to ensure that finite size effects are small. 

The \emph{Coupled Cluster} (or exp$S$) approach is also based on the use of a trial wave function, where the Slater determinant for the uncorrelated basis is multiplied by an exponential factor \citep{Coester1958,Coester1960}. The operator $S$ is formed by a set of creation-destruction operators that respectively add particles and holes on top of the Fermi sea to induce the correlated behavior on the wave function. The exponential $S$ ansazt, used in the coupled cluster equations, proves to be preferable with respect to other parametrizations of the trial wave function. In solving the energy expectation value, the coupled cluster equations can sum series of hole-hole and particle-particle scatterings, in some sense similar to the \emph{ladder approximation} which we will  use in the SCGF method \citep{Baardsen2013,Hag2013}.

A much implemented many-body method, which has gained high popularity in the field of nuclear many-body theory, is the one which was initiated by Brueckner in the mid-50's \citep{Brueckner1954,Brueckner1955a,Brueckner1955b}. This approach, known as the \emph{Brueckner-Goldstone} theory, is based on the Goldstone expansion \citep{Goldstone1957}, which uses \emph{Feynman diagrams} \citep{Feynman1949} to define a linked-diagram perturbation series for the ground-state energy of the many-body system. Most of the advances in the understanding of this theory were subsequently brought forward by Bethe and his collaborators \citep{Bethe1956,Rajaraman1967}, and for this reason one usually refers to this approach as the \emph{Bethe-Brueckner-Goldstone} theory. The diagrams appearing in the perturbation series give rise to the hole-line expansion, which in principle could be performed up to an arbitrary number of hole-lines \citep{Day1967,Day1981}. In practice, only calculations up to the two- and three-hole lines expansions have been performed up to now, which have proved the convergence of the series \citep{Baldo2001}. The \emph{Brueckner-Hartree-Fock} (BHF) approach corresponds to the lowest order in this \emph{Bethe-Brueckner-Goldstone} expansion. In the BHF approximation, only the sum of two-hole-line diagrams is considered, which is what accounts for the correlations present in the system. The potential is renormalized into an effective one, the $G$-matrix, which corresponds to the simultaneous particle-particle propagation on top of the Fermi momentum, iterated to all orders. The BHF method can be seen as a kind of mean-field approach, in that each SP momentum state is associated with a single SP energy, sum of a kinetic and a potential part obtained from the renormalized interaction. As far as the hole-line expansion goes, the contribution to the energy coming from the three-hole line diagrams has been computed \citep{Song1998}. This has proved the convergence of this kind of expansion, validating the calculations performed in the lowest-order (two hole-line) BHF approximation. Furthermore, a relativistic version of the BHF is also available, known as the \emph{Dirac-Brueckner-Hartree-Fock} approach \citep{Dalen2007}, in which the relativistic structure of the Dirac spinors is corrected to take into account the presence of the nuclear medium.

In the past decade, many-body approaches for the study of infinite nuclear matter have also been devised in the frame of chiral perturbation theory \citep{Lutz2000,Kaiser2002}. Observables are calculated with the help of an \emph{Effective Field Theory} formulated in terms of the Goldstone bosons, pions, and the low-lying baryons, nucleons. The energy per particle results in an expansion in terms of the Fermi momentum $p_\textrm F$ which, being twice the mass of the pion at saturation density, requires the pions to be kept as explicit degrees of freedom.

An innovative method was brought forward less than ten years ago which, by means of regularized potentials, avoids the problem related to strong correlations in the many-body wave function. This approach paved the way for the treating of nuclear matter with the use of \emph{Perturbation Theory} \citep{Bogner2005}. In these calculations, a potential obtained from $\chi$EFT is evolved to low-momentum via regularization techniques \citep{Nog2004,Bog2007}, preserving all the way the observables up to the cutoff value implemented in the regularization. Given the universality of low-momentum potentials, many different potentials, other than chiral ones, could be implemented in this approxiamation. Based on EFT insights, the concept which has been strongly stressed all along, in this kind of approaches, is that many-body forces are inevitable and must not be neglected in order to keep the calculation consistent \citep{Heb2011,Kru2013}. 

As already stressed in the introduction to this chapter, in this thesis we will perform calculations by means of the nonperturbative  SCGF theory. The SCGF approach has been continuously improved through the past decades to study infinite nuclear matter. This method is self-consistent in that the Green's function, which describes the propagating particle, is determined by the interaction of this particle with the surrounding ones. In turn this same interaction is described by the Green's function itself. An iterative self-consistent procedure is then due to find a stable solution for the propagator. In this approach, correlations are taken into account by summing up multiple particle-particle and hole-hole scatterings in the medium by means of the $T$-matrix. This treatment for nuclear matter is known as \emph{ladder approximation}. A diagrammatic approach by means of \emph{Feynman diagrams} is the preferred path to follow, to have a direct interpretation of the self-consistent equations one has to deal with. 

The first approach to many-particle systems from a Green's functions point of view was presented by the authors of Ref.~\citep{Mart1959}. A primitive attempt to carry out calculations for the ground-state properties of nuclear matter was presented not long after \citep{Puf1961,Rey1963}, with the use of simple separable potentials. First implementations with a hard-core realistic potential were performed by \citep{Foster1971,Fiset1972}. Later on, calculations with the use of certain realistic potentials were also presented by \citep{Web1985}. At this stage, the problem was solved only at the Hartree-Fock level. Subsequently, high improvements in the numerical implementation were introduced by the Barcelona and St. Louis groups \citep{Ram1988PhD}. With the use of realistic potentials, the problem was analyzed at zero temperature considering an all-order summation of particle-particle and hole-hole propagations. The calculations were performed in a quasi-particle approximation, meaning that the energy dependence of the spectral function was approximated to a delta-peak at the quasiparticle energy, and self-consistency was demanded only at the on-shell level \citep{Ramos1989a,Ram1989}. In addition, a careful analysis of the effects of short-range and tensor correlations on the SP self-energy and spectral function were analyzed in subsequent publications \citep{Vonderfecht1991a,Vonderfecht1991b,Vonderfecht1993}. Similar studies were presented at finite temperature by the Rostock group \citep{Alm1995,Sch1996}. Attention was especially drawn on the analysis of the occurrence of pairing instability, which appears below a critical temperature when the propagation of holes is considered \citep{Alm1993,Alm1996}. A substantial improvement was provided by the first attempts to consider fully dressed SP propagators, including the off-shell dependence of the spectral function. These were introduced by parametrizing the off-shell part with a set of Gaussian functions \citep{Dic1999}. The Ghent group tried to achieve a similar objective, using a discrete number of Dirac peaks \citep{Dew2002,Dew2003}. At the same time, complete off-shell results where presented both at zero and finite temperature by the Krakow group. First implementations in the off-shell part of the spectral function were done with simple separable potentials \citep{Bozek1999,Boz2001} and then, as an innovation, with separable versions of realistic NN potentials \citep{Boz2002,Boz2003}. The full off-shell dependency of the spectral function was then computed by the Krakow group with realistic potentials including both two- and, for the first time, three-body forces \citep{Som2008,Som2009,Som2009PhD}. A  milestone of the T{\"u}bingen and Barcelona groups has been that of achieving a detailed and extensive study of infinite matter at finite temperature. At zero temperature the $T$-matrix undergoes the so called \emph{pairing instability}, in which the formation of Cooper pairs in the hole-hole propagation part is favored and a superconducting phase appears. The definition of anomalous Green's functions would be necessary to solve this problem \citep{Dic2008}. Finite but low temperature automatically solves the problem related to this instability. A numerical procedure to calculate full off-shell structure of all relevant quantities at finite temperature was devised by Frick \emph{et al.} \citep{Fri2003,Fri2004PhD}. The effect of short range correlations was analyzed both for symmetric and asymmetric matter \citep{Frick2004,Fri2005,Rios2006}. Thermodynamical properties from a self-consistent Green's function approach were largely investigated using different kinds of realistic potentials \citep{Rio2006,Rio2007PhD,Rio2009Feb}.  The effect of correlations on the depletion of the SP momentum distribution is also still broadly investigated \citep{Rio2009Apr,Car2012,Rios2013}. Recently, a SCGF calculation of the nucleon mean-free path has been presented \citep{Rios2012}.

%%%%%%%%% The nuclear interaction %%%%%%%%

\section{The nuclear interaction}
\label{section:nuclear_int}

To solve the nuclear many-body problem, by means of any of the different approaches presented in the previous section, one needs to choose a specific definition of the strong interaction.

A potential is said to be \emph{realistic} if it is able to reproduce NN scattering data. After very primitive attempts of static local potentials \citep{Gammel1957} with additional spin-orbit terms included \citep{Eisenbud1941}, the most widely used potentials in the 60's were the Hamada-Johnston \citep{Hamada1962} and Yale \citep{Lassila1962} potentials. An improvement over these was presented later on by the authors of Ref.~\citep{Reid1968}. However, the Reid potential was mainly fit to reproduce the available $pp$ (proton-proton) data at the time. In a similar way, the Nijmegen \citep{Nagels1978}  and Paris \citep{Lacombe1980} potentials were also built to reproduce $pp$ phase shifts,  and hence provided a poor fit of the $np$ (neutron-proton) data. Fundamentally this problem was related to the charge-independence breaking that affects the strong interaction. Charge-independent potentials often implemented in calculations were the Argonne $v14$ \citep{Wiringa1984}, Urbana $v14$ \citep{Lag1981} and the Bonn potentials \citep{Machleidt1987,Machleidt1989}. The first high-precision NN microscopic potentials which were able to fit both $nn$ and $np$ scattering data up to an energy $E_\textrm{lab}=350$ MeV \citep{Sto1993}, were the Nijm93, Reid93 \citep{Sto1994} and Argonne $v18$ potentials \citep{Wir1995}. The Nijm93 and Reid93 potentials were built in momentum and spin/isospin space, while Argonne $v18$ was constructed in configuration space. These potentials were expanded in terms of all operators which respected the symmetries of the underlying behavior of the NN force. An extension of the Bonn potentials to include charge dependency was similarly presented a few years later, known as the CdBonn potential \citep{Mac2001}. Unlike other potentials, the Bonn and CdBonn forces were boson exchange kind potentials built in momentum space. 

Along with the formulation of realistic 2B potentials, 3BFs were constructed to be included in nuclear many-body calculations \citep{Coon1979,Day1978,Lag1981}. The first modern meson-exchange 3B potential was proposed by Fujita and Miyazawa \citep{Fuj1957}. It contained only a $2\pi$ exchange three nucleon interaction. This interaction was due to the scattering of a pion, which is exchanged between two nucleons, and a third nucleon via the $P$-wave $\Delta$ resonance. Other theoretical models, proposed later on, were the Tucson-Melbourne \citep{Coon1979} and Brazil \citep{Coelho1983} potentials, which furthermore included  $S$-wave scattering. This $2\pi$ exchange interaction was attractive in both nuclei and nuclear matter. While in nuclei this term would help to solve the underbinding obtained by the sole use of 2B forces (2BFs), it would increase the (already far too large) overbinding in nuclear matter. A phenomenological repulsive term, added to the Fujita-Miyazawa one, was introduced in the Urbana \citep{Pud1995} and Illinois \citep{Pie2001Jun} models for the 3B potential. This repulsive term mainly prevented nuclear matter from being too bound at high densities.

Another approach, which we will follow in this work, is inspired  by chiral perturbation theory. It exploits the structures of the spontaneous and explicit chiral symmetry breaking in quantum chromo-dynamics (QCD) \citep{Weinberg1990,Weinberg1991}. The theory of QCD describes the strong interaction within the standard model.  Its degrees of freedom are elementary particles, the quarks and gluons, whose coupling constant $\al_S$ runs to lower values as the energy increases. Perturbation theory is however inapplicable to a system described in terms of these fundamental particles, and for this reason one relies on the low-energy effective theory applied to QCD, \emph{chiral effective field theory}, $\chi$EFT. The separation of energy scales divides the family of quarks into two different sectors, the light sector with the \emph{u, d, s} quarks and the heavy sector with the \emph{c, b , t} quarks. This is exploited to integrate out the heavy degrees of freedom and define a low-energy theory which is based only on the light sector, where low refers to a typical hadronic scale of $\Lambda_{\rm QCD}=1$ GeV. In this new low-energy theory, the remaining particles can be considered as massless and the Lagrangian describing it exhibits an invariance under global transformations of the left- and right-hand quark fields, which goes under the name of \emph{chiral symmetry}. 

This invariance is spontaneously broken by the ground state of the theory. This breakdown is unveiled by the appearance of massless particles, the Goldstone bosons. In the specific sector that we are considering now, formed only by the $u$ and $d$ quarks, there are three Goldstone bosons with spin 0 and negative parity and they correspond to pions $\pi$. 

Chiral symmetry is also explicitly broken due to the non-zero mass of the light quarks. Nevertheless the effects of the explicit chiral symmetry breaking can still be analyzed in terms of $\chi$EFT. A further symmetry related to the $u$ and $d$ quark mass terms is observed in that hadrons appear in isospin multiplets, characterized by a splitting in their mass of a few MeV. This is generated by the small  mass difference $m_u-m_d$, and in the limit of $m_u=m_d$ we can define the theory as invariant under isospin transformations.

\begin{figure}[t!]
\begin{center}
\includegraphics[width=0.68\textwidth]{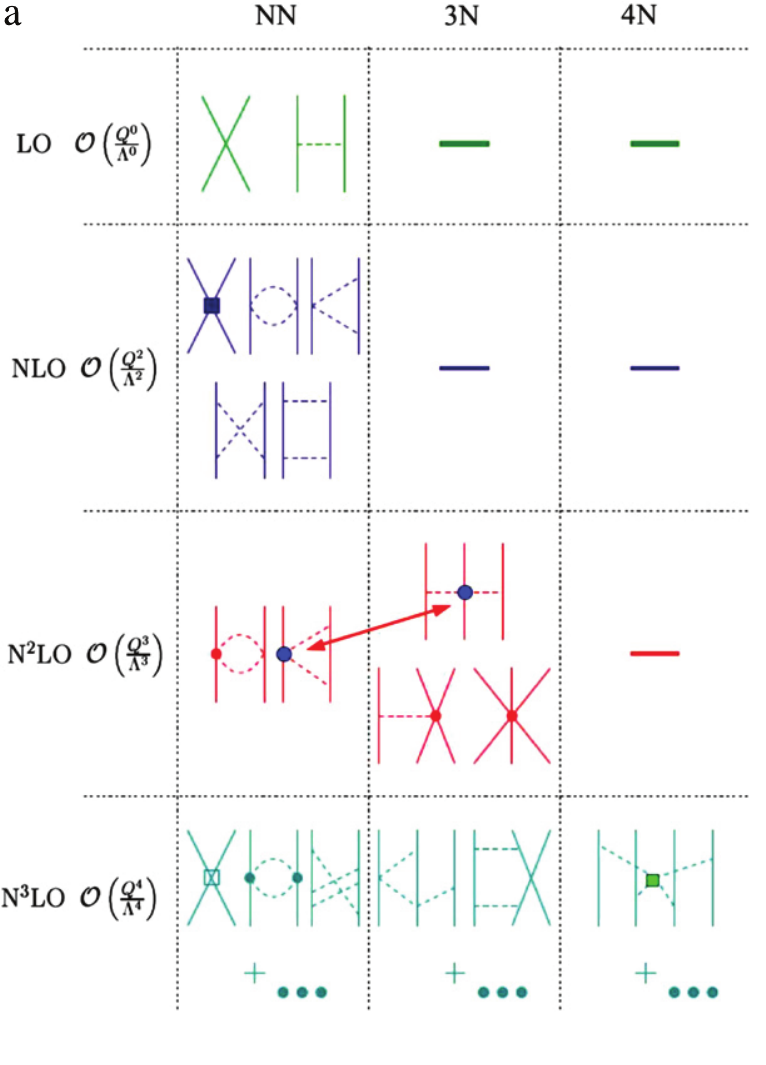}
\caption{Figure taken from Ref.~\citep{Bog2010}. 2NF, 3NF and 4NF chiral nuclear forces which appear up to fourth order in the chiral expansion, according to Weinberg's power counting. Solid lines represent nucleons, dashed lines represent pions. Different colors define the order in the expansion: green for leading-order LO; blue for next-to-leading-order NLO; red for next-to-next-to-leading-order N2LO and turquoise green for next-to-next-to-next-to-leading-order N3LO. The different colors and shapes for the nucleon-pion or contact interacting vertices are related to the number of nucleon field operators and to the number of derivatives and/or insertions of the pion mass $M_\pi$ \citep{Epe2009Oct}.}
\label{chiral_forces}
\end{center}
\end{figure}

In terms of $\chi$EFT, the nuclear force is constructed using nucleons and pions as degrees of freedom. Other light degrees of freedom, such as the $\Delta$, could be also taken into account. By construction, chiral nuclear forces are consistent with all symmetries of the underlying interaction. As long as the energy ${\cal Q}$ of the process lies below the hard scale $\Lambda_{\rm QCD}$ at which $\chi$EFT breaks down, the low-energy behavior of nuclear systems can be described in a perturbative expansion. This is done in orders of $({\cal Q}/\Lambda_{\rm QCD})^\nu$, where $\nu$ is the counting index which defines the so-called power counting \citep{Weinberg1990,Weinberg1991}. This counting allows to identify all contributions at a given order of expansion. The contributions with the lowest possible value of $\nu$ define the leading-order LO contribution; the first corrections with the second smallest allowed value of $\nu$ define the next-to-next-to-leading-order NLO terms, and so on. One sees a consistent appearance of 2B, 3B and up to many-body terms, according to the order in the perturbative expansion. The appearance of higher body interacting terms with the increasing of the order in the chiral expansion can be clearly seen in Fig.~\ref{chiral_forces}. The unknown high-energy physics is incorporated in low-energy constants (LECs) which necessarily have to be determined by a fit to experimental data. In order to exclude contributions of high-momentum components in intermediate states, the chiral potential is multiplied by a regulator, which suppresses momenta larger than a certain cut-off $\Lambda$, which is usually set at the $\rho$ mass value, 700 MeV.

The original approach based on this power counting was defined by Weinberg  \citep{Weinberg1990,Weinberg1991}. Pioneering work was presented shortly after by authors in Ref.~\citep{Ord1994,Ord1996}. Here the NLO contributions to the 2N potential were first considered. We must highlight that Weinberg's power counting is not necessarily the only way to evaluate the chiral expansion \citep{Kap1996,Kap1998Apr,Kap1998May,Nog2005,Epe2009May,Epe2009Oct}. 

Presently, the complete chiral 2NFs force have been presented up to the fourth order in Weinberg's power counting (N3LO). A $\Lambda=500$ MeV cutoff N3LO potential has been provided by authors in Ref.~\citep{Ent2003}. This is the potential which will be used in the 2B sector in this thesis. A 2NF potential at both third order, N2LO,  and at N3LO in the chiral expansion has been presented by authors in Refs.\citep{Epe2002Dec1,Epe2005}. In this formulation, a variation on the cutoff value has been applied, usually in ranges of $\Lambda=500-650$ MeV. Recently, an optimized version of the 2B chiral force at N2LO has been devised by authors in Ref~\citep{Eks2013}. In this thesis we will also present results using this force in the 2B sector.

The chiral 3NFs has been derived up to fifth order in the chiral expansion (N4LO). The N2LO 3NFs was first performed by author in Ref.~\citep{vKol1994} and then completed in Ref.\citep{Epe2002Dec2}. This potential is the one we will be using in the 3B sector of our calculations. The N3LO 3NFs have been computed by authors in Refs.\citep{Ishikawa2007,Bernard2008,Bernard2011}. A calculation of the 3NFs at N4LO has been also recently provided in Ref.~\citep{Krebs2012}. Evaluations of four nucleon forces (4NFs) at N3LO has been presented in Refs.\citep{Epelbaum2006,Epelbaum2007}. 

In conclusion, $\chi$EFT provides a consistent set of hamiltonians at the two-, three- and four-body level which can be systematically improved by including higher orders in an expansion of momenta over a large chiral scale. Furthermore, theoretical errors on observables estimations are expected to decrease as one increases the order in the chiral expansion. Benchmarking results testing the convergence of chiral interactions have been very recently presented by means of QMC methods \citep{Gezerlis2013,Roggero2014}.

%%%%%%%%%  program of the thesis %%%%%%%%

\section{Program of the thesis}
\label{section:program}

Following this brief overview of the different many-body approaches which have been used to solve the problem for the infinite nuclear matter system, in the next chapter we introduce the formalism for the many-body Green's functions theory at zero temperature \citep{Ram1988PhD,Dic2008}. As we have already discussed, at zero temperature the $T$-matrix undergoes the pairing instability. The specific formalism which we present in Chap.~\ref{chapter:formalism} does not take into account this instability. For this reason, the results that we present in Chap.~\ref{chapter:results} will be performed at finite but low temperature, to avoid the occurrence of this kind of instability. Nevertheless, by means of a simple extrapolation method, we provide an estimation of the zero temperature results.

After revising the general formalism of the Green's functions approach, in Chap.~\ref{chapter:formalism} we analyze the sum rule used to calculate the total energy of the many-body ground state. We will see how the Galistkii-Migdal-Koltun sum rule changes when 3BFs are included in the Hamiltonian. Subsequently, we describe the self-consistency approach achieved in the SCGF theory via solution of the Dyson equation, focusing especially on the use of specific classes of diagrams in the construction of the SP self-energy.

In Chap.~\ref{chapter:3BF_formalism} we extensively describe how the SCGF equations are altered with the inclusion of 3BFs. The full expansion of the SP self-energy up to third order will be presented. We introduce the concept of effective interactions, as it simplifies the expansion of the self-energy when including 3B interactions. We then solve the hierarchy of equations of motion (EOMs) up to the two-body propagator and perform specific truncations of the interacting vertex functions, putting special emphasis on the \emph{ladder approximation} for the $T$-matrix. We then also prove the up-to-third-order SP self-energy via a perturbative expansion of the interacting vertex functions.

We subsequently discuss in Chap.~\ref{chapter:eff_2b_int} the construction of a 2NF density-dependent (2NFdd). This is built by averaging over the third particle in the 3NFs appearing at N2LO in the chiral expansion. This is one of the crucial steps in order to incorporate 3BFs in the many-body calculations. We explore several strategies to perform the average, considering both a correlated and a free in-medium nucleon propagator. Moreover, the use of different regulator functions in the averaging procedure is also investigated. Finally, the partial wave matrix elements will be studied in detail, focusing on the effects of 3NFs and on the modifications due to the different types of averages.

In Chap.~\ref{chapter:results} we present bulk properties for symmetric nuclear and pure neutron matter. We compare the energy obtained with the sole use of the 2BF at N3LO \citep{Ent2003} with the one resulting from the inclusion of 3BFs. Discrepancies in the energy values due to the different averaging procedures are also presented. We then analyze the spread in results due to the use of different LECs in the 3B part for the case of SNM. Moreover, in the case of PNM, bands due to uncertainties in the LECs coming from the 2B part will be studied. By means of a newly optimized version of the 2B N2LO force \citep{Eks2013}, results using consistently 2N and 3N forces at the same order in the chiral expansion will be presented. As a conclusion to the chapter, we analyze the symmetry energy, focusing mostly on the different values obtained at saturation density.

A summary and conclusions drawn from the results obtained in the thesis will be outlined in the concluding Chap.~\ref{chapter:sum_concl}.

Moreover, three appendices are included at the end of the thesis.
App.~\ref{chapter:feynman_rules} reports the Feynman rules for diagrams. App.~\ref{chapter:int_irr_diag} provides the demonstration of the effective operators defined in Chap.~\ref{chapter:3BF_formalism}. App.~\ref{chapter:dens_dep_terms} gives complete details on the construction of the density-dependent potential from the 3NFs contributions at N2LO.

\clearpage{\pagestyle{empty}\cleardoublepage}

%%%%%%%%%%%%%%%%%%%%%%% FORMALISM %%%%%%%%%%%%%%%%%%%%%%%

\chapter{Many-body Green's functions}
\label{chapter:formalism}

%in the introduction conservation laws have to be introduced (thermodynamical consistency)

Green's functions (GFs) or \emph{propagators} can be considered as the \emph{heroes} of the many-body problem \citep{Mat1992}. These mathematical tools incorporate by definition the correlated behavior of an interacting many-body system. For this reason they are particularly useful to characterize the properties of complex many-particle systems, whether of nuclear, atomic or molecular nature. The great advantage of using GFs lies in the possibility to carry out the many-body problem in a general, systematic and graphical way through the use of \emph{Feynman diagrams}. A Feynman diagram can be seen as a graphical representation of a transition amplitude from an initial to a final state of the system. In systems where infinite series of transition amplitudes need to be summed up, as in the case of nuclear matter, diagrammatic techniques turn out to be indispensable. A graphical interface is then necessary to keep track of all the terms summed up in the infinite series. Furthermore, it is vital to control the approximations so far made via truncations of the series itself. For this reason the diagrammatic representation will be used throughout the entire thesis. This will lead to a direct interpretation of the specific processes under study.  

The original many-body Green's functions formalism dates back to the 1960s \citep{Mart1959,Kad1962,Noz1963, Abr1975}. At that time, it was impossible to implement it numerically, but provided a unique theoretical insight. In the past few decades, the rise of computational power has gradually improved to the point of allowing for \emph{ab-initio} studies which take into account the strong interacting characteristics of many-body systems. First principles calculations in the Green's functions framework are now routinely performed in solid state \citep{Ary1998,Oni2002}, atomic and molecular physics \citep{vNie1984,Bar2007,Deg2011,Bar2012,Ort2013} and nuclear structure \citep{Mut2000,Dic2004}.

In nuclear physics, GFs have been used to study the properties of finite nuclei. A variety of techniques, including the SCGF approach, have been exploited to study in detail the behavior of nuclei beyond the independent particle model approximation \citep{Bar2009Nov}. The Faddeev random phase approximation (FRPA), which includes both particle-particle and particle-hole phonons in the SP self-energy, has been used to describe closed-shell isotopes \citep{Bar2002PhD,Bar2002,Bar2009Jun}. Besides, within the Gorkov-Green's function method, semi-magic nuclei have been studied up to second order in the self-energy expansion \citep{Som2011,Som2013Nov}. Moreover, the problem of medium-mass nuclei with open shells has been also tackled down \citep{Som2013Jan}. Very recently, the effect of 3BFs has been addressed for medium-mass nuclei within an extended Gorkov-Green's function method \citep{Som2013Dec}. 

In addition to finite nuclei, as presented in the introduction to the thesis, SCGF formalism has also been used extensively to study infinite nuclear matter, mainly in the so-called \emph{ladder approximation}. In this approximation, an effective interaction in the medium, the $T$-matrix, is constructed in order to take into account the repeated scattering in between two nucleons. The necessity to implement this approximation comes from the strong short-range behavior which characterizes the nuclear force and which cannot be disregarded in high-density systems such as nuclear matter. This behavior cannot be treated perturbatively and therefore needs an infinite summation of terms to be handled correctly. Nevertheless, the sole use of two-body interactions doesn't provide the correct reproduction of empirical properties, such as the energy/density saturation point of nuclear matter. Saturation densities are too large and saturation energies too attractive, with calculations falling in the so-called Coester band \citep{Coester1970}. The inclusion of 3BFs is then necessary to avoid this trend. 

However, the many-body Green's functions framework has been originally developed with Hamiltonians containing up to 2B interactions in mind \citep{Fet1971,Rin1984,Bla1986,Dic2008}. The first implementations in infinite nuclear systems to include 3BFs were successfully brought forward by the Krakow group \citep{Som2009PhD}. A correlated average of the Urbana IX \citep{Pie2001Jun} force was included to define an effective 2B potential. This density-dependent force yielded strong improvements in properties of symmetric nuclear and pure neutron matter, for both zero \citep{Som2008} and finite temperature \citep{Som2009}. 

In this thesis we want to propose a more thorough analysis on how the SCGF method is extended to account for 3BFs \citep{Car2013Nov}. While the main motivation to do so are nuclear systems, the formalism can be easily applied to other many-body systems.  Such an approach is pivotal both to provide theoretical foundations to approximations made so far and to  advance the many-body formalisms for much-needed \emph{ab-initio} nuclear structure calculations. With a pedagogical perspective, the extension of the SCGF formalism to include 3BFs will be presented in Chap.~\ref{chapter:3BF_formalism}. This will be done by working out in full the first orders of the self-energy perturbative expansion and solving the self-consistent equations of motion for the propagators.

In this chapter, we will revise the basis of the Green's functions formalism, drawing attention to the self-consistency which is achieved through the solution of the Dyson equation. We will especially focus on the calculation of the total energy of the system which, when including 3BFs, requires a  corrected energy sum rule.

%%%%%%%%%%%%%%%%%% Self-consistent Green's functions %%%%%%%%%%%%%%%

\section{Green's functions formalism with three-body forces}
\label{section:GF_form_3bf}

The starting point for the analysis of a many-body system, in our case of a nuclear kind, is the description of the particles which compose it and how they interact in between each other. We consider a group of $N$ non-relativistic nucleons. The Hamiltonian $\h H$, in second quantization formalism,  is the sum of a one-, two- and a three-body term. It is useful to separate the Hamiltonian into two parts, $\h H = \h H_0 + \h H_1$. $\h H_0 = \h T + \h U$ is a 1B contribution, it is the sum of the kinetic term, $\h T$, and an auxiliary 1B potential, $\h U$. This potential defines the reference state for the perturbative expansion, $\rphizero$, on top of which correlations will be added\footnote{A typical choice in nuclear physics would be a Slater determinant of SP harmonic oscillator or Woods-Saxon wave functions.}. The second term of the Hamiltonian, $\h H_1 = -\h U + \h V + \h W$, describes the interactions. $\h V$ and $\h W$ denote, respectively, the two- and three-body interaction operators. The full Hamiltonian reads:
\beqn
\nn
\h H &=& \sum_{\al} \vep^0_\al\, a^\dg_\al a_\al 
%\\\nn &&
- \sum_{\al\be}U_{\al\be}\, a^\dg_\al a_{\be}
\\ &&
+\f 1 4 \sum_{\substack{\al\ga\\\be\de}}V_{\al\ga,\be\de}\, a_\al^\dg a_\ga^\dg a_{\de} a_{\be}
%\\ && \qquad\qquad
+\frac{1}{36}\sum_{\substack{\al\ga\ep \\ \be\de\eta}} W_{\al\ga\ep,\be\de\eta}\,
a_\al^\dg a_\ga^\dg a_\ep^\dg a_{\eta} a_{\de} a_{\be}\, .
\label{H}
\enqn
The greek indices $\al$,$\be$,$\ga$, etc. label a complete set of SP states which diagonalize the unperturbed Hamiltonian, $\h H_0$, with eigenvalues $\vep_\al^0$. $a^\dg_\al$ and $a_\al$ are creation and annihilation operators for a particle in state $\al$. The matrix elements of the 1B operator $\h U$ are given by $U_{\al\be}$. Equivalently, the matrix elements of the 2B and 3B forces are $V_{\al\ga,\be\de}$ and $W_{\al\ga\ep,\be\de\eta}$. In the following, we work with antisymmetrized matrix elements in both the 2B and the 3B sector, i.e.
\beqn
\nn
V_{\al\ga,\be\de}&=&\frac{1}{2}\, _A\langle\al\ga|V|\be\de\rangle_{A}=\langle\al\ga|V|\be\de\rangle_A
\\ &=&
\langle\al\ga|V(1-P_{12})|\be\de\rangle\,,
\label{v_asym}
\\ \nn
W_{\al\ga\ep,\be\de\eta}&=&\frac{1}{6}\,_A\langle\al\ga\ep|W|\be\de\eta\rangle_A=\langle\al\ga\ep|W|\be\de\eta\rangle_A
\\&=&
\langle\al\ga\ep|W(1-P_{12})(1-P_{13}-P_{23})|\be\de\eta\rangle\, .
\label{w_asym}
\enqn

The main ingredient of the formalism is the 1B GF, also called SP propagator or  2-point GF, which provides a complete description of one-particle and one-hole excitations of the many-body system. More specifically, the SP propagator is defined as the expectation value of the time-ordered product of an annihilation and a creation operators in the Heisenberg picture:
\beq
i\hbar\, G_{\al\be} (t_\al-t_\be)
=\langle\Psi_0^N|{\cal T}[a_\al(t_\al)a_{\be}^\dg(t_\be)]|\Psi_0^N\rangle \,,
\label{G}
\enq
where $|\Psi_0^N\rangle$ is the interacting $N$-body ground state of the system. The time ordering operator brings operators with earlier times to the right, with the corresponding fermionic permutation sign. For $t_\al-t_\be>0$, this results in the addition of a particle to the state $\be$ at time $t_\be$ and its removal from state $\al$ at time $t_\al$. Alternatively, for $t_\be-t_\al>0$, the removal of a particle from state $\al$ occurs at time $t_\al$ and its addition to state $\be$ at time $t_\be$. These correspond, respectively, to the propagation of a particle or a hole excitation through the system. We can also introduce the 4-point and 6-point GFs which will be useful in the following:
\beqn
\nn
&&
i\hbar \,G^{4-{\rm pt}}_{\al\ga,\be\de}(t_\al,t_\ga;t_{\be},t_{\de}) = 
\\ &&\quad\quad
\langle\Psi_0^N|{\cal T}[a_\ga(t_\ga)a_\al(t_\al)
a_{\be}^\dg(t_{\be})a_{\de}^\dg(t_{\de})]|\Psi_0^N\rangle \,,
\label{g4pt}
\enqn
\beqn
\nn
&&
i\hbar \,G^{6-{\rm pt}}_{\al\ga\ep,\be\de\eta}(t_\al,t_\ga,t_\ep;t_{\be},t_{\de},t_{\eta}) = 
\\ &&\quad
\langle\Psi_0^N|{\cal T}[a_\ep(t_\ep)a_\ga(t_\ga)a_\al(t_\al)
a_{\be}^\dg(t_{\be})a_{\de}^\dg(t_{\de})a_{\eta}^\dg(t_{\eta})]|\Psi_0^N\rangle \, .
\label{g6pt}
\enqn
Physically, the interpretation of Eq.~(\ref{g4pt}) and Eq.~(\ref{g6pt}) follows that of the 2-point GF in Eq.~(\ref{G}). In these cases, more combinations of particle and hole excitations are encountered depending on the ordering of the several time arguments. The extension to formal expressions for higher many-body GFs is straightforward.

In the following, we will consider propagators both in time representation, as defined above, or in energy representation. Note that, due to time-translation invariance, the \hbox{$m$-point} GF depends only on $m-1$ time differences or, equivalently, $m-1$ independent frequencies. Hence the Fourier transform to the energy representation is well-defined when the total energy is conserved:
\beqn
\nn
&& 2\pi \hbar\,\delta (\om_\al + \om_\ga + \ldots  - \om_\be - \om_\de - \ldots) 
%\\\nn && \qquad\qquad\qquad\times 
G^{m-{\rm pt}}_{\al \ga \ldots,  \be \de  \ldots} (\om_\al, \om_\ga, \ldots ; \om_\be , \om_\de ,  \ldots ) 
 \\\nn &&%\qquad\qquad
 =\int_{-\infty}^{+\infty} \hspace{-1mm} \d t_\al   \int_{-\infty}^{+\infty} \hspace{-1mm} \d t_\ga \; \ldots 
                \int_{-\infty}^{+\infty} \hspace{-1mm} \d t_\be  \int_{-\infty}^{+\infty} \hspace{-1mm} \d t_\de \ldots
\\ &&  \times e^{\frac{i}{\hbar}(\om_\al t_\al + \om_\ga t_\ga + \ldots)}
G^{m-{\rm pt}}_{\al \ga \ldots,  \be \de  \ldots} (t_\al, t_\ga, \ldots; t_\be , t_\de ,  \ldots ) \,
e^{-\frac{i}{\hbar}(\om_\be t_\be + \om_\de t_\de + \ldots)} \, .\,\,\,\,\,\,\quad
\label{Gmpt_ft}
\enqn
For the 1B GFs of Eq.~(\ref{G}) one also considers the Lehmann representation, that gives physical insight to the SP propagator: 
\beqn
\nn
G_{\al \be} (\om) &=& G^{2-pt} _{\al \be} (\om; \om) = \int_{-\infty}^{+\infty} d\t \; e^{\frac{i}{\hbar} \om \t} G_{\al \be}(\t) 
 \\\nn &=& 
 \sum_m\frac{\lPsizero a_\al|\Psi_m^{N+1}\rangle\langle\Psi_m^{N+1}|a^\dg_\be\rPsizero}
 {\om-(E_m^{N+1}-E_0^N)+i\eta} 
 \\ && 
 +\sum_n\frac{\lPsizero a^\dg_\be|\Psi_n^{N-1}\rangle\langle\Psi_n^{N-1}|a_\al\rPsizero}
 {\om-(E_0^N-E_n^{N-1})-i\eta} \, ,
 \label{G1B_Lehm}
\enqn
where $\tau=t_\al-t_\be$, and $m$ and $n$ label the many-body eigenstates of the system with $(N+1)$ or $(N-1)$ particles. The sum is intended over all discrete/continuum states of the energy spectrum. The poles in Eq.~(\ref{G1B_Lehm}) represent the energies of the excited states of the system. Namely, $\vep_m^+=E_m^{N+1}-E_0^N$ corresponds to the excitation energy of the system with an added particle $|\Psi_m^{N+1}\rangle$, with respect to the energy of the $N$-particle many-body ground state. $\vep_n^-=E_0^N-E_n^{N-1}$ corresponds to minus the excitation energy of the system with a removed particle $|\Psi_n^{N-1}\rangle$, with respect to the energy of the $N$-particle many-body  ground state. 

The information enclosed in Eq.~(\ref{G1B_Lehm}), about the transition amplitudes of an added or removed particle from the system, illustrates the potential power of the SP propagator and, for finite systems, its direct connection to experimental data. In fact, the poles in the denominators of Eq.~(\ref{G1B_Lehm}) signal the position in energy space of the $N+1$ or $N-1$ excited states of the system. Furthermore, the numerators define the corresponding transition amplitudes from the ground state with $N$ particles to the excited states of the $N\pm1$ systems. Experimentally, this information is related to the \emph{hole spectral function}, which can be extracted from the strength of the removed particle in (\emph{e,e'\,p}) knockout reactions \citep{Mou1980,Lap1993}. Mathematically this quantity corresponds to the residue of the poles of the 1B GF and is then related to the imaginary part of the 1B GF. It can be obtained from Eq.~(\ref{G1B_Lehm}) by means of the Plemejl identity:
\beq
\frac{1}{\om_{\pm}}=\frac{\cal P}{\om}\mp i\pi\delta(\om)\,,
\label{plemejl}
\enq
where $\om_{\pm}$ is the energy which corresponds respectively to the poles in the particle and hole part of Eq.~(\ref{G1B_Lehm}), and $\cal P$ defines the principal value. Hence one can write for the hole spectral function:
\beqn
\nn
\mathcal{S}_\alpha^h(\om) &=& \frac{1}{ \pi} \mathrm{Im}\, G_{\al\al}(\om) 
\\ 
&=& \sum_n \left| \left\langle \Psi_n^{N-1} | a_\al | \Psi_0^N \right\rangle \right|^2 
\delta \left( \om - \vep_n^-\right) \,,
\label{S_hole}
\enqn
for $\om\le\vep_\textrm F^-$, with $\vep_\textrm F^-=E_0^N-E_0^{N-1}$. $\vep_\textrm F^-$ corresponds to the minimum energy to remove a particle from the $N$-body ground state into the $N-1$-body ground state. In finite systems, the transition amplitude between the $N$ and the $N-1$ body systems is closely related to the definition of the theoretical spectroscopic factor \citep{Dic2004,Bar2009Nov}. This factor provides a measure of the correlations present in the system which induce its behavior to deviate form the independent-particle model description. 

The corresponding part of the hole spectral function in the particle domain is the \emph{particle spectral function}. This quantity represents the probability to add a particle to the many-body ground state and, analogously to Eq.~(\ref{S_hole}), it is defined as
\beqn
\nn
\mathcal{S}_\alpha^p(\om) &=& -\frac{1}{ \pi} \mathrm{Im}\, G_{\al\al}(\om) 
\\
&=& \sum_m \left| \left\langle \Psi_m^{N+1} | a^\dg_\al | \Psi_0^N \right\rangle \right|^2 
\delta \left( \om -\vep_m^+ \right) \,,
\label{S_part}
\enqn
for $\om\ge\vep_\textrm F^+$, with $\vep_\textrm F^+=E_0^{N+1}-E_0^N$. $\vep_\textrm F^+$ corresponds to the minimum energy necessary to add a particle to the $N$-body ground state into the $N+1$-body ground state. In finite systems there can be a considerable difference in between $\vep_\textrm F^-$ and $\vep_\textrm F^+$. In normal infinite systems, i.e. neither superfluid nor superconducting, which are the object of this thesis, the difference between $\vep_\textrm F^-$ and $\vep_\textrm F^+$ vanishes in the thermodynamical limit and approaches the value of the chemical potential.

For an infinite system, translational invariance suggests the use of momentum states for the SP basis. In this case the greek index identifies the quantum numbers of momentum, spin and isospin $\al=\{{\bf p},\bd\sigma,\bd\tau\}$. Consequently we can recast the Lehmann representation (\ref{G1B_Lehm}) for the 1B propagator as:
\beq
G({\bf p},\om)=\int_{\vep_\textrm F}^\infty\d\om'\frac{{\cal S}^p({\bf p},\om')}{\om-\om'+i\eta}
+\int_{-\infty}^{\vep_\textrm F}\d\om'\frac{{\cal S}^h({\bf p},\om')}{\om-\om'-i\eta}\,,
\label{Lehm_infty}
\enq
where for convenience we omit spin/isospin indices. Notice that we have introduced the Fermi energy $\vep_\textrm F$ which, in an uncorrelated system, defines the last filled energy level and hence corresponds to the energy needed to remove a particle from the many-body ground state.  In the present case of an interacting system, $\vep_\textrm F$ equals the chemical potential $\mu$, and corresponds to the minimum energy necessary to add or remove a particle to/from the system.
 
Furthermore, the knowledge of the hole spectral function gives access to the SP momentum occupation number in the many-body ground state. For an infinite system this can be written as:
\beq
\label{mom_dist}
n({\bf p})=\lPsizero a^\dg_\al a_\al\rPsizero= \int_{-\infty}^{\vep_\textrm F} \d \om {\cal S}^h({\bf p},\om)\,.
\enq
The study of the occupation number, especially at low-momentum, is a meaningful way to quantify the correlations present in the system. As a matter of fact, short-range and tensor correlations embodied in the nuclear potential increase the depletion observed at zero SP momentum. Consequently, they introduce high-momentum states in the distribution function, Eq.~(\ref{mom_dist}) \citep{Rio2009Apr,Car2012}. This can have  a direct effect on bulk properties of nuclear matter, such as the symmetry energy \citep{Vid2011}. 

As a concluding remark to this section, we would like to stress that the SCGF formalism aims at providing a reliable calculation of the SP propagator, and hence of the spectral function, in correlated systems. As we will see, an efficient way to achieve this purpose is by means of diagrammatic techniques. The use of a graphical interface can highly simplify the analysis of the dressed propagator and help to enumerate all contributions which are taken into account in the many-body approximation under hand.

%%%%%%%%%%%%%%%%% The GMK sumrule %%%%%%%%%%%%%%%%%%%%

\section{The Galistkii-Migdal-Koltun sum rule}
\label{section:gmk_sumrule}

The SP propagator provides access to expectation values of all 1B operators and hence is a useful tool to characterize a wide range of the properties of the system. Furthermore, if only up to 2B terms are considered in the Hamiltionian, the ground-state energy can be computed from the 1B GF itself. This is a crucial result that arises from the Galitskii-Migdal-Koltun (GMK) sum rule \citep{Gal1958,Kol1974}. In this section we will see how the inclusion of 3B interaction terms alter the form of the GMK sum rule. Hence we will need to derive a modified sum rule.

As outlined, the energy of an interacting system via 2BFs can be obtained within the SCGF approach by means of the GMK sum rule, which reads \citep{Gal1958,Kol1974}: 
 \beq
\label{gmk}
E_0^N=\frac{1}{2\pi}\int^{\epsilon^-_\textrm F}_{-\infty} \d\om \; \sum_{\al\be}(T_{\al\be}+
\om \de_{\al\be}) \mathrm{Im}\, G_{\be\al}(\om) \, .
\enq
Observing \eq{gmk}, we note that not all the information content from the propagator is needed to obtain the ground-state energy. The hole part (see \eq{S_hole}), which includes details about the transition amplitude for the removal of a particle from the many-body system, is enough for this purpose. To derive the GMK sum rule, one starts by considering the first moment of the hole spectral function:
\beq
\label{int_gmk}
I_\al = \frac{1}{\pi}\int^{\epsilon^-_\textrm F}_{-\infty}\mathrm{d} \om \, \om\, \mathrm{Im}\, G_{\al\al}(\om)= \int^{\epsilon^-_\textrm F}_{-\infty}\mathrm{d} \om \, \om\, \mathcal{S}_\alpha^h(\om)\,.
\enq
From the spectral representation of \eq{S_hole}, one can see that this integral is also the expectation value over the many-body ground state of the  commutator:
\beq
I_\al = \langle\Psi_0^N| \ad\al[\a\al,\h H]|\Psi_0^N\rangle\,.
\enq
Using the Hamiltonian in Eq.~(\ref{H}), one can evaluate the commutator to find:
\beqn
\nn
\label{I_alpha}
I_\al&=&\langle\Psi_0^N|\sum_{\be} T_{\al\be}\, \ad\al a_\be
+
\f 1 2\sum_{\ga\be\de} V_{\al\ga,\be\de}\, \ad\al\ad\ga a_\de a_\be
\\  &&\qquad
+ \frac{1}{12} \sum_{\substack{\ga\ep\be\de\eta}} W_{\al\ga\ep,\be\de\eta}\, 
\ad\al\ad\ga\ad\ep a_\eta a_\de a_\be |\Psi_0^N\rangle\,.
\enqn
Note that, in general, $T_{\al\be}$ represents the 1B part of the Hamiltonian which, in addition to the kinetic energy, might also contain a 1B potential. Summing over all the external SP states, $\alpha$, one finds,
\beq
\sum_\al I_\al=
\langle\Psi_0^N | \h T + 2 \h V + 3 \h W |\Psi_0^N\rangle\,.
\label{sumrule}
\enq
In other words, the sum over all SP states of the first moment of the spectral function yields a particular linear combination of the contributions  of the 1B, 2B and 3B potentials to the ground-state energy,
\beq
E_0^N = 
\langle\Psi_0^N| \h H | \Psi_0^N\rangle 
=
\langle\Psi_0^N | \h T + \h V + \h W |\Psi_0^N\rangle\,.
\label{gsenergy}
\enq

Since $\h T$ is a 1B operator, one can actually compute its expectation value from the SP propagator itself:
\beq
\langle\Psi_0^N|\h T|\Psi_0^N\rangle
 =\frac{1}{\pi}\int^{\epsilon^-_\textrm F}_{-\infty} \mathrm{d} \om \sum_{\al\be} T_{\al\be} \mathrm{Im}\, G_{\be\al}(\om) \, .
\enq
The energy integral on the right-hand side yields the reduced 1B density matrix:
\beq
\rho^{1B}_{\be \al}=\frac{1}{\pi}\int^{\epsilon^-_\textrm F}_{-\infty} \mathrm{d} \om \, \mathrm{Im}\, G_{\be\al}(\om)=\lPsizero a^\dag_\al a_\be\rPsizero \, ,
\enq
which can then be used to simplify the previous expression
\beq
\langle\Psi_0^N|\h T|\Psi_0^N\rangle=\sum_{\al\be}T_{\al\be}\rho^{1B}_{\be \al}\,.
\enq
For the 2B case, this is enough to provide an independent constraint and hence allows for the calculation of the total energy (see \eq{gmk}). The ground-state energy can then be computed from the 1B propagator alone. 

When 3BF are present, however, one needs a third independent linear combination of $\langle \h T\rangle$, $\langle \h V\rangle$ and $\langle \h W\rangle$. Knowledge of the 1B propagator is therefore not enough to compute the total energy, since either the 2B or the 3B propagators are needed to compute $\langle \h V\rangle$ or $\langle \h W\rangle$ exactly. Depending on which of the two operators is chosen, one is left with different expressions for the energy of the ground state. This freedom in choice could in principle be exploited to test the validity of different approximations. In practical applications, however, one should choose the combination that provides minimum uncertainty. 

Let us start by considering the case where the 3B operator is eliminated. Adding $2 \langle \h T\rangle$ and $\langle \h V\rangle$ to Eq.~(\ref{sumrule}), one finds the following exact expression for the total ground-state energy:
\beqn
E_0^N&=&\frac{1}{3\pi}\int^{\epsilon^-_\textrm F}_{-\infty} \mathrm{d} \om \; \sum_{\al \be}(2T_{\al\be}+
\om \de_{\al\be}) \mathrm{Im}\, G_{\be\al}(\om) 
\nn \\
&&+\f 1 3 \langle\Psi_0^N|\h{V}|\Psi_0^N\rangle \, .
\label{gmk_2b}
\enqn
The calculation of this expression requires the hole part of the 1B propagator and the two-hole part of the 2B propagator, which would appear in the second term. We note that this expression is somewhat equivalent to the original GMK, in that the ground-state energy is computed from 1B and 2B operators, even though the Hamiltonian itself is a 3B operator. This might prove advantageous in calculations where the 2B propagator is computed explicitly. 

Alternatively, one can eliminate the 2B contribution from the GMK sum rule by adding $\langle \h T\rangle$ and subtracting $\langle \h W\rangle$ to Eq.~(\ref{sumrule}). This leads to the expression:
\beqn
E_0^N&=&\frac{1}{2\pi}\int^{\epsilon^-_\textrm F}_{-\infty} \d\om \; \sum_{\al\be}(T_{\al\be}+
\om \de_{\al\be}) \mathrm{Im}\, G_{\be\al}(\om) \, 
\nn \\
&& - \f 1 2 \langle\Psi_0^N| \h W |\Psi_0^N\rangle\,.
\label{gmk_3b}
\enqn
The first term in this expression is formally the same as that obtained in the case where only 2BFs are present in the Hamiltonian. Note, however, that the 3BF does influence the 1B propagator in the first term and hence the correction should be applied at the very end of the self-consistent procedure. The second term in Eq.~(\ref{gmk_3b}) requires the knowledge of the three-hole part of the 3B propagator. If the 3B expectation value results as a small contribution to the total energy of the many-body ground state, Eq.~(\ref{gmk_3b}) can be favored compared to Eq.~(\ref{gmk_2b}) to compute the total energy of the system.

Eqs.~(\ref{gmk_2b}) and~(\ref{gmk_3b}) are both exact. Which of the two is employed in actual calculations will mostly depend on the accuracy associated with the evaluation of the expectation values, $\langle \h V \rangle$ and $\langle \h W \rangle$. If the 2B interaction is dominant with respect to the 3BF, for instance, the former will be a large contribution. Small errors in the calculation of the 2B propagator could eventually yield artificially large corrections in the ground-state energy. In nuclear physics, the 3BF expectation value is expected to provide a smaller contribution than the 2BF one \citep{Gra1989,Epe2009Oct}. Consequently, approximations in Eq.~(\ref{gmk_3b}) should lead to smaller absolute errors. This was the approach that was recently followed in both finite nuclei and infinite nuclear matter calculations \citep{Cip2013,Car2013Oct,Som2013Dec}. In finite nuclei, evaluating $\langle \h W \rangle$ at first order in terms of \emph{dressed} propagators leads to satisfactory results. However, accuracy is lost if free propagators, $G^{(0)}$, are used instead 	\citep{Cip2013}. Eq.~(\ref{gmk_2b}) may eventually be useful in calculations of infinite matter, in which the 2B propagator can be computed nonpertubatively by means of the $T$-matrix. Therefore, one could explicitly evaluate the expectation value of the 2BF in the many-body ground state and compute directly Eq.~(\ref{gmk_2b}). 

Even though Eq.~(\ref{gmk_2b}) can prove to be more accurate to study the energy of infinite matter, in the following we use the prescription presented in Eq.~(\ref{gmk_3b}). The main motivation for doing so is related to the approximation with which the second term in Eq.~(\ref{gmk_3b}) is computed. In the present calculations we estimate the 3BF expectation value computing it only at first order, i.e. in the HF approximation. Applying this same approximation to the 2BF expectation value in Eq.~(\ref{gmk_2b}) would lead to higher errors in the estimation of the total energy of the system (typically, $\langle\h W\rangle\approx\frac{1}{10} \langle\h V\rangle$ for nuclear interactions \citep{Epe2009Oct}). To avoid this drawback, and due to the fact that in the present state of the numerical code we only compute the 2B propagator at the HF level, we then rely  on Eq.~(\ref{gmk_3b}) to evaluate the energy of the many-body ground state.

%%%%%%%%%%%%%%%%%%%% Dyson's equation and self-consistency  %%%%%%%%%%%

\section{Dyson's equation and self-consistency}
\label{section:dyson_eq}

It has been pointed out that perturbation theory is helpless when confronted with the strong interaction in nuclear many-body systems. In the case of nuclear matter, where the strong short-range behavior of the force has to be treated, an infinite series of interacting terms must be summed up. For this reason, methods which perform all-order summations of Feynman diagrams are necessary, in order to correctly appreciate the correlated behavior of such systems.  

\begin{figure}
  \begin{center}
  \subfloat[]{\label{conn}\includegraphics[width=0.25\textwidth]{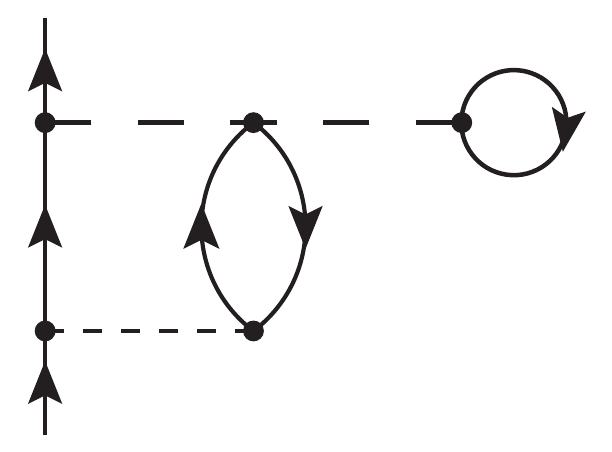}}
  \hspace{1.5cm}
  \subfloat[]{\label{disconn}\includegraphics[width=0.45\textwidth]{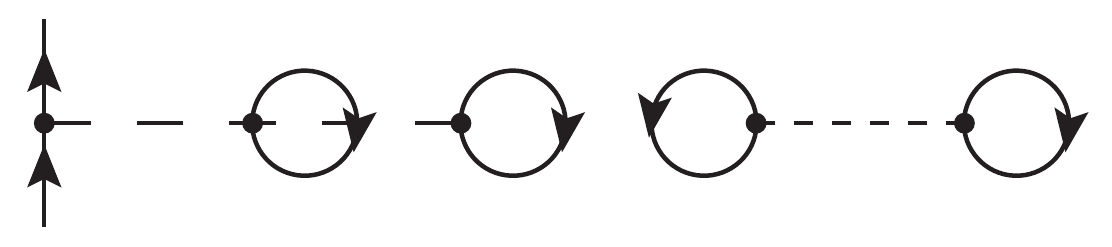}}
  \vfill
  \subfloat[]{\label{1pred}\includegraphics[width=0.25\textwidth]{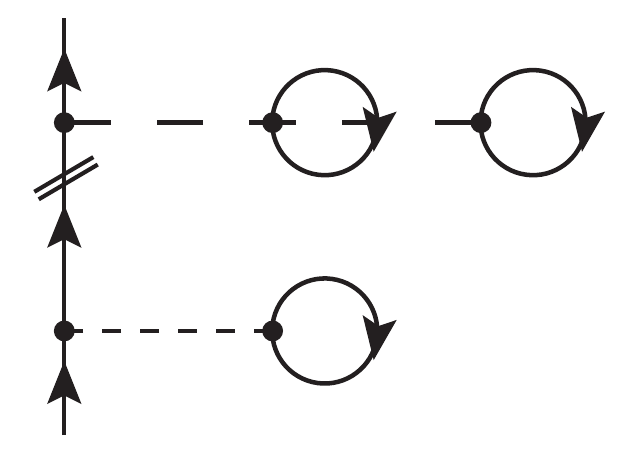}}
  \caption{Examples of diagrams contributing to the perturbative expansion of Eq.~(\ref{gpert}). \protect\subref{conn} is a connected and one-particle irreducible contribution. \protect\subref{disconn} is a disconnected contribution. \protect\subref{1pred} is a one-particle reducible contribution, where short double lines signals where the reduction can be performed. Single-arrowed lines describe an undressed SP propagator $G^{(0)}$, i.e. the first term of the expansion in Eq.~(\ref{gpert}). Short dashed lines describe a 2B interaction and long-dashed a 3B interaction.}
  \label{diagrams_types}
  \end{center}
\end{figure}

Let's start by defining the perturbative expansion of the 1B propagator. This reads \citep{Mat1992,Dic2008}:
\beqn
\nn
&&
G_{\al\be}(t_\al-t_\be) = -\f \ii \hbar \sum_{n=0}^\infty \left(-\f \ii\hbar\right)^n\frac{1}{n!}\int \hspace{-1mm} \d t_1 \; \ldots \int \hspace{-1mm} \d t_n 
\\ &&\qquad\qquad \times
\lphizero\T[\h H_1(t_1) \ldots \h H_1(t_n)a^I_\al(t_\al){a_{\be}^I}^\dg(t_\be)]\rphizero_\text{conn} \,,\,
\label{gpert}
\enqn
where $\rphizero$ is the unperturbed many-body ground state. $a^I_\al$, ${a_{\be}^I}^\dg$ and $\h H_1(t)$ are now operators in the interaction picture with respect to $H_0$.  The subscript ``conn'' implies that only \emph{connected} contributions (diagrams) have to be considered when performing the Wick contractions of the time-ordered product. As an example, a connected and a disconnected diagram are shown in Fig.~\subref*{conn} and \subref*{disconn}.

The interacting part of the Hamiltonian $H_1$ (see Eq.~(\ref{H})) contains contributions from 1B, 2B and 3B  interactions. Thus, the expansion in perturbation theory of the GF involves terms with individual contributions of each force, as well as combinations of these. Feynman diagrams are essential to keep track of such a variety of different contributions. The rules for Feynman diagrams used in this thesis are reported in App.~\ref{chapter:feynman_rules}.

\begin{figure}[t]
\begin{center}
\includegraphics[width=0.35\textwidth]{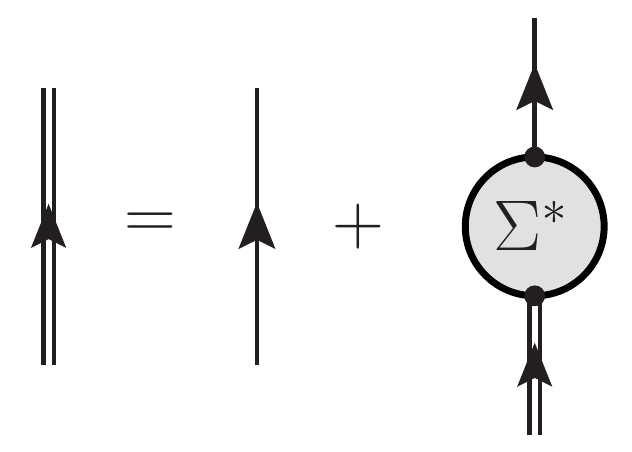}
\caption{Diagrammatic representation of the Dyson equation, Eq.~(\ref{Dyson}). Single-arrowed lines correspond to the unperturbed SP propagator $G^{(0)}$. Double-arrowed lines correspond to the fully dressed propagator $G$, result of the full expansion in Eq.~(\ref{gpert}). Dyson's equation introduces the concept of the irreducible self-energy $\Sigma^\star$ which groups together all 1PI diagrams.}
\label{dyson_fig}
\end{center}
\end{figure}

A first reorganization of all the contributions generated in the perturbative expansion of the SP propagator given in Eq.~(\ref{gpert}), is obtained by considering only those terms which are \emph{one-particle irreducible} (1PI) diagrams. These diagrams cannot be disconnected by cutting a single fermionic line. On the other hand, a one-particle reducible contribution is a diagram which can be split into two separated terms by cutting one single fermionic line. In Fig.~\ref{diagrams_types}, we depict with diagram Fig.~\subref*{conn} an example of an irreducible contribution to the 1B propagator expansion, and with Fig.~\subref*{1pred} an example of a reducible one. Reducible diagrams are generated by an all-order summation through Dyson's equation~\citep{Dys1949},
\beq
G_{\al\be}(\om)=G^{(0)}_{\al\be}(\om)+\sum_{\ga \de}G^{(0)}_{\al\ga}(\om)
\Sigma^\star_{\ga\de}(\om)G_{\de\be}(\om)\, ,
\label{Dyson}
\enq
where $\omega$ corresponds to the energy variable of the propagator. A diagrammatic representation of the Dyson equation is depicted in Fig.~\ref{dyson_fig}. The process of carrying out infinite summations is often called \emph{renormalization}. In this case, the renormalized quantity is the undressed propagator $G^{(0)}$, which is \emph{dressed up} through an infinite summation of contributions, which are grouped together in the so called \emph{irreducible self-energy}, $\Sigma^\star(\om)$. In fact, to sum properly all infinite diagrams by means of the irreducible self-energy, all reducible terms can be disregarded. The complete summation of these reducible contributions provides the definition of the \emph{reducible self-energy}, $\Sigma(\om)$. The diagrammatic definition of the reducible self-energy would correspond to the second term on the right-hand side in Fig.~\ref{dyson_fig}, where the dressed propagator $G$ entering the self-energy is replaced with a undressed one, $G^{(0)}$.

In Eq.~(\ref{Dyson}), the uncorrelated SP propagator, $G^{(0)}$, is associated with the system governed by the $H_0$ Hamiltonian and represents the $n=0$ order in the expansion of Eq.~(\ref{gpert}). For an infinite system, following Eq.~(\ref{Lehm_infty}), the unperturbed or free propagator can be written in the Lehmann representation as:
\beq
G^{(0)}({\bf p},\om)=\frac{\theta(p-p_\textrm F)}{\om-\vep_p+i\eta}+\frac{\theta(p_\textrm F-p)}{\om-\vep_p-i\eta}\,.
\label{G01B_Lehm}
 \enq
$p_\textrm F$ represents the Fermi momentum, which corresponds to the last occupied SP state in a system of non-interacting particles. $\vep_p=\hbar^2p^2/2m$  is the energy of the SP state above ($p>p_\textrm F$) and below ($p<p_\textrm F$) the Fermi level. In Eq.~(\ref{G01B_Lehm}) and in the following we omit explicitly spin/isospin indices.

In a infinite system we can restate Dyson's equation (see Eq.~\ref{Dyson}) as
\beq
G({\bf p},\om)=G^{(0)}({\bf p},\om)+G^{(0)}({\bf p},\om)
\Sigma^\star({\bf p},\om)G({\bf p},\om)\, ,
\enq
which can be solved algebraically:
\beq
G({\bf p},\om)=\frac{1}{\om-\vep_p-\Sigma^\star({\bf p},\om)}\,.
\label{G_algebraic}
\enq
The irreducible self-energy $\Sigma^\star({\bf p},\om)$ describes the kernel that includes all 1PI diagrams. This operator plays a central role in the GFs formalism and can be interpreted as the non-local and energy-dependent interaction that each particle feels due to the interaction with the medium. At positive energies, $\Sigma^\star({\bf p},\om)$ is also identified with the optical potential that the scattered particle feels from the many-body target \citep{Bla1986,Cap1996,Ced2001,Bar2005,Cha2006}. 

The knowledge of the self-energy, $\Sigma^\star({\bf p},\om)$, then provides direct access to the spectral functions. According to Eqs.~(\ref{S_hole}) and (\ref{S_part}), through the use of Eq.~(\ref{G_algebraic}), it can be demonstrated that \citep{Lut1961}:
\beqn
{\cal S}^p({\bf p},\om)&=&-\frac{1}{\pi}\frac{\mathrm{Im}\Sigma^\star({\bf p},\om)}
{(\om-\vep_p-\mathrm{Re}\Sigma^\star({\bf p},\om))^2+(\rm{Im}\Sigma^\star({\bf p},\om))^2} \quad \om>\vep_\textrm F\,,\qquad\,\,\,
\label{Sp_self}
\\ 
{\cal S}^h({\bf p},\om)&=&\frac{1}{\pi}\frac{\mathrm{Im}\Sigma^\star({\bf p},\om)}
{(\om-\vep_p-\mathrm{Re}\Sigma^\star({\bf p},\om))^2+(\rm{Im}\Sigma^\star({\bf p},\om))^2} \quad\,\,\,\, \om<\vep_\textrm F\,.\quad
\label{Sh_self}
\enqn
The numerator in Eqs.~(\ref{Sp_self}-\ref{Sh_self}) is related to the width of the spectral function in energy space. A vanishing width would result in a delta-like spectral function, and this corresponds to the case of a system in the non-interacting or mean-field picture. In this picture, each SP momentum is associated to one SP energy state. Conversely, in the correlated system, a finite width for the spectral function involves SP momentum states which are spread in the energy space, and therefore present a given lifetime when propagating through the system. The value of this life-time is then inversely proportional to the width of the spectral function. 

In Sec.~\ref{section:t_matrix} we will clearly see how the computation of the imaginary part of the self-energy requires the knowledge of the spectral function. This indeed demands a self-consistent iteration for the solution of the spectral function Eqs.~(\ref{Sp_self}-\ref{Sh_self}). We want to point out that, in the SCGF approach, the complete spectral decomposition of the propagator is kept throughout the entire self-consistent procedure, applying no approximations of the quasi-particle kind.

%In principle, the SCGF approach offers great advantages. It is intrinsically nonperturbative, and hence allows for massive summations of diagrams permitting the treatment of interactions of a nonperturbative nature. The SCGF formalism automatically satisfies the basic conservation laws \citep{Bay1961,Bay1962,Dic2008}. In practice, however, calculating diagrams with dressed propagators is computationally more expensive than using the plain $G^{(0)}$  in perturbation theory. Moreover, self-consistency requires an iterative solution for $\Sigma^\star$ and for $G$ via the Dyson equation, Eq.~(\ref{Dyson}). Therefore, the SCGF scheme is not always applied in full detail, but it is often employed to provide important guidance in developing working approximations to the self-energy.

%In the next chapter we will see in detail the effect of inclusion of 3BF analyzing the perturbative expansion of the irreducible self-energy, and we will then prove this expansion working out the EOM method for propagators.  

\subsection{The self-consistent propagator}
\label{section:self_cons}

In the previous section we saw how, by means of the iterative scheme defined by Dyson's equation, Eq.~(\ref{Dyson}), the unperturbed SP propagator, $G^{(0)}$, gets renormalized into a dressed one, $G$, via an infinite summation. In order to correctly iterate the equation through the use of the irreducible self-energy (see Fig.~\ref{dyson_fig}), the summations must only take into account the 1PI contributions. 

Nevertheless, one can further develop the description of the irreducible self-energy. This is achieved by including in its contributions the use of the same dressed propagator of which the self-energy provides the value through the solution of the Dyson equation. This is defined as a \emph{self-consistent renormalization} of the propagator, in which the dressed propagator is not only the solution of the iterative procedure, but furthermore generates the terms which are included in the irreducible self-energy. In some sense this leads to a further level of reorganization of the diagrams in the self-energy expansion. 

In practice, this self-consistent renormalization is obtained by means of the so called \emph{skeleton} diagrams \citep{Bla1986,Dic2008}. The use of these kind of diagrams ensures that the unperturbed propagators, $G^{(0)}$, in the internal fermionic lines of the irreducible self-energy can be safely replaced with dressed GFs, $G$. These are defined as 1PI diagrams that do not contain any portion that can be disconnected by cutting a fermion loop twice at two different points. This portion would be defined as a \emph{self-energy insertion}. Therefore a skeleton diagram is such if no self-energy insertions can be encountered. In fact, these insertions are already resummed into the dressed propagators which build the skeleton diagram.

\begin{figure}
  \begin{center}
  \subfloat[]{\label{skel}\includegraphics[width=0.17\textwidth]{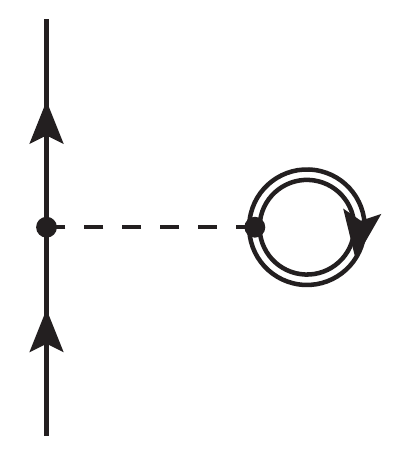}}
  \hspace{2cm}
  \subfloat[]{\label{nonskel}\includegraphics[width=0.3\textwidth]{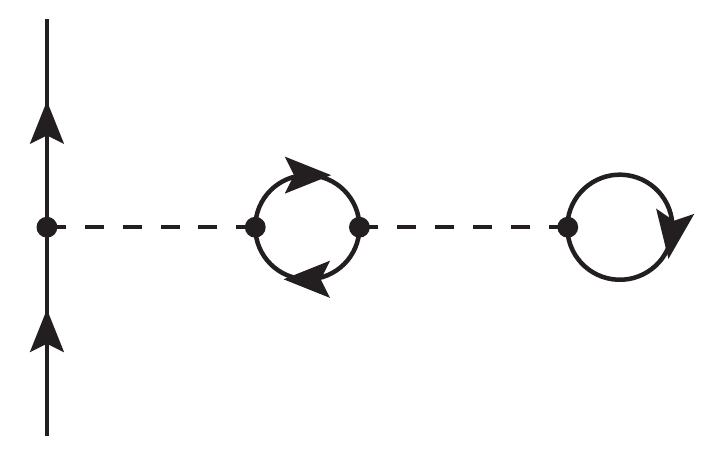}}
  \caption{Examples of a skeleton \protect\subref{skel} and a non skeleton \protect\subref{nonskel} diagram. The contribution depicted in \protect\subref{nonskel} contains a portion, a \emph{self-energy insertion}, which can be disconnected by cutting twice the middle fermion loop. This portion is contained by definition in the dressed fermionic loop of diagram \protect\subref{skel}.}
  \label{diagrams_skeleton}
  \end{center}
\end{figure}

An example of a skeleton and a non skeleton diagram is depicted in Fig.~\ref{diagrams_skeleton}. The SCGF approach is based on a diagrammatic expansion of such skeleton diagrams with renormalized propagators. When using skeleton diagrams in the irreducible self-energy, special care must be put when performing the iterative summation of the Dyson equation. As an example, a contribution like the one depicted in Fig.\subref*{nonskel}, if included in the irreducible self-energy as a skeleton diagram, i.e. with dressed propagators in both fermion loops, would be counted twice. In fact diagram Fig.\subref*{nonskel} is already accounted for by the contribution depicted in Fig.\subref*{skel}.

The number of diagrams which one sums up with the use of the skeleton irreducible self-energy is much larger than the number that one would obtain with the ordinary solution of the Dyson equation, Eq.~(\ref{Dyson}). Therefore, skeleton diagrams are not only a way to reorganize contributions in the self-energy, but above all a way to sum up big groups of diagrams hidden in the self-consistent dressed propagator.

\clearpage{\pagestyle{empty}\cleardoublepage}

%%%%%%%%%%%%%%%%%%% 3BF Formalism %%%%%%%%%%%%%%%%%%%%%

\chapter{Three-body forces in the SCGF approach}
\label{chapter:3BF_formalism}

In the introduction to this thesis we have underlined the necessity to consider 3BFs in order to obtain results both for finite and infinite nuclear systems which can resemble those obtained experimentally. As it has mostly been done in infinite nuclear matter calculations, 3BFs have been considered in the theoretical many-body approximation only by means of effective interactions. Nevertheless, even under these conditions, the 3B operators should be introduced in the formalism from first principles, i.e. from the definition of the Hamiltonian. This is crucial in order to avoid errors in the evaluation of the 3B contributions at the microscopic and macroscopic level.

Starting with the Hamiltonian described in Eq.~(\ref{H}), we want to derive in this chapter the perturbative expansion of the irreducible self-energy $\Sigma^\star$ up to third order. We will show how the description of the self-energy can be further reorganized thanks to the definition of a new class of irreducible contributions to the self-energy, the \emph{interaction-irreducible} diagrams. The comprehensive class of 1PI, skeleton and interaction-irreducible diagrams leads to a simplified description of the irreducible self-energy, which proves to be very useful in view of applying the hierarchy of equations of motion to the propagators. This hierarchy provides an all-order description not only for the SP self-energy but furthermore for the interacting vertex functions. Having control over these quantities allows us  to define specific truncations which lead to well defined many-body approximations. 

Such a formal approach to the problem is a much needed step in many-body calculations. In fact, the various $N$-body terms, which form the Hamiltonian, contribute with different factors to the single-particle or bulk properties of the system \citep{Bog2010,Heb2010Jul}. If not treated properly, these factors can be often mistaken in the process of the calculation. Special attention must be payed when dealing with effective interactions at a given $N$-body level. In fact, the physics of higher body operators included effectively in these quantities must preserve all the way its characteristics. In other words, the performance of contractions which turn higher body contributions into lower body ones must not alter the intrinsic multiplicity of the original operator.

For the specific purpose of the work presented in this thesis, the newly defined irreducible self-energy helps us to define in a consistent manner the \emph{ladder approximation} for infinite nuclear systems when including 3B forces, giving theoretical foundations to the calculations implemented numerically.

%%%%%%%%%%%%%%%%%%%% Self-energy perturbative expansion %%%%%%%%%%%%%%%

\section{Interaction-irreducible diagrams}
\label{section:self_en_pert}

\begin{figure}
  \begin{center}
  \subfloat[]{\label{direct}\includegraphics[width=0.2\textwidth]{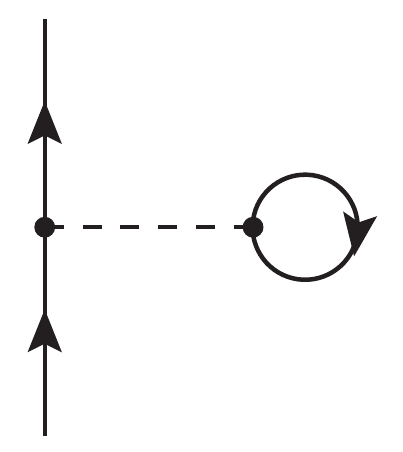}}
  \hspace{2cm}
  \subfloat[]{\label{exchange}\includegraphics[width=0.15\textwidth]{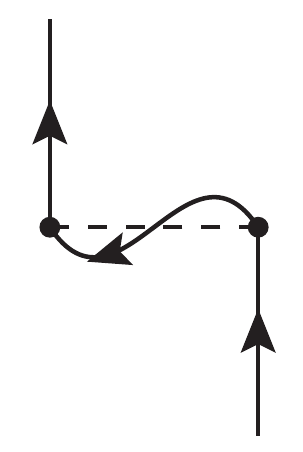}}
  \caption{Diagram \protect\subref{direct} and \protect\subref{exchange} correspond respectively to a direct and an exchange diagram. These diagrams arise from the two terms on the right-hand side of Eq.~(\ref{v_asym}). If antisymmetrized potential matrix elements are considered, \protect\subref{direct} and \protect\subref{exchange} are included in one another.}
  \label{dir_ex}
  \end{center}
\end{figure}

The best way to define the contributions which form the irreducible SP self-energy is by means of Feynman diagrams. All parts which compose a given diagram correspond to a specific mathematical expression and, once rules are defined, it is straightforward to associate a mathematical expression to a diagram. The rules used in this thesis are reported in App.~\ref{chapter:feynman_rules}. 

In the following, we will work mostly with \emph{unlabeled} Feynman diagrams, where unlabeled means that no time is specified for a given interaction vertex. According to Eq.~(\ref{gpert}), for a given order of expansion $n$, there are $n$ interacting terms $\h H_1(t_i)$, with $i$ going from 1 to $n$. Permutation of these operators gives rise to $n!$ equal diagrams, which can be considered as a unique diagram, a single ``time" unlabeled diagram. 

We also work with antisymmetrized potential matrix elements, as defined in Eq.~(\ref{v_asym}-\ref{w_asym}). This leads to consider as a unique diagram those contributions which come from the interchange of two incoming lines to a same interaction. An example of an antisymmetrized diagram is presented in Fig.~\ref{dir_ex}. Special attention has to be payed when considering  antisymmetrized matrix elements, in fact the correct counting of diagrams gives rise to specific multiplying factors which are usually not easy to define straightforwardly. This issue will be treated in detail in App.~\ref{chapter:int_irr_diag}.

\begin{figure}
\begin{center}
\includegraphics[width=0.9\textwidth]{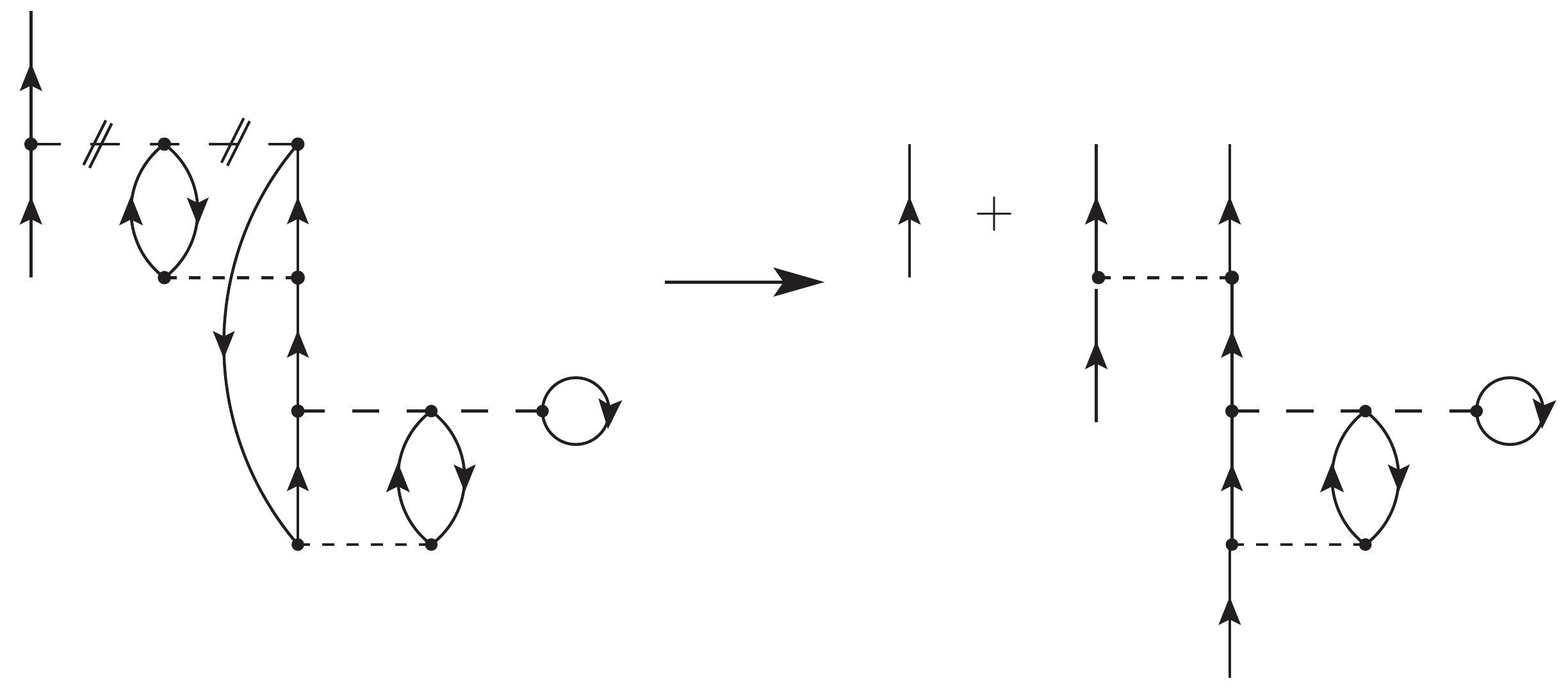}
\caption{Example of an articulation vertex: if the first 3B interaction is cut, where showed by the short double lines, the diagram is divided in two disconnected sections. One section corresponds to a group of $p=2$ lines which were previously entering and exiting a portion of the same interaction. This disconnected diagram is part of a 4-pt GF.}
\label{art_vertex}
\end{center}
\end{figure}

To go on and introduce the new class of \emph{interaction-irreducible diagrams} let us first consider an articulation vertex in a generic Feynman diagram. A 2B,  3B or higher interaction vertex is an articulation vertex if, when cut, it gives rise to a disconnected section where all propagating lines enter and exit a portion of this same interaction vertex\footnote{1B vertices cannot be split and therefore cannot be articulations.}. Formally, a diagram is said to be \emph{interaction-irreducible} if it contains no articulation vertices. Equivalently, a diagram is interaction-reducible if there exist a group of fermion lines (either interacting or not) that leave one interaction vertex and eventually all return to it. When an articulation vertex is cut, one is left with a cycle of fermion lines that all connect to the same interaction. If there were $p$ lines connected to this interaction vertex, this set of closed lines would necessarily be part of a $2p$-point GF \footnote{More specifically, these fermion lines contain an instantaneous contribution of the many-body GF that enters and exits the same interaction vertex, corresponding to a $p-$body reduced density matrix.}. An example of an articulation vertex  is shown in Fig.~\ref{art_vertex}. In this case, the second set of diagrams on the right-hand side of the figure belongs to a 4-point GF or a 2-body reduced density matrix. If this GF is computed explicitly in the calculation, one can use it to evaluate all these contributions straight away. This eliminates the need for computing all the diagrams looping in and out of the articulation vertex, at the expense of having to find the many-body propagator.

\begin{figure}[t!]
\begin{center}
\includegraphics[width=0.8\textwidth]{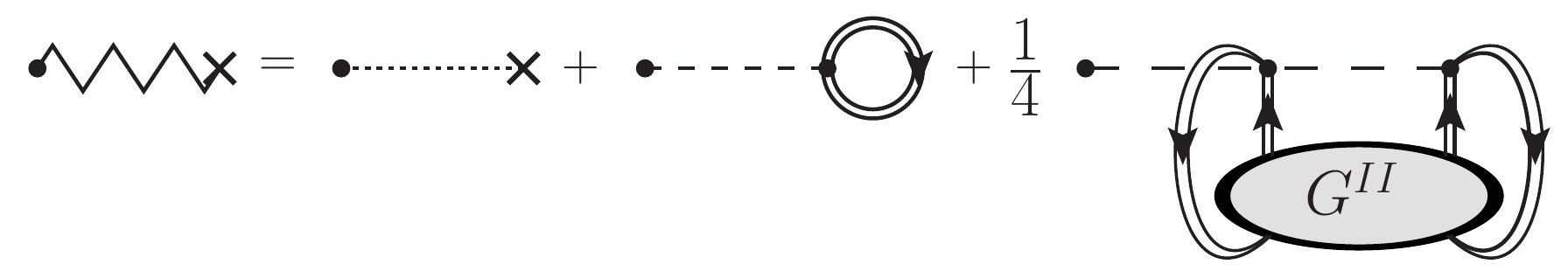}
\caption{Diagrammatic representation of the effective 1B interaction of Eq.~(\ref{ueff}). This is given by the sum of the original 1B potential (dotted line), the 2B interaction (dashed line) contracted with a dressed SP propagator, $G$ (double line with arrow), and the 3B interaction (long-dashed line) contracted with a dressed 2B propagator $G^{II}$. The correct symmetry factor of 1/4 in the last term is also shown explicitly. }
\label{ueffective}
\end{center}
\end{figure}

\begin{figure}
\begin{center}
\includegraphics[width=0.65\textwidth]{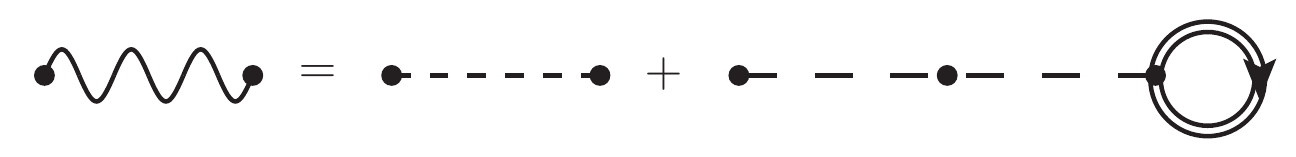}
\caption{Diagrammatic representation of the effective 2B interaction of Eq.~(\ref{veff}). This is given by the sum 
of the original 2B interaction (dashed line) and the
3B interaction (long-dashed line) contracted with a dressed SP propagator,  $G$.  }
\label{veffective}
\end{center}
\end{figure}

An $n$-body interaction vertex with $p$ fermion lines looping over it is an $n-p$ \emph{effective interaction} operator. Infinite sets of interaction-reducible diagrams can be sub-summed by means of effective contributions. The two cases of effective interactions which are of interest when 2B and 3B forces are present in the Hamiltonian are shown in Figs.~\ref{ueffective} and~\ref{veffective}. These give, respectively, the diagrammatic definition of  the 1B and 2B effective interactions. The 1B effective potential is obtained by adding up three contributions: the original 1B interaction; a 1B average over the 2B interaction; and a 2B average over the 3B force. The 1B and 2B averages are performed using fully dressed propagators. Similarly, the effective 2B force is obtained from the original 2B interaction plus a 1B average over the 3B force. Similar definitions would hold for higher-order forces and effective interactions beyond the 3B level. 

Hence, for a system with up to 3BFs, we define an effective Hamiltonian,
\beq
\widetilde H_1= {\widetilde U} + {\widetilde  V} + \h W \,,
\label{Heff}
\enq
where $\widetilde U$ and  $\widetilde V$ represent effective interaction operators. The diagrammatic expansion arising from  Eq.~(\ref{gpert}) with the effective Hamiltonian $\widetilde H_1$ is formed only of (1PI, skeleton) interaction-irreducible diagrams to avoid any possible double counting. Note that, in Eq.~(\ref{Heff}), the 3B interaction, $\h W$, remains the same as in Eq.~(\ref{H}) but enters only the interaction-irreducible diagrams with respect to 3B interactions. The explicit expressions for the 1B and 2B effective interaction operators are:
\beqn
\nn
\widetilde U& =&\sum_{\al\be}\Big[- U_{\al\be} 
- i\hbar \sum_{\ga\de}V_{\al\ga,\be\de} \, G_{\de \ga}(t-t^+) 
\\ &&  \qquad \qquad
\label{ueff} 
+ \f{i\hbar}{4} \sum_{\substack{\ga\ep \\ \de\eta}} W_{\al\ga\ep,\be\de\eta}
\,G^{II}_{\de\eta , \ga\ep}(t-t^+)\Big] a_\al^\dg a_{\be}\,, \\ 
\label{veff}
\widetilde V &=& \f 1 4\sum_{\substack{\al\ga\\\be\de}}\Big[V_{\al\ga,\be\de}
- i\hbar \sum_{\ep\eta}W_{\al\ga\ep,\be\de\eta} \,G_{\eta\ep}(t-t^+)\Big] a_\al^\dg a_\ga^\dg a_{\de}a_{\be} \, .
\enqn
We have introduced a specific component of the 4-point GFs of \eq{g4pt},
\beq
G^{II}_{\de\eta , \ga\ep}(t-t') = G^{4-\rm{pt}}_{\de\eta , \ga\ep}(t^+, t; t', t'^+) \, ,
\label{g2g4pt}
\enq
which involves two-particle and two-hole propagation. This is the so-called two-particle and two-time
Green's function. Let us also note that the contracted propagators in Eqs.~(\ref{ueff}) and (\ref{veff}) correspond to the full 1B and 2B reduced density matrices of the many-body system:
\beqn
\label{1B_densitymatrix}
\rho^{1B}_{\de\ga} &= ~ \langle\Psi_0^N| \, a_{\ga}^\dg a_{\de} \, |\Psi_0^N\rangle \quad &= ~ -i\hbar\, G_{\de\ga}(t-t^+) \; ,
\\
\rho^{2B}_{\de\eta , \ga\ep} &= ~\langle\Psi_0^N| \, a_{\ga}^\dg a_{\ep}^\dg a_{\eta} a_{\de} \, |\Psi_0^N\rangle \; &=
~i\hbar\,G^{II}_{\de\eta , \ga\ep}(t-t^+) \, . \qquad
\enqn
In a self-consistent calculation, effective interactions should be computed iteratively at each step, using correlated 1B and 2B propagators as input. All in all, the effective Hamiltonian of Eq.~(\ref{Heff})  not only regroups Feynman diagrams in a more efficient way, but also defines the effective 1B and 2B terms from higher-order interactions. Averaging the 3BF over one and two spectator particles in the medium is expected to yield the most important contributions to the many-body dynamics in nuclei \citep{Hag2007,Rot2012}. We note that Eqs.~(\ref{ueff}) and~(\ref{veff}) are exact and can be derived rigorously from the perturbative expansion. Details of the proof are discussed in App.~\ref{chapter:int_irr_diag}. As long as interaction-irreducible diagrams are used together with the effective Hamiltonian, $\widetilde{H}_1$, this approach provides a systematic way to incorporate many-body forces in the calculations and to generate effective in-medium interactions. More importantly, the formalism is such that symmetry factors are properly considered and no diagram is over-counted.

This approach can be seen as a generalization of the normal ordering of the Hamiltonian with respect to the reference state $\rphizero$, a procedure that has already been used in nuclear physics applications with 3BFs \citep{Hag2007,Bog2010,Rot2012}. The normal ordered interactions affect only excited configurations with respect to $\rphizero$, but not the reference state itself. Similarly, the effective operators discussed above only enter interaction-irreducible diagrams. In both the traditional normal ordering and our approach, the ${\widetilde{U}}$ and ${\widetilde{V}}$ operators contain contributions from higher-order forces, while $\hat{W}$ remains unchanged. Note however that in the latter we go beyond usual normal-ordering ``averages", in that they are performed over fully-correlated many-body propagators.  If the unperturbed 1B and 2B propagators were used in Eqs.~(\ref{ueff}) and~(\ref{veff}), the effective operators ${\widetilde{U}}$ and ${\widetilde{V}}$ would trivially reduce to the contracted 1B and 2B terms of normal ordering. In the present case the contraction is improved because it is performed with respect to the exact correlated density matrices. To some extent, one can think of the effective Hamiltonian, $\widetilde{H}_1$,  as being reordered with respect to the interacting many-body ground state $|\Psi_0^N\rangle$, rather than the non-interacting  $\rphizero$. This effectively incorporates correlations that, in the normal ordering procedure, must be instead calculated explicitly by the many-body approach. Calculations indicate that such correlated averages are important in both the saturation mechanism of nuclei and nuclear matter \citep{Cip2013,Car2013Oct}.

Note that a normal ordered Hamiltonian also contains a 0B term equal to the expectation value of the original Hamiltonian $\hat{H}$ with respect to $\rphizero$. Likewise, in our case, the full contraction of $\hat{H}$ in Eq.~(\ref{H}) according to the normal-ordering procedure of App.~\ref{chapter:int_irr_diag}, will yield a 0B term equal to the exact ground-state energy:
\beqn
\nn
E_0^N  &=&-i\hbar \,\sum_{\al\be} T_{\al\be} \, G_{\be\al}(t-t^+)
\\\nn &&+
\f{i\hbar}{4}\sum_{\substack{\al\ga\\\be\de}}V_{\al\ga,\be\de}\, G^{II}_{\be\de,\al\ga}(t-t^+) 
\\\nn && - \f{i\hbar}{36}
\sum_{\substack{\al\ga\ep \\ \be\de\eta}} W_{\al\ga\ep , \be\de\eta}  \, G^{III}_{\be\de\eta,\al\ga\ep}(t-t^+)
\\ &=& \langle\Psi_0^N| \, \hat H  \, |\Psi_0^N\rangle \,.
\label{e0}
\enqn
This is in accordance with our analogy between the effective Hamiltonian, $\widetilde{H}=\widetilde H_0+\widetilde{H}_1$, and the usual normal ordered one. 

\subsection{Self-energy expansion up to third order}

As a demonstration of the simplification introduced by the effective interactions approach, we want to derive all interaction-irreducible contributions to the proper self-energy up to third order in perturbation theory. We will discuss these contributions order by order, thus providing an overview of how the approach can be extended to higher-order perturbative and also to nonperturbative calculations. Among other things, this exercise will illustrate the amount and variety of new diagrams that need to be considered when 3BFs are used.

For a pure 2B Hamiltonian, the only possible interaction-reducible contribution to the self-energy is the generalized Hartree-Fock diagram. This corresponds to the second term on the right-hand side of Eq.~(\ref{ueff}) (see also Fig.~\ref{ueffective}). Note that this can go beyond the usual Hartree-Fock term in that the internal propagator is dressed. This diagram appears at first order in any SCGF expansion and it is routinely included in most GF calculations with 2B forces. Thus, regrouping diagrams in terms of effective interactions, such as Eqs.~(\ref{ueff}) and~(\ref{veff}), give no practical advantages unless 3BFs (or higher-body forces) are present. 

If 3BFs are considered, the only first-order, interaction-irreducible contribution is given by the one-body effective interaction depicted in Fig.~\ref{ueffective}, 
\beq
\Sigma^{\star , (1)}_{\al \be} = \widetilde{U}_{\al\be} \; . 
\label{eq:1ord}
\enq
Since $\widetilde{U}$ is in itself a self-energy insertion, it will not appear in any other, higher-order skeleton diagram. Even though it only contributes to Eq.~(\ref{eq:1ord}), the effective 1B potential is very important since it determines the energy-independent part of the self-energy. It therefore represents the (static) mean-field seen by every particle, due to both 2B and 3B interactions. As already mentioned, Eq.~(\ref{ueff}) shows that this potential incorporates three separate terms, including the Hartree-Fock potentials due  to both 2B and 3BFs and higher-order and interaction-reducible contributions due to the dressed $G$ and $G^{II}$ propagators. Thus, even the calculation of this lowest-order term $\Sigma^{\star , (1)}$ requires an iterative procedure to evaluate the internal many-body propagators self-consistently.

Note that, if one were to stop at the Hartree-Fock level, the 4-point GF would reduce to the direct and exchange product of two 1B propagators. In that case, the last term of Eq.~(\ref{ueff}) (or Fig.~\ref{ueffective}) would reduce to the pure 3BF Hartree-Fock contribution with the correct $1/2$ factor in front, due to the two equivalent fermionic lines (see Fig.~\ref{ueffective_hf} at the beginning of Chap.~\ref{chapter:results}). This approximate treatment of the 2B propagator in the 1B effective interaction has been employed in most nuclear physics calculations up to date, including both finite nuclei \citep{Ots2010,Rot2012,Cip2013} and nuclear matter \citep{Som2009,Heb2010Jul,Heb2011,Lov2012,Car2013Oct,Koh2013,Coraggio2014} applications.

Before we move on, let us mention a subtlety arising in the Hartree-Fock (or lowest-order) approximation to the two-body propagator.  If one were to insert $\widetilde{V}$ into the second term of the right-hand side of Eq.~(\ref{ueff}), one would introduce a double counting of the pure 3BF Hartree-Fock component. This is forbidden because the diagram in question would be interaction-reducible. The correct  3BF Hartree-Fock term is actually included as part of the last term of  Eq.~(\ref{ueff}) (see also Fig.~\ref{ueffective}). Consequently, there is no Hartree-Fock term arising from the effective interactions. Instead, this lowest-order contribution is fully taken into account within the 1B effective interaction.

\begin{figure}[t]
  \begin{center}
  \subfloat[]{\label{2ord_2B}\includegraphics[width=0.15\textwidth]{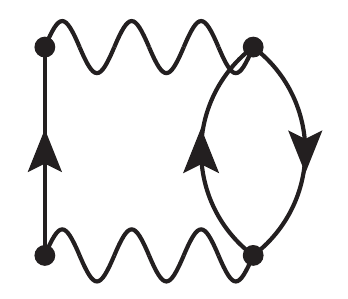}}
  \hspace{2cm}
  \subfloat[]{\label{2ord_3B}\includegraphics[width=0.25\textwidth]{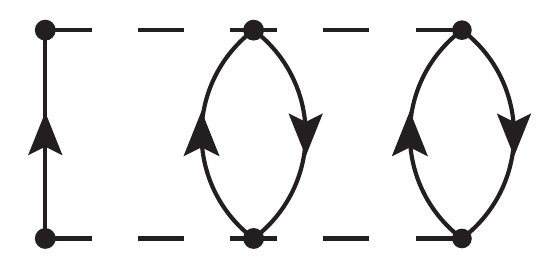}}
  \caption{1PI, \emph{skeleton} and \emph{interaction-irreducible} self-energy diagrams appearing at second order in the perturbative expansion of Eq.~(\ref{gpert}), using the effective Hamiltonian of Eq.~(\ref{Heff}).}
  \label{2ord}
  \end{center}
\end{figure}

\begin{figure}
  \begin{center}
  \subfloat[]{\label{2ord_2B_split_a}\includegraphics[width=0.17\textwidth]{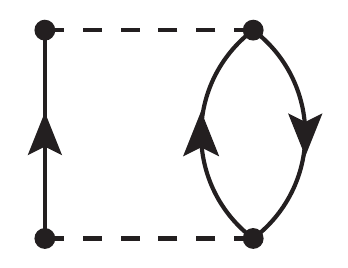}}
  \hspace{2cm}
  \subfloat[]{\label{2ord_2B_split_b}\includegraphics[width=0.3\textwidth]{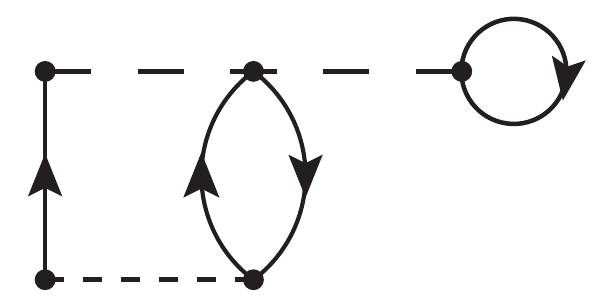}}
  \\
  \subfloat[]{\label{2ord_2B_split_c}\includegraphics[width=0.3\textwidth]{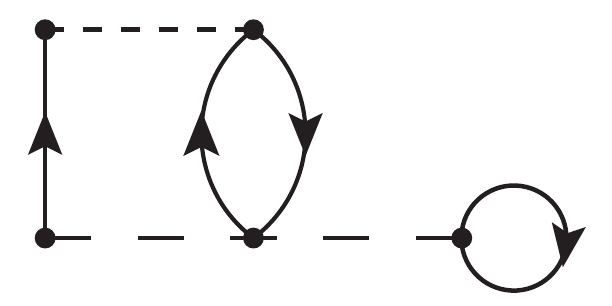}}
  \hspace{1cm} 
  \subfloat[]{\label{2ord_2B_split_d}\includegraphics[width=0.3\textwidth]{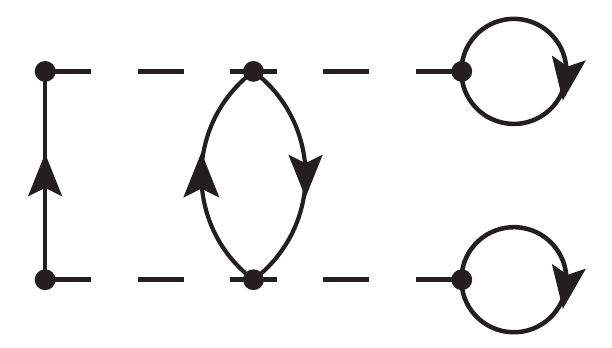}}
  \caption{These four diagrams are contained in diagram Fig.~\protect\subref*{2ord_2B}. 
 They correspond to one 2B \emph{interaction-irreducible} diagram, \protect\subref{2ord_2B_split_a}, and
 three \emph{interaction-reducible} diagrams, \protect\subref{2ord_2B_split_b}-\protect\subref{2ord_2B_split_d}. }
\label{2ord_2B_split}
\end{center}
\end{figure}

At second order, there are only two interaction-irreducible diagrams, that we show in Fig.~\ref{2ord}. Diagram Fig.\subref*{2ord_2B} has the same structure as the well-known contribution due to 2BFs only, involving two-particle--one-hole ($2p1h$) and two-hole--one-particle ($2h1p$) intermediate states. This diagram, however, is computed with the 2B effective interaction (notice the wiggly line) instead of the original 2B force and hence it corresponds to further interaction-reducible diagrams. 
By expanding the effective 2B interaction according to Eq.~(\ref{veff}), the contribution of Fig.~\subref*{2ord_2B} splits into the four diagrams of  Fig.~\ref{2ord_2B_split}.
 %~[see also a similar example in Fig.~\ref{rule9-2a}].
 
 The second interaction-irreducible diagram arises from explicit 3BFs and it  is given in Fig.~\subref*{2ord_3B}. One may expect this contribution to play a minor role due to phase space arguments, as it involves  $3p2h$ and $3h2p$ excitations at higher excitation energies. Moreover,  3BFs are generally weaker than the corresponding  2BFs (typically, $\langle\hat{W}\rangle\approx\f{1}{10}\langle\hat{V}\rangle$ for nuclear interactions~\citep{Gra1989,Epe2009Oct}). Summarizing, at second order in standard self-consistent perturbation theory, one would find a total of 5 skeleton diagrams. Of these, only 2 are interaction-irreducible and need to be calculated when effective interactions are considered.

\begin{figure}
  \begin{center}
  \subfloat[]{\label{3ord_2B_1}\includegraphics[width=0.13\textwidth]{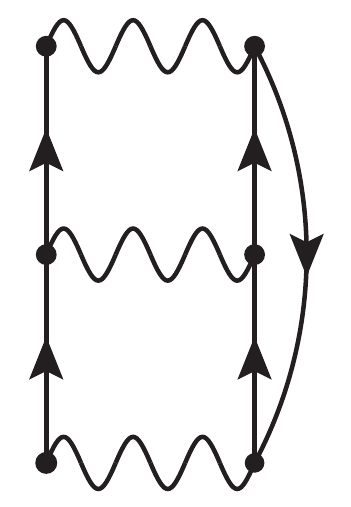}}
  \hspace{1cm}
  \subfloat[]{\label{3ord_2B_2}\includegraphics[width=0.21\textwidth]{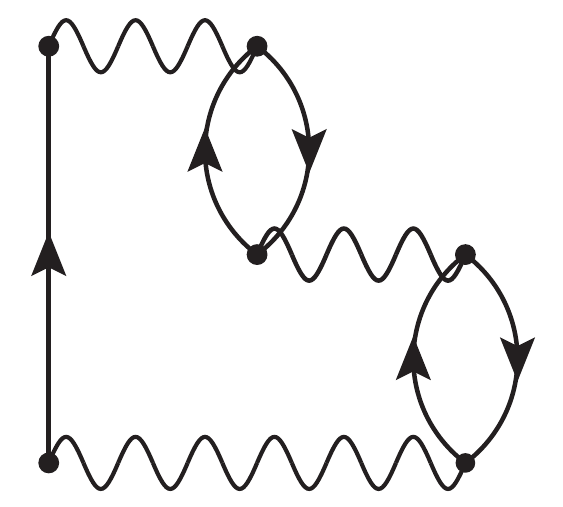}}
  \hspace{2cm}
  \subfloat[]{\label{3ord_232B}\includegraphics[width=0.19\textwidth]{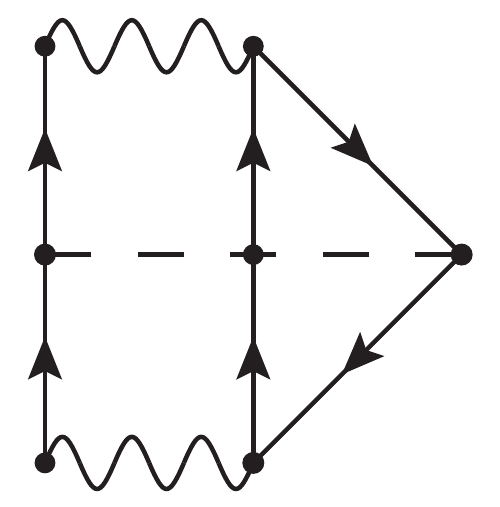}}
  \newline    \vskip .8cm
  \subfloat[]{\label{3ord_223B_1}\includegraphics[width=0.2\textwidth]{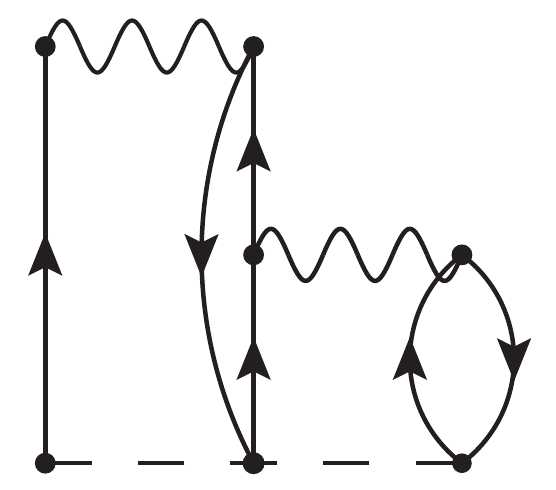}}
  \hspace{0.5cm}
  \subfloat[]{\label{3ord_223B_2}\includegraphics[width=0.2\textwidth]{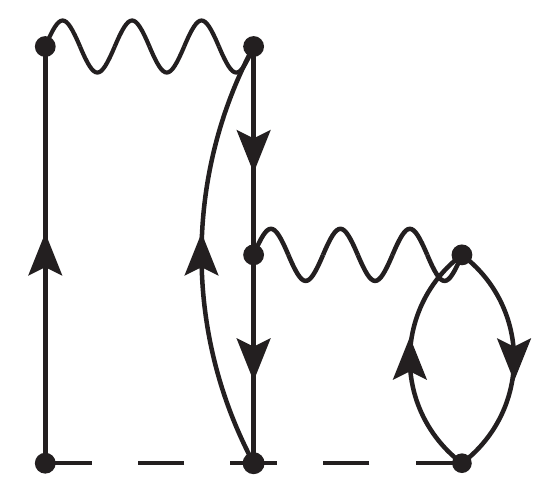}}
  \hspace{0.5cm}
   \subfloat[]{\label{3ord_322B_1}\includegraphics[width=0.2\textwidth]{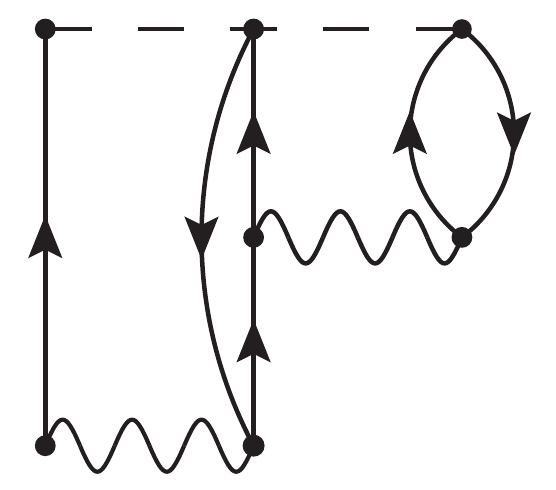}}
  \hspace{0.5cm}
  \subfloat[]{\label{3ord_322B_2}\includegraphics[width=0.2\textwidth]{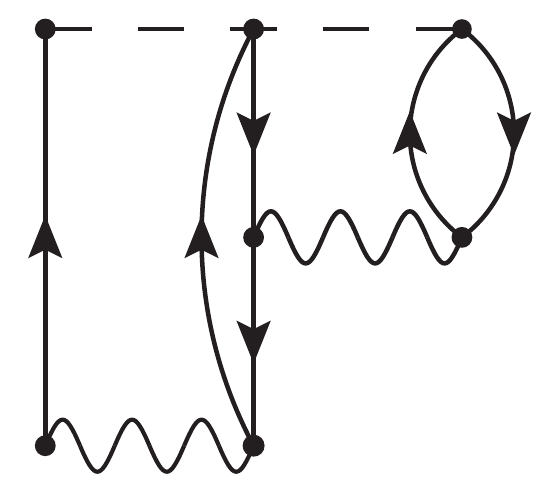}}
  \newline   \vskip .8cm
  \subfloat[]{\label{3ord_233B_1}\includegraphics[width=0.2\textwidth]{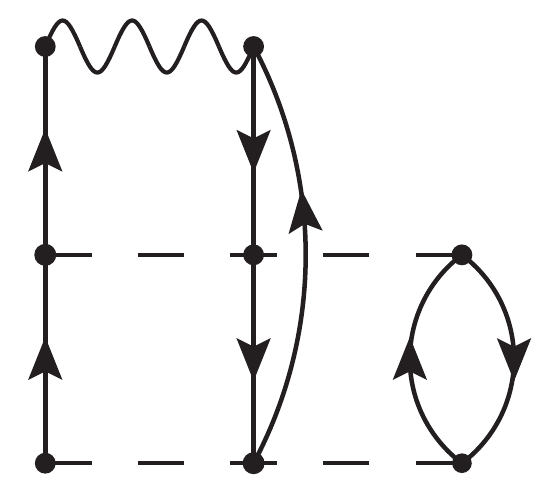}}
  \hspace{0.5cm}
  \subfloat[]{\label{3ord_233B_2}\includegraphics[width=0.2\textwidth]{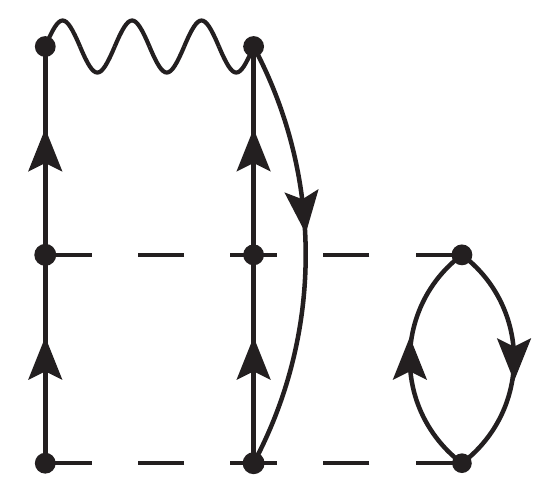}}
  \hspace{0.5cm}
  \subfloat[]{\label{3ord_332B_1}\includegraphics[width=0.2\textwidth]{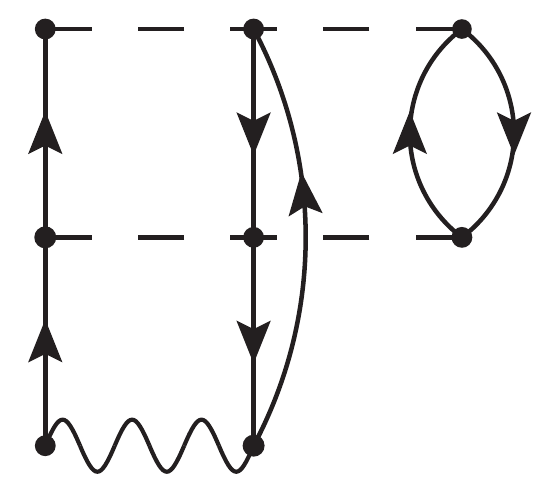}}
  \hspace{0.5cm}
  \subfloat[]{\label{3ord_332B_2}\includegraphics[width=0.2\textwidth]{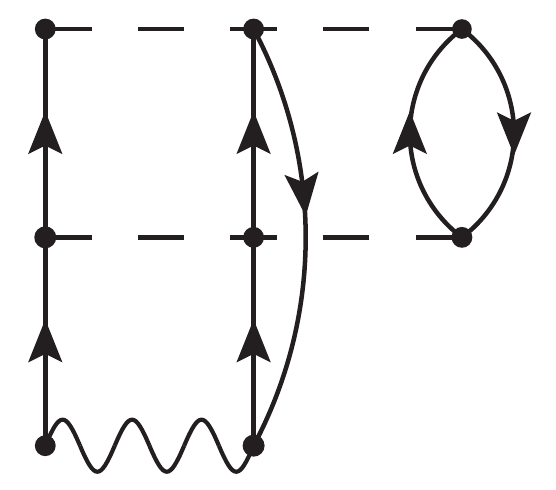}}
  \newline   \vskip .8cm
  \subfloat[]{\label{3ord_323B_1}\includegraphics[width=0.2\textwidth]{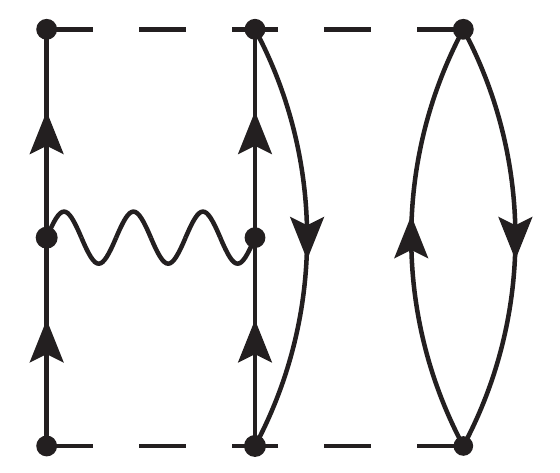}}
  \hspace{1.5cm}
  \subfloat[]{\label{3ord_323B_2}\includegraphics[width=0.2\textwidth]{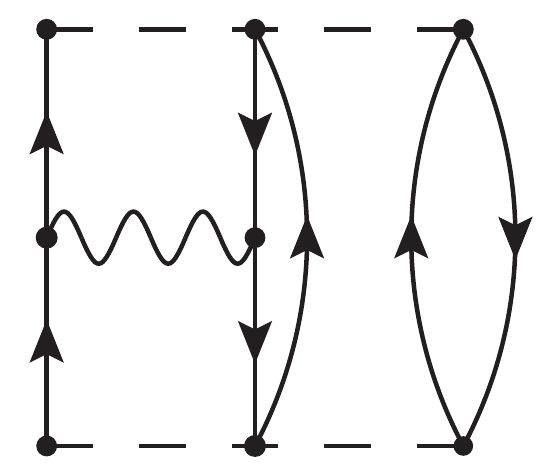}}
  \hspace{1.5cm}
  \subfloat[]{\label{3ord_323B_3}\includegraphics[width=0.2\textwidth]{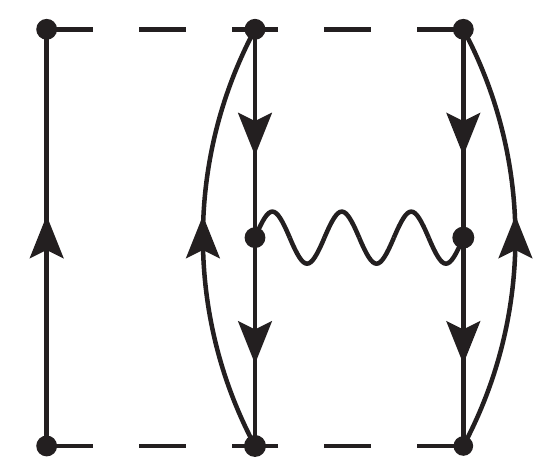}}
  \newline   \vskip .8cm
  \subfloat[]{\label{3ord_333B_1}\includegraphics[width=0.2\textwidth]{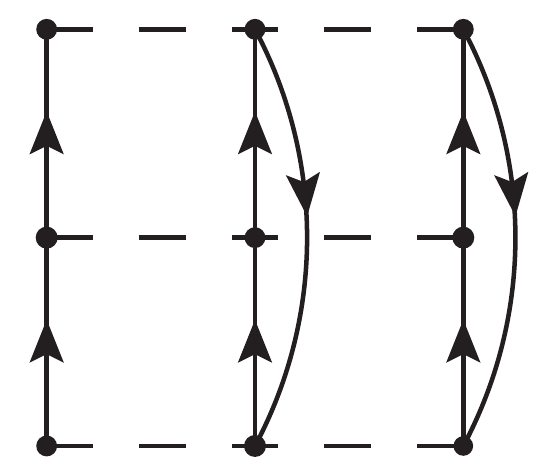}}
  \hspace{1.5cm}
  \subfloat[]{\label{3ord_333B_2}\includegraphics[width=0.2\textwidth]{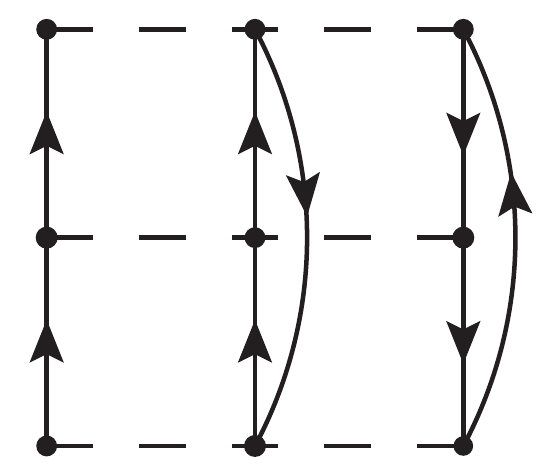}}
  \hspace{1.5cm}
  \subfloat[]{\label{3ord_333B_3}\includegraphics[width=0.2\textwidth]{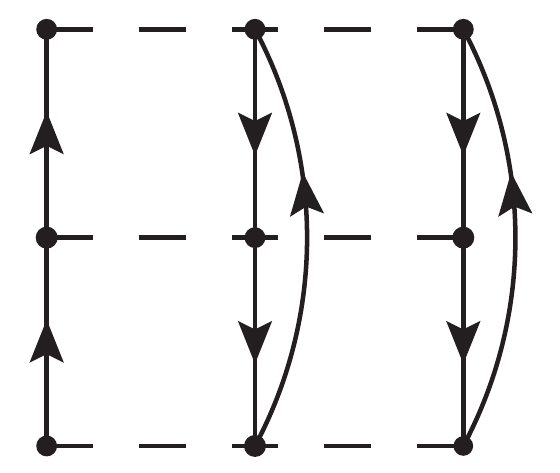}}
   \caption{1PI, \emph{skeleton} and \emph{interaction-irreducible} self-energy diagrams appearing at third order in the perturbative expansion of Eq.~(\ref{gpert}) using the effective Hamiltonian of Eq.~(\ref{Heff}).} 
  \label{3ord}
  \end{center}
\end{figure}

Fig.~\ref{3ord} shows all the 17 interaction-irreducible diagrams appearing at third order. Again, note that, expanding the effective interaction $\widetilde{V}$, would generate a much larger number of diagrams (53 in total). Diagrams Figs.~\subref*{3ord_2B_1} and~\subref*{3ord_2B_2} are the only third-order terms that would appear in the 2BF case. Numerically, these two diagrams only require evaluating Eq.~(\ref{veff}) beforehand, but can otherwise be dealt with using existing 2BF codes. They have already been exploited to include 3BFs in nuclear structure studies \citep{Som2008,Heb2010Jul,Heb2011,Cip2013,Car2013Oct}. 

The remaining 15 diagrams, from Figs.~\subref*{3ord_232B} to~\subref*{3ord_333B_3}, appear when irreducible 3BFs are introduced. These third-order diagrams are ordered in  Fig.~\ref{3ord} in terms of increasing numbers of 3B interactions and, within these, in terms of increasing number of particle-hole excitations. Qualitatively, one would expect that this should correspond to a decreasing importance of their contributions. Diagrams Figs.~\subref*{3ord_2B_1}-\subref*{3ord_232B}, for instance, only involve $2p1h$ and $2h1p$ intermediate configurations, normally needed to describe particle addition and removal energies to dominant quasiparticle peaks as well as total ground-state energies. Diagram Fig.~\subref*{3ord_232B} includes one 3B irreducible interaction term and still needs to be investigated within the SCGF method. Normal-ordered Hamiltonian studies \citep{Hag2007,Rot2012} clearly suggest that this brings in a small correction to the total energy with respect to diagrams Figs.~\subref*{3ord_2B_1} and~\subref*{3ord_2B_2}. This is in line with the qualitative analysis of the number of $\widetilde{V}$ and $\h{W}$ interactions entering these diagrams. 
%Diagrams Figs.~\subref*{3ord_2B_1}-\subref*{3ord_232B} all represent the first order term in an all order summation needed to account for configuration mixing between $2p1h$ or $2h1p$ excitations. Nowadays, resummations of these configurations are performed routinely for the first two diagrams in third-order algebraic diagrammatic construction, ADC(3), and FRPA calculations \citep{Bar2007,Ort2013,Bar2009Jun}.
The remaining diagrams of Fig.~\ref{3ord} all include $3p2h$ and $3h2p$ configurations. 
%These become necessary to reproduce the fragmentation patterns of shakeup configurations in particle removal and addition experiments, i.e. Dyson orbits beyond the main quasiparticle peaks. These contributions are computationally more demanding. 
Diagrams Fig.~\subref*{3ord_223B_1} to~\subref*{3ord_332B_2} all describe interaction between $2p1h$~($2h1p$) and $3p2h$~($3h2p$) configurations. These are split into two groups, four contributions containing two effective 2BFs and other four conatining two irreducible 3B interactions.  Similarly, diagrams Figs.~\subref*{3ord_323B_1} to~\subref*{3ord_333B_3} are the first contributions to the configuration mixing among $3p2h$  or $3h2p$ states.

App.~\ref{chapter:feynman_rules} provides the Feynman diagram rules to compute the contribution associated with these diagrams. Specific expressions for some diagrams in Fig.~\ref{3ord} are given. We note that the Feynman rules remain unaltered whether one uses the original, $\h U$ and $\h V$, or the effective, $\widetilde{U}$ or $\widetilde{V}$, interactions. Hence, symmetry factors due to equivalent lines remain unchanged. However, we provide in App.~\ref{chapter:feynman_rules} a few examples to illustrate the appearance of non-trivial symmetry factors when 3BFs are considered.  This complicates the rules of the symmetry factors and illustrates some of the difficulties associated with many-particle interactions.

%%%%%%%%%%%%%%%% Equation-of-motion method %%%%%%%%%%%%%%%%%%%

\section{Hierarchy of equations of motion}
\label{section:eom_method}

The perturbation theory expansion outlined in the previous section is useful to identify new contributions arising from the inclusion of 3B interactions. However, diagrams up to third order alone do not necessarily incorporate all the necessary information to describe strongly correlated quantum many-body systems. For example, the strong repulsive character of the nuclear force at short distances requires explicit all-order summations of ladder series. All-order summations of $2p1h$ and $2h1p$ are also required in finite systems to achieve accuracy for the predicted ground-state and separation energies, as well as to preserve the correct analytic properties of the self-energy beyond second order.

To investigate approximation schemes for all-order summations including 3BFs, we now develop the EOM method. This will provide special insight into possible self-consistent expansions of the irreducible self-energy, $\Sigma^\star$. For 2B forces only, the EOM technique defines a hierarchy of equations that link each $N$-body GF to the $(N-1)$- and the $(N+1)$-body GFs. When extended to include 3BFs, the hierarchy also involves the $(N+2)$-body GF. A truncation of this Martin-Schwinger hierarchy is necessary to solve the system of equations \citep{Mart1959} and can potentially give rise to physically relevant resummation schemes. Here, we will follow the footprints of Ref.~\citep{Mat1971} and apply truncations to obtain explicit equations for the 4-point (and 6-point, in the 3BF case) vertex functions. 

The EOM for a given propagator is found by taking the derivative of its time arguments. The time arguments are linked to the creation and annihilation operators in Eqs.~(\ref{G}) to (\ref{g6pt}), and hence the time dependence of these operators will drive that of the propagator \citep{Bla1986}. 
The unperturbed 1B propagator can be written as the $n=0$ order term of Eq.~(\ref{gpert}),
\beq
i\hbar \, \gz_{\al \be}(t_\al - t_\be) = \lphizero\T[a^I_\al(t_\al){a_{\be}^I}^\dg(t_\be)]\rphizero \, .
\label{G0}
\enq
Its time derivative will be given by the von Neumann equation for the operators in the interaction picture \citep{Abr1975}:
\beq
i\hbar\dev a^I_\al(t)= [a^I_\al(t),\h H_0] =  \vep^0_\al a^I_\al(t) \,. 
\label{eom_aI}
\enq
Taking the derivative of $\gz$ with respect to time and using Eq.~(\ref{eom_aI}), we find
\beq
\left\{i\hbar\frac{\partial}{\partial t_\al}-\vep^0_\al\right\} G^{(0)}_{\al\be}(t_\al-t_\be)=\delta(t_\al-t_\be) \de_{\al\be} \, .
\label{G0inv}
\enq
Note that the delta functions in time arise from the derivatives of the step-functions involved in the time-ordered product. 

\begin{figure}[t]
\begin{center}
\includegraphics[width=0.7\textwidth]{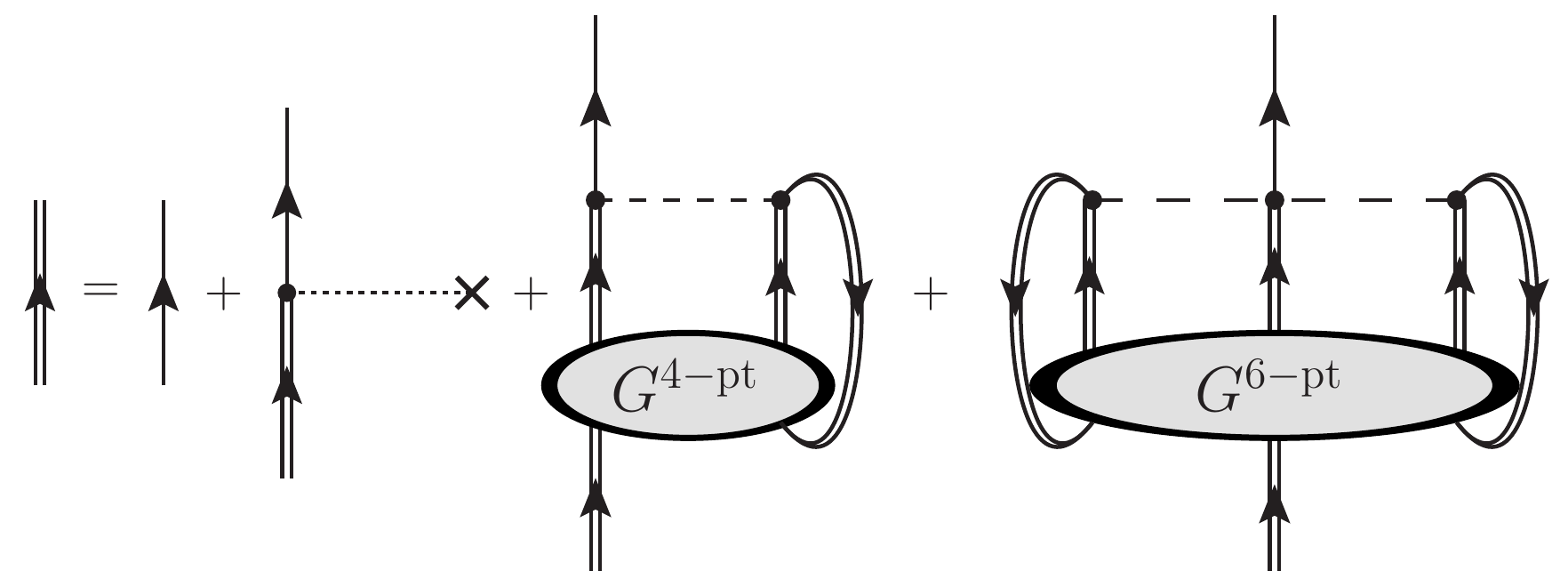}
\caption{Diagrammatic representation of the EOM, Eq.~(\ref{gexact}), 
for the dressed 1B propagator, $G$. The first term, given by a single line, defines the free 1B propagator, $\gz$. The second term denotes the interaction with a bare 1B potential, whereas the third and the fourth terms describe interactions involving the intermediate propagation of two- and three-particle configurations. }
\label{eq_g}
\end{center}
\end{figure}

The same procedure applied to the exact 1B propagator, $G$, requires the time-derivative of the operators in the Heisenberg picture. For the original Hamiltonian of Eq.~(\ref{H}), the EOM for the annihilation operator reads:

\beqn
\nn &&
i\hbar\dev a_\al(t) = [a_\al(t),\h H]  =
\vep^0_\al a_\al(t) - \sum_{\de} U_{\al\de} a_\de(t)  
\\\nn && \qquad
+ \f 1 2\sum_{\substack{\ep\\\de\mu}} V_{\al\ep,\de\mu} \ad\ep(t) a_\mu(t) a_\de(t) 
\\ && \qquad
+\frac{1}{12}\sum_{\substack{\ep\ta\\\de\mu\lm}}
W_{\al\ep\ta,\de\mu\lm} \ad\ep(t)\ad\ta(t) a_\lm(t)a_\mu(t)a_\de(t)\,. 
\label{ader}
\enqn
This can now be used to take the derivative of the full 1B propagator in Eq.~(\ref{G}):
\beqn
\nn &&
 \left\{i\hbar\frac{\partial}{\partial t_\al}-\vep^0_\al\right\}  G_{\al\be}(t_\al-t_\be) = \delta(t_\al-t_\be)\de_{\al\be} 
 \\\nn && \qquad\qquad 
- \sum_{\de}  U_{\al\de} G_{\de \be}(t_\al-t_\be) 
\\\nn &&  \qquad\qquad 
+\frac{1}{2} \sum_{\substack{\ep\\\de\mu}}  
V_{\al\ep,\de\mu} G^{4-{\rm pt}}_{\de\mu,\ep\be}(t_\al,t_\al;t_\al^+,t_\be) 
\\  && \qquad\qquad 
+\frac{1}{12}\sum_{\substack{\ep\ta\\\de\mu\lm}}
W_{\al\ep\ta,\de\mu\lm} G^{6-{\rm pt}}_{\de\mu\lm,\ep\ta\be}
(t_\al,t_\al,t_\al;t_\al^{++},t_\al^+,t_\be)\,. \qquad\qquad 
\label{devG}
\enqn
This equation links the 2-point GF to both the 4- and the 6-point GFs. Note that the connection with the latter is mediated by the 3BF and hence does not appear in the pure 2BF case. Regarding the time-arguments in Eq.~(\ref{devG}), the $t_\al^+$ and $t_\al^{++}$ in the 4- and 6-point GFs are necessary to keep the correct time-ordering in the creation operators when going from Eq.~(\ref{ader}) to Eq.~(\ref{devG}). An analogous equation can be obtained for the derivative of the time argument $t_\be$. After some manipulation, involving the Fourier transforms of Eqs.~(\ref{Gmpt_ft}) and (\ref{G1B_Lehm}), one obtains the equation of motion for the SP 
propagator in frequency representation: 
\beqn
\nn &&
G_{\al\be}(\om) = G^{(0)}_{\al\be}(\om)
- \sum_{\ga\de}G^{(0)}_{\al\ga}(\om) U_{\ga\de} G_{\de\be}(\om)
\\\nn && \quad
-\frac{1}{2} \sum_{\substack{\ga\ep\\\de\mu}}G^{(0)}_{\al\ga}(\om) V_{\ga\ep,\de\mu}
\int \frac{d \om_1}{2 \pi} \int \frac{d \om_2}{2 \pi} 
%\\\nn && \qquad
G^{4-{\rm pt}}_{\de\mu,\be\ep}(\om_1,\om_2;\om,\om_1+\om_2-\om) 
\\\nn&& \quad
+\frac{1}{12}\sum_{\substack{\ga\ep\ta\\\de\mu\lm}}G^{(0)}_{\al\ep}(\om)W_{\ga\ep\ta,\de\mu\lm}
\int \frac{d \om_1}{2 \pi}\int \frac{d \om_2}{2 \pi}\int \frac{d \om_3}{2 \pi}\int \frac{d \om_4}{2 \pi}
\\ && \qquad\qquad
\times G^{6-{\rm pt}}_{\de\mu\lm,\ga\be\ta}(\om_1,\om_2,\om_3;\om_4,\om,\om_1+\om_2+\om_3-\om_4-\om)\,.\quad
\label{gexact}
\enqn
Again, this involves both the 4- and the 6-point GFs, which appear due to the 2B and 3B interactions, respectively. The equation now involves $m-2$ frequency integrals for the $m$-point GFs. The diagrammatic representation of this equation is given in Fig.~\ref{eq_g}. 

The EOMs, Eqs.~(\ref{devG}) and~(\ref{gexact}), connect the 1B propagator to GFs of different $N$-body orders. In general, starting from an $N$-body GF, the derivative of the time-ordering operator generates a delta function between an incoming and outgoing particle, effectively separating a line and leaving an ($N-1$)-body propagator. Conversely, the 2B part of the Hamiltonian introduces an extra pair of creation and annihilation operators that adds another particle and leads to an ($N+1$)-body GF. For a 3B Hamiltonian, the ($N+2$)-body GF enters the EOM due to the commutator in Eq.~(\ref{ader}). This implies that higher-order GFs will be needed, at the same level of approximation, in the EOMs hierarchy when considering 3BFs.

Eq.~(\ref{gexact}) gives an exact equation for the SP propagator $G$ that, however, requires the knowledge of both the 4-point and 6-point GFs.  Full equations for the latter require applying the EOMs to these propagators as well.  Before that, however, it is possible to further simplify contributions in Eq.~(\ref{gexact}) by splitting the $m$-point GFs into two terms. The first one is relatively simple, involving the properly antisymmetrized independent propagation of $N$ dressed particles, with $N$ equal to two and three for respectively the 4-point and 6-point GFs. The second term will involve the interaction vertices, $\Gamma^{4-{\rm pt}}$ and $\Gamma^{6-{\rm pt}}$, 1PI vertex functions that include all interaction effects \citep{Bla1986}. These can be neatly connected to the irreducible self-energy.

\begin{figure}
\begin{center}
\includegraphics[width=0.65\textwidth]{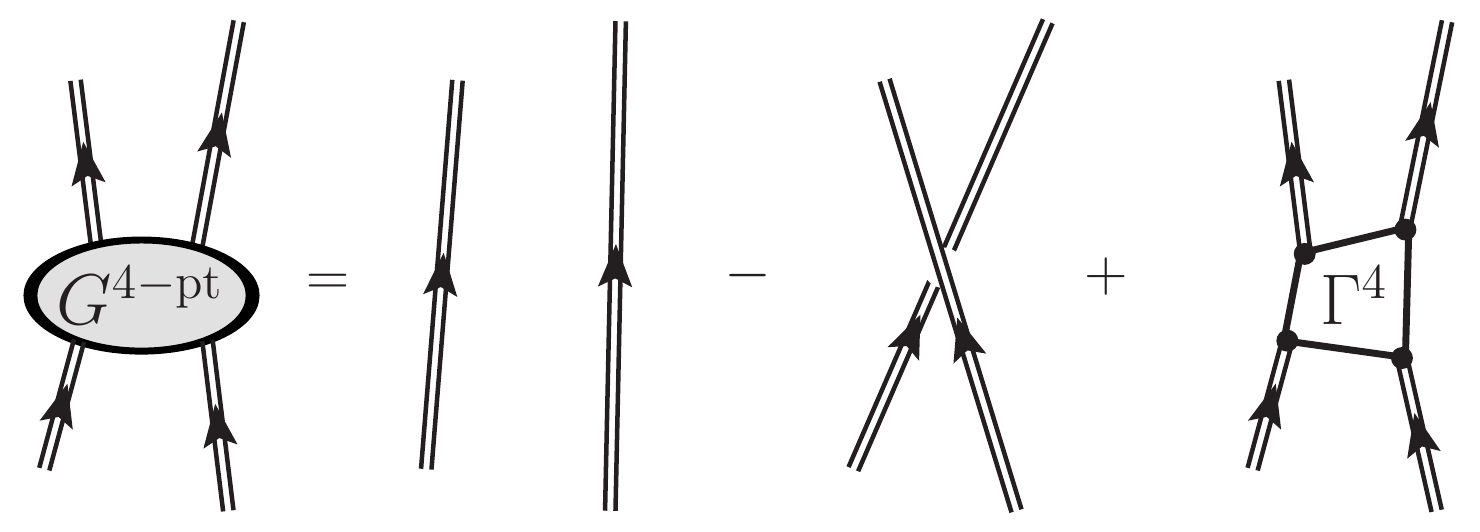}
\caption{Exact separation of the 4-point Green's function, $G^{4-{\rm pt}}$, in terms of non-interacting lines and a vertex function, as given in Eq.~(\ref{g4ptgamma}). The first two terms are the direct and exchange propagation of two non-interacting and fully dressed particles. The last term defines the 4-point vertex function, $\Gamma^{4-{\rm pt}}$, involving the sum of all 1PI diagrams.}
\label{g4point}
\end{center}
\end{figure}

For the 4-point GF, this separation is shown diagrammatically in Fig.~\ref{g4point}. The first two terms involve two dressed fermion lines propagating independently, and their exchange as required by the Pauli principle. The remaining part, stripped off its external legs, can contain only 1PI diagrams which are collected in a vertex function, $\Gamma^{4-{\rm pt}}$. This is associated with interactions and, at lowest level, it would correspond to a 2BF. As we will see in the following, however, 3B interactions also provide contributions to $\Gamma^{4-{\rm pt}}$. The 4-point vertex function is defined by the following equation:
\beqn
\nn &&
G^{4-{\rm pt}}_{\al\ga,\be\de}(\om_\al,\om_\ga;\om_\be,\om_\de) =
\\\nn && \,\,
 i \hbar \big[2\pi \de(\om_\al-\om_\be)G_{\al\be}(\om_\al)G_{\ga\de}(\om_\ga)
- 2\pi \de(\om_\ga-\om_\be)G_{\al\de}(\om_\al)G_{\ga\be}(\om_\ga)\big]
\\\nn && \,\, 
+(i\hbar)^2\sum_{\substack{\ta\mu\\\nu\lm}}G_{\al\ta}(\om_\al)G_{\ga\mu}(\om_\ga)
\Gamma^{4-{\rm pt}}_{\ta\mu,\nu\lm}(\om_\al,\om_\ga;\om_\be,\om_\de)
G_{\nu\be}(\om_\be)G_{\lm\de}(\om_\de)\,.\,\,\,\,
\\
\label{g4ptgamma}
\enqn 
Eq.~(\ref{g4ptgamma}) is \emph{exact} and is an implicit definition of $\Gamma^{4-{\rm pt}}$. Different many-body approximations arise when approximations are performed on this vertex function \citep{Dic2004,Dic2008}.

\begin{figure}
\begin{center}
\includegraphics[width=0.65\textwidth]{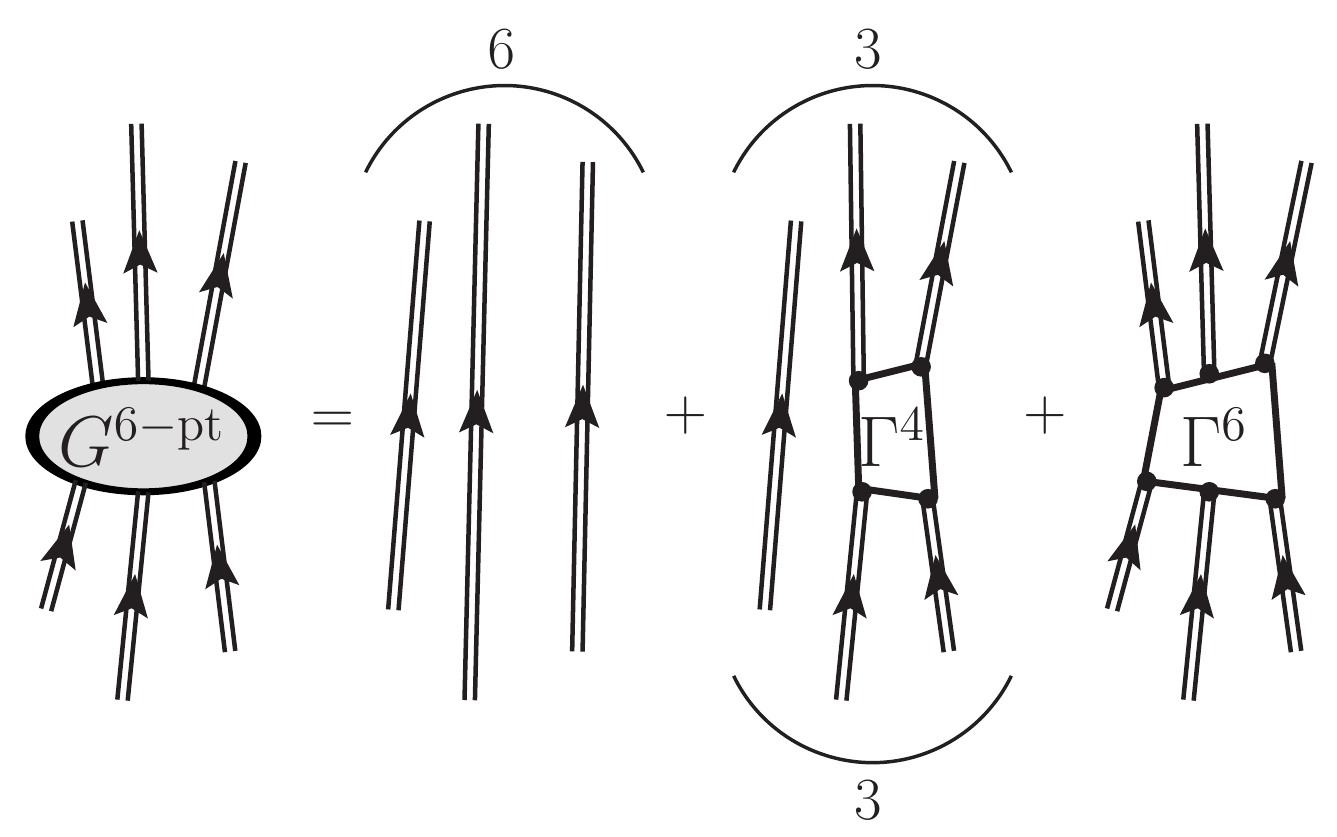}
\caption{Exact separation of the 6-point Green's function, $G^{6-{\rm pt}}$, in terms of non-interacting dressed fermion lines and vertex functions, as given in Eq.~(\ref{g6ptgamma}). The first two terms gather non-interacting dressed lines and subgroups of interacting particles that are fully connected to each other. Round brackets with numbers above (below) these diagrams indicate the numbers of permutations of outgoing (incoming) legs needed to generate all possible diagrams. The last term defines the 6-point 1PI vertex function~$\Gamma^{6-{\rm pt}}$.
}
\label{g6point}
\end{center}
\end{figure}

A similar expression holds for the 6-point GF. In this case, the diagrams that involve non interacting lines can contain either all 3 dressed propagators moving independently from each other or groups of two lines interacting through a 4-point vertex function. The remaining terms are collected in a 6-point vertex function, $\Gamma^{6-{\rm pt}}$, which contains terms where all 3 lines are interacting. This separation is demonstrated diagrammatically in Fig.~\ref{g6point}. The Pauli principle requires a complete antisymmetrization of these diagrams. For the ``free propagating" term, this implies all $3!=6$ permutations of the 3 lines. The second term, involving $\Gamma^{4-{\rm pt}}$, requires $3^2=9$ cyclic permutations within both incoming and outgoing legs. The 6-point vertex function is already antisymmetrized and hence no permutations are needed. The  equation corresponding to Fig.~\ref{g6point} is exact and provides an implicit definition of the $\Gamma^{6-{\rm pt}}$ vertex function:
\beqn
\label{g6ptgamma}
\nn &&
G^{6-{\rm pt}}_{\al\ga\ep,\be\de\eta}(\om_\al,\om_\ga,\om_\ep;\om_\be,\om_\de,\om_\eta) =
(2\pi)^2 (i\hbar)^2\,{\cal A}_{[\{\al\om_\al\},\{\ga\om_\ga\},\{\ep\om_\ep\}]} 
\\\nn &&\qquad\qquad\qquad
\times\big[\de(\om_\al-\om_\be)\,\de(\om_\ga-\om_\de)\, G_{\al\be}(\om_\al)
G_{\ga\de}(\om_\ga)G_{\ep\eta}(\om_\ep)
 \big]
\\\nn && 
+ 2\pi(i\hbar)^3\, {\cal P}^{\rm cycl.}_{[\{\al\om_\al\},\{\ga\om_\ga\},\{\ep\om_\ep\}]}
                           {\cal P}^{\rm cycl.}_{[\{\be\om_\be\},\{\de\om_\de\},\{\eta\om_\eta\}]}
                           \Big[
                           \de(\om_\al-\om_\be) G_{\al\be}(\om_\al) 
\\\nn && 
\qquad  \times
\sum_{\substack{\ta\mu\\\nu\lm}}G_{\ga\ta}(\om_\ga)
G_{\ep\mu}(\om_\ep)\Gamma^{4-{\rm pt}}_{\ta\mu,\nu\lm}(\om_\ga,\om_\ep;\om_\de,\om_\eta) 
G_{\nu\de}(\om_\de)G_{\lm\eta}(\om_\eta) \Big]
\\\nn && 
+ (i\hbar)^4 
\sum_{\substack{\ta\mu\chi\\\nu\lm\xi}}G_{\al\ta}(\om_\al)G_{\ga\mu}(\om_\ga)
G_{\ep\chi}(\om_\ep)
\Gamma^{6-{\rm pt}}_{\ta\mu\chi,\nu\lm\xi}(\om_\al,\om_\ga,\om_\ep;\om_\be,\om_\de, \om_\eta)
\\ && \qquad\qquad\qquad
 \times G_{\nu\be}(\om_\be)G_{\lm\de}(\om_\de)G_{\xi\eta}(\om_\eta) \, .
\enqn 
Here, we have introduced the antisymmetrization operator, ${\cal A}$, which sums all possible permutations of pairs of indices and frequencies, $\{\al\om_\al\}$, with their corresponding sign. Likewise, ${\cal P^{\rm cycl.}}$ sums all possible cyclic permutations of the index-frequency pairs. Again, let us stress that both $\Gamma^{4-{\rm pt}}$ and $\Gamma^{6-{\rm pt}}$ are formed of 1PI diagrams only, since they are defined by removing all external {\em dressed} legs from the $G^{4-{\rm pt}}$ and $G^{6-{\rm pt}}$ propagators. However, they can be still two-particle reducible, since they include diagrams that can be split by cutting two lines. In general, $\Gamma^{4-{\rm pt}}$ and $\Gamma^{6-{\rm pt}}$ are solution of  all-order summations analogous to the Bethe-Salpeter equation, in which the kernels are 2PI and 3PI vertices (see Eqs.~(\ref{GmLadd}) to
(\ref{GmParquet}) below).

\begin{figure}[t]
\begin{center}
\includegraphics[width=0.75\textwidth]{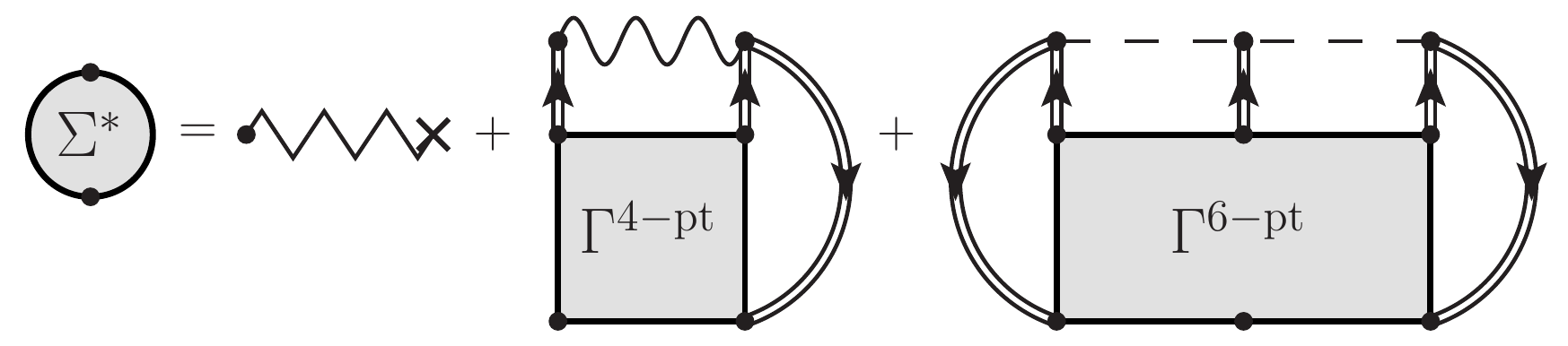}
\caption{Diagrammatic representation of the irreducible self-energy $\Sigma^\star$ by means of effective 1B and 2B potentials and 1PI vertex functions, as given in Eq.~(\ref{irrself}).  The first term is the energy independent part of $\Sigma^\star$ and contains all diagrams depicted in Fig.~\ref{ueffective}. The second and third terms are dynamical terms consisting of excited configurations generated through 2B and 3BFs. This is an exact equation for Hamiltonians including 3BFs and it is not derived from perturbation theory. }
\label{self_en}
\end{center}
\end{figure}

Inserting Eqs.~(\ref{g4ptgamma}) and~(\ref{g6ptgamma}) into Eq.~(\ref{gexact}), and exploiting the effective 1B and 2B operators  defined in Eqs.~(\ref{ueff}) and (\ref{veff}), one recovers the Dyson equation, Eq.~(\ref{Dyson}). One can therefore identify the exact expression of the irreducible self-energy $\Sigma^\star$ in terms of 1PI vertex functions:
\beqn
\nn
 \Sigma^\star_{\ga\de}(\om)&=& 
\widetilde U_{\ga\de}  
- \frac{(i\hbar)^2}{2}\sum_{\substack{\mu\\\nu\lm}}\sum_{\substack{\xi\ta\\\ep}}
\widetilde V_{\ga\mu,\nu\lm}\int\frac{\d\om_1}{2\pi}\int\frac{\d\om_2}{2\pi}
G_{\nu\xi}(\om_1)G_{\lm\ta}(\om_2)
\\\nn && 
\times
\Gamma^{4-{\rm pt}}_{\xi\ta,\de\ep}(\om_1,\om_2;\om,\om_1+\om_2-\om)
G_{\ep\mu}(\om_1+\om_2-\om)
\\\nn &&
+\frac{(i\hbar)^4}{12}\sum_{\substack{\mu\phi\\\lm\nu\chi}}\sum_{\substack{\ta\xi\eta\\\ep\sig}}
W_{\mu\ga\phi,\lm\nu\chi}
\int\frac{\d\om_1}{2\pi}\int\frac{\d\om_2}{2\pi}\int\frac{\d\om_3}{2\pi}\int\frac{\d\om_4}{2\pi}
\\\nn && \times
G_{\lm\ta}(\om_1)G_{\nu\xi}(\om_2)G_{\chi\eta}(\om_3) 
\\\nn && 
\times\Gamma^{6-{\rm pt}}_{\ta\xi\eta,\ep\de\sig}
(\om_1,\om_2,\om_3;\om_4,\om,\om_1+\om_2+\om_3-\om_4-\om)
\\ && \times
G_{\ep\mu}(\om_4)G_{\sig\phi}(\om_1+\om_2+\om_3-\om_4-\om)\,.
\label{irrself}
\enqn
The diagrammatic representation of Eq.~(\ref{irrself}) is shown in Fig.~\ref{self_en}. We note that, as an irreducible self-energy, this should include all the connected, 1PI diagrams. These can be regrouped in terms of \emph{skeleton} and \emph{interaction-irreducible} contributions, as long as $\Gamma^{4-{\rm pt}}$ and $\Gamma^{6-{\rm pt}}$ are expressed that way. Note that effective interactions are used here. The interaction-reducible components of $\widetilde U$, $\widetilde{V}$ and $\h W$ are actually generated by contributions involving partially non-interacting propagators contributions inside $G^{4-{\rm pt}}$ and $G^{6-{\rm pt}}$. The first two terms in both Eqs.~(\ref{g4ptgamma}) and~(\ref{g6ptgamma}) only contribute to generate effective interactions. Note, however, that the 2B effective interaction does receive contributions from both $\Gamma^{4-{\rm pt}}$ and $\Gamma^{6-{\rm pt}}$ in the self-consistent procedure, as will be clear in the following.

The first term entering  Eq.~(\ref{irrself}) is the energy-independent contribution to the irreducible self-energy, already found in Eq.~(\ref{eq:1ord}). This includes the subtraction of the auxiliary field, $\hat{U}$, as well as the 1B interaction-irreducible contributions due to the 2B and 3BFs. Once again, we note that the definition of this term, shown in Fig.~\ref{ueffective}, involves fully correlated density matrices. Consequently, even though this is a static contribution, it goes beyond the Hartree-Fock approximation. The dispersive part of the self-energy is described by the second and third terms on the right side of Eq.~(\ref{irrself}). These account for all higher-order contributions and incorporate correlations on a 2B and 3B level associated with the vertex functions $\Gamma^{4-{\rm pt}}$ and $\Gamma^{6-{\rm pt}}$, respectively. 

\subsection{Interaction vertices: the $\Gamma^{4-\textrm{pt}}$}

\begin{figure}[t]
\begin{center}
\includegraphics[width=0.6\textwidth]{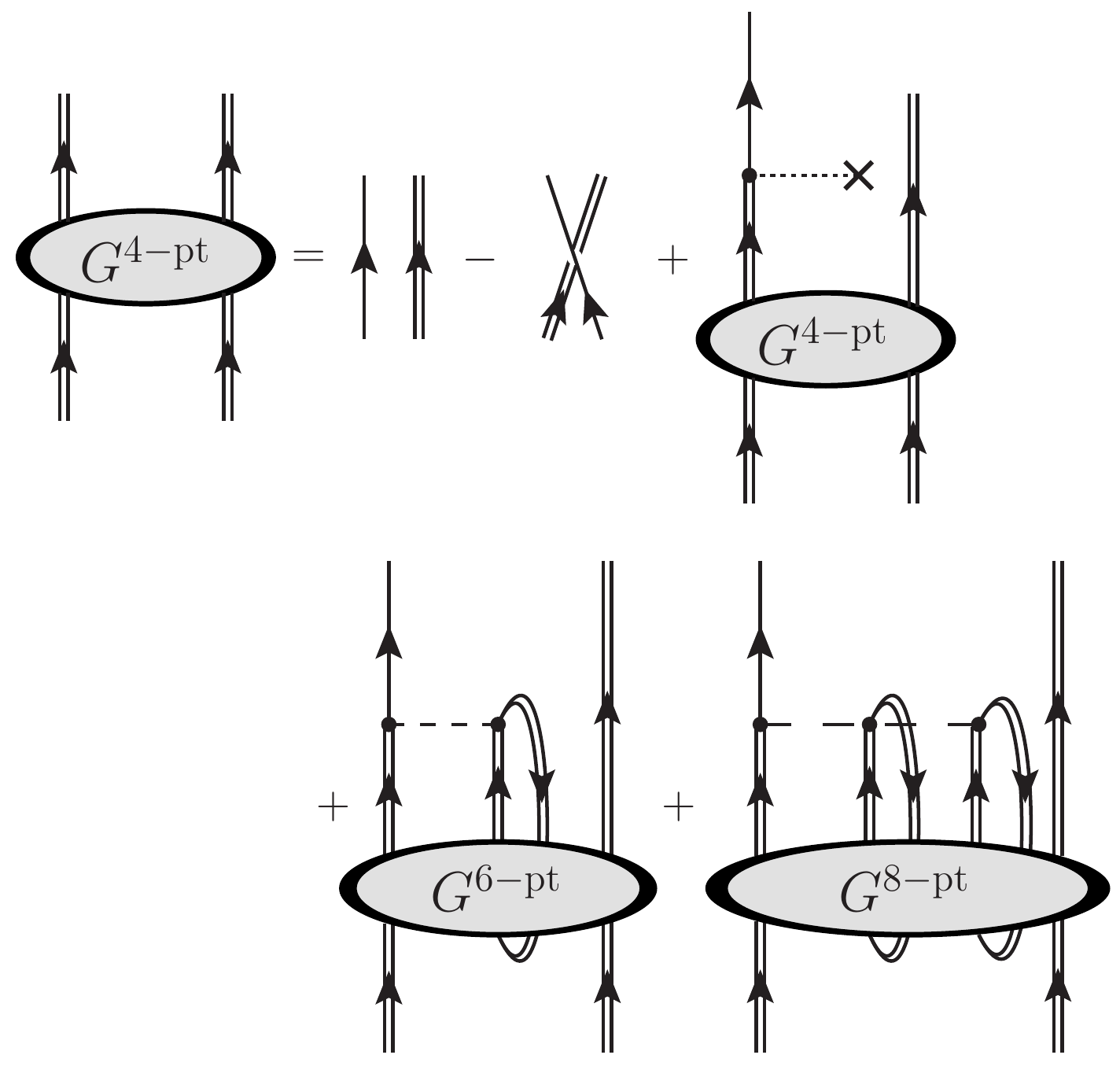}
\caption{Diagrammatic representation of the EOM for the 4-point propagator, $G^{4-{\rm pt}}$, given in Eq.~(\ref{gIIexact}). The last term, involving an 8-point GF, arises due to the presence of 3B interactions. }
\label{eq_g2}
\end{center}
\end{figure}

We now apply the EOM method to the 4-point GF. This will provide insight into approximation schemes that involve correlations at or beyond the 2B-level. Let us stress that our final aim is to obtain generic nonperturbative approximation schemes in the many-body sector. Taking the time derivative of the first argument in Eq.~(\ref{g4pt}) and following the same procedure as for the 2-point GF, we find:
\beqn
\label{gIIexact}
\nn &&
G^{4-{\rm pt}}_{\al\ga,\be\de}(\om_\al,\om_\ga;\om_\be,\om_\de) = 
\\\nn &&
i\hbar\,[ 2 \pi \de(\om_\al-\om_\be) G^{(0)}_{\al\be}(\om_\al) G_{\ga\de}(\om_\ga)-
2 \pi \de(\om_\ga-\om_\be) G^{(0)}_{\al\de}(\om_\al)G_{\ga\be}(\om_\ga)]
\\\nn && 
+ \sum_{\mu\lm}G^{(0)}_{\al\mu}(\om_\al) U_{\mu\lm} G^{4-{\rm pt}}_{\lm\ga,\be\de}(\om_\al,\om_\ga;\om_\be,\om_\de)
\\\nn &&
-\frac{1}{2} \sum_{\substack{\mu\ep\\\lm\ta}}G^{(0)}_{\al\mu}(\om_\al)
 V_{\mu\ep,\lm\ta}
\int\frac{d \om_1}{2 \pi}\int\frac{d \om_2}{2 \pi}
\\\nn && \quad
\times G^{6-{\rm pt}}_{\lm\ta\ga,\be\ep\de}(\om_1,\om_2,\om_\ga;\om_\be,\om_1+\om_2-\om_\al,\om_\de) 
\\\nn&&
+\frac{1}{12}\sum_{\substack{\mu\ep\chi\\\lm\ta\eta}}G^{(0)}_{\al\mu}(\om_\al)
W_{\mu\ep\chi,\lm\ta\eta}
\int\frac{d \om_1}{2 \pi}\int\frac{d \om_2}{2 \pi}\int\frac{d \om_3}{2 \pi}\int\frac{d \om_4}{2 \pi}
\\ && \quad
\times G^{8-pt}_{\lm\ta\eta\ga,\be\ep\chi\de}(\om_1,\om_2,\om_3,\om_\ga;\om_\be,\om_4,\om_1+\om_2+\om_3-\om_\al-\om_4,\om_\de), \quad\,\,\,\,\,\,\,\,\,\,
\enqn
which is the analogous, in the case of the 2-body GF, of Eq.~(\ref{gexact}) for the SP propagator. As expected, the EOM connects the 2-body (4-point) GF to other propagators. The 1B propagator term just provides the non-interacting dynamics, with the proper antisymmetrization. The interactions bring in admixtures with the 4-point GF itself, via the one-body potential, but also with the 6- and 8-point GFs, via the the 2B and the 3B interactions, respectively. Similarly to what we observed in Eq.~(\ref{gexact}), the dynamics involve $m-4$ frequency integrals for the $m$-point GFs. The diagrammatic representation of this equation is given in Fig. \ref{eq_g2}.

\begin{figure}[t!]
\begin{center}
\includegraphics[width=0.6\textwidth]{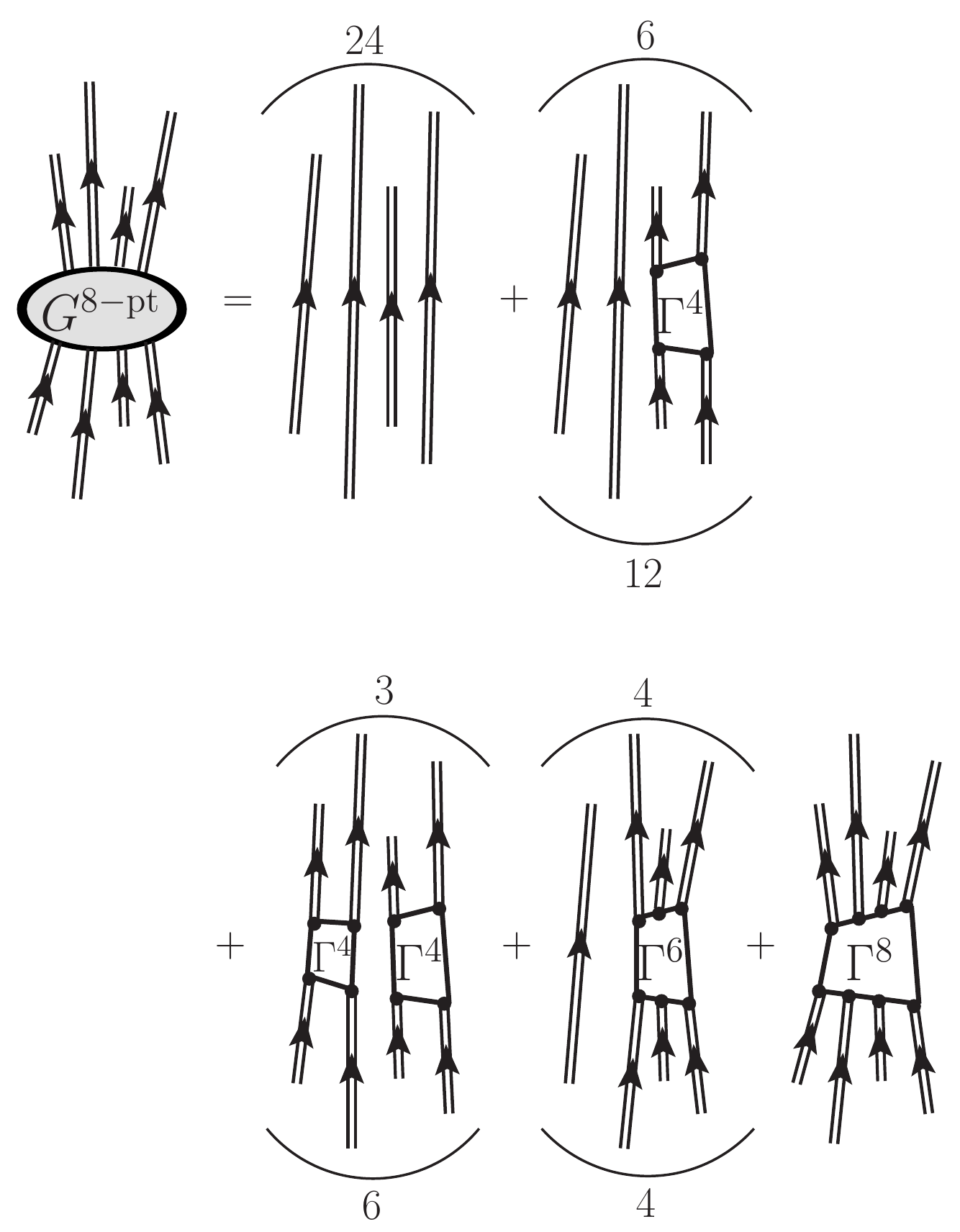}
\caption{Exact separation of the 8-point Green's function, $G^{8-{\rm pt}}$, in terms of non-interacting lines and vertex functions. The first four terms gather non-interacting dressed lines and subgroups of interacting particles that are fully connected to each other. Round brackets with numbers above (below) these diagrams indicate the numbers of permutations of outgoing (incoming) legs needed to generate all possible diagrams. The last term defines the 8-point 1PI vertex function~$\Gamma^{8-{\rm pt}}$.
\label{g8point} }
\end{center}
\end{figure}

To proceed further, as done for the analysis of the 1B GF and following the steps of Ref.~\citep{Mat1971}, we split the 8-point GF into free dressed propagators and 1PI vertex functions. This decomposition is shown in Fig.~\ref{g8point}. In addition to the already-defined vertex functions, one needs 1PI objects with 4 incoming and outgoing indices. To this end, we introduce the 8-point vertex function $\Gamma^{8-{\rm pt}}$ in the last term. Note that due care has to be taken in enumerating all antisymmetrization possibilities when groups of fermion lines that are not connected by $\Gamma^{8-{\rm pt}}$ are considered. The first term in Fig.~\ref{g8point}, for instance, involves 4 non-interacting but dressed fermion lines, and there are $4!=24$ possible combinations. There are $\binom{4}{2}\binom{4}{2}\f 1 2=72$ equivalent terms involving two non-interacting lines and a single $\Gamma^{4-{\rm pt}}$, as in the second term of Fig.~\ref{g8point}. The double $\Gamma^{4-{\rm pt}}$ contribution (third term) can be obtained in $6 \times 3=18$ equivalent ways. For the term with an independent line and three interacting lines through a $\Gamma^{6-{\rm pt}}$ vertex function there are $4\times4=16$ equivalent terms. 

\begin{figure}[t]
\begin{center}
\includegraphics[width=1.\textwidth]{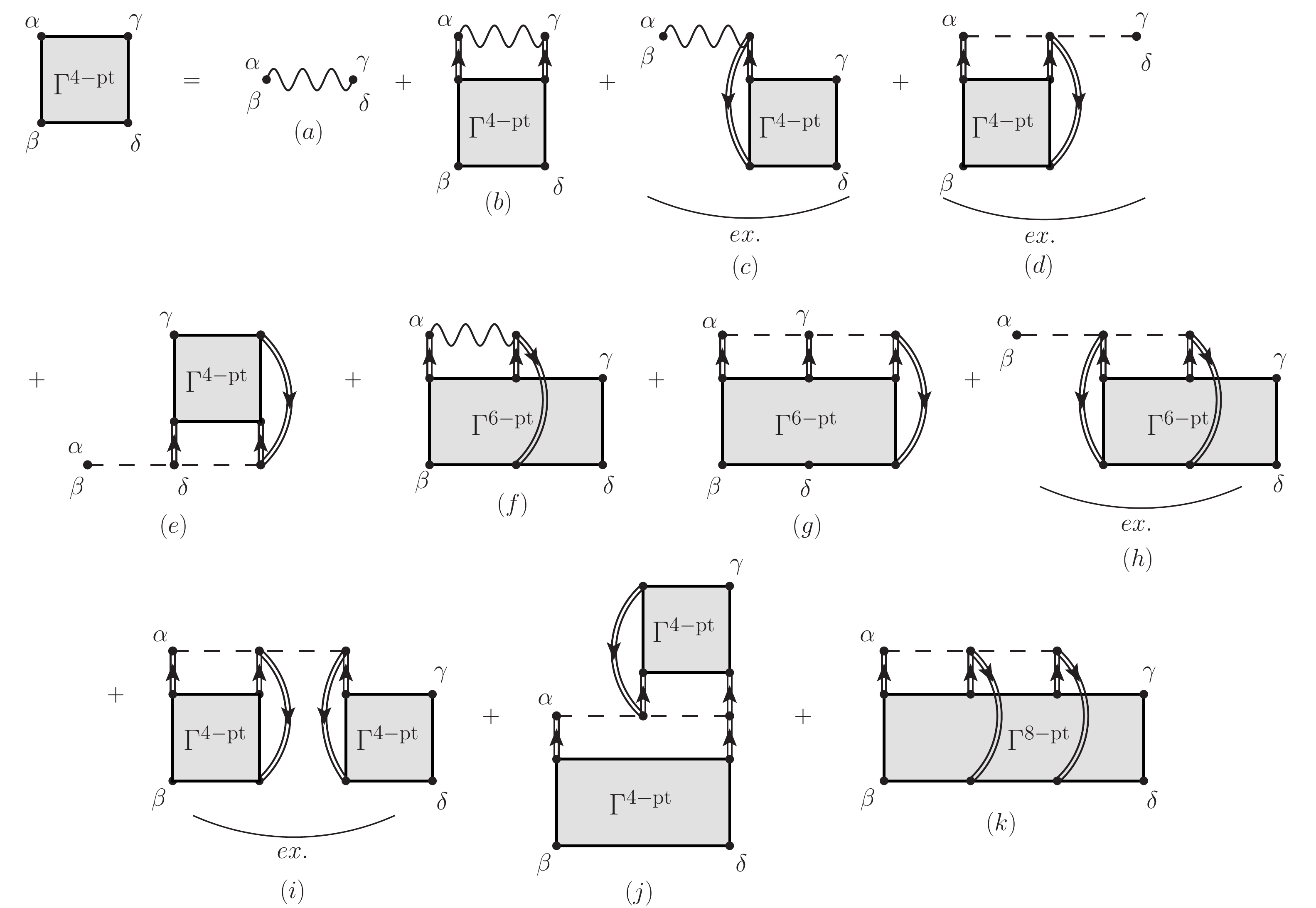}
\caption{Self-consistent expression for the $\Gamma^{4-{\rm pt}}$ vertex function derived from the EOM for $G^{4-{\rm pt}}$. The round brackets underneath some of the diagrams indicate that the term obtained by exchanging the $\{\be\om_\be\}$ and $\{\de\om_\de\}$ arguments must also be included. Diagrams (a), (b), (c) and (f) are the only ones present for 2B Hamiltonians, although (f) also contains some intrinsic 3BF contributions such as the $\{\al\om_\al\} \leftrightarrow \{\ga\om_\ga\}$ exchange of (e). All other diagrams arise from the inclusion of 3B interactions. Diagram (b) is responsible for generating the ladder summation,  the direct part of (c) generates the series of antisymmetrized rings, and the three sets together [(b), (c) and the exchange of (c)] would give rise  to a Parquet-type resummation.}
\label{gamma4pt}
\end{center}
\end{figure}

With this decomposition at hand, one can now proceed and find an equation for the 4-point vertex function, $\Gamma^{4-{\rm pt}}$. Inserting the exact decompositions of the 4-, 6- and 8-point GFs, given respectively by Figs.~\ref{g4point}, \ref{g6point} and~\ref{g8point}, into the EOM of the 2B propagator, Eq.~(\ref{gIIexact}), one obtains an equation with $\Gamma^{4-{\rm pt}}$ on both sides. The diagrammatic representation of this self-consistent equation is shown in Fig.~\ref{gamma4pt}.

A few comments are in order at this point. The left-hand side of Eq.~(\ref{gIIexact}) in principle contains two dressed and non interacting propagators, as shown in the first two terms of Fig.~\ref{g4point}. On the right-hand side of Eq.~(\ref{gIIexact}), however, one of the 1B propagators is not dressed. However, when expanding the GFs in terms of the $\Gamma^{2n-{\rm pt}}$ vertex functions,  the remaining contributions to the Dyson equation arise automatically (see Fig.~\ref{eq_g}). The free unperturbed line, therefore, becomes dressed. As a consequence, the pair of dressed non-interacting propagators cancel out exactly on both sides of Eq.~(\ref{gIIexact}). This dressing procedure of the $\gz$ propagator happens only partially in the last three terms of Eq.~(\ref{gIIexact}) and has been disregarded in our derivation. In this sense, Fig.~\ref{gamma4pt} should be taken as an approximation to the exact EOM  for $G^{4-{\rm pt}}$.

Eq.~(\ref{gIIexact})  links  1B, 2B, 3B and 4B propagators. Correspondingly, Fig.~\ref{gamma4pt} involves higher-order vertex functions, such as $\Gamma^{6-{\rm pt}}$ and $\Gamma^{8-{\rm pt}}$, which are in principle coupled, through their own EOMs, to more complex GFs. The hierarchy of these equations has to be necessarily truncated. In Ref.~\citep{Mat1971}, truncation schemes were explored by neglecting the $\Gamma^{6-{\rm pt}}$ vertex function at the level of Fig.~\ref{gamma4pt} ($\Gamma^{8-{\rm pt}}$ did not appear in the 2BF-only case). This level of truncation is already sufficient to retain physically-relevant subsets of diagrams, such as ladders and rings. Let us note, in particular, that the summation of these infinite series leads to nonperturbative many-body schemes. For completeness, we show in Fig.~\ref{gamma4pt} all contributions coming also from the $\Gamma^{6-{\rm pt}}$ and $\Gamma^{8-{\rm pt}}$ vertices, many of them arising from 3BFs. 
%\begin{landscape}
%\end{landscape}

We have ordered the diagrams in Fig.~\ref{gamma4pt} in terms of increasing contributions from 3BFs, and in the order of expansion in perturbation theory at which they start contributing to $\Gamma^{4-{\rm pt}}$. Intuitively, we expect that this should arrange them in decreasing importance. Diagrams Figs.~\ref{gamma4pt}(a), \ref{gamma4pt}(b), \ref{gamma4pt}(c) and \ref{gamma4pt}(f) are those that are also present in the 2BF-only case. Diagram Fig.~\ref{gamma4pt}(f), however, is of a mixed nature: it can contribute only at third order with effective 2BFs, but does contain interaction-irreducible 3BF contributions at second order that are similar to diagrams Figs.~\ref{gamma4pt}(d) and \ref{gamma4pt}(e). Diagrams Figs.~\ref{gamma4pt}(d)-(h) all contribute to $\Gamma^{4-{\rm pt}}$ at second order, although the first three require a combination of a $\widetilde{V}$ and a $\h W$ term. The remaining diagrams in this group, Figs.~\ref{gamma4pt}(g) and \ref{gamma4pt}(h), require two 3B interactions at second order and are expected to be subleading. Note that diagram Fig.~\ref{gamma4pt}(d) is antisymmetric in $\al$ and $\ga$, but it must also be antisymmetrized with respect to $\be$ and $\de$. Its conjugate contribution, diagram Fig.~\ref{gamma4pt}(e), should not be further antisymmetrized in $\al$ and $\ga$, because such exchange term is already included in Fig.~\ref{gamma4pt}(f). All the remaining terms, diagrams Figs.~\ref{gamma4pt}(i)-(k), only contribute from the third order on.

In this section we saw how the hierarchy of EOMs links different $N$-body GFs to each other, increasing the complexity of the solution together with the growing of the $N$-body interacting terms which appear in the description of the system. We demonstrated how solution of the EOM for the 2-point GF yields the formulation of the irreducible self-energy, and how following the same pattern for the 4-point GF provides an expression for the $\Gamma^{4-{\rm pt}}$ vertex function.

\subsection{Interaction vertices truncations}

\begin{figure}[t!]
\begin{center}
\includegraphics[width=.75\textwidth]{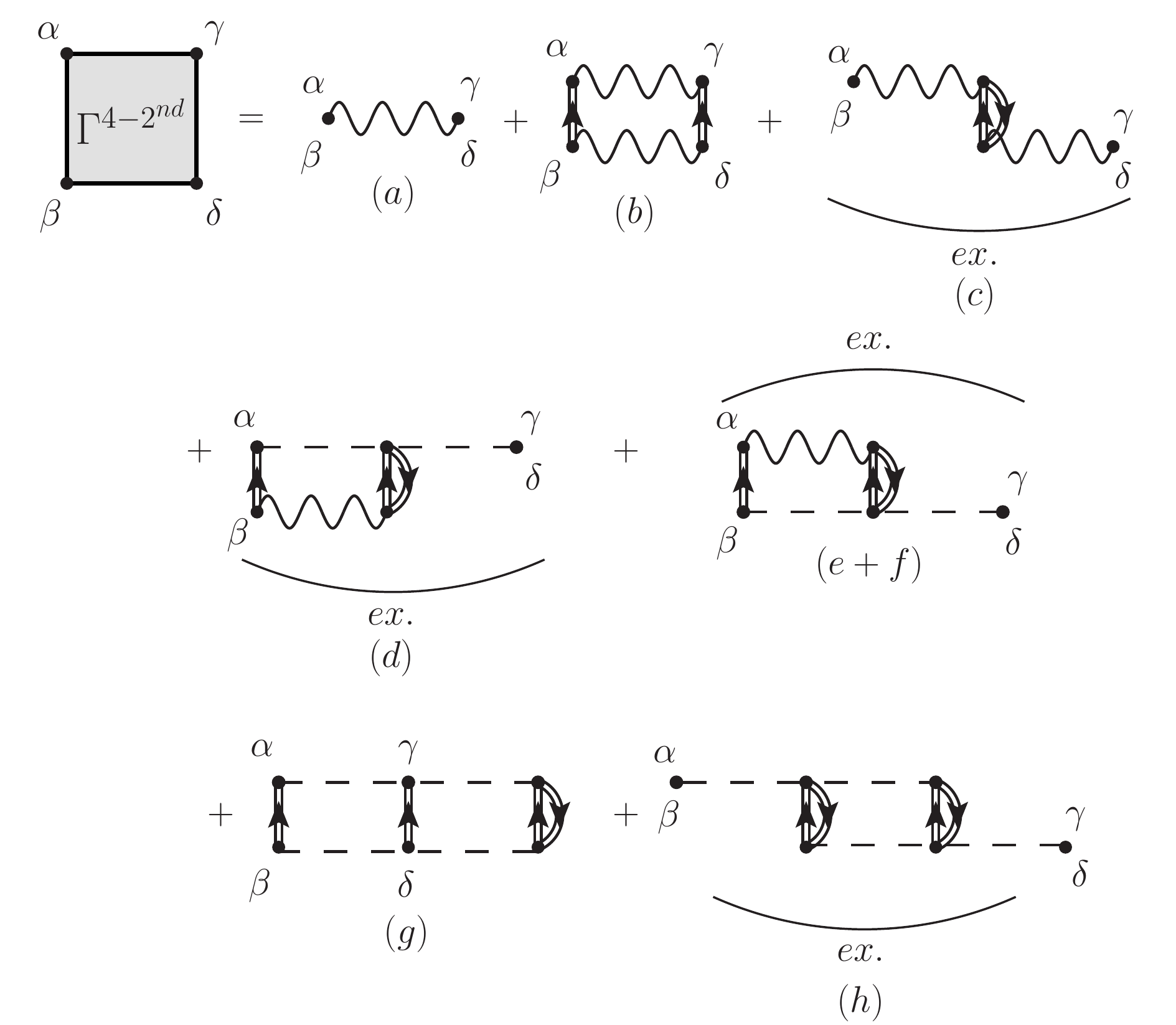}
\caption{Skeleton and interaction-irreducible diagrams contributing to the $\Gamma^{4-{\rm pt}}$ vertex function up to second order. The round brackets above (below) some diagrams indicate that the exchange diagram between the $\{\al\om_\al\}$ and $\{\ga\om_\ga\}$ \hbox{($\{\be\om_\be\}$ and $\{\de\om_\de\}$)} arguments must also be included. }
\label{g4pt_2nd}
\end{center}
\end{figure}

We now want to demonstrate the correspondence between the techniques derived in Sec.~\ref{section:self_en_pert} and the EOM method described in Sec.~\ref{section:eom_method}. In particular, we want to show how  the self-energy obtained with the EOM expression, Eq.~(\ref{irrself}), leads to the perturbative expansion of Eq.~(\ref{gpert}). We will do this by expanding the self-energy up to third order and showing the equivalence of both approaches at this order. To this end, we need to expand  the vertex functions in terms of the effective Hamiltonian, $\widetilde{H}_1$. The lowest-order terms entering $\Gamma^{4-{\rm pt}}$ can be easily read from Fig.~\ref{gamma4pt}. We show these second-order, skeleton and interaction-irreducible diagrams in Fig.~\ref{g4pt_2nd}. Only the first three terms would contribute in a 2BF calculation. There are two terms involving mixed 2BFs and 3BFs, whereas the final two contributions come from two independent 3BFs. Note that, to get the third-order expressions of the self-energy, we expand the vertex functions to second order, i.e. one order less. 

\begin{figure}[t!]
\begin{center}
\includegraphics[width=.75\textwidth]{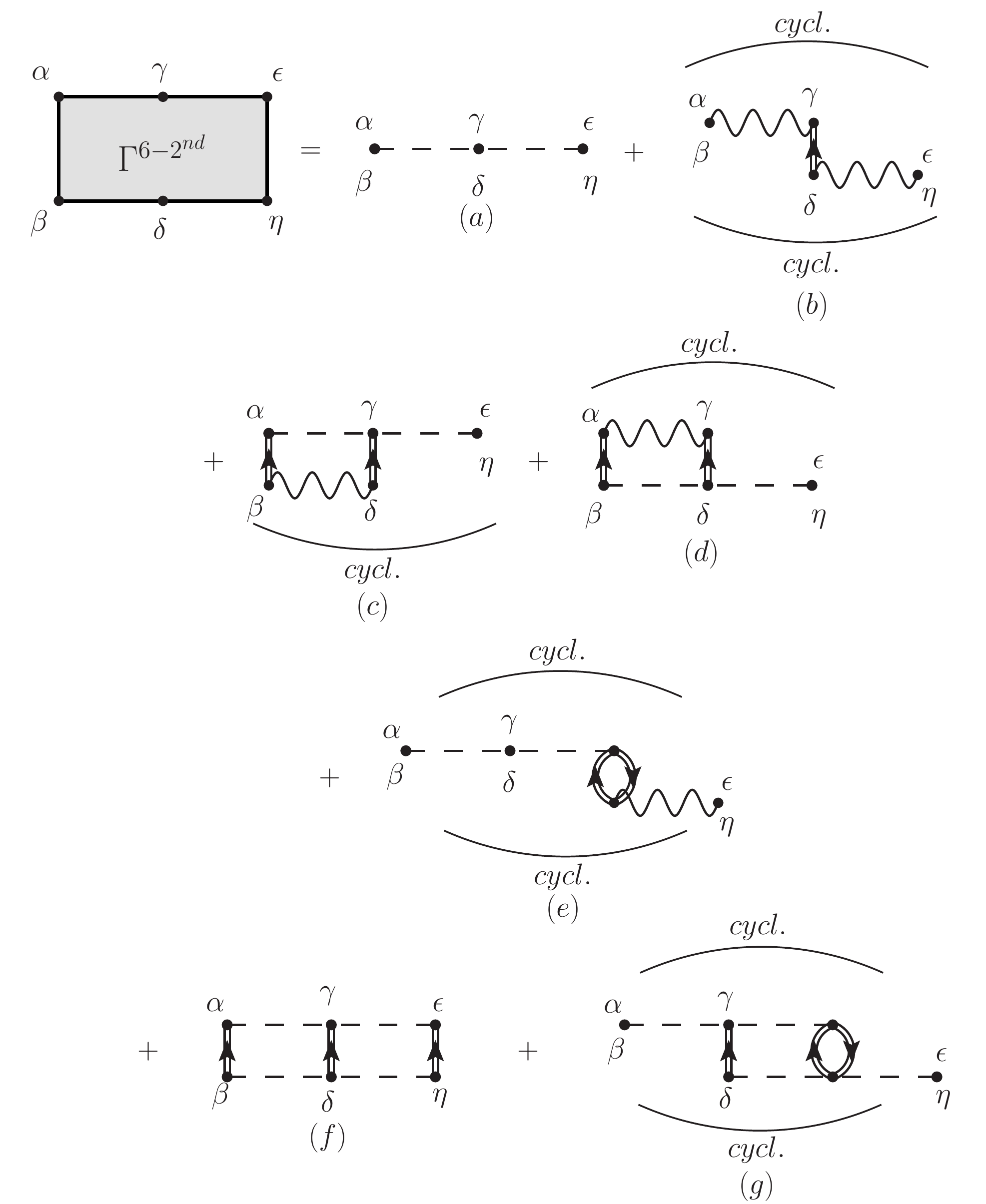}
\caption{ The same as Fig.~\ref{g4pt_2nd} for the  $\Gamma^{6-{\rm pt}}$ vertex function. The round brackets above (below) some diagrams indicate that cyclic permutations of the $\{\al\om_\al\}$, $\{\ga\om_\ga\}$ and $\{\ep\om_\ep\}$ ($\{\be\om_\be\}$, $\{\de\om_\de\}$ and $\{\eta\om_\eta\}$) arguments must also be included. }
\label{g6pt_2nd}
\end{center}
\end{figure}

Analogously, we display the expansion up to second order of $\Gamma^{6-{\rm pt}}$ in Fig.~\ref{g6pt_2nd}. Most contributions to this vertex function contain 3BFs. The lowest-order term, for instance, is given by the 3B interaction itself. Note, however, that second-order terms formed only by 2B effective interactions are possible, such as the second term on the right-hand side of Fig.~\ref{g6pt_2nd}. These will eventually be connected with a 3BF to give a mixed self-energy contribution (see Eq.~(\ref{irrself}) and Fig.~\ref{self_en}). 

If one includes the diagrams in Figs.~\ref{g4pt_2nd} and \ref{g6pt_2nd} into the irreducible self-energy $\Sigma^\star$ of Fig.~\ref{self_en}, all the diagrams discussed in Eq.~\ref{eq:1ord}, Fig.~\ref{2ord} and Fig.~\ref{3ord} of Sec.~\ref{section:self_en_pert} are recovered. This does prove, at least up to third order, the correspondence between the perturbative expansion approach and the EOM method for the GFs. Proceeding in this manner to higher orders, one should obtain equivalent diagrams all the way through. 

It is important to note that diagrams representing conjugate contributions to $\Sigma^\star$ are generated by different, not necessarily conjugate, terms of $\Gamma^{4-{\rm pt}}$ and~$\Gamma^{6-{\rm pt}}$. For instance, diagram Fig.~\subref*{3ord_223B_1} is the result of the term \ref{g4pt_2nd}($e+f$) and its exchange, on the right-hand side of Fig.~\ref{g4pt_2nd}. Its conjugate self-energy diagram Fig.~\subref*{3ord_322B_1}, however, is generated by the second contribution to $\Gamma^{6-{\rm pt}}$, Fig.~\ref{g6pt_2nd}(b). This term is also related to diagram Fig.~\subref*{3ord_322B_2}. More specifically, the term in Fig.~\ref{g6pt_2nd}(b) has 9 cyclic permutations of its indices, of which 6 contribute to diagram Fig.~\subref*{3ord_322B_1} and 3 to diagram Fig.~\subref*{3ord_322B_2}. On the other hand, the conjugate of diagram Fig.\subref*{3ord_322B_2} is diagram Fig.~\subref*{3ord_223B_2}, which is entirely due to the exchange contribution of the Fig.~\ref{g4pt_2nd}(d) term in $\Gamma^{4-{\rm pt}}$. The direct contribution of this same term leads to diagram Fig.~\subref*{3ord_232B}, which is already self-conjugate. 

More importantly, nonperturbative self-energy expansions can be obtained by means of other hierarchy truncations at the level of $\Gamma^{4-{\rm pt}}$ and $\Gamma^{6-{\rm pt}}$. Translating these into self-energy expansions is then just an issue of introducing them in Eq.~(\ref{irrself}). According to the approximation  chosen for the vertex functions appearing in Fig.~\ref{self_en}, we will be summing specific sets of diagrams when solving Dyson's equation, Eq.~(\ref{Dyson}). However, from the above discussion it should be clear that extra care must be taken to guarantee that the truncations lead to physically coherent results. In particular, it is not always possible to naively neglect~$\Gamma^{6-{\rm pt}}$. The last two terms of the self-energy equation, Eq.~(\ref{irrself}), generate conjugate contributions. Hence, neglecting one term or the other will spoil the analytic properties of the self-energy which require an Hermitian real part and an anti-Hermitian imaginary part. In the examples discussed above, diagrams Fig.~\subref*{3ord_322B_1} and Fig.~\subref*{3ord_322B_2} would be missing if $\Gamma^{6-{\rm pt}}$ had not been considered. 

When no irreducible 3B interaction terms are present in the hierarchy truncation, only the $\Gamma^{4-{\rm pt}}$ term contributes to Eq.~(\ref{irrself}). The ladder and the ring truncations, (see further down in Eqs.~(\ref{GmLadd}) and (\ref{GmRing})), generate their own conjugate diagrams and can be used on their own to obtain physical approximations to the self-energy. However, this need not be true in general. A counterexample is actually provided by the truncation of Eq.~(\ref{DeltaGm}) (see further down in Sec.~\ref{section:t_matrix}) which, if inserted in Eq.~(\ref{irrself}) without the corresponding contributions to $\Gamma^{6-{\rm pt}}$, cannot generate a correct self-energy.   Because of its diagrammatic content, Eq.~(\ref{DeltaGm}) can only be used as a correction to~$\Gamma^{4-{\rm pt}}$.

%%%%%%%%%%%%%%% The T-matrix %%%%%%%%%%%%%%%%%%%

\section{Ladder and other approximations}
\label{section:t_matrix}

Let us now introduce truncations on the $\Gamma^{4-{\rm pt}}$ derived in the previous section, which will lead us to define specific approximations in the many-body calculation. The simplest truncation schemes to $\Gamma^{4-{\rm pt}}$ come from considering the first three terms of Fig.~\ref{gamma4pt}, which involve effective 2BFs only. In the pure 2B case, these have already been discussed in the literature \citep{Mat1971}. Retaining diagrams Figs.~\ref{gamma4pt}(a) and \ref{gamma4pt}(b) leads to the ladder resummation to include effectively 3BFs, used in recent studies of infinite nucleonic matter \citep{Som2008,Car2013Oct}: 
\beqn
\nn
&&\Gamma^{4_\text{ladd}}_{\al\ga,\be\de}(\om_\al,\om_\ga;\om_\be,\om_\al+\om_\ga-\om_\be)=
\\\nn && \qquad
\widetilde V_{\al\ga,\be\de} 
+ \frac{\ii\hbar}{2}\int\frac{\d\om_1}{2\pi}\sum_{\ep\mu\ta\lm}
\widetilde V_{\al\ga,\ep\mu}G_{\ep\ta}(\om_1)G_{\mu\lm}(\om_\al+\om_\ga-\om_1)
\\ &&\qquad\qquad
\times\Gamma^{4_\text{ladd}}_{\ta\lm,\be\de}(\om_1,\om_\al+\om_\ga-\om_1;\om_\be,\om_\al+\om_\ga-\om_\be) \, ,
\label{GmLadd}
\enqn
where we have explicitly used the fact that $\Gamma^{2p-{\rm pt}}$, for a vertex function connecting $p$ incoming and outgoing lines, is only defined when incoming and outgoing energies are conserved. Likewise, diagrams Figs.~\ref{gamma4pt}(a) and the direct contribution of \ref{gamma4pt}(c) generate a series of ring diagram which correspond to the antisymmetrized version of the random phase approximation (RPA): 
\beqn
\nn
&&\Gamma^{4_\text{ring}}_{\al\ga,\be\de}(\om_\al,\om_\ga;\om_\be,\om_\al+\om_\ga-\om_\be)= 
\\\nn &&\qquad
\widetilde V_{\al\ga,\be\de}
- \ii\hbar\int\frac{\d\om_1}{2\pi}\sum_{\ep\mu\ta\lm} 
\widetilde V_{\al\ep,\be\mu}G_{\mu\lm}(\om_1)G_{\ta\ep}(\om_1-\om_\al+\om_\be)
\\ &&\qquad\qquad
\times\Gamma^{4_\text{ring}}_{\lm\ga,\ta\de}(\om_1,\om_\ga;\om_1-\om_\al+\om_\be,\om_\al+\om_\ga-\om_\be)
 \, .
 \label{GmRing}
\enqn
Adding up the first three contributions together, \ref{gamma4pt}(a)-(c), and including the exchange, will generate a Parquet-type of resummation, with ladders and rings embedded into each other:
\beqn
\nn&&
\Gamma^{4_\text{Parquet}}_{\al\ga,\be\de}(\om_\al,\om_\ga;\om_\be,\om_\al+\om_\ga-\om_\be)=
\\\nn && \quad 
\widetilde V_{\al\ga,\be\de} + \ii\hbar\int\frac{\d\om_1}{2\pi}\sum_{\ep\mu\ta\lm}
%\\\nn &&  
\left[ \frac{1}{2}\widetilde V_{\al\ga,\ep\mu}
G_{\ep\ta}(\om_1)G_{\mu\lm}(\om_\al+\om_\ga-\om_1)\right.
\\\nn && \qquad\qquad\left.
\times\Gamma^{4_\text{Parquet}}_{\ta\lm,\be\de}(\om_1,\om_\al+\om_\ga-\om_1;\om_\be,\om_\al+\om_\ga-\om_\be) \right.
\\ \nn && \qquad \left.
- \widetilde V_{\al\ep,\be\mu}G_{\mu\lm}(\om_1)G_{\ta\ep}(\om_1-\om_\al+\om_\be)\right.
\\ \nn && \qquad \qquad\left.
\times\Gamma^{4_\text{Parquet}}_{\lm\ga,\ta\de}(\om_1,\om_\ga;\om_1-\om_\al+\om_\be,\om_\al+\om_\ga-\om_\be)\right.
\\\nn &&\qquad\left.
+\widetilde V_{\al\ep,\de\mu}G_{\mu\lm}(\om_1)G_{\ta\ep}(\om_1+\om_\ga-\om_\be)\right.
\\ && \qquad \qquad\left.
\times\Gamma^{4_\text{Parquet}}_{\lm\ga,\ta\be}(\om_1,\om_\ga;\om_1+\om_\ga-\om_\be,\om_\be)\right]   \, .
\label{GmParquet}
\enqn

Eqs.~(\ref{GmLadd}) and~(\ref{GmRing}) can be solved in a more or less simple fashion because the corresponding vertex functions effectively depend on only one frequency ($\Omega=\om_\al+\om_\ga$ and $\Omega=\om_\al-\om_\be$, respectively). Hence these two resummation schemes have been traditionally used to study extended systems \citep{Dic2004,Ary1998,Oni2002}. The simultaneous resummation of both rings and ladders within the self-energy is possible for finite systems, and it is routinely used in both quantum chemistry and nuclear physics \citep{Sch1983,Dan2011,Bar2007,Cip2013}. The Parquet summation, as shown in Eq.~(\ref{GmParquet}), does require all three independent frequencies and it is difficult to implement numerically. Specific approximations to rewrite these in terms of two-time vertex functions have been recently attempted \citep{Ber2011}, but further developments are still required. 

The next approximation to $\Gamma^{4-{\rm pt}}$ would involve diagrams Figs.~\ref{gamma4pt}(d), \ref{gamma4pt}(e), and the exchange part included in \ref{gamma4pt}(f). All these should be added together to preserve the antisymmetry and conjugate properties of the vertex function. The resulting contributions still depend on all three frequencies and cannot be simply embedded in all-order summations such as the ladder, Eq.~(\ref{GmLadd}), or the ring, Eq.~(\ref{GmRing}), approximations. However, these diagrams could be used to obtain corrections, at first order in the interaction-irreducible $\hat W$, to the previously calculated 4-point vertices. The explicit expression for these terms is:
\beqn
\nn&&\Delta\Gamma^{4_{d+e+e'}}_{\al\ga,\be\de}
(\om_\al,\om_\ga;\om_\be,\om_\al+\om_\ga-\om_\be)=
\\\nn && \qquad
\frac{(\ii\hbar)^2}{2}\int\frac{\d\om_1}{2\pi}\int\frac{\d\om_2}{2\pi}
\sum_{\substack{\ep\mu\xi\\ \ta\lm\nu}}
\\\nn && \qquad\qquad
\times\left[- W_{\al\nu\ga,\ep\mu\de} \, G_{\ep\ta}(\om_1)G_{\mu\lm}(\om_2)
G_{\xi\nu}(\om_1+\om_2-\om_\be) \right. \, 
\\\nn && \left. \qquad\qquad\qquad
\times\Gamma_{\ta\lm,\be\xi}(\om_1,\om_2; \om_\be,\om_1+\om_2-\om_\be) \, \right.
\\\nn && \left.  \qquad\qquad
+ W_{\al\nu\ga,\ep\mu\be} \, G_{\ep\ta}(\om_1)G_{\mu\lm}(\om_2) 
G_{\xi\nu}(\om_1+\om_2-\om_\al-\om_\ga+\om_\be)\right. \,
\\\nn && \left. \qquad\qquad\qquad
\times\Gamma_{\ta\lm,\de\xi}
(\om_1,\om_2;\om_\al+\om_\ga-\om_\be, \om_1+\om_2-\om_\al-\om_\ga+\om_\be) \,\right.
\\\nn && \left. \qquad\qquad
- \Gamma_{\ga\nu,\ep\mu}(\om_\ga,\om_1+\om_2-\om_\ga;\om_1,\om_2) \right.\,
\\\nn && \left.  \qquad\qquad\qquad
\times G_{\ep\ta}(\om_1)G_{\mu\lm}(\om_2)   \,  W_{\al\ta\lm,\be\de\xi} \,
G_{\xi\nu}(\om_1+\om_2-\om_\ga)\right.
\\\nn && \left. \qquad\qquad \,
+ \Gamma_{\al\nu,\ep\mu}(\om_\al,\om_1+\om_2-\om_\al;\om_1,\om_2)\right. \,
\\ && \left. \qquad\qquad \qquad
\times G_{\ep\ta}(\om_1)G_{\mu\lm}(\om_2)   \,  W_{\ga\ta\lm,\be\de\xi} \,
G_{\xi\nu}(\om_1+\om_2-\om_\al)\right] \; .
\label{DeltaGm}
\enqn
Eq.~(\ref{DeltaGm}) has some very attractive features. First, it should provide the dominant contribution beyond those associated with the effective 2B interaction, $\widetilde{V}$. Perhaps more importantly, this contribution can be easily calculated in terms of one of the two-time vertex functions,  $\Gamma^{4_{ladd}}$ and~$\Gamma^{4_{ring}}$. This could then be inserted in Eq.~(\ref {g4ptgamma}) to generate corrections of expectation values of 2B operators stemming from purely irreducible 3B contributions. 
%A similar correction for the irreducible self-energy is also discussed further in the section.

Once a truncation scheme is chosen at the level of the vertex functions, one can immediately derive a diagrammatic approximation for the self-energy \citep{Dic2008}. In the case of the present work, where we deal with infinite nuclear matter, we want to focus on the dominant effect of short-range correlations which characterizes the repulsion at small distances between nucleons. To take this effect into account we choose to work within the ladder summation of diagrams described by Eq.~(\ref{GmLadd}). To analyze more in detail the analytic structure of the ladder vertex function, let's introduce the Lehman representation, in the specific case of an infinite system, of the 2B propagator of Eq.~(\ref{g2g4pt}). We focus on the specific case of particle-particle $(pp)$ and hole-hole $(hh)$ propagation for the $G^{4-\rm{pt}}$:
\beqn
\nn
G_{\al\ga,\be\de}^{II}(\Omega)&=&\int_{2\vep_\textrm F}^\infty\d\Omega_m^{'\,N+2}\frac
{\lPsizero a_\ga a_\al|\Psi_m^{N+2}\rangle\langle\Psi_m^{N+2}|a^\dg_\be a^\dg_\de\rPsizero}
{\Omega-\Omega_m^{'\,N+2}+i\eta}
\\\nn && -\int^{2\vep_\textrm F}_{-\infty}\d\Omega_n^{'\,N-2}\frac
{\lPsizero a^\dg_\be a^\dg_\de|\Psi_n^{N-2}\rangle\langle\Psi_n^{N-2}|a_\ga a_\al\rPsizero}
{\Omega-\Omega_n^{'\,N-2}-i\eta}\,;
\\
\label{G2B_Lehm}
\enqn
$\Omega$ defines the energy variable for a 2B state. We must point out that for the specific purpose of the ladder approximation we only need the first order approximation of the expectation values that appear in the numerators of Eq.~(\ref{G2B_Lehm}). This corresponds to considering the HF approximation for the $G^{II}$, namely the independent propagation of two correlated SP states (see  the first two terms on the right-hand side of Fig.~\ref{g4point}). Let's define relative incoming and outgoing momenta of the two particles as ${\bf k}=({\bf p}_1-{\bf p}_2)/2$ and ${\bf k'}=({\bf p'}_1-{\bf p'}_2)/2$, and the momenta of the center of mass as ${\bf P}={\bf p}_1+{\bf p}_2={\bf p'}_1+{\bf p'}_2$. ${\bf p}_1,{\bf p}_2$ are momenta of each single incoming nucleon, and ${\bf p'}_1,{\bf p'}_2$ of each outgoing nucleon. In the following equations we omit the use of spin/isospin indices for convenience. Making use of Eq.~(\ref{S_hole}) and Eq.~(\ref{S_part}), we can rewrite the first order approximation of Eq.~(\ref{G2B_Lehm}) as:
\beqn
\nn
&&G^{II,f}_{pp,hh}({\bf q},{\bf P};\om)=\int_{\vep_\textrm F}^\infty \d\om'\int_{\vep_\textrm F}^\infty\d\om''
\frac{S^p({\bf q}+{\bf P}/2;\om')S^p({\bf P}/2-{\bf q};\om'')}{\om-\om'-\om''+i\eta}
\\ &&\qquad\qquad -
\int^{\vep_\textrm F}_{-\infty} \d\om'\int^{\vep_\textrm F}_{-\infty}\d\om''
\frac{S^h({\bf q}+{\bf P}/2;\om')S^h({\bf P}/2-{\bf q};\om'')}{\om-\om'-\om''-i\eta}\,,
\label{2BG_Lehm_f}
\enqn
where ${\bf q}={\bf k'}-{\bf k}$ is the transferred momentum. Diagrammatically the previous expression corresponds to the first two diagrams in Fig.~\ref{g4point}.

Let's now restate Eq.~(\ref{GmLadd})  as:
\beq
\Gamma^{4_\mathrm{ladd}}_{pphh}({\bf k},{\bf k'},{\bf P};\Omega)=V({\bf k},{\bf k'})
+\Delta\Gamma^{4_\mathrm{ladd}}_{pphh}({\bf k},{\bf k'},{\bf P};\Omega)\,,
\label{ladder_eq}
\enq
where we have introduced the dispersive energy-dependent part of the ladder vertex function, $\Delta\Gamma^{4_\mathrm{ladd}}_{pphh}$. Eq.~(\ref{ladder_eq}) is a Lippman-Schwinger-type equation that defines the ladder vertex function $\Gamma^{4^\mathrm{ladd}}_{pphh}$ as an in-medium $T$-matrix. This quantity can be seen as the analogous to the scattering $T$-matrix in free space. But, in the present context, it takes into account the medium through the repeated scattering to all orders of two independent but fully dressed particles. This is described by the 2B propagator $G^{II,f}_{pp,hh}$ included in its dispersive energy-dependent part, $\Delta\Gamma^{4_\mathrm{ladd}}_{pphh}$. Making use of Eq.~(\ref{2BG_Lehm_f}) and combining Eqs.~(\ref{GmLadd}) and (\ref{ladder_eq}), we can write this quantity as
\beqn
\nn
&&\Delta\Gamma^{4_\mathrm{ladd}}_{pphh}({\bf k},{\bf k'},{\bf P};\Omega)= 
\\\nn && \qquad
-\frac{1}{\pi}\int_{2\vep_\textrm F}^\infty\d\Omega'\frac{\rm{Im}\Delta\Gamma^{4_\mathrm{ladd}}_{pphh}({\bf k},{\bf k'},{\bf P};\Omega')}
{\Omega-\Omega'+i\eta}
\\\nn && \qquad\quad
+\frac{1}{\pi}\int^{2\vep_\textrm F}_{-\infty}\d\Omega'\frac{\rm{Im}\Delta\Gamma^{4_\mathrm{ladd}}_{pphh}({\bf k},{\bf k'},{\bf P};\Omega')}
{\Omega-\Omega'-i\eta}\,
\\&& \,\,\,= 
\Delta\Gamma_{\downarrow}^{4_\mathrm{ladd}}({\bf k},{\bf k'},{\bf P};\Omega)
+\Delta\Gamma_{\uparrow}^{4_\mathrm{ladd}}({\bf k},{\bf k'},{\bf P};\Omega)\,.
\label{ladd_disp}
\enqn
The notation with $\downarrow$ and $\uparrow$ signals that the corresponding contributions to $\Delta\Gamma^{4^\mathrm{ladd}}_{pphh}$ has poles in the lower or upper half of the complex energy plane. Poles in the lower half plane are associated with forward $pp$ propagation, i.e. energies above $2\vep_\textrm F$, while poles in the upper half plane apply for backward $hh$ propagation, i.e. energies below $2\vep_\textrm F$. This distinction is helpful to allow the proper construction of the ladder self-energy which can be defined as:
\beq
\Sigma^\star_\mathrm{ladd}({\bf p};\om)=\Sigma^\star_\mathrm{\widetilde V}({\bf p})+
\Sigma^\star_\mathrm{\Delta\Gamma}({\bf p};\om)\,,
\label{ladd_self}
\enq
where ${\bf p}$ defines SP momentum. The first term corresponds to the HF self-energy, which is included in the first term of Eq.~(\ref{irrself}). In the present work, the HF self-energy is calculated as
\beq
\Sigma^\star_\mathrm{\widetilde V}({\bf p})=\int\frac{\d{\bf p}_1}{(2\pi)^3}\widetilde V({\bf p},{\bf p}_1)n({\bf p}_1)\,,
\label{self_en_hf}
\enq
where the SP momentum distribution function has been introduced in Eq.~(\ref{mom_dist}). Diagrammatically this contribution to the self-energy corresponds to the first order approximation of Fig.~\ref{ueffective} stripped off the 1B term. The second term in Eq.~(\ref{ladd_self}) can be obtained by considering the pole structure defined in Eq.~(\ref{ladd_disp}), and can be written as:
\beqn
\nn
&&\Sigma^\star_{\Delta\Gamma}({\bf p};\om)=\int\frac{\d{\bf p}_1}{(2\pi)^3}\int_{-\infty}^{\vep_\textrm F}\d\om'
\Delta\Gamma_{\downarrow}^{4_\mathrm{ladd}}({\bf p},{\bf p}_1;\om+\om')S^h({\bf p}_1;\om')
\\ &&\qquad\qquad-
\int\frac{\d{\bf p}_1}{(2\pi)^3}\int^\infty_{\vep_\textrm F}\d\om'
\Delta\Gamma_{\uparrow}^{4_\mathrm{ladd}}({\bf p},{\bf p}_1;\om+\om')S^p({\bf p}_1;\om')\,.\,\,\,\,\,\quad
\label{self_en_disp}
\enqn
Note that conversely to Eq.~(\ref{ladd_disp}), the ladder self-energy is now defined in terms of SP momenta. Eq.~(\ref{self_en_disp}) tells us that a self-consistent solution for the system is indeed required, given that the spectral functions are functions of the self-energy itself through Dyson's equation (see Eqs.~(\ref{Sp_self})-(\ref{Sh_self})). A diagrammatic representation of the previous equation is shown by the second term in Fig.~\ref{self_en}. Consequently, it is clear how the dispersive part of the ladder self-energy, Eq.~(\ref{self_en_disp}), is connected to the in-medium $T$-matrix. 

In Sec.~\ref{section:dyson_eq} we saw how the spectral functions are given by a combination of the real and imaginary part of the self-energy (see Eqs.~(\ref{Sp_self})-(\ref{Sh_self})). For this reason, it is useful to introduce the expression for Eq.~(\ref{self_en_disp}) in which the imaginary and real part are defined separately. Plugging the spectral decomposition of the $\Gamma^{4_\mathrm{ladd}}_{pphh}$ vertex function, given in Eq.~(\ref{ladd_disp}), into Eq.~(\ref{self_en_disp}), leads us to write the imaginary part of the ladder self-energy: 
\beqn
&& \nn
\mathrm{Im}\Sigma^\star_{\Delta\Gamma}({\bf p};\om)=\int\frac{\d{\bf p}_1}{(2\pi)^3}\int_{-\infty}^{\vep_\textrm F}\d\om'
\mathrm{Im}\Delta\Gamma_{\downarrow}^{4_\mathrm{ladd}}({\bf p},{\bf p}_1;\om+\om')S^h({\bf p}_1;\om')
\\ && \qquad +
\int\frac{\d{\bf p}_1}{(2\pi)^3}\int^\infty_{\vep_\textrm F}\d\om'
\mathrm{Im}\Delta\Gamma_{\uparrow}^{4_\mathrm{ladd}}({\bf p},{\bf p}_1;\om+\om')S^p({\bf p}_1;\om')\,.\,\,
\label{imself_eq}
\enqn
The real part of the ladder self-energy can then be obtained through a dispersion relation:
\beq
\mathrm{Re}\Sigma^\star_{\Delta\Gamma}({\bf p};\om)=
-\frac{\cal P}{\pi}\int^{\vep_\textrm F}_{-\infty}\d\om'\frac{\mathrm{Im}\Sigma^\star_{\Delta\Gamma_{\downarrow}}({\bf p};\om')}{\om-\om'}
%\\\nn && \qquad\qquad
+\frac{\cal P}{\pi}\int_{\vep_\textrm F}^{\infty}\d\om'\frac{\mathrm{Im}\Sigma^\star_{\Delta\Gamma_\uparrow}({\bf p};\om')}
{\om-\om'}\,,
\label{reself_eq}
\enq
to which the HF self-energy, Eq.~(\ref{self_en_hf}),  must be summed up to obtain the total real part of the self-energy.

To conclude this chapter, we would like to stress the fact that extensions to include 3BFs beyond effective 2B interactions, like $\widetilde{V}$, are a completely virgin territory. To our knowledge, these have not been evaluated for nuclear systems (or any other system, for that matter) with diagrammatic formalisms. Truncation schemes, like those proposed here, should provide insight on in-medium 3B correlations. The advantage that the SCGF formalism provides is the access to nonperturbative, conserving approximations that contain pure 3B dynamics without the need for \emph{ad hoc} assumptions.

\clearpage{\pagestyle{empty}\cleardoublepage}

%%%%%%%%%% Effective two-body chiral interaction %%%%%%%%%%%%%%%

\chapter{Effective two-body chiral interaction}
\label{chapter:eff_2b_int}

Exploiting the extended SCGF formalism to include 3B forces, we want to obtain results in the ladder approximation for infinite nuclear matter, as described in Sec.~\ref{section:t_matrix}. To achieve this goal, we calculate an effective 2B potential, sum of a 2NF and a contracted 3NF, as the one defined in Eq.~(\ref{veff}) and depicted in Fig.~\ref{veffective}. This effective potential will then be used in the dispersive part of the SP self-energy, Eq.~(\ref{self_en_disp}). In order to compute the HF part of the self-energy, Eq.~(\ref{self_en_hf}), we furthermore need to calculate a 1B effective potential. In the present calculations, however, we don't compute the complete 1B effective potential, as described in Eq.~(\ref{ueff}) and graphically shown in Fig.~\ref{ueffective}, but only its HF approximation, stripped off the 1B part (see Fig.~\ref{ueffective_hf} in the introduction to Chap.~\ref{chapter:results}). In this case, as we have already pointed out in Sec.~\ref{section:self_en_pert}, the factor in front of the contracted 3B term in Eq.~(\ref{ueff}) must change from a 1/4 to a 1/2. Hence, to calculate the 1B effective potential we can resume to use the same 2B effective potential defined for the dispersive part, but carefully correct the contracted 3NF term with a factor 1/2. The use of these effective interactions will then lead to a correct evaluation of the complete SP self-energy, Eq.~(\ref{ladd_self}). 

Consequently, the necessary step, which we will perform in this chapter, is the calculation of an averaged 3NF via the contraction with a 1B propagator (see second term on right-hand side in Fig.~\ref{veffective}). The advantage of using a density-dependent 2NF constructed from three-body physics is clear. The definition of an effective potential is the \emph{easiest} way to include, at the lowest order in 3B correlations, the effect of 3NFs. In fact this, \emph{quick fix}, provides the effect of 3B forces without the need, given a specific many-body approximation, to alter the overall construction of the theory built for the 2B sector only. Nevertheless, as explained in detail in Sec.~\ref{section:self_en_pert}, caution must be kept when defining the effective terms at the 1B or 2B level.

The Krakow group paved the way for calculations in the SCGF approach including three-body forces. In their works they built a contracted version of the Urbana IX force \citep{Pie2001Jun}, obtained with an average that included the effect of correlations. \citep{Som2008,Som2009,Som2009PhD}. This density-dependent potential was then summed to different 2NFs, such as the Nijmegen \citep{Sto1994}, CdBonn\citep{Mac2001} and Argonne $v18$\citep{Wir1995} potentials, to define an effective 2B interaction as the one described in Eq.~(\ref{veff}). 

In this thesis we want to employ interactions at the 2B and 3B level obtained from the same approach, $\chi$EFT. We choose to construct our density-dependent 2B force from a 3NF obtained at N2LO in $\chi$EFT \citep{vKol1994,Epe2002Dec2}. This order corresponds to the leading order for 3B chiral forces. We reduce the 3NF to a density-dependent effective interaction through a correlated SP average over the third particle. This represents in some sense the lowest-order contribution of the three-body force in the many-body system. This is formally defined by the second term in Eq.~(\ref{veff}) and diagrammatically represented by the second contribution in Fig.~\ref{veffective}. As we will see in the following sections, three of the five LECs appearing in the three leading-order 3B terms are known from the corresponding two-body potential. The remaining two low-energy constants need to be obtained by fits in the few-body sector \citep{Nog2006,Nav2007Jul,Heb2011,Mar2013}.

%First principles calculations should be based on a consistent hamiltonian, with 2NF and 3NF that include the same ingredients and are computed within the same model space. This has not been necessarily the case in the past, where forces, particularly at the three-body level, had phenomenological ingredients \citep{Pie2001Jun}. As discussed in the introduction to this work, a way out of \emph{ad hoc} parametrizations has been provided by EFTs applied to low-energy QCD. 

In the past, most calculations for infinite matter where performed using forces for the 2B and 3B sector derived in different model spaces, and at the three-body level often included phenomenological ingredients \citep{Car1983}. A density-dependent 2NF, arising from three-body physics, was added to the original 2NF to provide an interaction that effectively included both two- and three-body effects, in a similar fashion to what is presented in this work. Within the CBF theory, phenomenological density-dependent 3NFs \citep{Lag1981,Frie1981} have been used for over 30 years \citep{Fan1984,Fan1987,Fab1989,Ben1989,Ben1992,Ben2007,Car2011pr}. Recently, further static and dynamical correlations have been introduced in the density-dependent force within the FHNC approach \citep{Lov2011,Lov2012,Lov2012PhD}, for Urbana IX, a chiral inspired revision of the Tucson Melbourne potential \citep{Coo1981} and for a local version of the N2LO chiral 3NF \citep{Nav2007Nov}. Urbana IX has also been used extensively within the BHF approach \citep{Bal1999,Zuo2002Aug,Zuo2002b,Vid2009}. Within this approach, most results have been derived from a density-dependent 2NF following the prescription of Ref.~\citep{Gra1989}. Modern calculations also include a consistent defect function in the averaging procedure for other 3NF \citep{Li2008Mar,Li2008Aug}. 

Density-dependent effective 2NF have been recently constructed from chiral 3NFs at N2LO in Refs.~\citep{JWHol2010,Heb2010Jul}. In both cases, the average is performed using a non-interacting propagator. This averaged force has been used in perturbative calculations of infinite \citep{Heb2011,Heb2013Jul,Coraggio2014} and finite nuclear systems \citep{JWHol2009}. BHF calculations, based on the same chiral 3NF  but with an alternative averaging procedure, indicate a very strong overbinding of nuclear matter \citep{Li2012}. In contrast, within the same approach, a recent calculation which follows a similar procedure for the construction of the density-dependent chiral force to the one presented in Ref.~\citep{JWHol2010}, obtains good saturation properties for nuclear matter \citep{Koh2013}.

In this chapter we describe the construction of the density-dependent force from 3NF at N2LO in the chiral expansion. We then analyze in detail its effect on the partial wave matrix elements.

%%%%%%%%%%%%%%%% Density dependent N2LO interaction %%%%%%%%%%%%%%%

\section{Density-dependent potential at N2LO}
\label{section:dd_n2lo}

The 3NF at third order in chiral perturbation theory is given by three terms: a two-pion-exchange (TPE) contribution, which corresponds to the Fujita-Miyazawa original $2\pi$ exchange term \citep{Fuj1957}; an iterated one-pion-exchange (OPE); and a contact (cont) term \citep{vKol1994,Epe2002Dec2}. These three contributions are diagrammatically represented in Fig.~\ref{3NF_terms}.  Their analytical expressions are given by:

\begin{figure}[t]
  \begin{center}
  \subfloat[]{\label{TPE-3B}\includegraphics[width=0.24\textwidth]{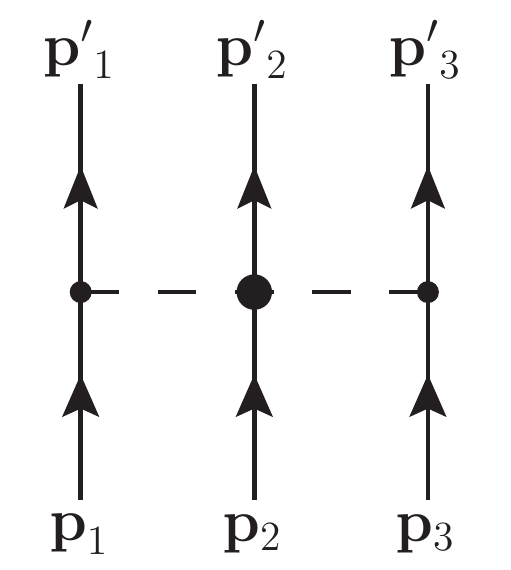}}
  \hspace{1cm}
  \subfloat[]{\label{OPE-3B}\includegraphics[width=0.23\textwidth]{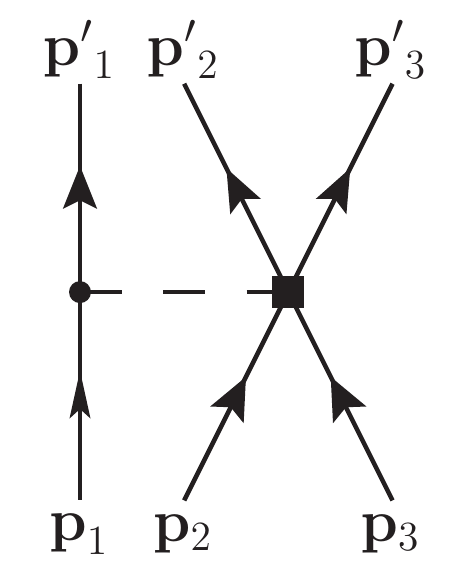}}
  \hspace{1cm}
  \subfloat[]{\label{cont-3B}\includegraphics[width=0.18\textwidth]{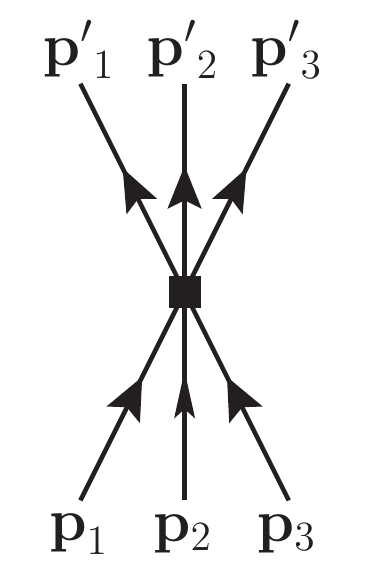}}
  \caption{Three 3B interaction terms appearing at N2LO in the chiral expansion. Diagram \protect\subref{TPE-3B} corresponds to the TPE term of Eq.~(\ref{tpe}). Diagram \protect\subref{OPE-3B} to the OPE term in Eq.~(\ref{ope}). Diagram \protect\subref{cont-3B} is the contact term defined in Eq.~(\ref{cont}). Dashed lines describe the 3B interaction. Small and big dots, and squares define the nature of the vertices as described in Fig~\ref{chiral_forces}.}
  \label{3NF_terms}
  \end{center}
\end{figure}

\beq
W_\mathrm{TPE} =\sum_{i\neq j\neq k}  \frac{g_A^2}{8F_\pi^4}
\frac{(\bd\sigma_i\cdot{\bf q}_i)(\bd\sigma_j\cdot{\bf q}_j)}{({\bf q}_i^2 + M_\pi^2)
({\bf q}_j^2 + M_\pi^2)}
F_{ijk}^{\alpha\beta}\tau_i^{\alpha}\tau_j^{\beta}\,,
\label{tpe}
\enq
\beq
W_\mathrm{OPE} = -\sum_{i\neq j\neq k} \frac{c_D g_A}{8F_\pi^4\Lambda_\chi}
\frac{\bd\sigma_j\cdot{\bf q}_j}{{\bf q}_j^2 + M_\pi^2}(\bd\tau_i\cdot\bd\tau_j)
(\bd\sigma_i\cdot{\bf q}_j)\,;
\label{ope}
\enq
\beq
W_\mathrm{cont} =  \sum_{j\neq k} \frac{c_E}{2F_\pi^4\Lambda_\chi}
\bd\tau_j \cdot \bd\tau_k \,.
\label{cont}
\enq
In the TPE 3B contribution of Eq.~(\ref{tpe}), the quantity $F_{ijk}^{\alpha\beta}$ is
\beq
F_{ijk}^{\alpha\beta}=\delta^{\alpha\beta} [-4M_\pi^2c_1+2 c_3{\bf q}_i\cdot{\bf q}_j]
+\sum_\gamma c_4\epsilon^{\alpha\beta\gamma}\tau^\gamma_k\bd\sigma_k\cdot[{\bf q}_i\times{\bf q}_j]\,.
\label{tpe_tensor}
\enq
In the previous expressions, ${\bf q}_i={\bf p'}_i- {\bf p}_i$ is the transferred momentum between incoming particle and outgoing particle $i$; ${\bf p}_i$ and ${\bf p'}_i$ are SP initial and final momentum of nucleon $i=1,2,3$ as depicted in Fig.~\ref{3NF_terms}. $\bd\tau_i$ and $\bd\sigma_i$ define the isospin and spin matrices for particle $i$. $g_A=1.29$ is the axial-vector coupling constant, $M_\pi=138.04$ MeV the pion mass, $F_\pi=92.4$ MeV the weak pion decay constant and $\Lambda_\chi=700$ MeV is the chosen chiral symmetry breaking constant, of the order of the $\rho$ meson mass.

The LECs appearing in the two-pion exchange term, $c_1,\,c_3,\,c_4$, are the same as those appearing in the original 2NF at N3LO. These are fixed by experimental NN phase-shifts and deuteron properties \citep{Ent2002,Epe2002Dec1}. In contrast, the two LECs appearing in the one-pion and contact 3NF terms, $c_D$ and $c_E$, remain undetermined and have to be fit to further experimental values in the few-body sector. Their value can be obtained from fits to the triton binding energy and to the value of the \emph{nd} scattering lenght \citep{Epe2002Dec2}. Fits to both the binding energies of $^{4}$He and $^3$H can be exploited \citep{Nog2004,Nog2006}, or in some cases even to 3B systems binding energies alone \citep{Nav2007Jul,Nav2007Nov}. Fitting procedures to $^3$H binding energy and the radii of the $\al$ particle have also been implemented \citep{Heb2011}. Furthermore, the triton binding energy together with magnetic moments and the Gamow-Teller in tritium $\beta$-decay have also been used to constrain the value of these LECs \citep{Mar2013}. In the next chapter we will analyze the dependence of our results on different couples of LECs $c_D$ and $c_E$.

It must be noted that the leading-order 3NF contributions appearing in Eqs.~(\ref{tpe}-\ref{cont}) are antisymmetrized terms \citep{Epe2002Dec2}. In other words, the 3B antisymmetrization operator $A_{123}$ is applied to the potential term which, being a symmetric operator, leaves unaltered the following quantity:
\beq
V^\mathrm{3NF}|\Psi\rangle_A=A_{123}V^\mathrm{3NF}|\Psi\rangle_A
=V^\mathrm{3NF}A_{123}|\Psi\rangle_A=V^\mathrm{3NF}|\Psi\rangle_A\,.
\enq
Hence one can work equally well with forces which are antisymmetrized or not. $|\Psi\rangle_A$ represents an antisymmetric many-body state. In the specific case of a 3-particle normalized antisymmetric state we have
\beq
\frac{1}{\sqrt 3!}|\Psi_3\rangle_A= \sqrt{3!}A_{123}|\Psi_3\rangle\,,
\enq
where the three-particle antisymmetrization operator reads
\beq
\label{antisymm}
A_{123}=\frac{(1-P_{12})}{2}\frac{(1-P_{13}-P_{23})}{3}\,.
\enq
$P_{12}$ is the permutation operator of momentum and spin/isospin of particles 1 and 2, which in spin/isospin space reads
\beq
P_{12}=\frac{1+\bd\sigma_1\cdot\bd\sigma_2}{2}\frac{1+\bd\tau_1\cdot\bd\tau_2}{2}\,.
\label{perm_op}
\enq

The density-dependent 2NF is obtained by tracing over spin/isospin indices and integrating over the correlated momentum occupation $n({\bf p}_3)$, Eq~(\ref{mom_dist}), of the averaged particle, here identified as the third particle:
\beqn
\nn
&&\langle {\bf p}_1'\bd\sigma_1'\bd\tau_1';{\bf p}_2'\bd\sigma_2'\bd\tau_2'|\tilde V^\mathrm{3NF}
|{\bf p}_1\bd\sigma_1\bd\tau_1;{\bf p}_2\bd\sigma_2\bd\tau_2\rangle_A =
\\\nn && \quad\quad  \sum_{\sigma_3\sigma'_3}\delta_{\sigma_3\sigma'_3} \sum_{\tau_3\tau'_3}
\delta_{\tau_3\tau'_3}
\int \frac{{\mathrm d}{\bf p}_3}{(2\pi)^3}
\int\frac{{\mathrm d}{\bf p}_3'}{(2\pi)^3}n({\bf p}_3) 
\delta({\bf p}_3'-{\bf p}_3)
\\ && \quad\quad
\langle {\bf p}_1'\bd\sigma_1'\bd\tau_1';{\bf p}_2'\bd\sigma_2'\bd\tau_2';{\bf p}_3'\bd\sigma_3'\bd\tau_3'|W
|{\bf p}_1\bd\sigma_1\bd\tau_1;{\bf p}_2\bd\sigma_2\bd\tau_2;{\bf p}_3\bd\sigma_3\bd\tau_3\rangle_A\,.\,\,\,\,\,\quad
\label{dd3bf}
\enqn
In the previous equation we used the antisymmetric potential matrix elements which have been already introduced in Eq.~(\ref{v_asym}-\ref{w_asym}). For practical reasons from now on we will write $|{\bf p}_1\bd\sigma_1\bd\tau_1\rangle = |{\bf 1}\rangle$. When calculating Eq.~(\ref{dd3bf}), only part of the exchange which characterizes the 3B state $|123\rangle_A$ has to be considered in the averaging procedure, i.e. on the right-hand side of Eq.~(\ref{dd3bf}). The exchange concerning particles 1 and 2 should in fact be included only in the final result, i.e. on the left-hand side of Eq.~(\ref{dd3bf}), because this interchange doesn't affect the contraction over the third particle. Namely, only the $(1-P_{13}-P_{23})$ exchange part of the antisymmetrization operator $A_{123}$, see Eq.~(\ref{antisymm}), has to be considered in the average. We can then rewrite Eq.~(\ref{dd3bf}) as:
\beqn
\nn &&
\langle {\bf 1' 2'}|\tilde V^\mathrm{3NF}
|{\bf 1 2}\rangle_A =
\mathrm{Tr}_{\sigma_3}\mathrm{Tr}_{\tau_3}
\int \frac{{\mathrm d}{\bf p}_3}{(2\pi)^3}n({\bf p}_3)
\\ && \quad\quad\quad\quad\quad
\langle {\bf 1' 2' 3'}|W
(1-P_{13}-P_{23})
|{\bf 1 2 3}\rangle_{A_{12}} \,,
\label{dd3bf_new}
\enqn
where we have written $\sum_{\sigma_3\sigma'_3}\delta_{\sigma_3\sigma'_3} =\mathrm{Tr}_{\sigma_3}$, and the same for the isospin space. $A_{12}$ on the right-hand side matrix element means that the three-particle ket is antisymmetrized only with respect to particles 1,2. The application of the interchange $(1-P_{13}-P_{23})$ when calculating Eq.~(\ref{dd3bf_new}) is a must in order to obtain a density-dependent force which takes into account correctly all possible permutations. If neglected, it leads to the definition of an incorrect density-dependent force which has lost part of the permutation which characterized the original 3B matrix element. 

In this work, the in-medium NN contributions have been derived in the approximation of zero centre of mass momentum, i.e. ${\bf P}={\bf p}_1+{\bf p}_2=0$, which has been shown to drive small errors on bulk properties of infinite matter \citep{Heb2010Jul}. In the process of integration given in Eq.~(\ref{dd3bf_new}), the 3NF has been regularized with a function that reads
\beq
f(p,q)=\exp{\left[-\frac{(p^2+3q^2/4)^2}{\Lambda^4_\textrm{3NF}}\right]}\,,
\enq
where ${\bf p}=({\bf p}_1-{\bf p}_2)/2$ and ${\bf q}=2/3({\bf p_3}-({\bf p}_1+{\bf p}_2)/2)$ are, only in this expression, identified as the Jacobi momenta. $\Lambda_\textrm{3NF}$ defines the cutoff value applied to the 3NF in order to obtain a 3B contribution which dies down similarly to the 2B part one. The regulator function is applied both on incoming $({\bf p},{\bf q})$ and outgoing $({\bf p'},{\bf q'})$ Jacobi momenta. In the approximation of ${\bf P}=0$, we have for the complete regulator function:
\beqn
\nn 
f(k,k',p_3)&=&\exp{\left[-\left(\frac{k}{\Lambda_\textrm{3NF}}\right)^4
-\left(\frac{k'}{\Lambda_\textrm{3NF}}\right)^4\right]}
\\ && \qquad
\exp{\left[-\frac{2}{3}\frac{p_3^2}{\Lambda^4_\textrm{3NF}}\left(\frac{p_3^2}{3}+(k^2+k'^2)\right)\right]}\,.
\label{reg_fun}
\enqn
$k=|{\bf k}|=|{\bf p}_1-{\bf p}_2|/2$ and $k'=|{\bf k'}|=|{\bf p'}_1-{\bf p'}_2|/2$ are the modules of the relative incoming and outgoing momenta. $p_3$ is the module of the SP momentum of the averaged particle. The first exponential term on the right-hand side of Eq.~(\ref{reg_fun}) only affects external relative incoming and outgoing momenta and is similar to the regulator function applied on the 2B part of the chiral potential \citep{Ent2003}. The second term in Eq.~(\ref{reg_fun}) also affects the momentum of the averaged particle. In Ref.~\citep{JWHol2010} the authors use a regulator function which is equal only to the first exponential term in Eq.~(\ref{reg_fun}). For this reason they obtain semi-analytical expressions for the integrals of Eq.~(\ref{dd3bf_new}). The authors of Ref.~\citep{Heb2010Jul} use a regulator function equal to the full Eq.~(\ref{reg_fun}). We will see in the following section the effect of using these two different regulator functions in the averaging procedure. Note that the function in Eq.~(\ref{reg_fun}) is symmetric in the interchange of the three particles, hence it is not affected by the permutations performed in the average (see Eq.~(\ref{dd3bf_new})).

In the following we present  the in-medium density-dependent contributions in the specific case of diagonal matrix elements, i.e. equal relative incoming and outgoing momentum $k=|{\bf k}|=|{\bf k'}|$. We calculate these contributions for both cases of symmetric nuclear and pure neutron matter. We want to underline that the expressions that will be presented are formally equivalent to those obtained in Ref.~\citep{JWHol2010}, whose method we followed to calculate our density-dependent contributions. We will see however, that our expressions differ with respect to those in Ref.~\citep{JWHol2010}, because of a different treatment of regulators and correlations in the construction of the in-medium contributions. We must also point out that in our present calculations, the off-diagonal elements of the density-dependent potential are extrapolated from the on-diagonal ones following the prescription given in Ref.~\citep{JWHol2010}. We follow this approach because it greatly simplifies the evaluation of the density-dependent terms, and avoids the inclusion of additional operatorial structures in the definition of the general NN potential \citep{Erk1971} (see Eq.~(\ref{on-shell_vnn}) further down). We will comment more on this extrapolation at the end of this chapter. Nevertheless, we write complete off-diagonal potential elements for the density-dependent 2NF at N2LO in App.~\ref{chapter:dens_dep_terms}.

The most general form for a two-nucleon potential, in diagonal momenta, which is charge independent, Hermitian and invariant under translation, particle exchange, rotation, space reflection and time reversal is given by \citep{Erk1971}:
\beqn
\nn 
V({\bf k},{\bf q})&=&V^s_c+\bd\tau_1\cdot\bd\tau_2V^v_c
\\\nn 	&&
+[V^s_\sigma+\bd\tau_1\cdot\bd\tau_2V^v_\sigma]\bd\sigma_1\cdot\bd\sigma_2
\\\nn &&
+[V^s_{\sigma q}+\bd\tau_1\cdot\bd\tau_2V^v_{\sigma q}]\bd\sigma_1\cdot{\bf q}\bd\sigma_2\cdot{\bf q}
\\\nn &&
+[V^s_{SL}+\bd\tau_1\cdot\bd\tau_2V^v_{SL}]i(\bd\sigma_1+\bd\sigma_2)\cdot({\bf q}\times{\bf k})
\\ &&
+[V^s_{\sigma L}+\bd\tau_1\cdot\bd\tau_2V^v_{\sigma L}]\bd\sigma_1\cdot({\bf q}\times{\bf k})\bd\sigma_2\cdot({\bf q}\times{\bf k})\,.
\label{on-shell_vnn}
\enqn
We highlight this expression because it will be useful in identifying the different contributions of the density-dependent interaction which arise from contraction of the 3NF terms written in Eqs.(\ref{tpe}-\ref{cont}). Furthermore, following Refs. \citep{Erk1971,Kaiser1997,JWHol2010}, the choice to express the density-dependent force in the form presented in Eq.~(\ref{on-shell_vnn}), turns out to be helpful when defining the partial wave decomposition of the matrix elements. In Eq.~(\ref{on-shell_vnn}), the subscripts identify respectively: $c$ for the central term; $\sigma$ the spin-spin term; $\sig q$ the tensor term; $SL$ spin-orbit and $\sigma L$ quadratic spin-orbit terms. All contributions are present in an isoscalar $V^s$ and isovector $V^v$ form. For non-diagonal momentum matrix elements, Eq.~(\ref{on-shell_vnn}) would include a further operatorial structure \citep{Erk1971}, leading to a more complicated treatment when performing the partial wave decomposition (see App.~\ref{chapter:dens_dep_terms} for a detailed description for off-diagonal momentum elements).

%%%%%%%%%%%%%%%%% nuclear matter %%%%%%%%%%%%%%%%%

\subsubsection{Symmetric nuclear matter}

\begin{figure}[t]
  \begin{center}
  \subfloat[]{\label{TPE-1}\includegraphics[width=0.22\textwidth]{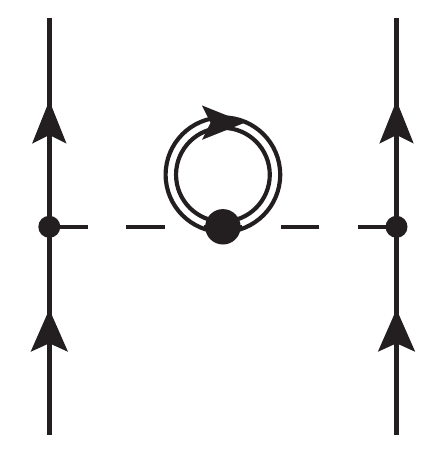}}
  \hspace{1cm}
  \subfloat[]{\label{TPE-2}\includegraphics[width=0.16\textwidth]{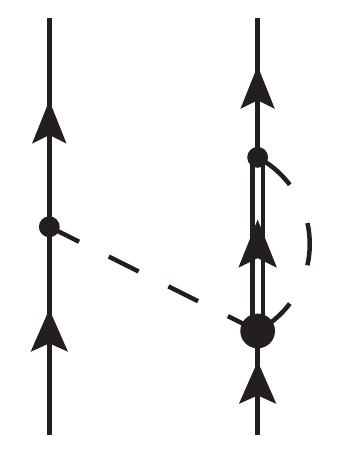}}
  \hspace{1cm}
  \subfloat[]{\label{TPE-3}\includegraphics[width=0.14\textwidth]{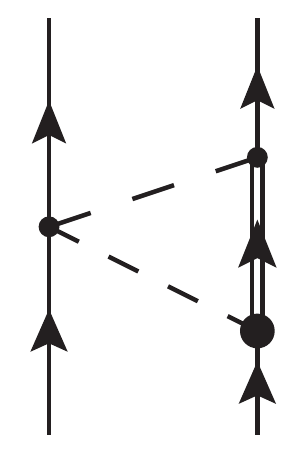}}
  \newline    \vskip .8cm
  \subfloat[]{\label{OPE-1}\includegraphics[width=0.22\textwidth]{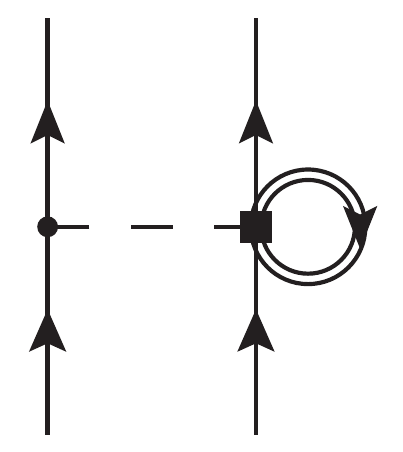}}
  \hspace{1cm}
  \subfloat[]{\label{OPE-2}\includegraphics[width=0.13\textwidth]{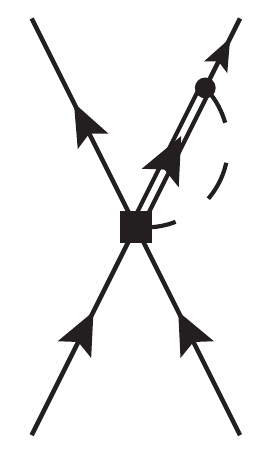}}
  \hspace{1cm}
  \subfloat[]{\label{contact}\includegraphics[width=0.14\textwidth]{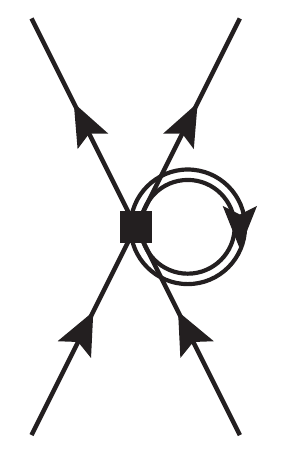}}
  \caption{Six density-dependent contributions arising from contractions of the three 3NF terms appearing at N2LO in the chiral expansion. Dashed lines define the 3B interaction; double arrowed lines correspond to a dressed single particle propagator. Diagrams \protect\subref{TPE-1}, \protect\subref{TPE-2} and \protect\subref{TPE-3} arise form contraction of the long-range 3NF two-pion-exchange term given in Eq.~(\ref{tpe}), and correspond respectively to Eqs.~(\ref{tpe_dd_1}), (\ref{tpe_dd_2}) and (\ref{tpe_dd_3}). Diagrams \protect\subref{OPE-1} and \protect\subref{OPE-2} are obtained from averaging the medium-range 3NF one-pion-exchange term, Eq.~(\ref{ope}), and correspond respectively to Eqs.~(\ref{ope_dd_1}) and (\ref{ope_dd_2}). Diagram \protect\subref{contact} is the result of contracting the contact 3NF contribution given in Eq.~(\ref{cont}), and is defined in Eq.~(\ref{cont_dd}). Small and big dots, and squares define the nature of the vertices in the chiral expansion.}
  \label{2NF_dd_terms}
  \end{center}
\end{figure}

Let's start with the isospin-symmetric case of nuclear matter. Evaluating Eq.~(\ref{dd3bf_new}) for the TPE term of Eq.~(\ref{tpe}), depicted in Fig.~\subref*{TPE-3B}, leads to three contracted in-medium 2B interactions. These are represented in Figs.~\subref*{TPE-1}-\subref*{TPE-3}. 

The first term, Fig.~\subref*{TPE-1}, is an isovector tensor term, i.e. will only contribute to $V^v_{\sigma q}$ in Eq.~(\ref{on-shell_vnn}).  This corresponds to a 1$\pi$ exchange contribution with an in-medium pion propagator:
\beq
\tilde V_\mathrm{TPE-1}^\mathrm{3NF}=\frac{g_A\,\rho_f}{2 F_\pi^4}
\frac{(\bd\sigma_1\cdot{\bf q})(\bd\sigma_2\cdot{\bf q})}{[q^2 + M_\pi^2]^2}
\bd\tau_1\cdot\bd\tau_2[2 c_1M_\pi^2+ c_3\,q^2]\,.
\label{tpe_dd_1}
\enq
$\rho_f$ defines the integral of the correlated momentum distribution function weighed by the regulator function $f(k,k,p_3)$
\beq
\frac{\rho_f}{\nu}=\int \frac{{\mathrm d}{\bf p}_3}{(2\pi)^3}n({\bf p}_3)f(k,k,p_3)\,,
\label{rho_f}
\enq
where $\nu$ is the degeneracy of the system, $\nu=4$ in the isospin symmetric case. If the regulator function included in Eq.~(\ref{rho_f}) were not dependent on the internal integrated momentum $p_3$, the integral would reduce to the value of the total density of the system, $\rho$, divided by the degeneracy and multiplied by an external regulator function. Consequently, the expression in Eq.~(\ref{rho_f}) would exactly equal the one presented in Ref.~\citep{JWHol2010}, even though in our case we use a correlated momentum distribution function, while a step function up to the Fermi level is used in the latter.

The second term, Fig.~\subref*{TPE-2}, is also a tensor contribution to the in-medium NN interaction. It adds up to the previous term and contributes to $V^v_{\sigma q}$ in Eq.~(\ref{on-shell_vnn}). This term includes vertex corrections to the 1$\pi$ exchange due to the presence of the nuclear medium:
\beqn
\nn
&& \tilde V_\mathrm{TPE-2}^\mathrm{3NF}= \frac{g_A^2}{8\pi^2F_\pi^4}
\frac{\bd\sigma_1\cdot{\bf q}\bd\sigma_2\cdot{\bf q}}{q^2 + M_\pi^2} \bd\tau_1\cdot\bd\tau_2
\\\nn && \qquad \times
\Big\{-4c_1M_\pi^2\left[\Gamma_1(k)+\Gamma_0(k)\right]
\\\nn &&  \qquad\qquad
-(c_3+c_4)\left[q^2(\Gamma_0(k)+2\Gamma_1(k)+\Gamma_3(k))+4\Gamma_2(k)\right]
\\ && \qquad\qquad
+4c_4{\cal I}(k)\Big\}\,.
\label{tpe_dd_2}
\enqn
We have introduced the functions $\Gamma_0(k), \Gamma_1(k), \Gamma_2(k), \Gamma_3(k), {\cal I}(k)$, which are integrals over a single pion propagator:
\beq
\label{gamma0}
\frac{\Gamma_0(k)}{(2\pi)^2}=\int\frac{{\mathrm d}{\bf p}_3}{(2\pi)^3}n({\bf p}_3)
\frac{1}{[{\bf k}\pm{\bf p}_3]^2 + M_\pi^2}f(k,k,p_3)\,;
\enq
\beq
\label{gamma1}
\frac{\Gamma_1(k)}{(2\pi)^2}=\frac{1}{k^2}\int\frac{\d{\bf p}_3}{(2\pi)^3}n({\bf p}_3)
\frac{\pm{\bf k}\cdot{\bf p_3}}{[{\bf k}\pm{\bf p}_3]^2 + M_\pi^2}f(k,k,p_3)\,;
\enq
\beq
\label{gamma2}
\frac{\Gamma_2(k)}{(2\pi)^2}=\frac{1}{2k^2}\int\frac{\d{\bf p}_3}{(2\pi)^3}n({\bf p}_3)
\frac{p_3^2k^2-({\bf k}\cdot{\bf p_3})^2}{[{\bf k}\pm{\bf p}_3]^2 + M_\pi^2}f(k,k,p_3)\,;
\enq
\beq
\label{gamma3}
\frac{\Gamma_3(k)}{(2\pi)^2}=\frac{1}{2k^4}\int\frac{\d{\bf p}_3}{(2\pi)^3}n({\bf p}_3)
\frac{3({\bf k}\cdot{\bf p_3})^2-p_3^2k^2}{[{\bf k}\pm{\bf p}_3]^2 + M_\pi^2}f(k,k,p_3)\,;
\enq
\beq
\label{i_integral}
\frac{{\cal I}(k)}{(2\pi)^2}=\int \frac{{\mathrm d}{\bf p}_3}{(2\pi)^3}n({\bf p}_3)
\frac{[{\bf p}_3\pm{\bf k}]^2}{[{\bf p}_3\pm{\bf k}]^2 + M_\pi^2}f(k,k,p_3)\,.
\enq
These integrals are formally equal to those presented in Ref.~\citep{JWHol2010}, but differ in that a dressed propagator is used in our average, and the expressions are weighed with an internal regulator function.

The last TPE contracted term, depicted in Fig.~\subref*{TPE-3}, includes in-medium effects for a 2$\pi$ exchange 2B term. This expression contributes to all operatorial structures of Eq.~(\ref{on-shell_vnn}). Specifically it contributes to the scalar central term $V^s_c$,  to the isovector spin-spin $V^v_\sigma$ and tensor term $V^v_{\sigma q}$, to the spin-orbit in both isoscalar $V^s_{SL}$ and isovector form $V^v_{SL}$, and to the isovector quadratic spin-orbit term $V^v_{\sigma L}$:
\beqn
\nn &&
\tilde V_\mathrm{TPE-3}^\mathrm{3NF}=\frac{g_A^2}{16\pi^2F_\pi^4}\Big\{
-12c_1M_\pi^2\big[2\Gamma_0(k)-G_0(k,q)(2M_\pi^2+q^2)\big]
\\\nn &&   %\quad
-c_3\big[8k_F^3-12(2M_\pi^2+q^2)\Gamma_0(k)
-6q^2\Gamma_1(k)+3(2M_\pi^2+q^2)^2G_0(k,q)\big] 
\\\nn &&  %\qquad
+ 4c_4 \bd\tau_1\cdot\bd\tau_2(\bd\sigma_1\cdot\bd\sigma_2\, q^2-\bd\sigma_1\cdot{\bf q}\bd\sigma_2\cdot{\bf q})
G_2(k,q)
\\\nn &&  %\qquad
-(3c_3+c_4\bd\tau_1\cdot\bd\tau_2)\,i(\bd\sigma_1+\bd\sigma_2)\cdot({\bf q}\times{\bf k})
\\\nn &&  \qquad \times
\big[2\Gamma_0(k)+2\Gamma_1(k)-(2M_\pi^2+q^2)G_0(k,q)+2G_1(k,q)\big]
\\\nn &&  %\qquad 
-12c_1M_\pi^2\,  i(\bd\sigma_1+\bd\sigma_2)\cdot({\bf q}\times{\bf k})
\big[G_0(k,q)+2G_1(k,q)\big]
\\\nn &&  %\qquad
+4c_4\bd\tau_1\cdot\bd\tau_2\bd\sigma_1\cdot({\bf q}\times{\bf k})\bd\sigma_2\cdot({\bf q}\times{\bf k})
\\ && \qquad\times
\big[G_0(k,q)+4G_1(k,q)+4G_3(k,q)\big]\Big\}\,.
\label{tpe_dd_3}
\enqn
Here we have introduced the function $G_0(k,q)$, which is an integral over the product of two different pion propagators:
\beqn
&& \nn
\frac{G_{0,\star,\star\star}}{(2\pi)^2}(k,q)=
\\ && \,\,
\int \frac{{\mathrm d}{\bf p}_3}{(2\pi)^3}n({\bf p}_3)
\frac{\{p_3^0,p_3^2,p_3^4\}}{\big[[{\bf k}+{\bf q}+{\bf p}_3]^2+M_\pi^2\big]\big[[{\bf p}_3+{\bf k}]^2+M_\pi^2\big]}f(k,k,p_3)\,.\qquad\,\,\,\,\,
\label{G_0} 
\enqn
The functions $G_{\star}(k,q), G_{\star\star}(k,q)$ have been introduced to define the rest of the functions, $G_1(k,q), G_2(k,q)$ and $G_3(k,q)$:
\beq
\label{G_1}
G_1(k,q)=\frac{\Gamma_0(k)-(M_\pi^2+k^2)G_0(k,q)-G_\star(k,q)}{4k^2-q^2}\,,
\enq
\beq
\label{G_1star}
G_{1\star}(k,q)=\frac{3\Gamma_2(k)+k^2\Gamma_3(k)-(M_\pi^2+k^2)G_\star(k,q)-G_{\star\star}(k,q)}{4k^2-q^2}\,,
\enq
\beq
\label{G_2}
G_2(k,q)=(M_\pi^2+k^2)G_1(k,q)+G_\star(k,q)+G_{1\star}(k,q)\,,
\enq
\beq
\label{G_3}
G_3(k,q)=\frac{\Gamma_1(k)/2-2(M_\pi^2+k^2)G_1(k,q)-2G_{1\star}(k,q)-G_\star(k,q)}{4k^2-q^2}\,.
\enq
Note that $G_{1\star}(k,q)$ is needed only to define $G_2(k,q),\,G_3(k,q)$.

Integrating Eq.~(\ref{dd3bf_new}) for the OPE 3NF term, given in Eq.~(\ref{ope}), leads to two contributions, shown in Figs.~\subref*{OPE-1}-\subref*{OPE-2}. The first one, Fig.~\subref*{OPE-1}, is a tensor contribution which defines a vertex correction to a 1$\pi$ exchange NN term. It is proportional to the quantity $\rho_f$, similar to what was obtained for the TPE 3NF contracted term $\tilde V_\mathrm{TPE-1}^\mathrm{3NF}$ (see Eq.~\ref{tpe_dd_1}):
\beq
\tilde V_\mathrm{OPE-1}^\mathrm{3NF}=-\frac{c_D\,g_A\,\rho_f}{8\,F_\pi^4\,\Lambda_\chi}
\frac{(\bd\sigma_1\cdot{\bf q})(\bd\sigma_2\cdot{\bf q})}{q^2 + M_\pi^2}
(\bd\tau_1\cdot\bd\tau_2)\,.
\label{ope_dd_1}
\enq
As for the $\tilde V_\mathrm{TPE-1}^\mathrm{3NF}$ term, $\tilde V_\mathrm{OPE-1}^\mathrm{3NF}$ contributes uniquely to the isovector tensor term $V^v_{\sigma q}$ of Eq.~(\ref{on-shell_vnn}).

The second term derived from the 3NF OPE is depicted in Fig.~\subref*{OPE-2}. It defines a vertex correction to the short-range contact NN interaction.  This in-medium interaction contribution is formed of terms of various kinds: a central scalar $V^s_c$, a  spin-spin $V^v_\sigma$, a tensor $V^v_{\sigma q}$ and quadratic spin-orbit $V^v_{\sigma L}$ terms, all contributions in the isovector form (see Eq.~(\ref{on-shell_vnn})). It reads:
\beqn
\nn &&
\tilde V_\mathrm{OPE-2}^\mathrm{3NF}=\frac{c_Dg_A}{16\pi^2F_\pi^4\Lambda_\chi}\Big\{
\big(\Gamma_0(k)+2\Gamma_1(k)+\Gamma_3(k)\big)
%\\\nn &&
\left[\bd\sigma_1\cdot\bd\sigma_2\Big(2k^2-\frac{q^2}{2}\Big)\right.
\\\nn && \left.\qquad
+(\bd\sigma_1\cdot{\bf q}\,\bd\sigma_2\cdot{\bf q})\Big(1-\frac{2k^2}{q^2}\Big)-\frac{2}{q^2}\bd\sigma_1\cdot({\bf q}\times{\bf k})
\bd\sigma_2\cdot({\bf q}\times{\bf k})\frac{1}{q^2}\right]
\\ && \qquad
+2\Gamma_2(k)(\bd\sigma_1\cdot\bd\sigma_2)\Big](\bd\tau_1\cdot\bd\tau_2)
+6{\cal I}(k)\Big\}\,.
\label{ope_dd_2}
\enqn

The last density-dependent term, shown in Fig.~\subref*{contact}, arises from a contraction of the contact 3NF term given in Eq.~(\ref{cont}). This yields a scalar central contribution to the in-medium NN interaction, i.e. a $V^s_c$ term, proportional to $\rho_f$. Hence, being momentum independent it will contribute only to $S$ partial waves, as we will see in the following section. Its formal expression is:
\beq
\tilde V_\mathrm{cont}^\mathrm{3NF}=-\frac{3 c_E\rho_f}{2 F_\pi^4\Lambda_\chi}\,.
\label{cont_dd}
\enq

We would like to underline once more that the obtained in-medium NN interaction terms, Eqs.~(\ref{tpe_dd_1}), (\ref{tpe_dd_2}), (\ref{tpe_dd_3}), (\ref{ope_dd_1}), (\ref{ope_dd_2}) and (\ref{cont_dd}) are formally the same to those obtained by the authors in Ref.~\citep{JWHol2010}. The difference lies in the integrals over a single and a product of pion propagators and in the function $\rho_f$. These integrals differ not only in that our averaging procedure is performed using the correlated momentum distribution function $n({\bf p}_3)$, but furthermore in the weighing of the integrand with the full regulator function $f(k,k',p_3)$ presented in Eq.~(\ref{reg_fun}). In the next section we will test the discrepancies obtained on the effective 2B potential given by different averaging procedures.

%%%%%%%%%%%% neutron matter %%%%%%%%%%%%%%%%%%%%%%
\subsubsection{Pure neutron matter}

In the case of pure neutron matter, the evaluation of Eq.~(\ref{dd3bf_new}) is simplified. In fact,
the trace over isospin is trivial, neutron matter can only be in total isospin $T=1$, i.e. $\bd\tau_1\cdot\bd\tau_2=1$. Consequently the exchange operator of Eq.~(\ref{perm_op}) reduces only to the momentum and spin part, i.e. in spin space it reads:
\beq
P_{12}=\frac{1+\bd\sigma_1\cdot\bd\sigma_2}{2}\,.
\label{perm_op_2}
\enq
It can then be proved that in-medium terms proportional to $c_4, c_D, c_E$ go to zero \citep{Tolos2008,Heb2010Jul,JWHol2010}. 

In fact, for the term proportional to $c_E$, the contact contribution in Eq.~(\ref{cont}), the permutation of spin indices leads to equal direct and exchange terms which directly cancel one another. Physically this is a consequence of the Pauli principle which neutrons, being fermions, must respect. In other words, $W_\textrm{cont}$ vanishes between antisymmetrized three-nucleon states.

For the OPE term proportional to $c_D$, Eq.~(\ref{ope}), it can be demonstrated that the spin-momentum structure of this contribution leads to a vanishing quantity when the trace over spin is applied (for further details see App.~\ref{chapter:dens_dep_terms}). As explained in \citep{Heb2010Jul}, the physical reason lies in the fact that both spin and relative momentum state of the two-neutron system interacting via the contact term in Eq.~(\ref{ope}) is symmetric, which cannot be the case for an overall wave function for a two-fermion state. 

The other vanishing term, the $c_4$ part of the TPE contribution in Eq.~(\ref{tpe}), contains an operatorial structure of the kind $(\bd\tau_1\times\bd\tau_2)\cdot\bd\tau_3$, and further permutations of particles 1,3 and 2,3. This structure is zero in a three neutron system. 

Therefore the only density-dependent contributions, which are non-zero in neutron matter, are those proportional to LECs $c_1$ and $c_3$ in Eqs.~(\ref{tpe})-(\ref{tpe_tensor}). The density-dependent interacting terms obtained in neutron matter will only differ with respect to the symmetric case ones by different pre-factors. This is due to the fact that the only part which changes from the symmetric to the pure neutron matter case is the trace over isospin indices. 

In order to obtain the correct degeneracy for neutron matter, i.e.  $\nu=2$, we need to replace $\rho_f \rightarrow 2\rho_f$ in the $\tilde V_\mathrm{TPE-1}^\mathrm{3NF}$ contribution of Eq.~(\ref{tpe_dd_1}), (see also Eq.~(\ref{rho_f})). The isovector tensor terms $\tilde V_\mathrm{TPE-1}^\mathrm{3NF}$ and  $\tilde V_\mathrm{TPE-2}^\mathrm{3NF}$, given in Eqs.~(\ref{tpe_dd_1})-(\ref{tpe_dd_2}), which contribute to $V_{\sigma q}^v$ in Eq.~(\ref{on-shell_vnn}), must then change prefactor according to:
\beq
\tilde V_\mathrm{TPE-1}^\mathrm{3NF}: \bd\tau_1\cdot\bd\tau_2 \rightarrow \frac{1}{2}\bd\tau_1\cdot\bd\tau_2\,, 
\label{pnm_tpe_1}
\enq
\beq
\tilde V_\mathrm{TPE-2}^\mathrm{3NF}: \bd\tau_1\cdot\bd\tau_2 \rightarrow 
\frac{1}{4}(\bd\tau_1\cdot\bd\tau_2-2)\,.
\label{pnm_tpe_2}
\enq
The isoscalar part of the density-dependent potential appearing in $\tilde V_\mathrm{TPE-3}^\mathrm{3NF}$, which contributes to both the central $V_c^s$ and the spin-orbit $V_{SL}^s$ terms in Eq.~(\ref{on-shell_vnn}), must change prefactor according to:
\beq
\tilde V_\mathrm{TPE-3}^\mathrm{3NF}: 1 \rightarrow \frac{1}{3}\,.
\label{pnm_tpe_3}
\enq

Neutron matter represents a very interesting and unique system for chiral forces. All many-body forces up to N3LO in the chiral expansion are predicted. Complete calculations including sub-leading three-nucleon forces and leading four-nucleon forces can then be performed without the need to adjust further LECs, except for those obtained in the 2B sector, i.e. $c_1$ and $c_3$. Recent results presented by the Darmstadt group \citep{Tew2013,Kru2013} found significant contributions from sub-leading chiral 3NFs, i.e. 3NF at N3LO. Results obtained at complete N3LO level are indeed a major step forward in providing further constrains for the definition of the neutron matter equation of state.

%%%%%%%%%%%% Derivation of terms %%%%%%%%%%%%%%%%%%%

\section{Partial-wave matrix elements }
\label{section:pw_analysis}

We now want to analyze the different partial wave components of the six in-medium NN interacting contributions obtained in the previous section. The partial wave analysis is performed following the prescription presented in Ref.~\citep{JWHol2010}. Given the general formulation for the NN interaction of Eq.~(\ref{on-shell_vnn}), each of the 5 operatorial structures is expanded into eigenstates of the $|LSJ\rangle$ basis. To do this, one first evaluates the matrix elements in the helicity basis and then rotates with Wigner $d$-functions to the $|LSJ\rangle$ basis (for details see Sec.~4 in Ref.~\citep{Erk1971}). To follow this approach, we first define the quantities $U_i=V^s_i+(4T-3)V^v_i$, where $i=c,\,\sigma,\,\sigma\,q,\,SL,\,\sigma L$ (see Eq.~(\ref{on-shell_vnn})) and $T=0,1$ is the total isospin of the pair (for details see Sec.~3 of Ref.~\citep{Kaiser1997}). We then apply strictly the projection formulas given in Sec.~IVA of Ref.~\citep{JWHol2010}. 

In the $|LSJ\rangle$ basis we consider six different matrix elements: 
\begin{itemize}
\item single matrix element with $S=0$ and $L=0$;
\item triple matrix element with $S=1$ and $L=J$;
\item triple matrix elements with $S=1$ and $L=J\pm1$; 
\item triple mixing matrix elements with $S=1$ and  $L'=J-1$, $L=J+1$ and its reverse. 
\end{itemize}
Note that the reverse mixing matrix element, i.e. $L'=J+1$, $L=J-1$, is equal to the former one only for diagonal momentum elements, but it differs in the off-diagonal ones.

The effect of the density-dependent 2NF is tested on top of the N3LO NN force of Ref.~\citep{Ent2003}. Low-energy constants coming from the 2B sector are chosen accordingly, $c_1=-0.81 \rm{GeV}^{-1}, c_3=-3.2\rm{GeV}^{-1}, c_4=5.4\rm{GeV}^{-1}$. The remaining LECs, $c_D=-1.11$ and $c_E=-0.66$, are taken from fits to ground-state properties of $^3$H and $^4$He \citep{Nog2006}, obtained with the same chiral 2NF and 3NF used in this thesis. The cutoff on the 3NF included in the regulator function Eq.~(\ref{reg_fun}) is also chosen accordingly, i.e. $\Lambda_\textrm{3NF}=2.5\textrm{fm}^{-1}$ . Now and in the following, unless specified otherwise, calculations are performed with full regulator function given in Eq.~(\ref{reg_fun}).

\subsection{The six terms of the density-dependent 2NF at N2LO}

\begin{figure}
\begin{center}
\subfloat[]{
\includegraphics[width=0.85\textwidth]{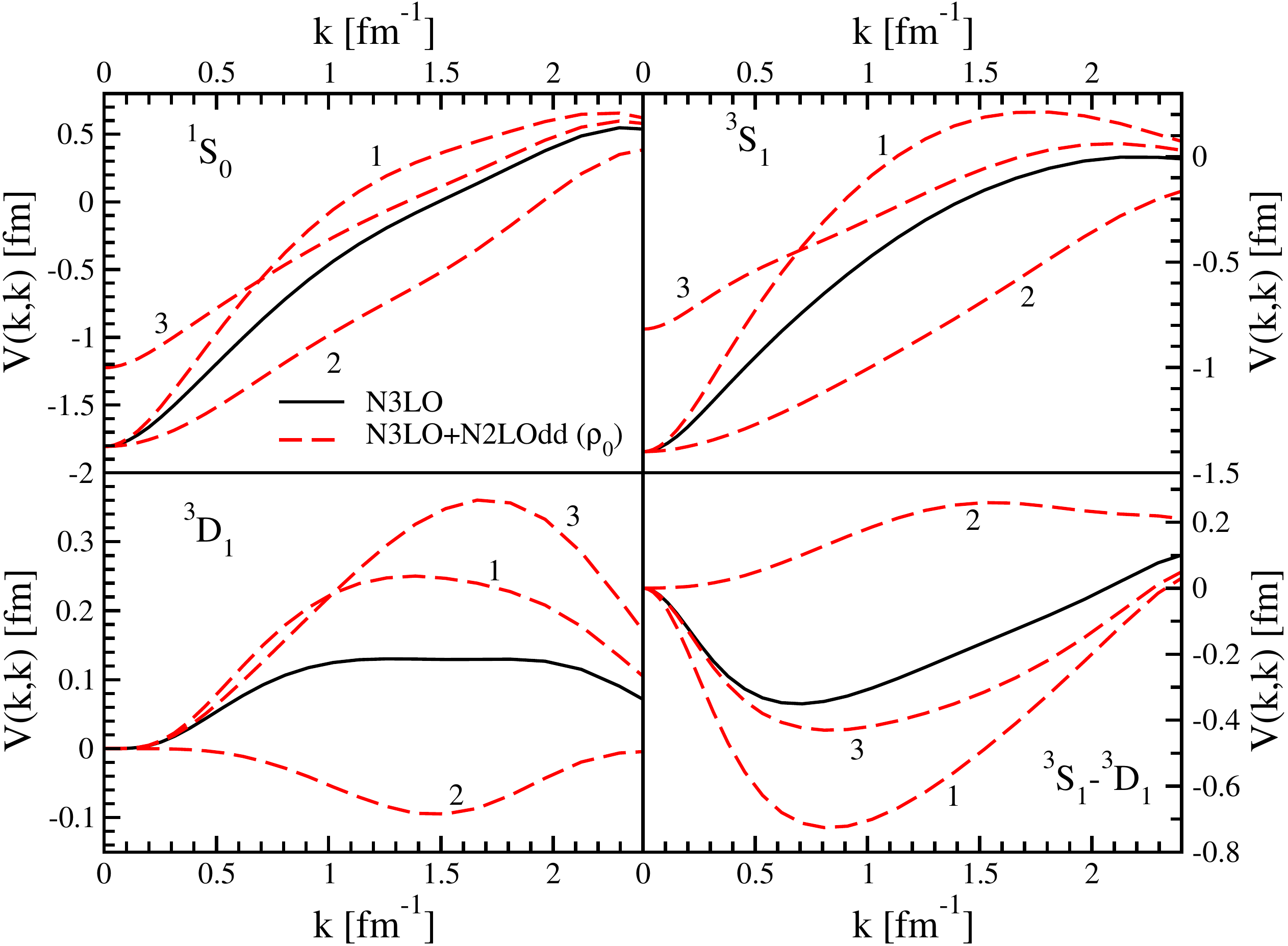}
\label{pot_123_S}
}
\hfill
\subfloat[]{
\includegraphics[width=0.85\textwidth]{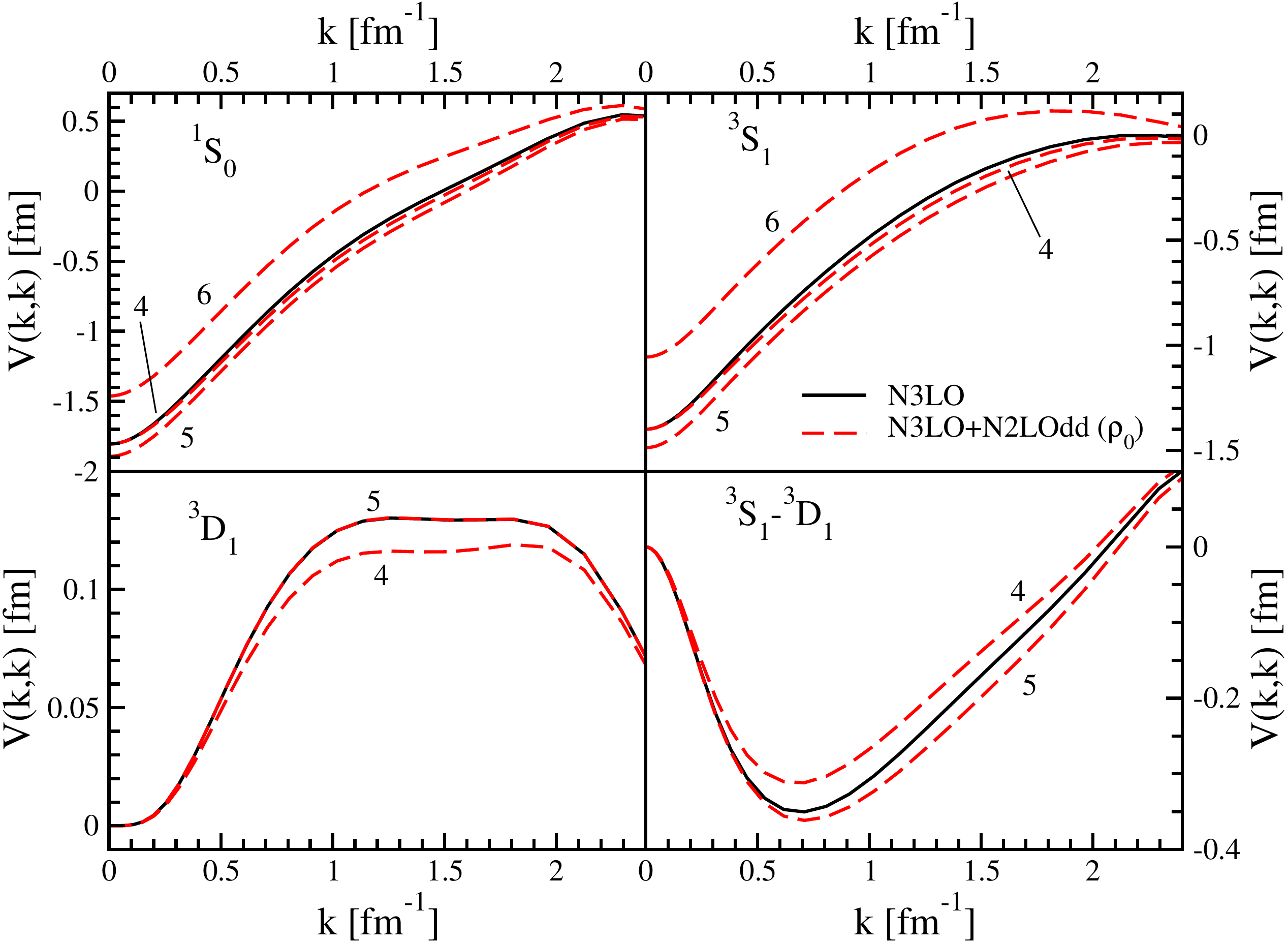}
\label{pot_456_S}
}
\caption{$S$, $D$ and $S-D$ mixing partial waves modifications due to the six in-medium components derived from the 3NFs at N2LO. Black-solid lines correspond to the bare NN N3LO potential. Red-dashed lines depict the contribution coming from the sum of the NN N3LO plus: 1 for the $\tilde V_\textrm{TPE-1}^\textrm{3NF}$, Eq.~(\ref{tpe_dd_1}); 2 for the $\tilde V_\textrm{TPE-2}^\textrm{3NF}$,  Eq.~(\ref{tpe_dd_2}); 3 for the $\tilde V_\textrm{TPE-3}^\textrm{3NF}$, Eq.~(\ref{tpe_dd_3}); 4 for the $\tilde V_\textrm{OPE-1}^\textrm{3NF}$, Eq.~(\ref{ope_dd_1}); 5 for the $\tilde V_\textrm{OPE-2}^\textrm{3NF}$, Eq.~(\ref{ope_dd_2}); and 6 for the $\tilde V_\textrm{cont}^\textrm{3NF}$, Eq.~(\ref{cont_dd}). Density-dependent terms are obtained at empirical saturation density of symmetric nuclear matter, $\rho_0=0.16$ fm$^{-3}$.}
\label{pot_S}
\end{center}
\end{figure}

Let us start by analyzing the effect of each of the six components of the in-medium interaction $\tilde V^\textrm{3NF}$. We do it for the specific case of symmetric nuclear matter at empirical saturation density $\rho=0.16$ fm$^{-3}$. Figs.~\ref{pot_S} reports the effect for $S$, $D$ and $S-D$ mixing partial waves. 

For $S$-waves (see upper left and right panels of Fig.~\subref*{pot_123_S}), the tensor components derived from the long-range chiral 3NF,  $\tilde V_\textrm{TPE-1}^\textrm{3NF}$ and $\tilde V_\textrm{TPE-2}^\textrm{3NF}$, present in all waves a similar modification but with opposite sign. On one side, $\tilde V_\textrm{TPE-1}^\textrm{3NF}$ adds repulsion, reaching an approximate maximum value of $\sim0.4-0.5$ fm at intermediate relative momentum. On the other side, $\tilde V_\textrm{TPE-2}^\textrm{3NF}$ contributes attractively showing similar values. Therefore their combined modification to the NN potential will be small, usually providing less than $\sim0.1$ fm of attraction due to the slightly higher values of $\tilde V_\textrm{TPE-2}^\textrm{3NF}$. These components go to zero for zero transferred momentum, hence vanish for $k\rightarrow0$. The other contribution derived from the 3NF long-range term,  $\tilde V_\textrm{TPE-3}^\textrm{3NF}$, has a repulsive effect for all momenta. It contributes strongly at low momentum, weighing up to $\sim30\%$ of the value of the NN N3LO force at $k=0$. This modification is the main cause for the suppression of the attraction of the bare 2B potential. Repulsion goes down as the relative momentum increases. 3NF medium-range contracted components (see upper left and right panels of Fig.~\subref*{pot_456_S}),  $\tilde V_\textrm{OPE-1}^\textrm{3NF}$ and  $\tilde V_\textrm{OPE-2}^\textrm{3NF}$, give a small constant attractive contribution for all momenta, of the order of $\sim 0.15$ fm for the sum of the two. Conversely, $\tilde V_\textrm{cont}^\textrm{3NF}$, coming from the contraction of the contact 3NF, weakens the attraction of the NN N3LO potential, adding an average constant value of 0.35 fm in repulsion, a bit higher for the $^3S_1$ wave. Its effect diminishes only at high relative momentum. Combined with $\tilde V_\textrm{TPE-3}^\textrm{3NF}$, these two contributions to the in-medium potential have the dominant effect, providing an overall strong repulsion. We want to point out that the contracted contact term,  $\tilde V_\textrm{cont}^\textrm{3NF}$, is proportional to the low-energy constant $c_E$, and changes sign accordingly to the sign of $c_E$. Hence it can give an attractive or a repulsive effect. %Fitting procedures to obtain this constant must be precise in order to obtain a reliable value to be used in nuclear structure calculations.

For the $D$-wave (see lower left panel of Fig.~\subref*{pot_123_S}) the combined effect of $\tilde V_\textrm{TPE-1}^\textrm{3NF}$ and $\tilde V_\textrm{TPE-2}^\textrm{3NF}$ is also small. At intermediate relative momenta, the former adds an 0.1 fm of repulsion, while the latter a 0.2 fm of attraction, leading to a small combined attractive effect similar to what is observed for $S$-waves.
The repulsive effect of the $\tilde V_\textrm{TPE-3}^\textrm{3NF}$ appears mainly in the range of $1.5-2$ fm$^{-1}$ relative momenta, reaching a maximum repulsion of $0.36$ fm on top of the NN N3LO. For terms derived from the medium-range 3NF (see lower left panel of Fig.~\subref*{pot_456_S}), $\tilde V_\textrm{OPE-1}^\textrm{3NF}$ adds a more or less constant attraction of 0.02 fm at intermediate momentum ranges, diminishing the repulsion of the $\tilde V_\textrm{TPE-3}^\textrm{3NF}$. $\tilde V_\textrm{OPE-2}^\textrm{3NF}$ has a negligible but non zero effect. 

The mixing $S-D$-wave (see lower right panel of Fig.~\subref*{pot_123_S}-\subref*{pot_456_S}) is the result of a strict balance in between the different components. This balance results in repulsion at low momenta, and attraction at intermediate and high momenta, with respect to the NN N3LO force. As can be observed from Fig.~\subref*{pot_123_S}-\subref*{pot_456_S} (lower right panels), the attractive sum of the density-dependent contributions $\tilde V_\textrm{TPE-1}^\textrm{3NF}$ and $\tilde V_\textrm{TPE-3}^\textrm{3NF}$ balances the repulsion coming from the  $\tilde V_\textrm{TPE-2}^\textrm{3NF}$ term. A similar behavior is observed for the contributions derived from the medium-range 3NF, where the attraction of $\tilde V_\textrm{OPE-2}^\textrm{3NF}$ is opposed by the repulsion of $\tilde V_\textrm{OPE-1}^\textrm{3NF}$. 

\begin{figure}
\begin{center}
\subfloat[]{
\includegraphics[width=0.85\textwidth]{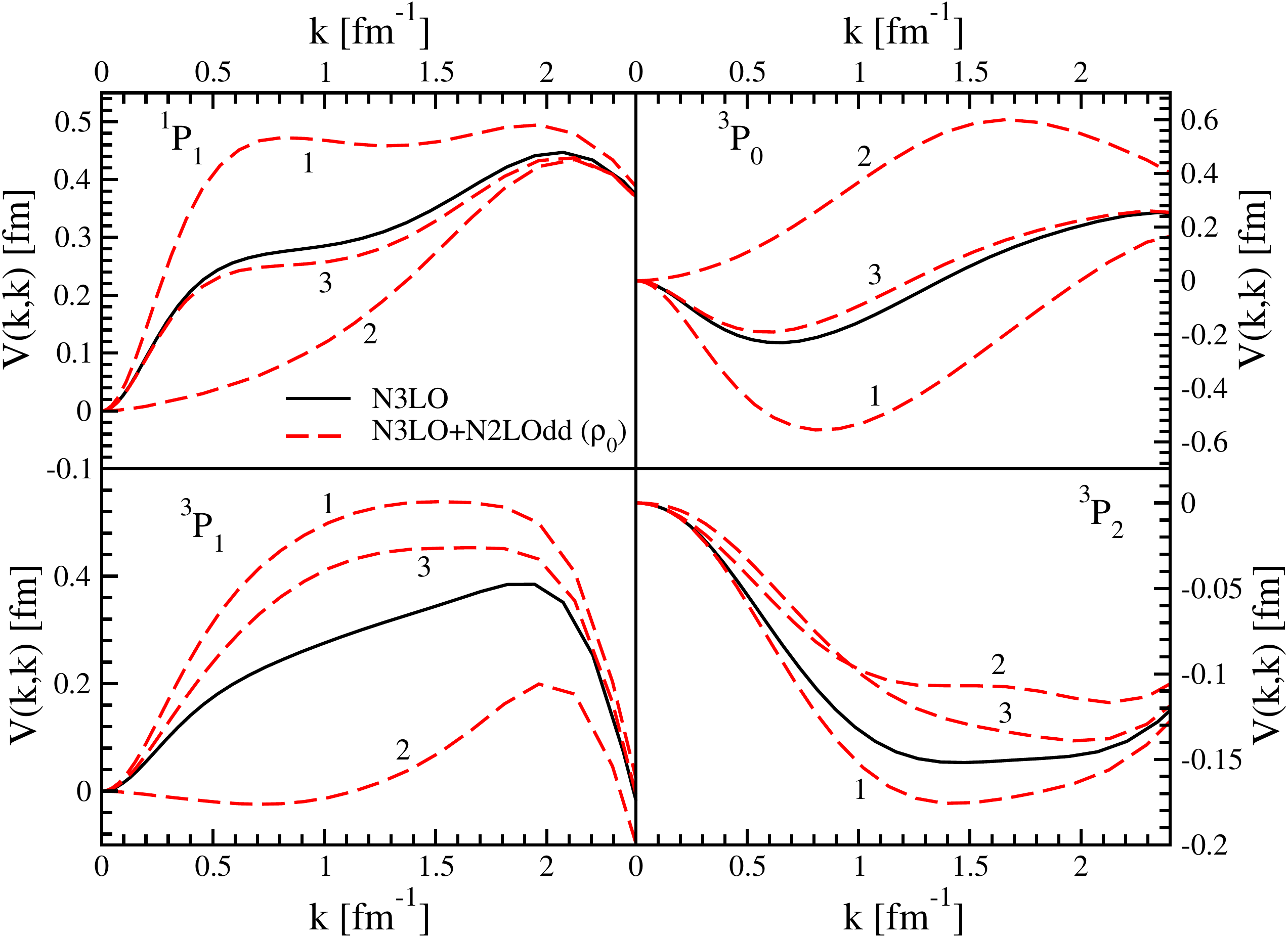}
\label{pot_123_P}
}
\hfill 
\subfloat[]{
\includegraphics[width=0.85\textwidth]{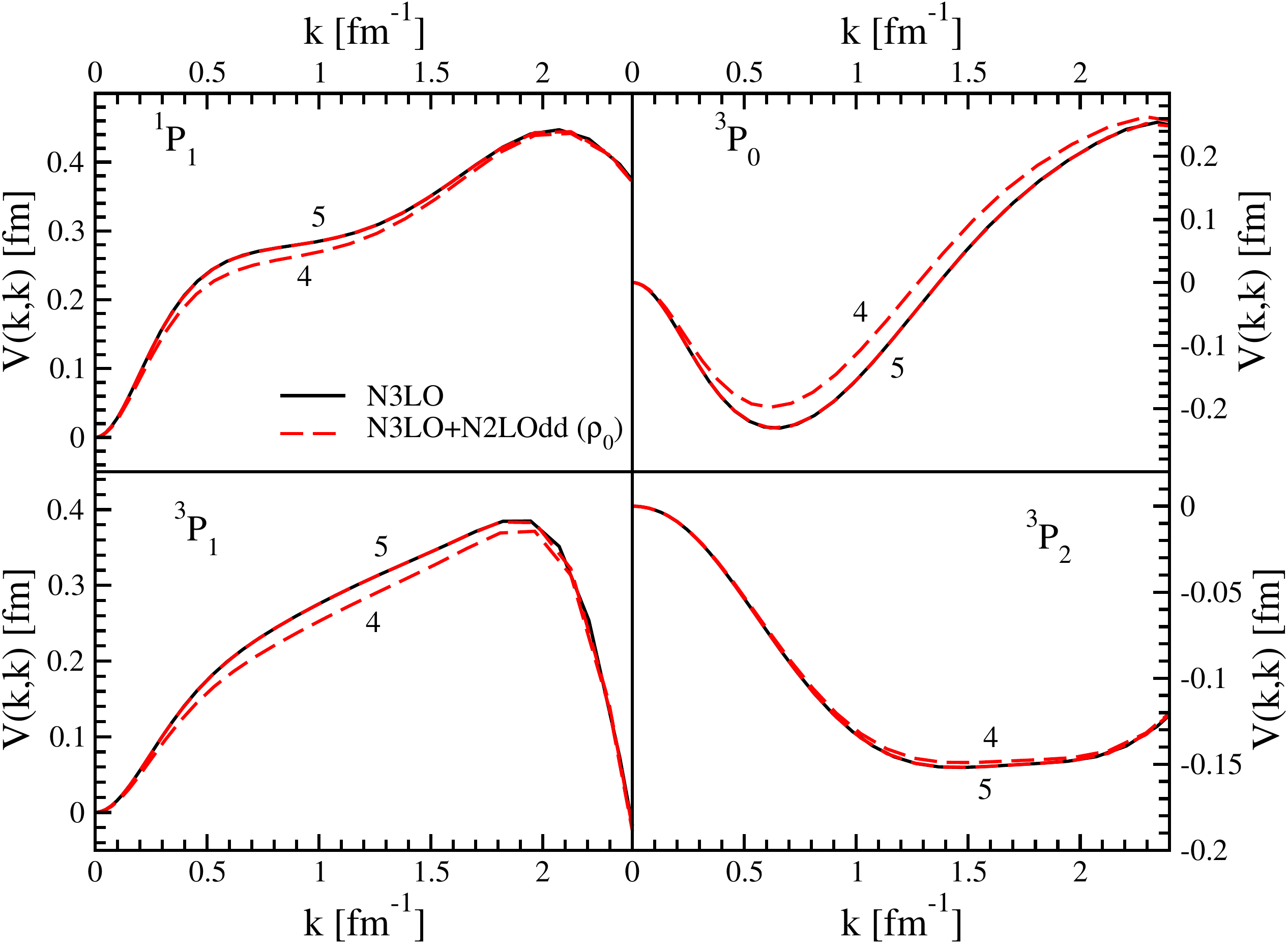}
\label{pot_456_P}
}
\caption{$P$ partial waves modifications due to the six in-medium components derived from the 3NFs at N2LO. Black-solid lines correspond to the bare NN N3LO potential. Red-dashed lines depict the contribution coming from the sum of the N3LO plus: 1 for the $\tilde V_\textrm{TPE-1}^\textrm{3NF}$, Eq.~(\ref{tpe_dd_1}); 2 for the $\tilde V_\textrm{TPE-2}^\textrm{3NF}$,  Eq.~(\ref{tpe_dd_2}); 3 for the $\tilde V_\textrm{TPE-3}^\textrm{3NF}$, Eq.~(\ref{tpe_dd_3}); 4 for the $\tilde V_\textrm{OPE-1}^\textrm{3NF}$, Eq.~(\ref{ope_dd_1}); 5 for the $\tilde V_\textrm{OPE-2}^\textrm{3NF}$, Eq.~(\ref{ope_dd_2}); and 6 for the $\tilde V_\textrm{cont}^\textrm{3NF}$, Eq.~(\ref{cont_dd}). Density-dependent terms are obtained at empirical saturation density of symmetric nuclear matter, $\rho_0=0.16$ fm$^{-3}$.}
\label{pot_P}
\end{center}
\end{figure}

Let's now analyze the effect of each in-medium contribution on $P$ partial waves. In all waves (see Fig.~\ref{pot_P}), the major change with respect to the bare NN potential is given by a combined effect of the three 3NF TPE contracted in-medium contributions. To this combined effect, the 2NFdd medium-range term, $\tilde V_\textrm{OPE-1}^\textrm{3NF}$, adds small modifications. The effect due to $\tilde V_\textrm{OPE-2}^\textrm{3NF}$ is negligible but non zero in all $P$ partial waves. 

For the uncoupled singlet $^1P_1$, see upper left panel in Fig.~\subref*{pot_123_P},  we have that the $\tilde V_\textrm{TPE-1}^\textrm{3NF}$ adds a repulsion of $\sim0.2$ fm around relative momentum of 1 fm$^{-1}$. This is compensated by the joint attraction of $\tilde V_\textrm{TPE-2}^\textrm{3NF}$ and $\tilde V_\textrm{TPE-3}^\textrm{3NF}$ which is approximately $\sim 0.2$ fm. $\tilde V_\textrm{OPE-1}^\textrm{3NF}$, in upper left panel of  Fig.~\subref*{pot_456_P}, is then the contribution which provides a small visible attraction, $\sim 0.04$ fm. This behavior is maintained up to  relative momenta of 1.5 fm$^{-1}$, subsequently the repulsive effect of $\tilde V_\textrm{TPE-1}^\textrm{3NF}$ wins over the other contributions.

In the triplet $^3P_1$ partial wave the changes due to each component are similar to the ones observed for the former one, except for the 3NF long-range contracted contribution $\tilde V_\textrm{TPE-3}^\textrm{3NF}$, see lower left panel of Fig.~\subref*{pot_123_P}. This terms presents in this case a reversed behavior, with respect to the $^1P_1$ wave, being repulsive for all momenta. At intermediate momenta, the repulsion obtained by the sum of $\tilde V_\textrm{TPE-1}^\textrm{3NF}$ and $\tilde V_\textrm{TPE-3}^\textrm{3NF}$ is stronger than the attraction coming from $\tilde V_\textrm{TPE-2}^\textrm{3NF}$ and $\tilde V_\textrm{OPE-1}^\textrm{3NF}$, see lower left panel in Fig.~\subref*{pot_456_P} for the latter contribution. This provides an overall enhancement for the NN N3LO at $k=1$ fm$^{-1}$ of $\sim0.05$ fm. In this range of relative momenta is where the main modifications for the  $^3P_1$ partial wave are observed.

In the remaining $P$ partial waves the changes provided by each contribution are mainly similar. In both $^3P_0$ and $^3P_2$, right panels in Fig.~\subref*{pot_123_P}, the averaged TPE contribution, $\tilde V_\textrm{TPE-2}^\textrm{3NF}$ and $\tilde V_\textrm{TPE-3}^\textrm{3NF}$, sum up to add repulsion to the bare NN N3LO. For intermediate relative momentum,  the repulsion can grow up to maximum values of $\sim 0.6$ fm in the $^3P_0$ wave. In the $^3P_2$ wave this is much less, of around $\sim 0.06$ fm for $k=1.5$ fm$^{-1}$ relative momenta. This effect is enhanced in both partial waves by the OPE in-medium term $\tilde V_\textrm{OPE-1}^\textrm{3NF}$,  right panels in Fig.~\subref*{pot_456_P}, especially in the $^3P_0$, where it adds a further 0.02 fm repulsion to the NN potential. The joint effect of these terms is proportionally balanced in each partial wave by the attraction of the density-dependent contribution $\tilde V_\textrm{TPE-1}^\textrm{3NF}$.  However, the attraction provided by this contribution is not enough, and both partial waves result less attractive at all momenta when considering the inclusion of the complete density-dependent interaction. 

As a concluding remark, we would like to underline that the terms dependent on the LECs $c_D$ and $c_E$ can change substantially depending  on the value of these constants. While the effect of the vertex correction to the short-range contact NN term, $\tilde V_\textrm{OPE-2}^\textrm{3NF}$ in Eq.~(\ref{ope_dd_2}), is always negligible, the other terms can greatly influence the total modification of the partial waves. In fact, according to the sign of $c_D$ and $c_E$, the vertex correction to the $1\pi$ exchange term, $\tilde V_\textrm{OPE-1}^\textrm{3NF}$ in Eq.~(\ref{ope_dd_1}), and the vertex correction to the contact NN interaction,  $\tilde V_\textrm{cont}^\textrm{3NF}$ in Eq.~(\ref{cont_dd}), can yield an attractive or repulsive effect, whose strength varies with the modules of the LECs,  $c_D$ and $c_E$ respectively. Therefore fitting procedures to obtain these constants must be precise, in order to obtain reliable results for nuclear structure calculations.

\subsection{The effect of correlations on the density-dependent 2NF}

\begin{figure}
\begin{center}
\subfloat[]{
\includegraphics[width=0.85\textwidth]{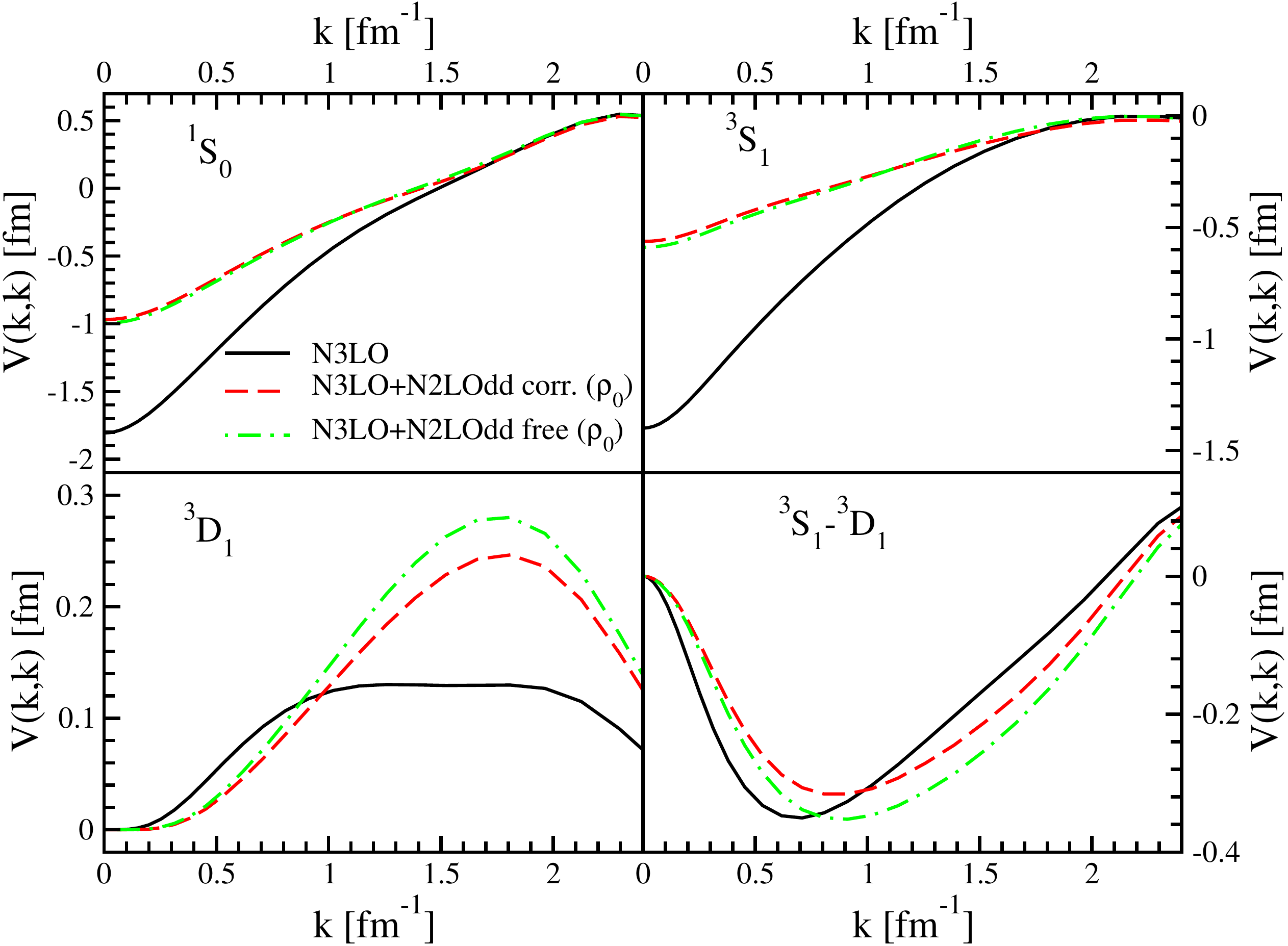}
\label{S_freevsmed_reg2}
}
\hfill 
\subfloat[]{
\includegraphics[width=0.85\textwidth]{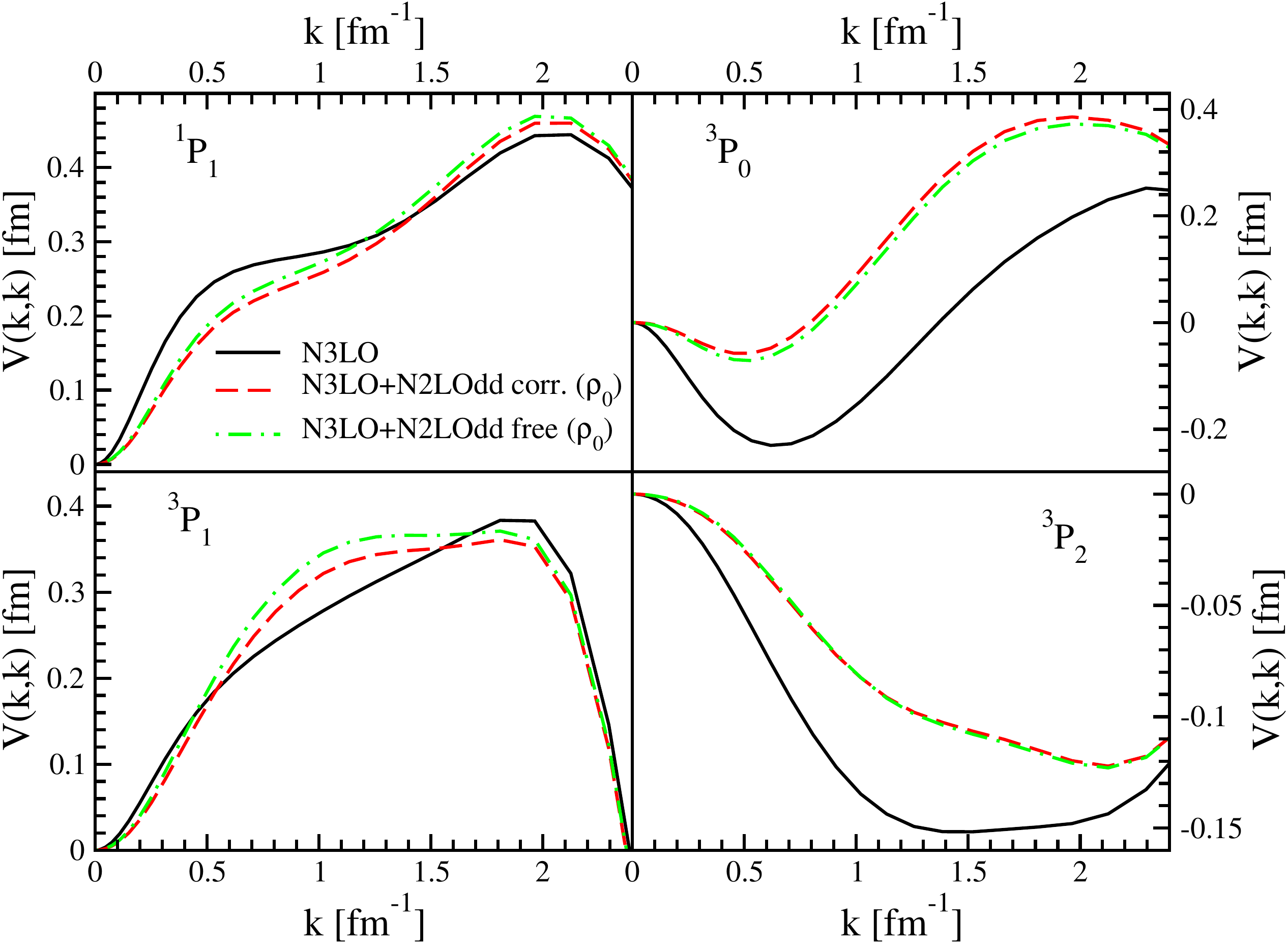}
\label{P_freevsmed_reg2}
}
\caption{$S$, $D$, $S-D$ mixing (upper panel \protect\subref{S_freevsmed_reg2}) and $P$ (lower panel \protect\subref{P_freevsmed_reg2}) partial waves modifications due to complete in-medium potential $\tilde V^\textrm{3NF}$. Black-solid lines depict the bare NN N3LO potential. Red-dashed line correspond to the sum of NN N3LO to the complete in-medium NN potential calculated with a dressed nucleon propagator (corr.). Green-dot-dashed correspond to the case obtained with a free nucleon propagator in the averaging procedure (free). Density-dependent terms are obtained at empirical saturation density of symmetric nuclear matter, $\rho_0=0.16$ fm$^{-3}$.}
\label{freevsmed}
\end{center}
\end{figure}

In Fig.~\ref{freevsmed} we plot at empirical saturation density the complete effect of the in-medium NN interaction on top of the NN N3LO chiral force. Curves obtained using both a correlated and a free distribution function in Eq.~(\ref{dd3bf_new}) are presented. The correlated distribution function is the self-consistent momentum distribution obtained at each iteration in the solution of the Dyson equation, Eq.~($\ref{Dyson}$), considering both 2NFs and 3NFs in the calculation. The use of a free in-medium propagator corresponds to replacing $n({\bf p}_3)$ by a step function up to the Fermi momentum $p_\textrm F$ in Eq.~(\ref{rho_f}), Eq.~(\ref{gamma0}-\ref{i_integral}) and Eq.(\ref{G_0}). 

Focusing on the $S$ partial waves, upper panels in Fig.~\subref*{S_freevsmed_reg2},  the NN N3LO potential is pushed to less attractive values for all momenta. As we previously analyzed, this effect is mainly provided by the term including medium effects in a $2\pi$ exchange 2B term, $\tilde V_\textrm{TPE-3}^\textrm{3NF}$ of Eq.~(\ref{tpe_dd_3}), and by the contact term, $\tilde V_\textrm{cont}^\textrm{3NF}$ given in Eq.~(\ref{cont_dd}). The repulsion provided by the density-dependent force is as large as $50\%$ of the value of the bare NN force at zero momentum. If we now turn to the $D$-wave, lower left panel in Fig.~\subref*{S_freevsmed_reg2},  we observe a small attraction at low-momenta, $\sim0.02$ fm, which evolves in a stronger repulsion at intermediate momenta, $\sim0.1$ fm. Also in this case, the $\tilde V_\textrm{TPE-3}^\textrm{3NF}$ term plays a substantial role in providing this repulsion. 

The modifications on $S$ and $D$ waves are dominant. We have checked that the repulsive change which characterizes these partial waves increases with the density. This promotes the total energy of the system to more repulsive values, providing the mechanism for nuclear matter saturation. While for the 3NF TPE contracted term, $\tilde V_\textrm{TPE-3}^\textrm{3NF}$ in Eq.~(\ref{tpe_dd_3}), this enhancement is difficult to see from its formal expression, for the contact term, $\tilde V_\textrm{cont}^\textrm{3NF}$, it's easy to verify. As a matter of fact, this contribution is proportional to the quantity $\rho_f$, which directly grows with density (see Eq.~(\ref{cont_dd})).

The behavior of the mixing $S-D$ wave is reversed with respect to the one of the $D$ wave. A small repulsion of the order of $\sim0.02$ fm is observed at momenta up to $1$ fm$^{-1}$, and a similar value but in attraction is seen for higher momenta.  This is given by the combined effect of all in-medium terms which contribute to this partial wave.
 
We can observe that the main effect coming from the use of a free or dressed nucleon propagator appears in the $D$ and $S-D$ mixing waves. Due to the use of the free propagator, the change is repulsive in the previous and attractive in the latter, i.e. it enhances in both cases the absolute value of the potential matrix element. For the $S$ waves the modifications are negligible, less than 0.03 fm in attraction with the use of the free in-medium propagator. From an analysis of each of the six density-dependent components, it arises that the major effect, due to the use of different in-medium propagators, is observed for the $\tilde V_\textrm{TPE-3}^\textrm{3NF}$ contribution. This term increases in absolute value when going from the correlated to the free average. This is a consequence of the availability of momenta below the Fermi momentum. In other words, the more momentum states are available below $p_\textrm F$, i.e. free in-medium propagator, the more the partial waves increase in absolute values. This will be clear from the analysis that we will perform in the next section on the different internal regulators used in the average.

%Given that the attraction in the $S-D$-wave is a bit larger than the repulsion coming from the $D$-wave, this will result in a slighter more attractive total energy value when using a free in-medium nucleon propagator.

In Fig.~\subref*{P_freevsmed_reg2} we show for $P$ waves, the sum of the six in-medium NN contributions on top of the NN N3LO potential. The effect is mainly given by the relative effect of the three contributions derived from averaging the TPE 3NF, $\tilde V_\textrm{TPE-1}^\textrm{3NF}$, $\tilde V_\textrm{TPE-2}^\textrm{3NF}$  and $\tilde V_\textrm{TPE-3}^\textrm{3NF}$. As expected from the previous analysis performed on each contribution, in waves $^3P_0$ and $^3P_2$ this effect is to add repulsion at all momenta with respect to the bare NN N3LO potential. In the $^3P_0$ wave, the average enhancement is $\sim0.1$ fm, which dies at small and high relative momenta. In the $^3P_2$ this value is a bit smaller, reaching its maximum at $\sim0.05$ fm at  intermediate momenta.  In $^1P_1$ partial wave, the effect is attractive at low momenta and repulsive for higher; the behavior is reversed in wave $^3P_1$. In both cases, the modification is never higher than $\sim0.04$ fm in absolute value.

In $P$ partial waves, results obtained performing the calculation with a free or dressed in-medium nucleon propagator present small differences. In $^1P_1$ and $^3P_1$ waves, the change due to the use of a step function up to the Fermi momentum adds a repulsion for all momenta of the size of $\sim0.02$ fm. In the $^3P_0$ the effect is slightly smaller but with the opposite sign. 

Finally, according to Fig.~\ref{freevsmed}, we can conclude that, except for the $^3P_0$ partial wave, the effect of using a free propagator in the averaging procedure is to enhance the absolute value of the potential matrix elements in all partial waves. Similar conclusions as for the $S$, $D$ and $S-D$ waves can be drawn. The global effect will be visible in the next chapter, where we will analyze the bulk properties of nuclear matter. We will see how, at saturation density, the effect of using the correlated average, with respect to the free one, lowers the attraction provided by the 3BF, while at double saturation energy it lowers its repulsion. 

\subsection{The effect of the regulator function}

\begin{figure}
\begin{center}
\subfloat[]{
\includegraphics[width=0.85\textwidth]{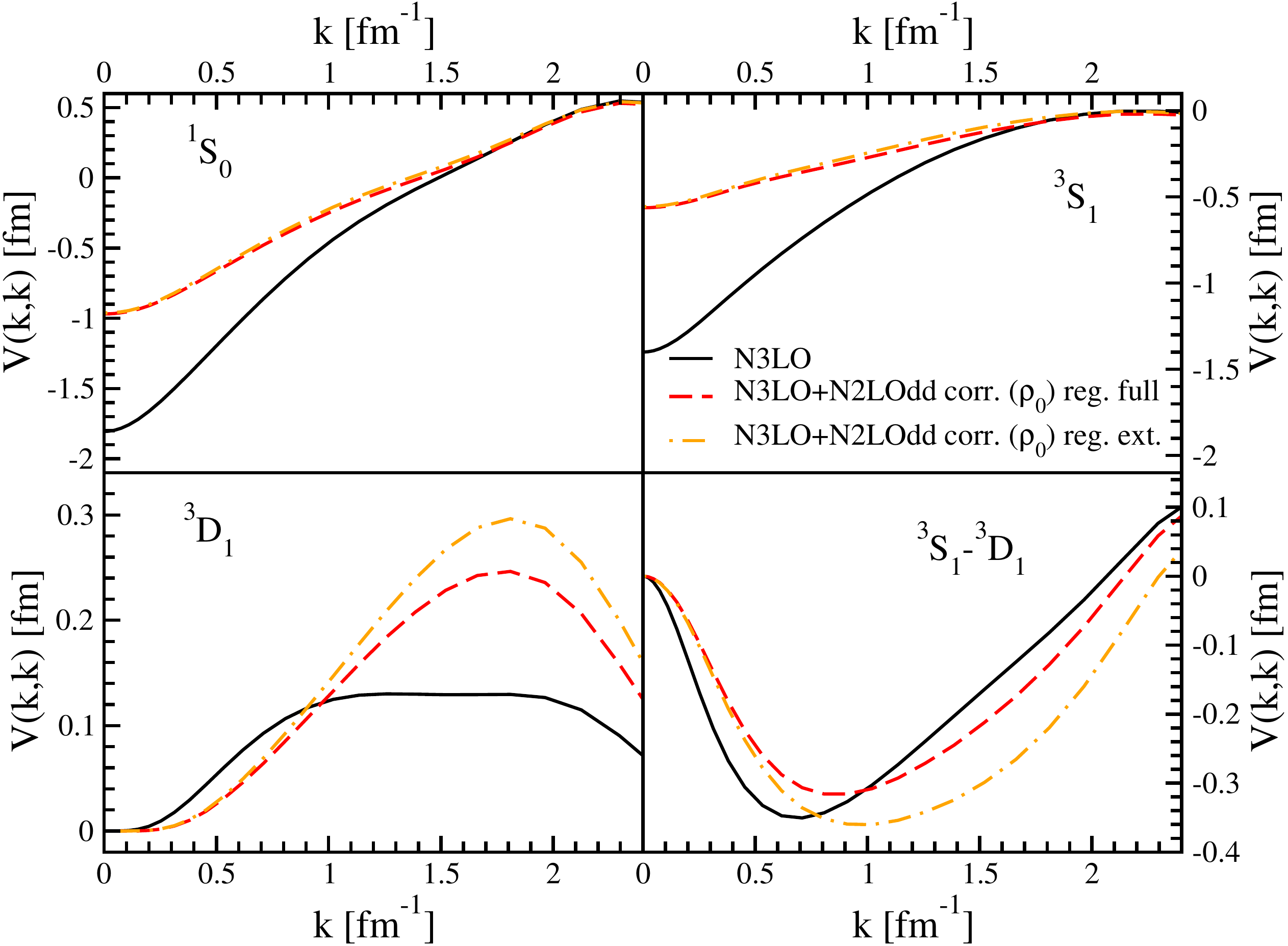}
\label{S_reg1vsreg2_med}
}
\hfill
\subfloat[]{
\includegraphics[width=0.85\textwidth]{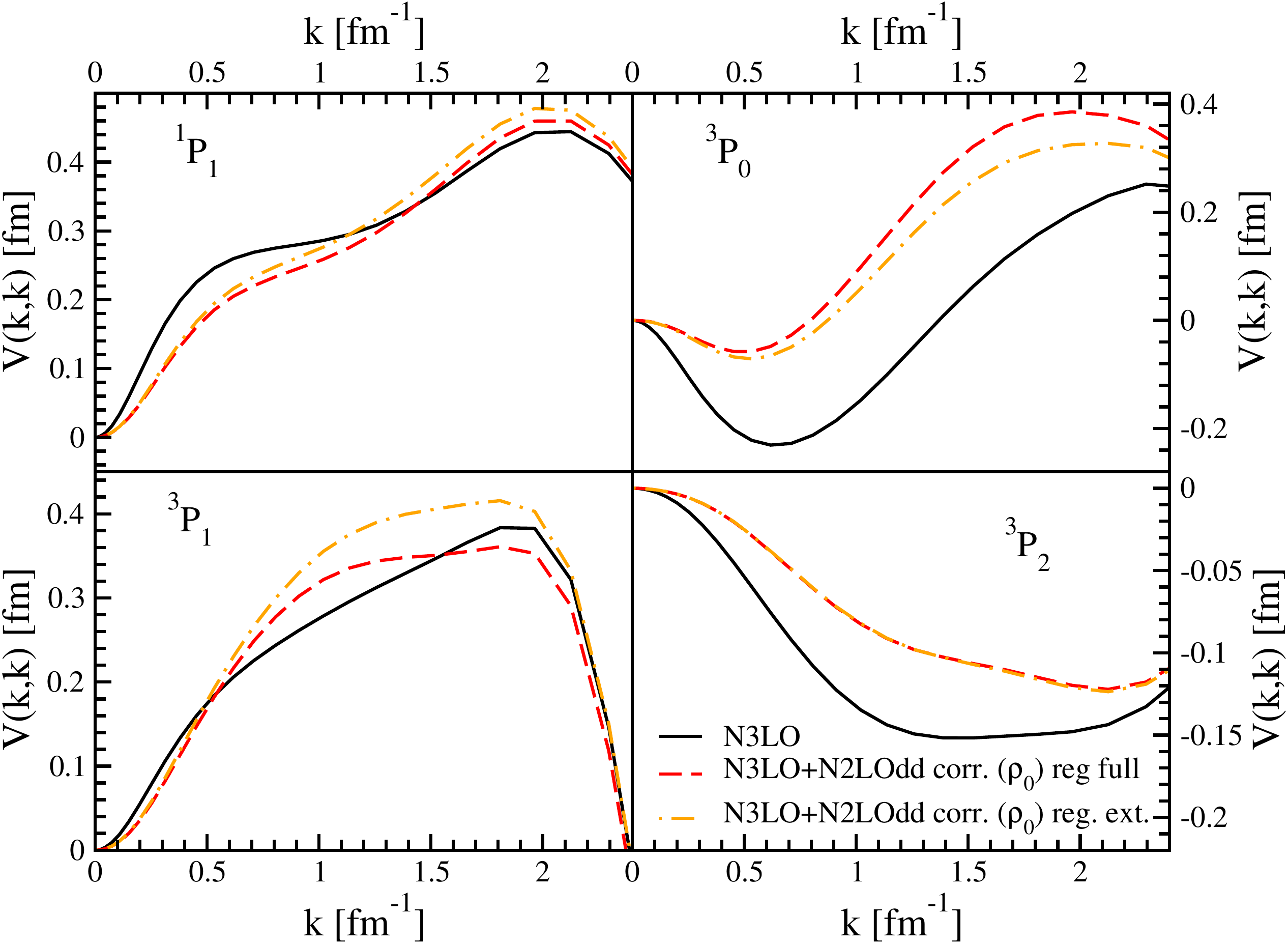}
\label{P_reg1vsreg2_med}
}
\caption{$S$, $D$, $S-D$ mixing (upper panel \protect\subref{S_reg1vsreg2_med}) and $P$ (lower panel \protect\subref{P_reg1vsreg2_med}) partial waves modifications due to complete in-medium potential $\tilde V^\textrm{3NF}$. Black solid lines depict the bare NN N3LO potential. Red-dashed lines correspond to the sum of the NN N3LO to the complete in-medium NN potential calculated with the full regulator function Eq.~(\ref{reg_fun}). Orange-dot-dashed lines correspond to the case obtained with an external regulator function (first term in Eq.~(\ref{reg_fun}). Density-dependent terms are obtained at empirical saturation density of symmetric nuclear matter, $\rho_0=0.16$ fm$^{-3}$.}
\label{reg1vsreg2}
\end{center}
\end{figure}

We now want to analyze the effect of using different regulator functions when performing the average over the third particle, as explained in the previous section (see Eq.~(\ref{reg_fun}) and paragraph thereafter). We will refer to full regulator when the calculation is computed using the complete function given in Eq.~(\ref{reg_fun}). Conversely, we will refer to external regulator when the average is computed using only the first exponential term in Eq.~(\ref{reg_fun}). In both cases the self-consistent correlated momentum distribution function is used. We present in Fig.~\ref{reg1vsreg2} the curves obtained with the use of the external regulator compared to those obtained with the full function. From a global perspective, it is interesting to observe the similarities with the modifications which were obtained when performing the average with a free in-medium propagator (see Fig.~(\ref{freevsmed})). This is a consequence of how momenta below $p_\textrm F$ are available in the integration of 3NFs. In the present case, momenta are not internally regulated. They appear to be more available with respect to the other averaging procedures. As a matter a fact, an even stronger enhancement, with respect to the other cases, is observed on the absolute values of the partial waves.

If we look at partial waves depicted in Fig.~\subref*{S_reg1vsreg2_med}, we see a negligible effect for the $S$ waves (upper panels), while an appreciable change is observed in $D$ and $S-D$ mixing waves (lower panels). For the latter waves, the modifications observed with respect to the average with full regulator are similar to those seen in Fig.~\subref*{S_freevsmed_reg2} (lower panels) when using the free propagator. With respect to the values in Fig.~\subref*{S_freevsmed_reg2}, in the present case we observe a further enhancement of the absolute quantities at intermediate momenta of approximately $\sim0.02$ fm. For $P$ waves, Fig.~\subref*{P_reg1vsreg2_med}, a similar behavior is also observed with respect to changes seen in Fig.~\subref*{P_freevsmed_reg2}. For the $^1P_1$ and $^3P_1$ waves, the use of an external regulator in the average provides repulsion for all momenta, while it has the opposite effect for the $^3P_0$ partial wave. The effect is negligible in the $^3P_2$ wave. With respect to Fig.~\subref*{P_freevsmed_reg2}, the change observed in Fig.~\subref*{P_reg1vsreg2_med} on the $^3P_1$ and $^3P_0$ is higher. Not weighing the internal integrated momentum results in a repulsion in the  $^3P_1$ wave which can reach values of $\sim0.5$ fm at relative momenta $k=1.5-2.0$ fm$^{-1}$. In the $^3P_0$, for similar relative momenta, an attractive shift of the same strength as the previous one is observed.  

We can affirm that the use of an external regulator function drives an overall enhancement in the absolute values of the potential partial waves. Using an external regulator results in a less depleted momentum distribution function in the integral of Eq.~(\ref{dd3bf_new}). We can conclude that the non regulation of internal momenta below the Fermi momentum is what provides an enhanced effect of the density-dependent force. In other words, the fact that momenta below $p_\textrm F$ are more available in the averaging procedure, provides a stronger attraction for negative values and a stronger repulsion for positive values of the partial waves. The effect will be clearer when we will look at the total energy of nuclear matter in the next chapter. We will see that, when using the external regulator, stronger attraction is observed at saturation density, while higher repulsion is observed at double saturation energy, with respect to the full regulator calculation.  This behavior is similar to what we will observe for results obtained with the free in-medium propagator with respect to the correlated average with internal regulator. Conversely, the free in-medium propagator calculations will yield less attractive negative energy values, and less repulsive positive energy values with respect to the ones obtained with the correlated average with external regulator. As momenta become more and more depleted below $p_\textrm F$, from the free average to the correlated average both internally regulated, the effect of 3B forces is less attractive for negative energy values and less repulsive for positive energy values. This once again leads us to the conclusion that the most important role in the integrated functions is provided by momenta below $p_\textrm F$.
 
\begin{figure}[t]
\begin{center}
\includegraphics[scale=0.45]{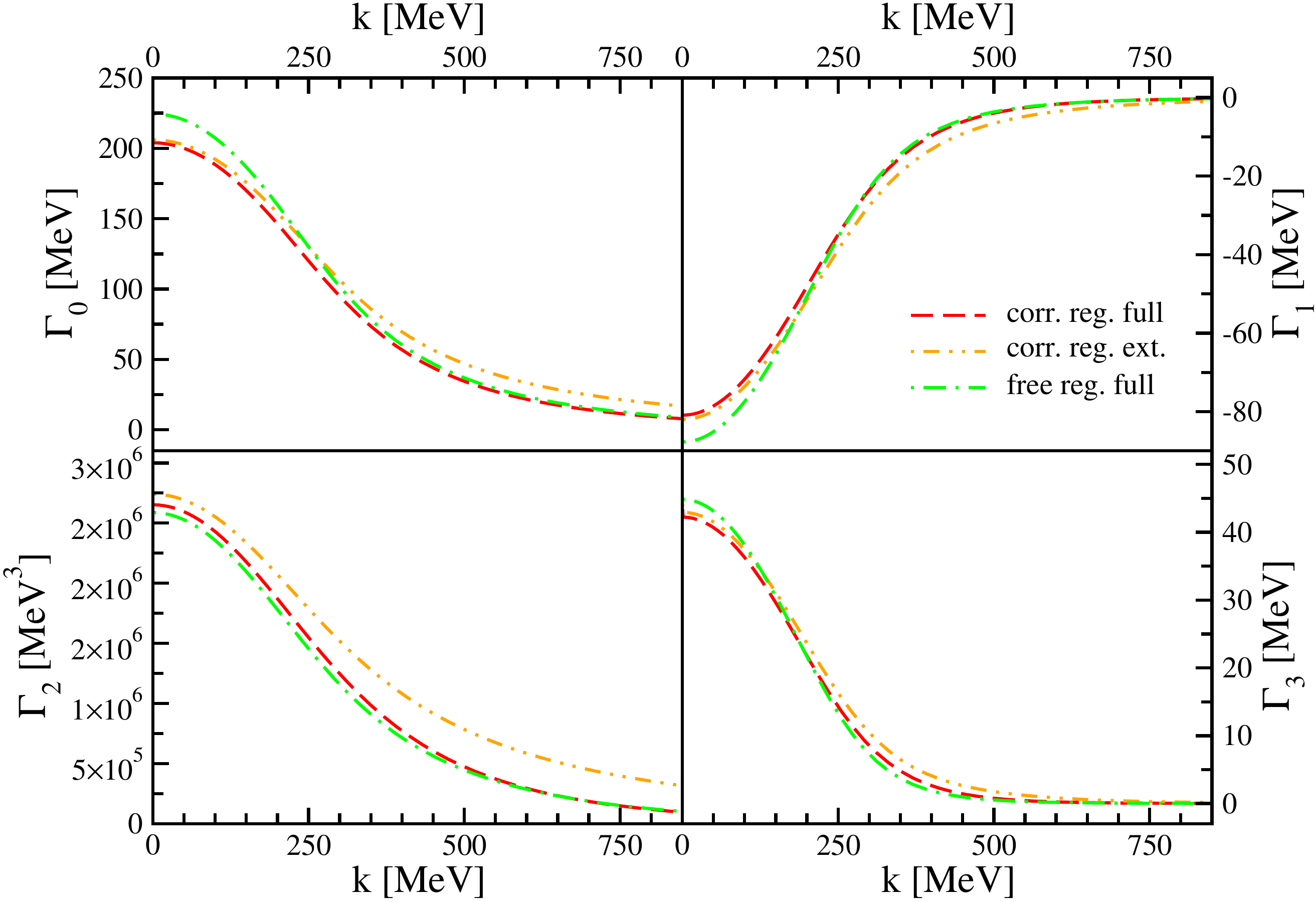}
\caption{External relative momentum dependence for the integral gamma functions of Eqs.~(\ref{gamma0})-(\ref{gamma3}). Red-dashed lines correspond to calculation performed with a correlated momentum distribution function weighed with the full regulator function, Eq.~(\ref{reg_fun}) (corr. reg. full). Orange double-dot-dashed lines correspond to the case with a correlated distribution function and an external regulator (corr. reg. ext.). Green dot-dashed lines depict the case with a free momentum distribution function and the full regulator (free reg. full).}
\label{gamma_funct}
\end{center}
\end{figure}

The explanation for this behavior could be drawn from the analysis of the integrated functions which are present in the averaging procedure, Eqs.~(\ref{gamma0}-\ref{i_integral}). 
%Following the previous conclusions, we expect a similar behavior for these functions when looking on one side at the correlated calculation with an external regulator, and on the other at the free propagator average but with full regulator. As can be observed 
In Fig.~\ref{gamma_funct} we plot, as functions of relative momenta, the quantities $\Gamma_0(k)$, $\Gamma_1(k)$, $\Gamma_2(k)$ and $\Gamma_3(k)$, given respectively in Eqs.~(\ref{gamma0}), (\ref{gamma1}), (\ref{gamma2}) and (\ref{gamma3}). 

For all the depicted functions, the main effect of using the external regulator is to enhance the absolute value of the integrals for all relative momenta. Consequently the effect observed on the partial waves, Fig.~\ref{reg1vsreg2}, is due to this enhancement. This increased value for the functions plotted in Fig.~\ref{gamma_funct} is to be expected given that the external regulator doesn't weigh the internal integrated momenta, providing larger values for these integrated quantities. 

In contrast, the effect of using a free propagator in the integrating procedure results in a non unique behavior. For the $\Gamma_0(k)$ the effect is to increase the value of the integral for small momenta, while the effect is the inverse for the $\Gamma_2(k)$. This modification dies for both functions with increasing relative momenta. For the $\Gamma_1(k)$ and $\Gamma_3(k)$, we observe equal modifications in the curves absolute values:  as a consequence of using a free propagator, higher values are seen for small  relative momenta and slight lower for intermediate ones. In this case, it is then not directly predictable how the interplay of these and the other integrated functions, (see for example Eqs.~(\ref{rho_f}), (\ref{i_integral}) and (\ref{G_0})), provides the effect observed on the partial waves, Fig.~\ref{freevsmed}. 

All in all, we can conclude that the effect of the density-dependent 2NFs is mostly given by the weight of momenta below the Fermi momentum. This implies variations on all the integrated quantities presented in the previous section and, consequently, the interplay of all of these provides the effect observed on the partial waves.

\subsection{Partial waves in neutron matter}

\begin{figure}
\begin{center}
\subfloat[]{
\includegraphics[width=0.85\textwidth]{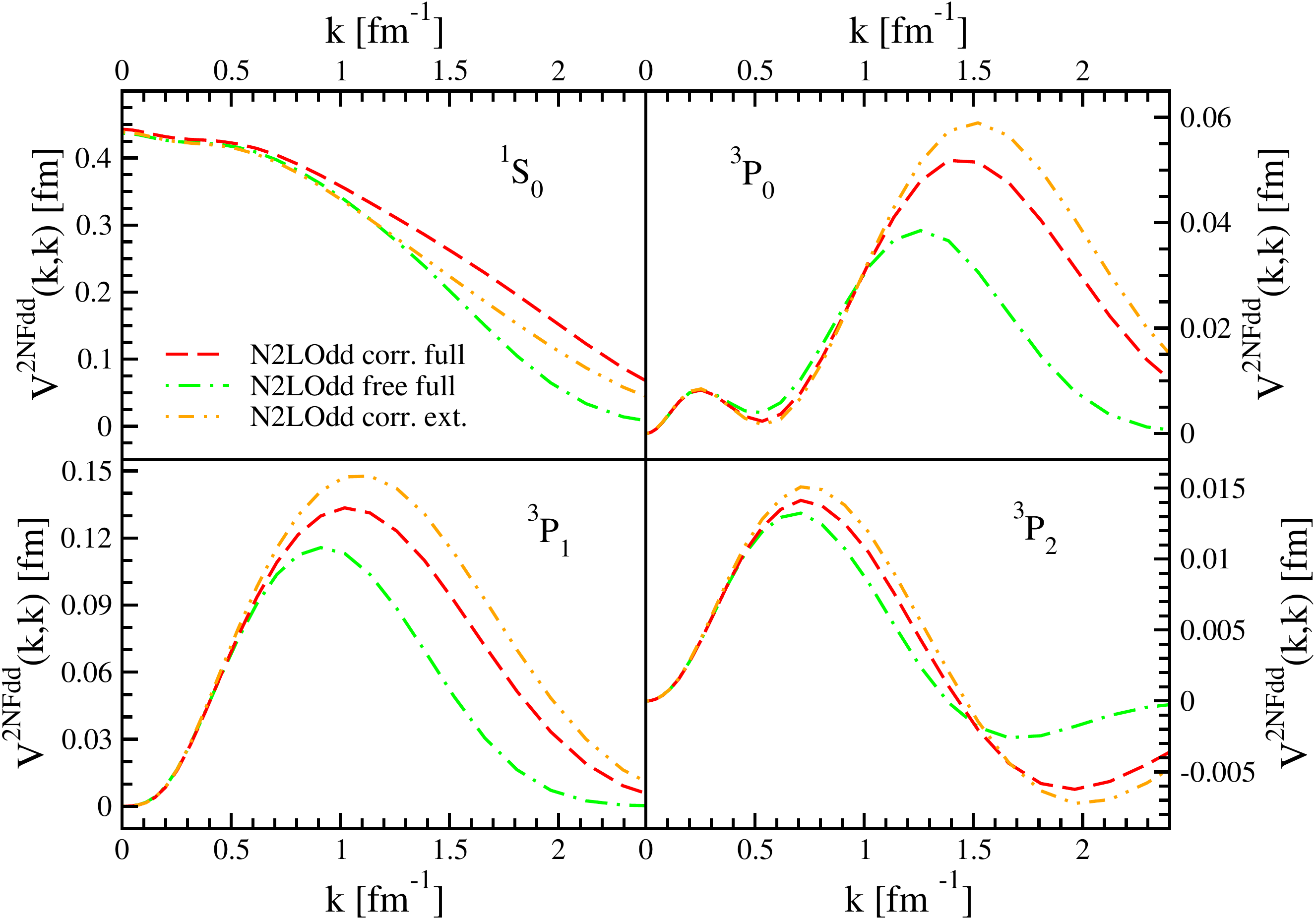}
\label{3b_pnm}
}
\hfill
\subfloat[]{
\includegraphics[width=0.85\textwidth]{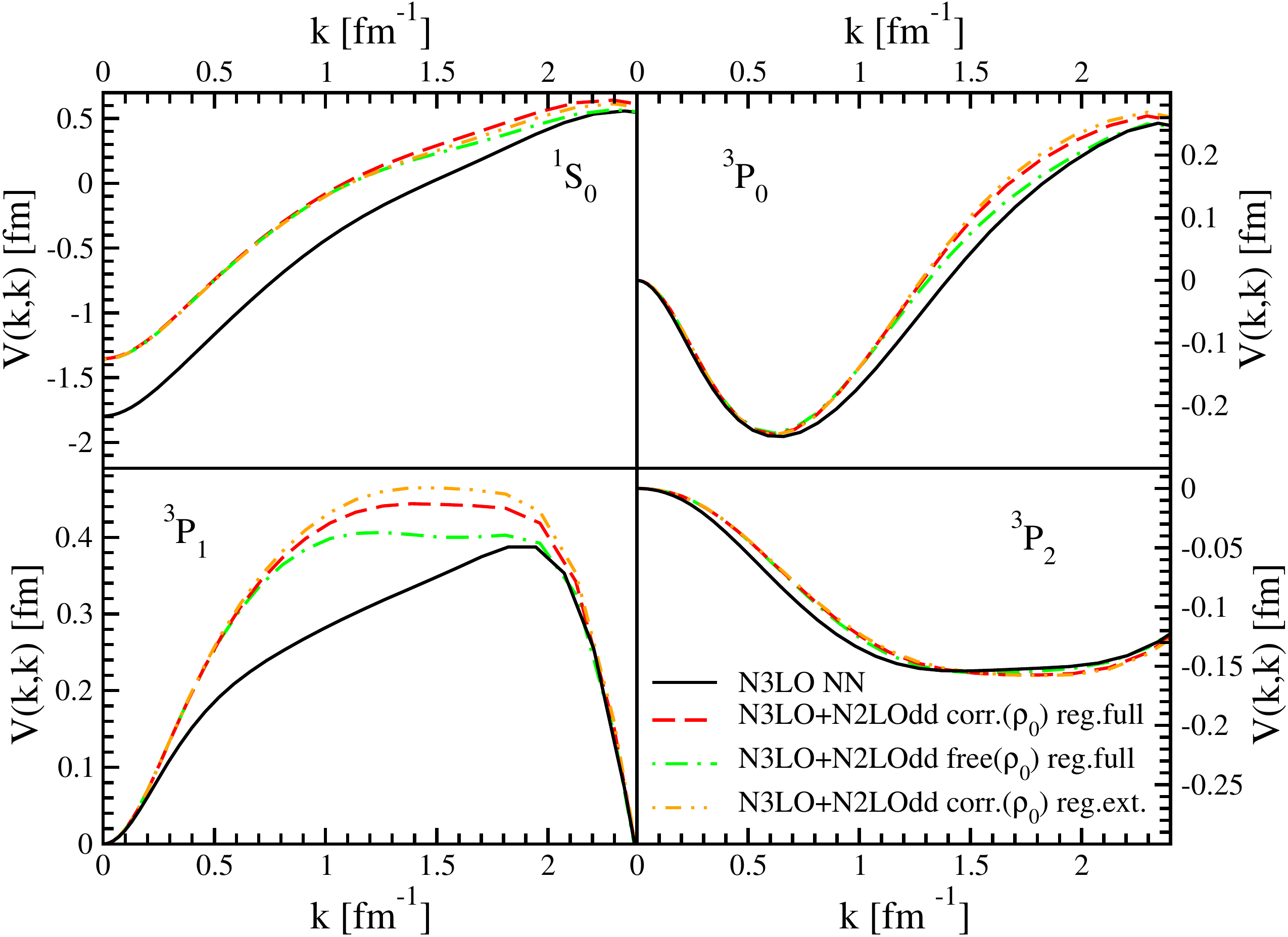}
\label{2b+3b_pnm}
}
\caption{$S$ and $P$ waves available in PNM for the total in-medium potential $\tilde V^\textrm{3NF}$ (upper panel \protect\subref{3b_pnm}), and for the sum of this to the NN N3LO (lower panel \protect\subref{2b+3b_pnm}). Red-dashed lines correspond to the in-medium NN contribution calculated with a dressed nucleon propagator and full regulator (corr. reg. full). Green-dot-dashed lines correspond to the case with a free propagator and full regulator (free reg. full). Orange double-dot-dashed lines correspond to the case with a dressed propagator and external regulator (corr. reg. ext). Density-dependent terms are obtained at empirical saturation density of SNM, $\rho_0=0.16$ fm$^{-3}$.}
\label{pw_pnm}
\end{center}
\end{figure}

We will now proceed with the analysis of the in-medium interaction in the specific case of totally asymmetric isospin matter. In the case of pure neutron matter, some partial waves are blocked in the antisymmetrization procedure of the two-neutron wave function, i.e. they can only be in total $T=1$ isospin state. We plot in Fig.~\ref{pw_pnm} the dominant partial wave matrix elements in PNM.
In Fig.~\subref*{3b_pnm} we present the curves for the complete density-dependent potential $\tilde V^\textrm{3NF}$ calculated at the empirical saturation density value, $\rho_0=0.16\,\textrm{fm}^{-3}$. In a similar fashion to the analysis we followed for the case of SNM, we present all cases of averaging procedures: the ones obtained with a correlated or a free momentum distribution function with full regulator and the correlated average but with external regulator. As a starting remark, we observe that for all averaging cases the contribution of the contracted 3NFs is a positive value, meaning that repulsion will be added to the bare NN N3LO potential. The only distinction is presented by the $^3P_2$ partial wave at high momenta. 

The dominant repulsive effect is observed for the $^1S_0$ partial wave, with values up to $0.44$ fm at zero relative momentum. This repulsion decreases with increasing momenta. The enhancement provided by the other waves decreases in order starting from the effect of the $^3P_1$, to the $^3P_0$ and finally to the $^3P_2$ partial wave, whose maximum effect is an order of magnitude less than the $^1S_0$ wave. 

It is interesting to observe that for all partial waves, except for the singlet $^1S_0$, the modifications due to the different averaging procedures seem to be a consequence of the availability of the complete range of SP momenta, not only of those below $p_\textrm F$. In other words, if states over $p_\textrm F$ are present, the 3NF has a higher repulsive effect, if furthermore states are not internally regulated the effect is even stronger. Nevertheless the dominant effect for the potential modification comes from the sinlget $^1S_0$, in which this behavior is not respected. In this partial wave, on the contrary, the biggest modification is observed in repulsion, at high relative momenta, for the correlated average with full regulator. The modification decreases for the same average but with external regulator, and decreases even more with the free average. We can derive the same conclusion already stated in the previous section, whereby the effect is due to an interplay of all the integrated quantities playing in the averaging of 3NFs. However, when we will look at the total energy of the system, we will see that the comprehensive repulsive effect of 3NFs increases according to the same analysis that was drawn in the previous section, i.e. it depends on the availability of momentum states below the Fermi momentum.

In Fig.~\subref*{2b+3b_pnm}, we plot the effect of the density-dependent force summed to the NN N3LO potential. As expected, the major change is observed in the $^1S_0$ partial wave, where an increase in the 2B potential matrix element is observed, with values that range from 0.45 fm to 0.1 fm  in a spread of relative momenta of $k=0-2.5$ fm$^{-1}$. The other visible effect is in the $^3P_1$ wave, where the strongest repulsion is observed around $k=1$ fm$^{-1}$ with a positive value of 0.15 fm, more or less the 30\% of the effect of the previous partial wave. The effect is orders of magnitude smaller in the other two waves.

The overall repulsive effect provided by the density-dependent force in all partial waves is what accounts for the enhancement of the total energy of PNM. In other words, what drives the stiffening of the equation of state of the isospin asymmetric system. As in the symmetric matter case, changes in pure neutron matter provided by the inclusion of the density-dependent 2NF will be of a repulsive kind for all densities. 

%Furthermore we will see how the small spread in energy curves due to the different averaging procedures increases with density. It will be interesting to see that the high density behavior for these different curves in PNM, will reflect the same one already observed at high densities in SNM. In this density range, we will see that the correlated average with full regulator provides the less repulsive curve. This lead us to the assumption that the less repulsion is not a consequence of the available phase space, but rather depends mostly on the regulator function used in the averaging procedure. 

%From Fig.~\ref{pw_pnm}, it is also clear that calculations performed with an in-medium correlated nucleon propagator will provide more repulsive results, with respect to the ones obtained with a free propagator in the averaging procedure. This will be directly observable in the results for the total energy of pure neutron matter in the next chapter. We can conclude that the overall effect of using a dressed propagator provides opposite effect in symmetric nuclear matter and pure neutron matter.

\subsection{Non-diagonal momentum matrix elements}

All results presented in this chapter where obtained for the specific case of diagonal potential matrix elements in relative momentum space, that is equal relative incoming and outgoing momenta, i.e. $k=|{\bf k}|=|{\bf k'}|$. Complete off-diagonal expressions for the six density-dependent potential contributions are nevertheless necessary to perform nuclear structure calculations. We present in App.~\ref{chapter:dens_dep_terms} the complete expressions in momentum space for the six density-dependent terms, Eqs~(\ref{tpe_dd_1}), (\ref{tpe_dd_2}), (\ref{tpe_dd_3}), (\ref{ope_dd_1}), (\ref{ope_dd_2}) and (\ref{cont_dd}) . For the case of  non diagonal potential matrix elements in relative momentum space, $|{\bf k}|\neq|{\bf k'}|$, the general formulation given in Eq.~(\ref{on-shell_vnn}) for the NN potential acquires a further operatorial structure \citep{Erk1971}.  
Instead of performing the partial wave decomposition of the potential for all non-diagonal matrix elements, in the present thesis we decided to follow the approach presented in Refs.~\citep{Kaiser1997,JWHol2009,JWHol2010}. We have performed the average only for diagonal momentum matrix elements. The off-diagonal elements are then extrapolated from the diagonal ones, with the substitution $k^2\rightarrow(k^2+k'^2)/2$. 

In Ref.~\citep{JWHol2009}, it is observed that the strongest off-diagonal dependency is provided by the  vertex correction to the $1\pi$ exchange 2B term which includes the presence of the medium, i.e. $\tilde V_\textrm{TPE-2}^\textrm{3NF}$ in Eq.~(\ref{tpe_dd_2}). This term decreases the value of the bare 2B $1\pi$ exchange in $S$ waves, it has hence an attractive modification (see also Fig.~\subref*{pot_123_S}, upper panels). We can then consider the bare NN $1\pi$ exchange contribution:
\beq
V^\textrm{2NF}_\textrm{OPE}=-\frac{g_A}{4F_\pi^4}\bd\tau_1\cdot\bd\tau_2
\frac{\bd\sigma_1\cdot{\bf q}\bd\sigma_2\cdot{\bf q}}{q^2+M_\pi^2}\,,
\label{2nf_ope}
\enq 
and compare its off-diagonal dependency in $S$ waves to the one obtained for  the $\tilde V_\textrm{TPE-2}^\textrm{3NF}$ contribution, applying the outlined off-diagonal extrapolation. When performing the off-diagonal partial wave analysis of the NN OPE term of Eq.~(\ref{2nf_ope}) no approximation is performed \citep{Erk1971,JWHol2010}.

\begin{figure}
\begin{center}
\includegraphics[width=0.7\textwidth]{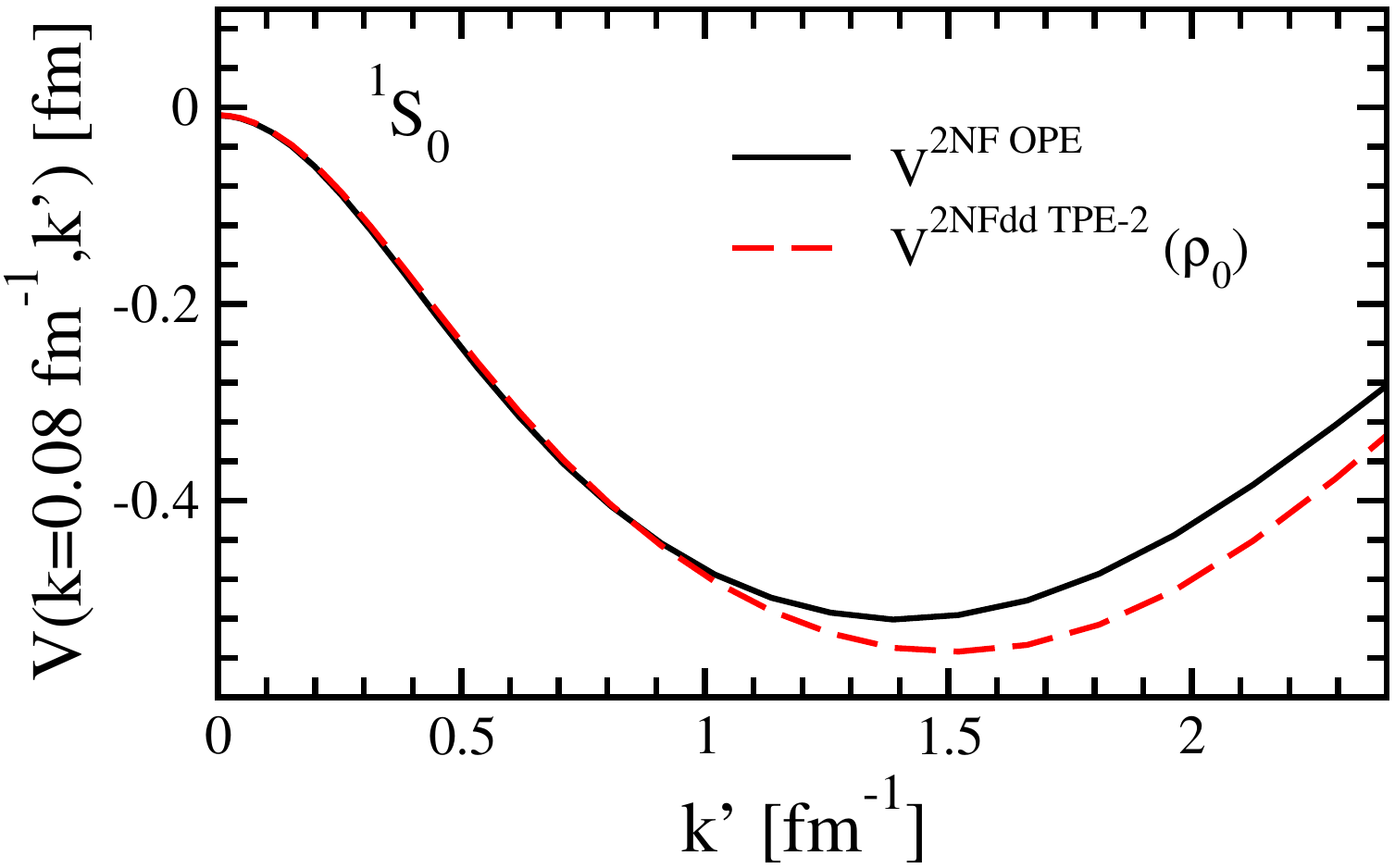}
\caption{Off-diagonal potential matrix elements for the $^1S_0$ partial wave, for a fixed incoming relative momentum $k=0.08$ fm$^{-1}$.  The black-solid line corresponds to the bare OPE NN potential. The red-dashed line corresponds to the $\tilde V_\textrm{TPE-2}^\textrm{3NF}$ obtained at density  $\rho_0=0.16\,\textrm{fm}^{-3}$. }
\label{nodiag_pw}
\end{center}
\end{figure}

As depicted in Fig.~\ref{nodiag_pw}, the qualitative behavior of the two matrix elements, presented for a fixed incoming relative momentum $k=0.08$ fm$^{-1}$ and as a function of the outgoing momenta $k'$, is in good agreement (note that the NN OPE term is plotted with an overall negative sign). As already discussed previously, we observe that the vertex correction to the 1$\pi$ exchange term,  $\tilde V_\textrm{TPE-2}^\textrm{3NF}$, has an attractive effect (see also Fig.~\subref*{pot_123_S}). 

Therefore, in the results of the next chapter, we employ this off-diagonal momentum extrapolation for all potential matrix elements.

\clearpage{\pagestyle{empty}\cleardoublepage}

%%%%%%%%%%%%%%% Results %%%%%%%%%%%%%%%%%%

\chapter{Nuclear and Neutron matter with three-body forces}
\label{chapter:results}

In this chapter we present the results obtained for microscopic and macroscopic properties of SNM and PNM. We will use the extended SCGF approach to include consistently 3BFs and analyze the effect of this inclusion in the properties of the many-body system.

Using a variety of many-body approaches \citep{Mut2000,Dew2003,Bal2012}, it has been demonstrated that saturation properties of nuclear matter fail to be reproduced whatever realistic 2NF is used in the calculation. This deficiency can be cured by the addition of 3NF in the Hamiltonian. In fact, it is proved that the inclusion of three-body forces has a crucial effect on bulk properties of both symmetric nuclear and pure neutron matter \citep{Akm1998,Zuo2002Aug,Li2006,Som2008,Heb2010Jul,Heb2011,Lov2012,Car2013Oct,Koh2013,Hag2013,Coraggio2014}.

Following the partial wave analysis of the potential matrix elements presented in the previous chapter, we will observe the striking effect of the inclusion of the density-dependent force for the total energy of the many-body ground state. Exploiting the extended SCGF approach presented in this thesis, we will see how the modifications induced by the 3B force are larger as the density increases. This density dependence is what provides the saturation mechanism for nuclear matter. In pure neutron matter, 3NFs are the main cause for the stiffening of the equation of state. In view of astrophysical studies for neutron star masses, this stiffening is a major ingredient for the achievement of theoretical results which can better match recent astrophysical observations \citep{Dem2010,Ant2013}. We will furthermore analyze the variation of the symmetry energy due to the inclusion of 3B forces. The symmetry energy determines to a large extent the composition of $\beta$-stable  matter and therefore the structure and mass of neutron stars \citep{Sch2006}. Its density dependence is a crucial ingredient to understand many properties of isospin-rich nuclei and neutron stars \citep{BALi2008,Ste2005}. A detailed study of this quantity from fully microscopic approaches \citep{Vid2009,Car2012,Heb2013Jul}, can be seen as a safe and necessary alternative to the variety of phenomenological methods which yield a large dispersion in the symmetry energy values.

As a starting point, we will first analyze in this chapter the microscopic properties. In the Green's functions method, the knowledge of the spectral function gives direct access to the calculation of microscopic quantities, such as the momentum distribution function. We study the behavior of the self-energy, separating the real and imaginary parts. Looking at the real part of the self-energy, it will be clear how the inclusion of the density-dependent 2NF influences the momentum dependency of the quasi-particle spectra. Conversely, we will observe how the inclusion of the density-dependent force has very little effect on the spectral function itself and hence on the momentum distribution. 

\begin{figure}
  \begin{center}
  \subfloat[]{\label{snmtdep}\includegraphics[width=0.47\textwidth]{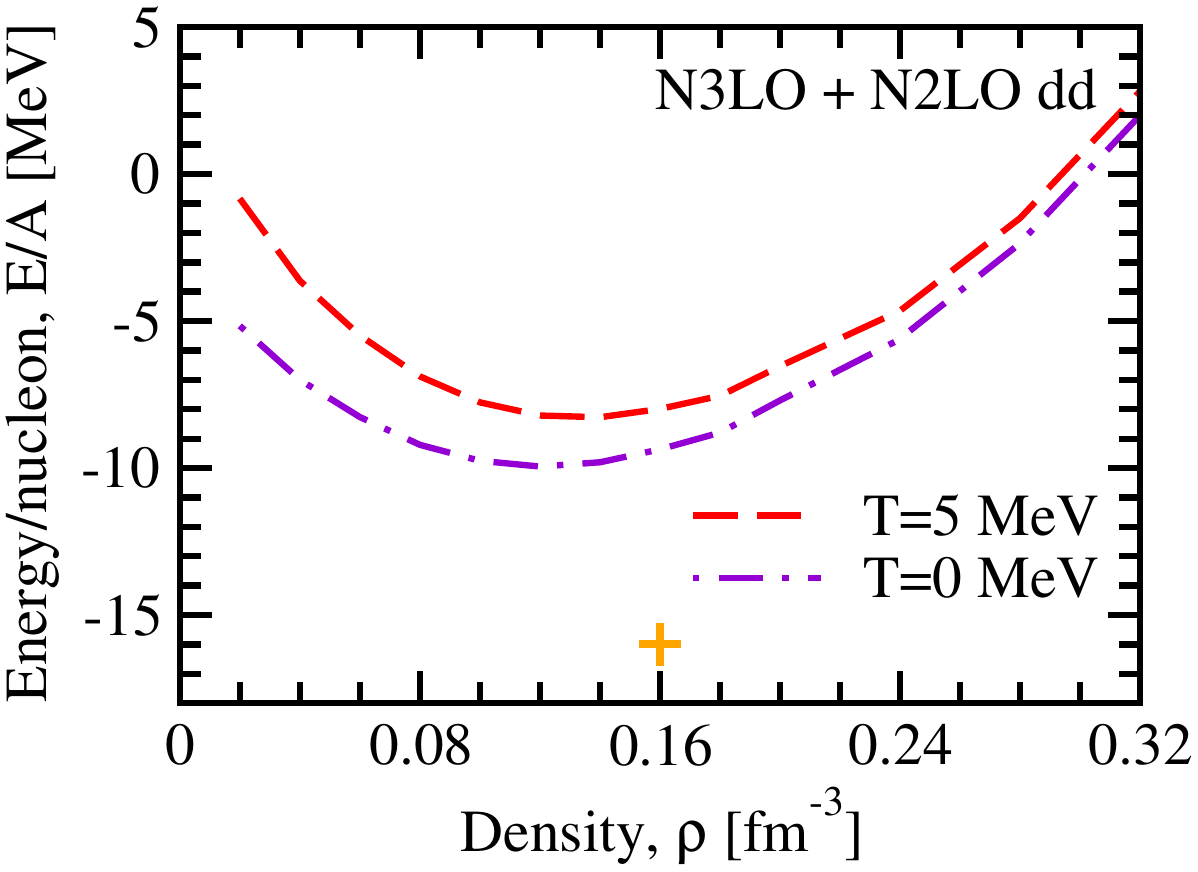}}
  \hspace{0.5cm}
  \subfloat[]{\label{pnmtdep}\includegraphics[width=0.46\textwidth]{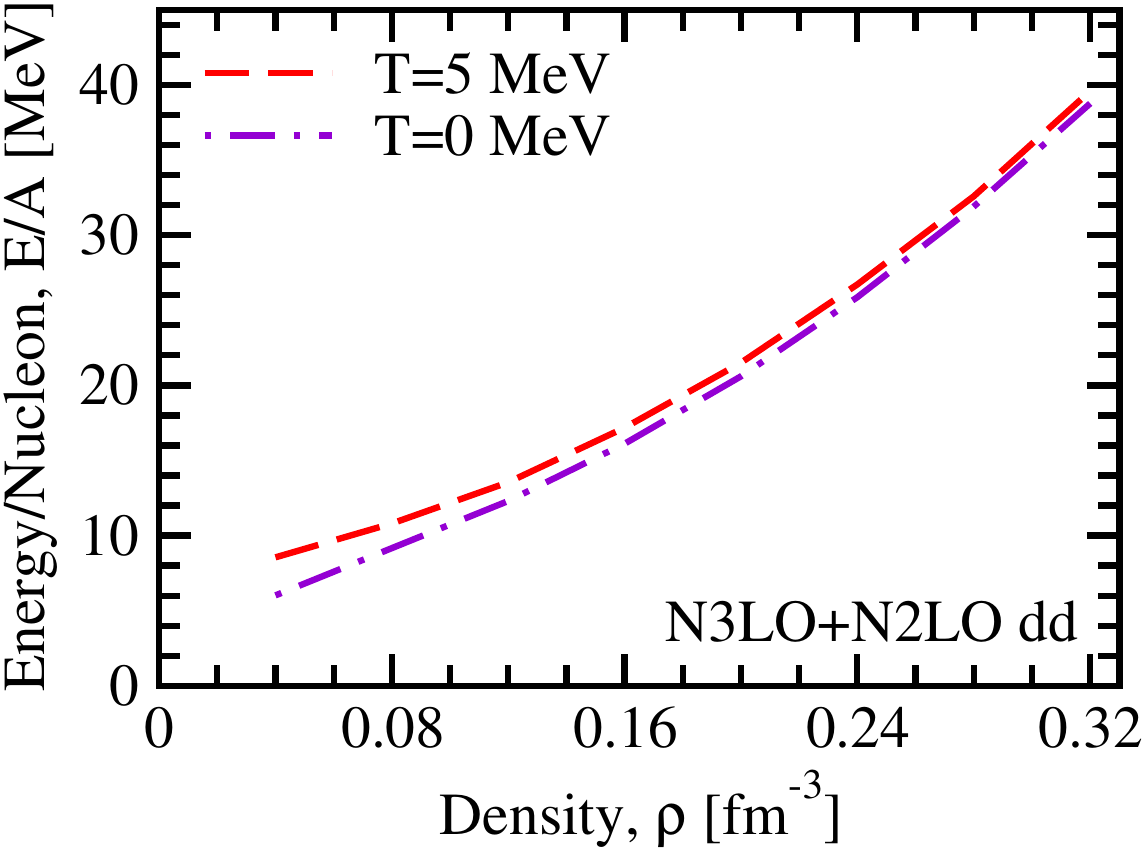}}
  \caption{Energy per nucleon for SNM, panel \protect\subref{snmtdep}, and PNM, panel \protect\subref{pnmtdep}. Results include NN N3LO and density-dependent N2LO, in the correlated average version and with full regulator. Red-dashed lines display results obtained at $T=5$ MeV. Dot-dashed-violet lines display the extrapolated $T=0$ MeV results as explained in the text. The orange cross in panel \protect\subref{snmtdep} defines the empirical energy/density saturation point, $E_0=-16$ MeV and $\rho_0=0.16$ fm$^{-3}$.}
  \label{T_dep}
  \end{center}
\end{figure}

Before going into the details of our calculations, let us first comment on the temperature dependence of the results we present. As explained in the introduction to the thesis, calculations are performed at finite temperature to avoid pairing instabilities. Thermal effects can be estimated using the Sommerfeld expansion \citep{Rio2009Feb}. At low temperatures, the Sommerfeld expansion indicates that the effect of temperature is quadratic for both the energy and the free energy but with opposite sign. These read respectively $e\sim e_0+a_eT^2$ and $f\sim f_0+a_fT^2$, where $a_e=-a_f$ and $e_0=f_0$ \citep{Rio2009Feb}. Consequently, the semi-sum of both quantities is an estimate of the zero-temperature energy. This expansion is only valid if $T/\vep_\textrm F\ll 1$, which is the case under study. 
%It can be demonstrated than under these conditions, the energy per nucleon of the system can be expressed in the Sommerfeld expansion as \citep{Rio2007PhD}:
%\beq
%\frac{E}{A}\sim\left(\frac{T}{\vep_F}\right)^2+C
%\label{somm}
%\enq
In other words, as the density gets higher for a fixed $T$, the energy temperature dependence gets milder \citep{Rio2007PhD}.

In Fig.~\ref{T_dep} we plot, for symmetric nuclear and pure neutron matter, the density dependence of the total energy per nucleon for two different temperatures. The $T=5$ MeV corresponds to the calculation of the energy via the GMK sum rule, once self-consistency for the spectral function is achieved. The $T=0$ MeV curve is given by the semi-sum of total energy and total free energy per nucleon both obtained at $T=5$ MeV. Calculations are performed with the use of the 2B N3LO force and including the density-dependent force presented in the previous chapter, performing the correlated average with full regulator. For SNM, temperature drives an enhancement of less than 5 MeV at very low densities for results calculated at $T=5$ MeV. As expected from the Sommerfeld expansion, the thermal repulsion is reduced with density, approaching approximately 1 MeV around empirical saturation density, and further diminishes as the density increases. A similar behavior is observed for pure neutron matter, here the spread of the curves due to the temperature is already small at low densities, $\sim2$ MeV, and quickly decreases as density increases. 

We want to stress once again that all expressions, for both microscopic and macroscopic properties, discussed in chapter 2 and 3, were derived at zero temperature. In the following, in order to prevent the pairing instability and also due to practical advantages in the construction of the numerical code, all microscopic and bulk results will be derived at temperature $T=5$ MeV, unless otherwise stated.

\subsubsection{Numerical inclusion of 3BFs in the SCGF approach}

\begin{figure}[t]
\begin{center}
\includegraphics[width=0.95\textwidth]{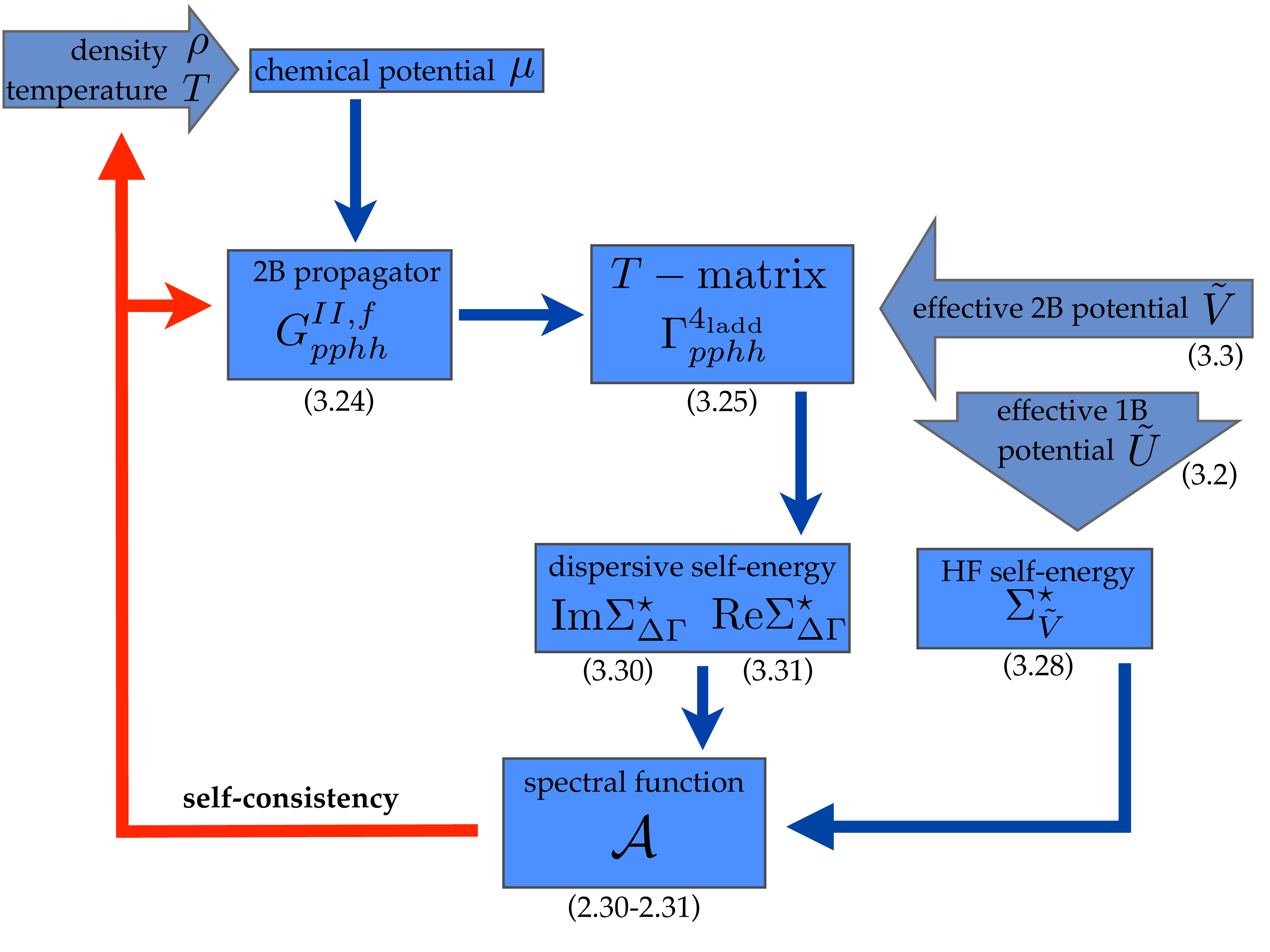}
\caption{The structure of a ladder SCGF calculation with 3BFs. The numbers below each box represent the relevant equations associated to it.}
\label{num_impl}
\end{center}
\end{figure}

A detailed description of the numerical implementation of the complete SCGF method at finite temperature when considering only 2NFs can be found in \citep{Rio2007PhD}. The code has been written in FORTRAN programming language. 

We show in Fig.~\ref{num_impl} a schematic representation of how the code works when considering both 2B and 3B forces. The thick external arrows correspond to the input quantities: the density and temperature of the system on one side, and the nuclear potential on the other. Further on, we will explain how the potential has to be included when considering 3BFs. The external density and temperature of the system, together with the spectral function obtained from the previous iterative step, are needed to calculate the chemical potential \citep{Rio2007PhD}. Once this is known, the first-order 2B GF can be calculated. This quantity, together with the nuclear potential chosen, are used to obtain the $T$-matrix, i.e. the ladder approximation for the $\Gamma^{4-\rm{pt}}$ vertex function. Via the knowledge of the $T$-matrix, the dispersive part of the imaginary and real part of the SP self-energy can be evaluated. The HF part of the SP self-energy is calculated directly from the potential. Finally, the spectral function can be obtained and the procedure starts again until a consistent result is achieved. Some numerical details on the calculation the 2B GF, the $T$-matrix and the self-energy are given at the end of App.~\ref{chapter:dens_dep_terms}. 

\begin{figure}
\begin{center}
\includegraphics[width=0.7\textwidth]{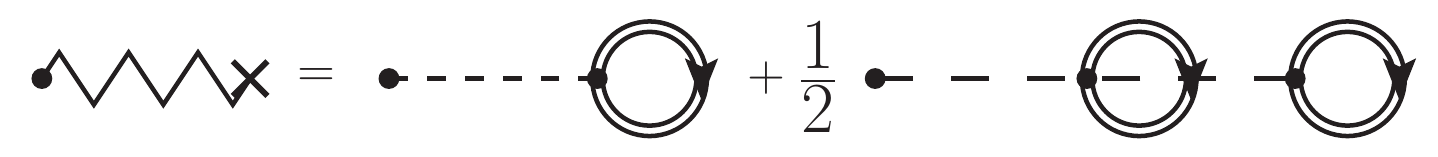}
\caption{Hartree-Fock term of the 1B effective potential depicted in Fig.~\ref{ueffective}, without considering the bare 1B potential.}
\label{ueffective_hf}
\end{center}
\end{figure}

Following the extension of the formalism presented in Chap.~\ref{chapter:3BF_formalism} and the evaluation of the density-dependent force $\tilde V^\textrm{3NF}$ given in Chap.~\ref{chapter:eff_2b_int}, we would like to focus here mostly on the inclusion of 3BFs in the existing code. Numerical details for the calculation of the density-dependent contributions are reported at the end of App.~\ref{chapter:dens_dep_terms}. 

In order to include correctly the 3B force, we add its contributions to the bare 2B potential, $V^\textrm{2NF}$, in two different forms:
\begin{itemize}
\item we define a $V^\textrm{NN}_{T}=V^\textrm{2NF}+\tilde V^\textrm{3NF}$ for the calculation of the $T$-matrix, defined in Eq.~(\ref{ladder_eq}). This potential corresponds to the effective 2B potential depicted in Fig.~\ref{veffective}. In Fig.~\ref{num_impl} this is represented by the external right arrow specifying ``effective 2B potential $\tilde V$".
\item we define a $V^\textrm{NN}_\textrm{HF}=V^\textrm{2NF}+\frac{1}{2}\tilde V^\textrm{3NF}$ for the calculation of the HF part of the self-energy, defined in Eq.~(\ref{self_en_hf}). The HF self-energy corresponds to the first-order term of the effective 1B potential depicted in Fig.~\ref{ueffective}, dropping the bare 1B part which, in the case under study, falls out of the calculation. We show this first-order term in Fig.~\ref{ueffective_hf}. In Fig.~\ref{num_impl} this is represented by the external right arrow specifying ``effective 1B potential $\tilde U$".
\end{itemize}
According to what was explained in Chap.~\ref{chapter:3BF_formalism}, the inclusion of the 3B contributions by means of this strategy ensures that all diagrams included in the self-energy are counted properly. 

\begin{figure}
\begin{center}
\includegraphics[width=0.3\textwidth]{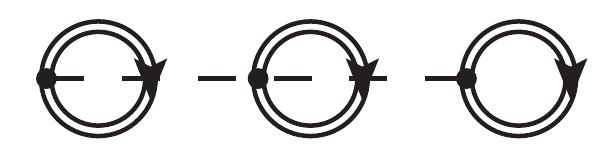}
\caption{Hartree-Fock approximation of the expectation value of the 3B interaction.}
\label{3b_hf}
\end{center}
\end{figure}

For the evaluation of the total energy via the modified GMK sum rule given in Eq.~(\ref{gmk_3b}), we need to evaluate the expectation value of the 3B operator. As already explained previously, in our calculations this term is obtained only a the HF level. We depict it in Fig.~\ref{3b_hf}. To obtain the 3B expectation value, we then perform an integration of the kind:
\beq
\frac{\langle W\rangle_\textrm{HF}}{A}=\frac{\nu}{\rho}\int \frac{{\mathrm d}{\bf p}}{(2\pi)^3}n({\bf p}) \frac{\Sigma^\star_{\tilde V^\textrm{3NF}}({\bf p})}{3}\,.
\label{3B_exp}
\enq
In the previous expression, the HF self-energy for the 3B part, i.e. $\Sigma^\star_{\tilde V^\textrm{3NF}}$, which is also shown in the second term on the right-hand side of Fig.~\ref{ueffective_hf}, has been computed from the 3B part of $V^\textrm{NN}_\textrm{HF}$, i.e. $1/2\tilde V^\textrm{3NF}$. Numerical details for the determination of the momentum distribution in Eq.~(\ref{3B_exp}) are given at the end of App.~\ref{chapter:dens_dep_terms}.

%%%%%%%%%%%%%% The self-energy %%%%%%%%%%%%%%%%

\section{The self-energy}
\label{section:self_energy}

As defined by Dyson's equation, Eq.~(\ref{Dyson}), the self-energy incorporates all interacting terms, for a specific approximation, which \emph{dress up} the SP propagator. For the ladder self-energy, we calculate  separately the imaginary and the real part.  We derived these contributions at zero temperature in Eq.~(\ref{imself_eq}-\ref{reself_eq}). The real part is derived from the imaginary part by means of a dispersion relation. As a matter of fact, as shown in Eq.~(\ref{Sp_self}-\ref{Sh_self}), a combination of the real and imaginary part of the self-energy is needed to obtain the spectral function.

\begin{figure}
\begin{center}
\subfloat[]{\label{imselfk0}\includegraphics[width=0.9\textwidth]{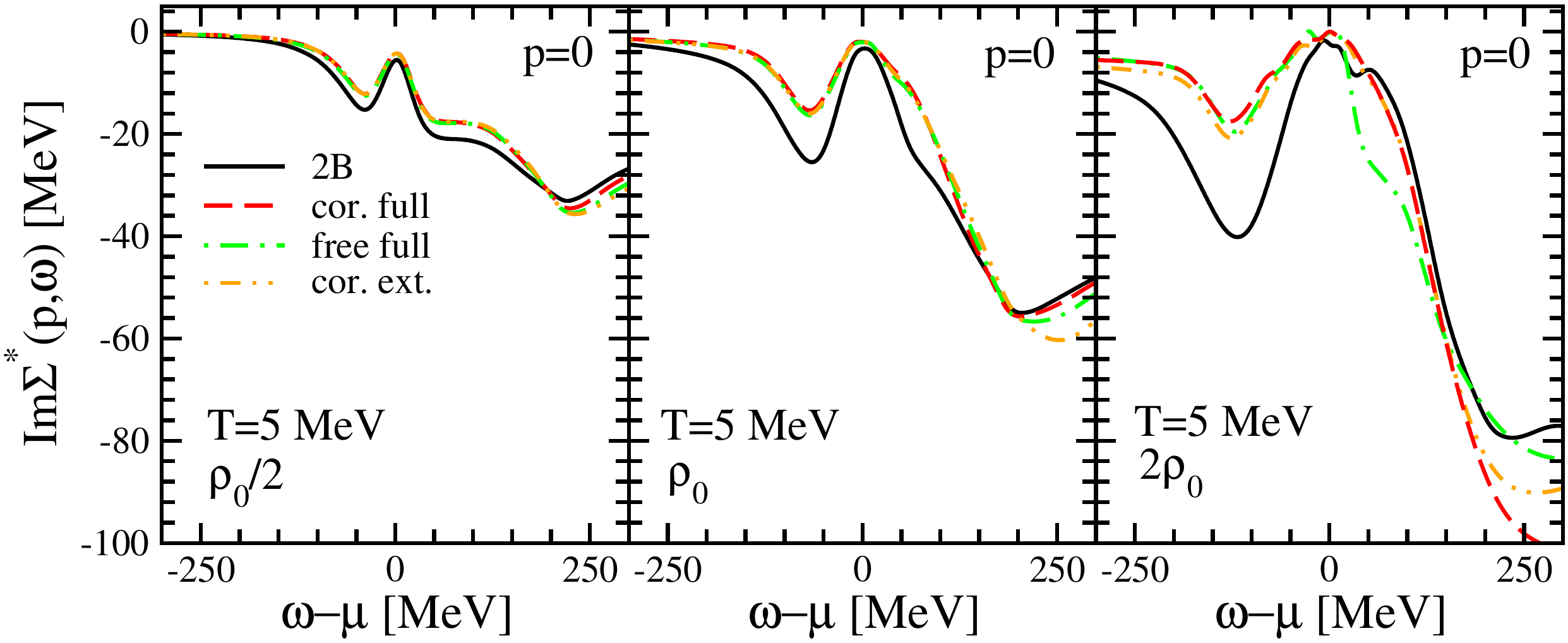}}
\hfill    \vskip .5cm
\subfloat[]{\label{imselfkF}\includegraphics[width=0.9\textwidth]{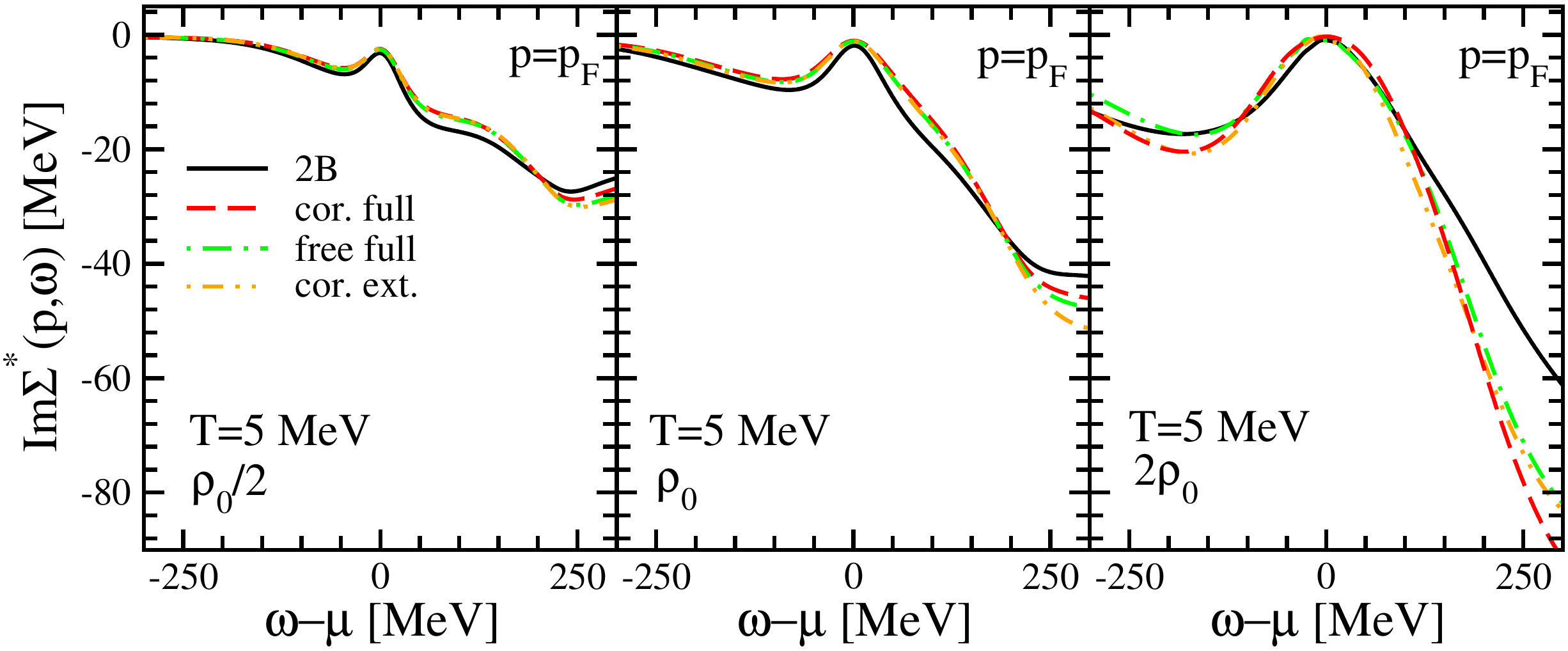}}
\hfill    \vskip .5cm
\subfloat[]{\label{imself2kF}\includegraphics[width=0.9\textwidth]{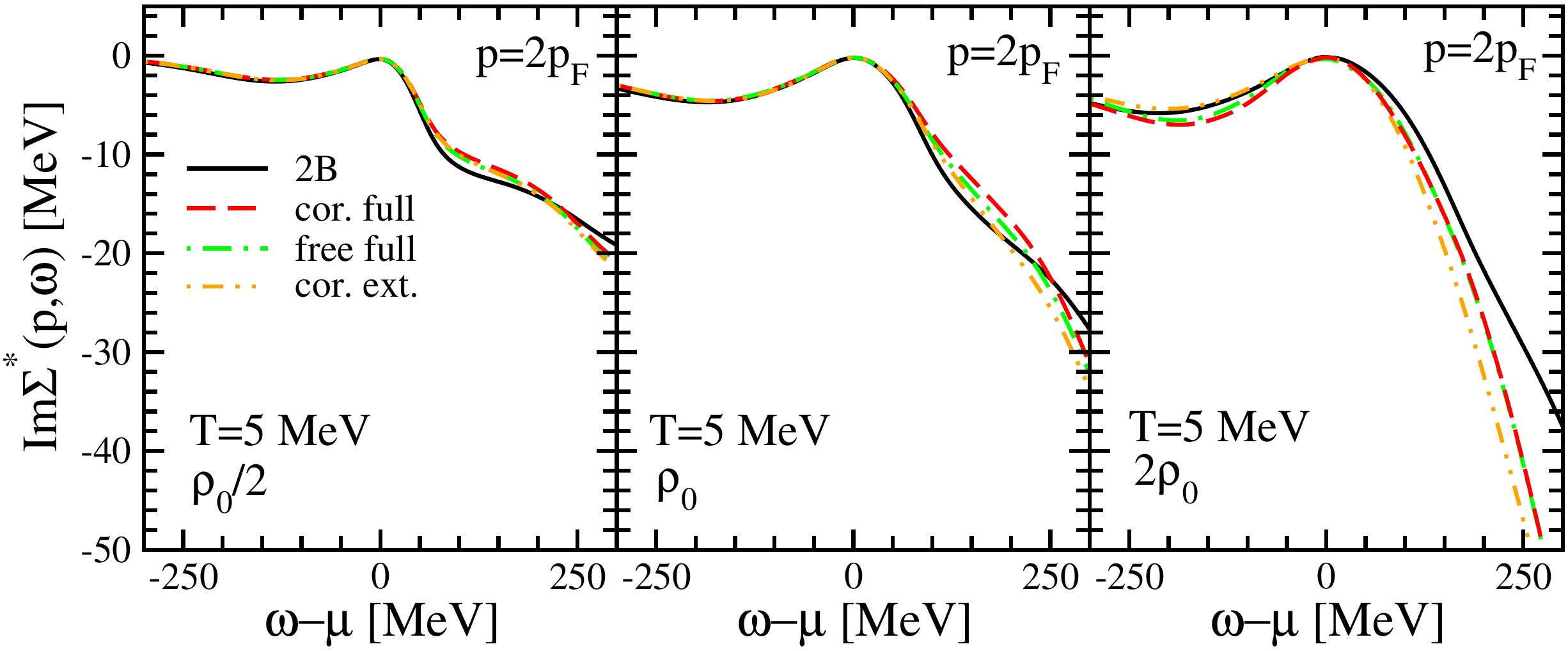}}
\caption{Imaginary part of the self-energy in SNM for SP momentum $p=0, p_\textrm{F}, 2p_\textrm{F}$ in the first, second and third rows respectively. In each row, panels going from left to right show the self-energies at densities $\rho$ = 0.08 - 0.16 - 0.32 fm$^{-3}$. Black-solid lines represent the NN N3LO calculation. Red-dashed, green-dot-dashed and orange-double-dot-dashed curves correspond to the inclusion of the 2NFdd, obtained, respectively, in the correlated and free version with full regulator, and in the correlated version with external regulator.}
\label{imself}
\end{center}
\end{figure}

In Fig.~\ref{imself}, we present  the imaginary part of the ladder self-energy obtained at $T=5$ MeV. For a deeper comprehension of the qualitative and quantitative behavior of the imaginary part of the self-energy, its mathematical expression at finite temperature must be analyzed in detail \citep{Fri2004PhD,Rio2007PhD}. However, a general qualitative explanation can be given. While the imaginary part is mostly connected to absorption and to the available phase space, the real part of the self-energy is also related to mean-field effects. We analyze in Fig.~\ref{imself} its behavior for three different SP momenta, $p=0, p_\textrm{F}, 2p_\textrm{F}$, and for increasing density, $\rho=0.08, 0.16, 0.32$ fm$^{-3}$. In the figure we compare the self-energy when only the 2B N3LO force is considered, to the case when the density-dependent N2LO interaction is included. The inclusion of the density-dependent force is presented in the three different cases: correlated and uncorrelated average with the use of the full regulator as well as the correlated average with the external regulator, as explained in the previous chapter.

We note  that the inclusion of the contracted 3B force doesn't strongly affect the overall behavior of the imaginary part of the self-energy. If we look at momentum $p=0$, we recognize, around energies close to the chemical potential $\mu$, the parabolic shape predicted by Luttinger for the zero temperature case, i.e. $\textrm{Im}\Sigma(p,\om)\sim a(\om-\mu)^2$ \citep{Luttinger1960}. Actually, the imaginary part of the self-energy approaches zero for $\om=\mu$ as the density increases. The fact that the system resembles the zero temperature behavior is because the higher the density the stronger the degeneracy of the system, and temperature effects are washed out. For a fixed density, if we go to higher momenta, $p=p_\textrm{F}$ or even more at $p=2p_\textrm{F}$, the cusp induced by the parabolic behavior decreases. Here we see that the imaginary part of the self-energy flattens for energies close and below $\mu$. As Luttinger's argument is independent of the force, the inclusion of the density-dependent force generates no variation in the position of the cusp. On the contrary, mainly at zero momentum, 3NFs generate a smaller absolute value in the minima observed in the imaginary self-energy, for energies close and below the chemical potential, with respect to the 2B results. As we go to higher densities, the difference between the values at the minima obtained with the two cases increases. This is caused by the fact that the energy values at the minima in the 2B case become more negative as the density increases, while the ones including 3NFs remain approximately unaltered for all densities. For $p=0$, this spread goes from $\sim 2$ MeV at $\rho_0/2$ to approximately $20$ MeV at double saturation density. For higher momenta the behavior is somewhat different, with both minima, obtained with and without 3NF, increasing to more negative values as the density grows. At $2\rho_0$ the density-dependent force generates a few MeV more negative results for the imaginary part of the self-energy.  In general, the minima flattens for all densities as the momenta increases. Nevertheless, the fact that the value of the minima changes is a consequence of the combination of both phase space availability and the force itself. The overall behavior of the imaginary part of the self-energy at energies $\om<\mu$, below the minima, is left unaltered by the inclusion of 3BFs, for all densities and momenta, showing a negligible value at energies already around $250$ MeV below the chemical potential. 

\begin{figure}
\begin{center}
\includegraphics[width=0.9\textwidth]{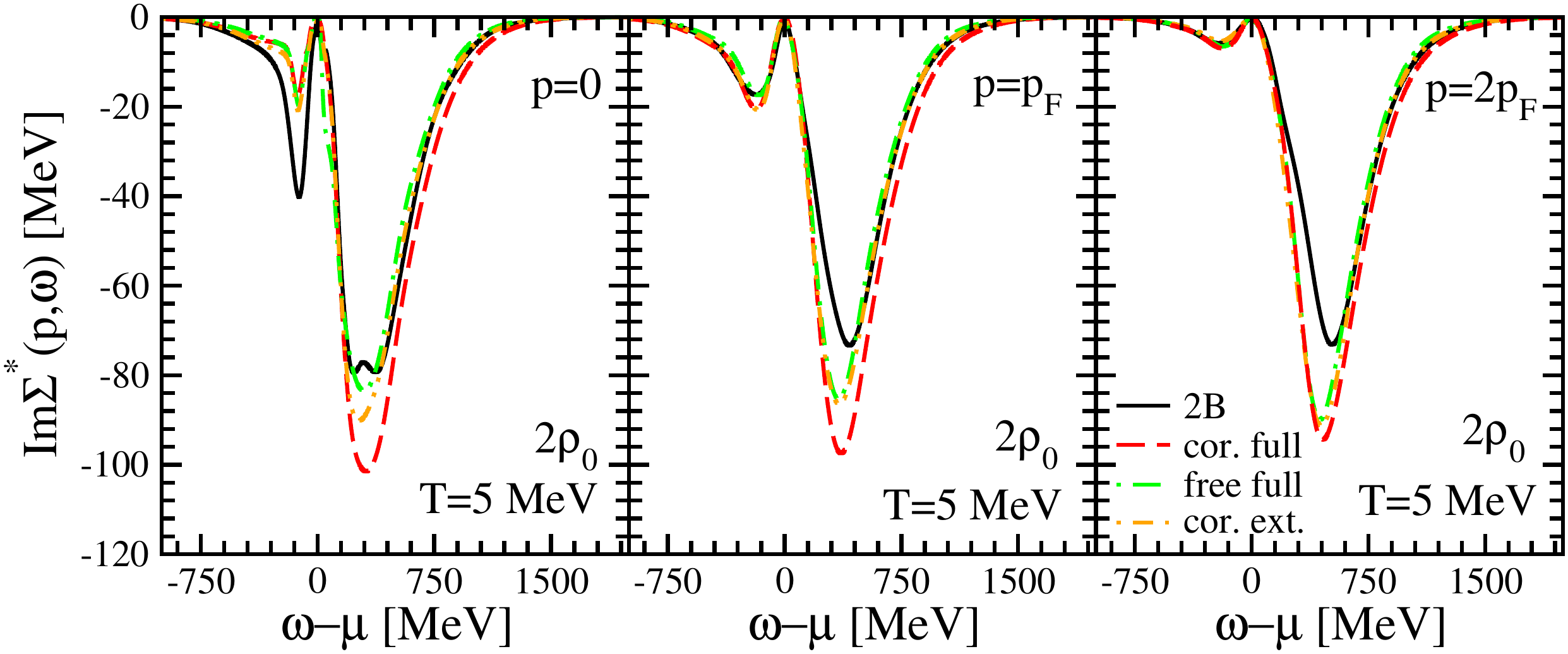}
\caption{Imaginary part of the self-energy in SNM for density $\rho=0.32$ fm$^{-3}$, for momentum $p=0, p_\textrm{F}, 2p_\textrm{F}$ in panels going from left to right. Black-solid lines depict the NN N3LO calculation. Red-dashed, green-dot-dashed and orange-double-dot-dashed curves represent the inclusion of the 2NFdd, obtained, respectively, in the correlated and free version with full regulator, and in the correlated version with external regulator.}
\label{imself_hightail}
\end{center}
\end{figure}

Looking at energies close but higher than the chemical potential, the imaginary self-energy presents a bump around $\sim100$ MeV, for all momenta, which disappears  with increasing density. The three-body contracted force induces smaller absolute values around the bump, but conversely results fall below the 2B-only calculation when the self-energy reaches the minimum observed in the region $\om>\mu$. As the density increases, for a fixed momenta, this falling to more negative results, for calculations performed with the 2NF density-dependent, becomes steeper. In the region of energies $\sim200-400$ MeV, especially for density $2\rho_0$, is where we observe the strongest dependence on the averaging/regulator procedure for the construction of the density-dependent force. As it can be seen in Fig.~\ref{imself_hightail}, where we focus on this high density region, we obtained higher absolute values for the minimum when the correlated average procedure with full regulator is performed, with respect to the other procedures. This calculation shows at the minimum a somewhat stable separation of $\sim 20$ MeV, observed for all momenta, with respect to the two-body only calculation. Conversely, results including the density-dependent force obtained with a correlated average but with the external regulator or with a free average with the use of the full regulator, present a variation in the position of the energy minimum with increasing momenta. In these two latter cases, the self-energy value at the minimum is at first close to the 2B calculation one, and departs from it with increasing momentum. Simultaneously, these minima approach the more negative result obtained with the complete calculation of the density-dependent 2NF. We will see in the next section that it is exactly in this high-density region where the main variation for the spectral function appears. All in all, except for this high-density high-energy region, we can conclude that the modifications induced by the different averaging procedures are negligible. From a qualitative point of view, the fact that modifications due to the different averaging procedures appear mostly at high densities, is a direct consequence of what we concluded from the partial wave analysis. In fact, we stated that the difference in between the averages is mainly due to how momenta below the Fermi momenta are treated. As the density increases, the region of momenta below $p_\textrm F$ is bigger, consequently the effect is enhanced. 

Furthermore, the low and high energy tails of the imaginary part of the self-energy are related to off-shell effects, which are determined by the short-range part of the potential. The inclusion of three-body forces seems to alter the behavior of this part of the potential especially at high densities, where we observe a somehow enhanced, in absolute values, high-energy tail. In some sense, we will observe this effect when analyzing the depletion of the momentum distribution in the next section, which will result more depleted at high densities when considering 3BFs. This is in agreement with a stronger off-shell behavior when including 3BFs.

\begin{figure}
\begin{center}
\includegraphics[width=0.9\textwidth]{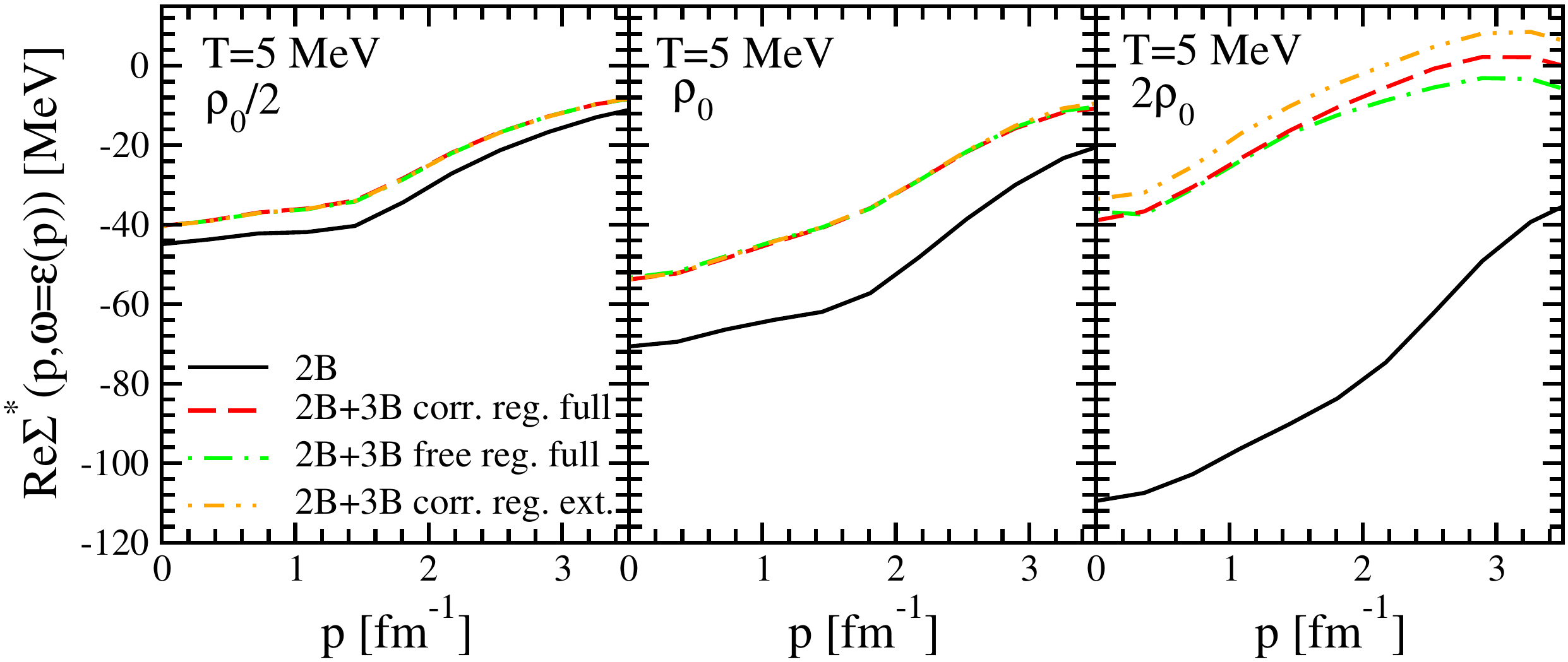}
\caption{Quasi-particle potential in SNM calculated at densities $\rho$ = 0.08 - 0.16 - 0.32 fm$^{-3}$ respectively in panels going from left to right. Black-solid lines represent the NN N3LO calculation. Red-dashed, green-dot-dashed and orange-double-dot-dashed curves correspond to the inclusion of the 2NFdd, obtained, respectively, in the correlated and free version with full regulator, and in the correlated version with external regulator.}
\label{qp_pot}
\end{center}
\end{figure}

A much more visible consequence of the density-dependent 2NF can be detected at the quasi-particle energy. This is obtained from the real part of the self-energy. The quasi-particle energy is obtained at each momentum as the solution of the following self-consistent equation:
\beq
\vep_{qp}(p)=\frac{p^2}{2m}+\textrm{Re}\Sigma^\star(p,\vep_{qp}(p))\,.
\label{qp_spectra}
\enq
The second term on the right-hand side of Eq.~(\ref{qp_spectra}) is shown in Fig.~\ref{qp_pot}. This corresponds to the potential part of the quasi-particle energy. We show it for three different densities, $\rho=0.08, 0.16, 0.32$ fm$^{-3}$. For all momenta, the contracted 3NF shifts the quasi-particle energy to more repulsive values. At zero momentum, the shift amounts to $\sim 5$ MeV for $\rho_0/2$, increases to $\sim 20$ MeV at saturation density $\rho_0$, and grows as high as $70$ MeV at double saturation density $2\rho_0$. For all densities, this shift gradually diminishes as momentum increases. At low-momenta this modification leaves mostly unaltered the qualitative behavior of the SP potential at $\rho_0/2$ and $\rho_0$, except for a repulsive effect in the entire momenta range. Conversely, at $\rho=0.32$ fm$^{-3}$, the strong repulsion at low-momenta, induced by the density-dependent 2NF, also affects the qualitative momentum dependence of the SP potential. At high densities, the quasi-particle potential shows an even stronger repulsion becoming positive at intermediate momenta. This strong density-dependent repulsive effect drives the saturation of the total energy of symmetric nuclear matter, as we will see in the following sections. Furthermore, the modification of the momentum dependence of the quasi-particle potential affects the value of the nucleon effective mass \citep{Som2009PhD}.

Variations caused by the different averaging procedures in the construction of the density-dependent force are negligible at densities $\rho_0/2$ and $\rho_0$. As expected from the analysis of the high density behavior of the imaginary part of the self-energy, modifications are mainly visible at $2\rho_0$. Here the use of the external regulator in the correlated average drives results for the quasi-particle spectrum which are $\sim 8$ MeV more repulsive at zero momentum with respect to the other averaging procedures. This repulsion is mainly maintained for all momentum with respect to the correlated average performed with the full regulator. The main qualitative variation is observed for the curve performed in the free average with the use of full regulator. The effect of weighing internal contracted momenta from 0 to $p_\textrm{F}$ and having no presence of momenta higher than the Fermi momentum, causes a more rapid decreasing of the quasi-particle potential at high momenta. This could be justified in a HF picture, where a sharp disappearance of momenta higher than $p_\textrm F$ increases the damping of the SP potential.  

%%%%%%%%%%%%%%%%% self-energy and spectral function%%%%%%%%%%%%%%%%%

\section{Spectral function and momentum distribution}
\label{section:spec_fun}

The SP momentum distribution, defined in Eq.~(\ref{mom_dist}) at zero temperature, provides a measure of the correlations embedded in the nuclear wave function. The spectral function, as shown in Chap.~\ref{chapter:formalism} (see Eqs.~(\ref{S_hole}-\ref{S_part})), describes the spread in energy of the probability to find single-particle states at energies other than the quasi-particle energy, and hence tests the presence of correlations in the system. This behavior can be experimentally seen in ($e,e'p$) knockout reactions on finite nuclei \citep{Mou1980,Ben1989,Dic2004}. 

While we presented in Eq.~(\ref{S_hole}) and Eq.~(\ref{S_part}) the formal expressions for the spectral functions at zero temperature of respectively the hole and particle part, at finite temperature the spectral function can be described by a unique quantity, the $A(p,\om)$ \citep{Bla1986}. The spectral function $A(p,\om)$ is related to the discontinuity of the SP propagator along the real axis (see Eq.~(\ref{G1B_Lehm})), and takes into account both effects coming from Eq.~(\ref{S_hole}) for the hole contribution and from Eq.~(\ref{S_part}) for the particle contribution. For a non-interacting system, this quantity would result in a delta peaked at an energy equal to the SP energy \citep{Bla1986}. 

\begin{figure}
\begin{center}
\subfloat[]{\label{sf_k0}\includegraphics[width=0.9\textwidth]{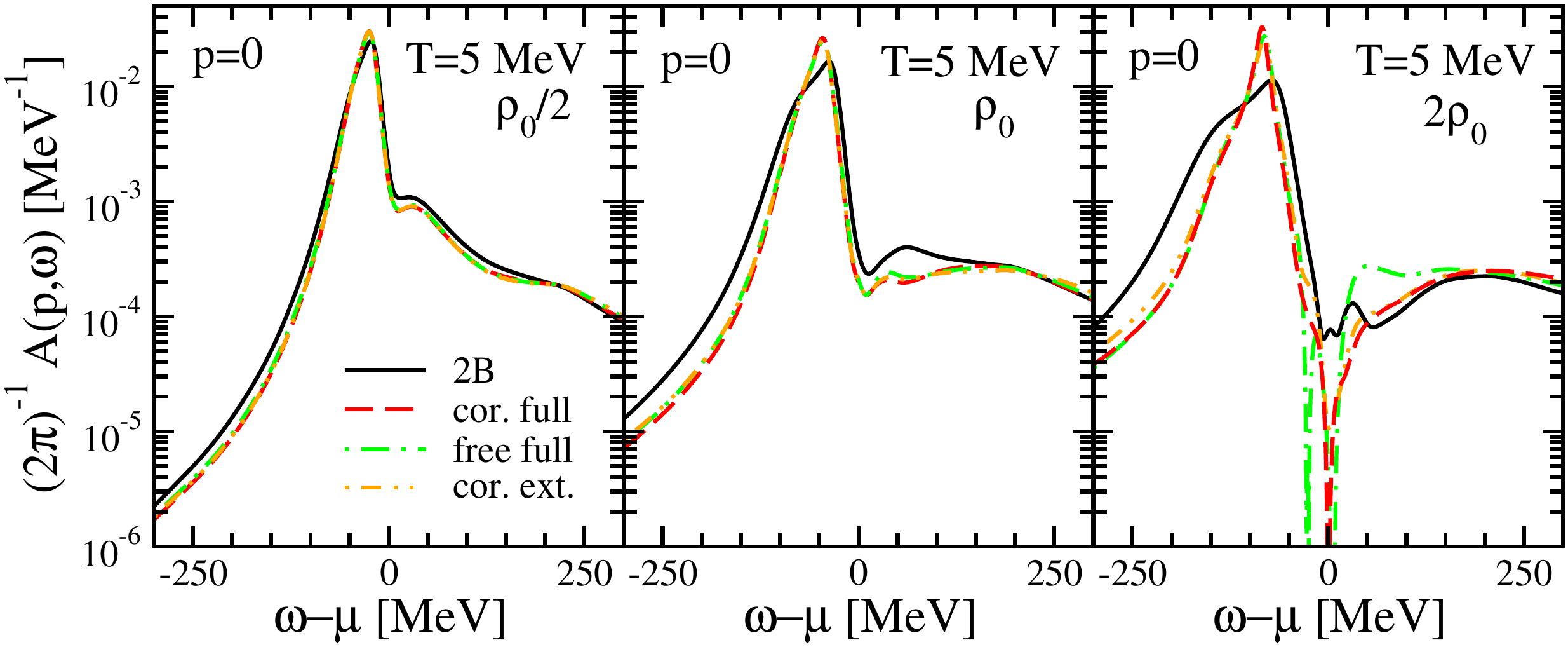}}
\hfill    \vskip .5cm
\subfloat[]{\label{sf_kF}\includegraphics[width=0.9\textwidth]{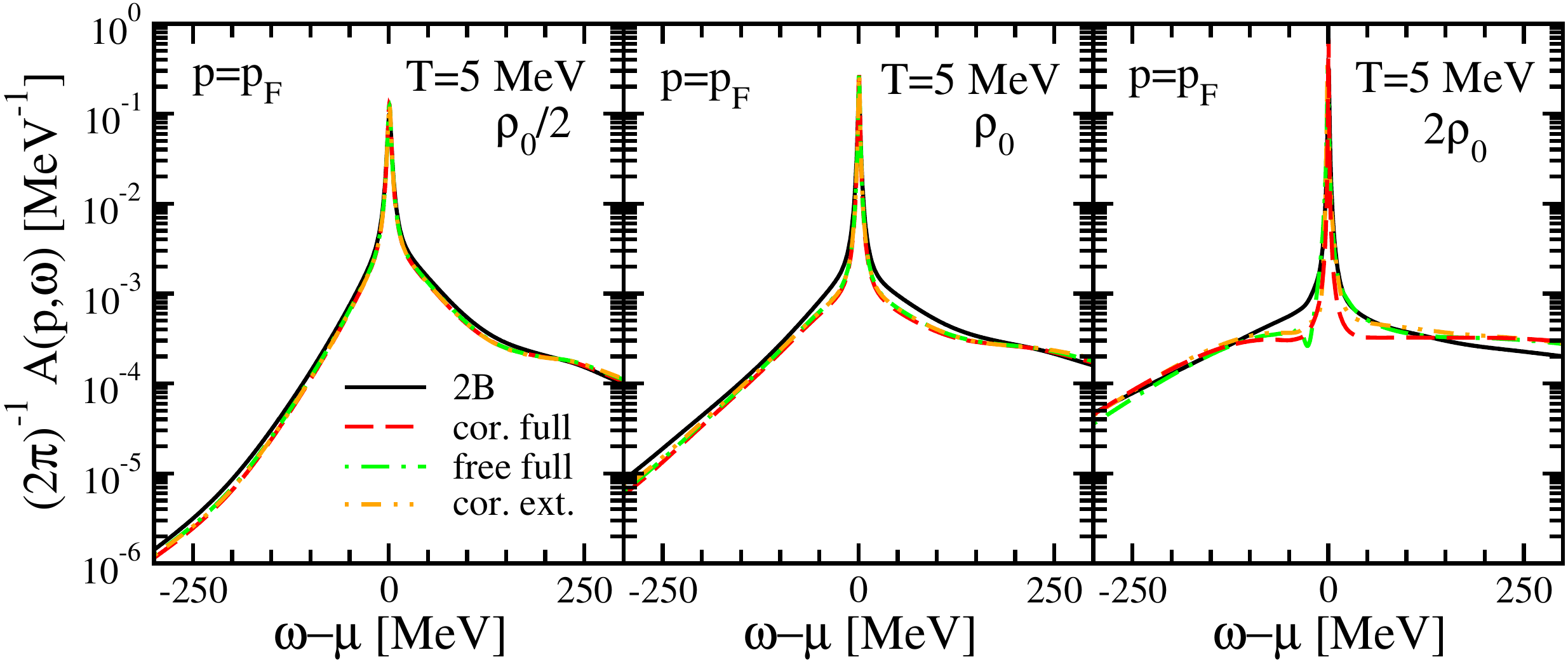}}
\hfill    \vskip .5cm
\subfloat[]{\label{sf_2kF}\includegraphics[width=0.9\textwidth]{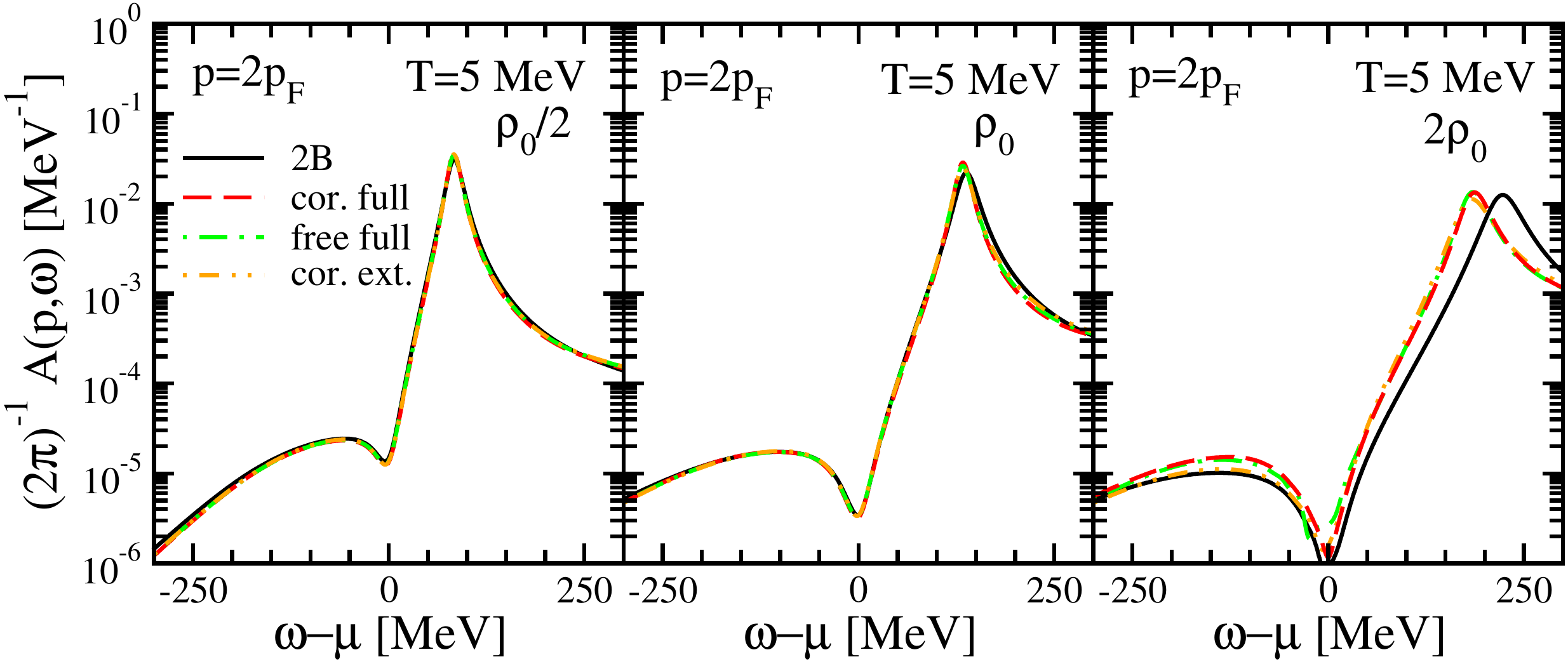}}
\caption{SP spectral function for SNM for SP momentum $p=0, p_\textrm{F}, 2p_\textrm{F}$ in first, second and third row respectively. In each row, panels going from left to right show the spectral function at densities $\rho$= 0.08 - 0.16 - 0.32 fm$^{-3}$. Black-solid lines depict the NN N3LO calculation. Red-dashed, green-dot-dashed and orange-double-dot-dashed curves correspond to the inclusion of the 2NFdd, obtained, respectively, in the correlated and free version with full regulator, and in the correlated version with external regulator.}
\label{spec_fun}
\end{center}
\end{figure}

We plot $A(p,\om)$ in Fig.~\ref{spec_fun} for SNM at temperature $T=5$ MeV for different densities and SP momentum. Spectral functions at three characteristic momenta, $p=0, p_\mathrm F, 2p_\mathrm F$, are presented. The left to right panels correspond, for a specific momentum, to a density of $\rho=0.08, 0.16, 0.32$ fm$^{-3}$. For low and high momenta, the quasi-particle peak positioned in either the hole, $\om<\mu$, or particle, $\om>\mu$, part can be clearly distinguished in the energy dependency of the spectral function. Actually at zero temperature, for $p<p_\textrm{F}$ and $p>p_\textrm{F}$, the spectral function $A(p,\om)$ must go to zero at $\om=\mu$, separating into two distinct sections which correspond exactly to the hole and particle part given in Eq.~(\ref{S_hole}-\ref{S_part}). At finite but  low temperature, as presented in Fig.~(\ref{spec_fun}), this behavior is strongly visible especially at low momentum and high density,  given the increased degeneracy of the system under these conditions. 

The energy of the spectral function peak is well described by the quasi-particle potential analyzed in the previous section. For $p=0$, panel \subref{sf_k0} of Fig.~\ref{spec_fun}, the quasi-particle peak is located in the hole region of energies. This peak moves to more attractive energies as the density increases. As a matter of fact, the higher the density, the more bound the zero momentum state becomes, as observed in the quasi-particle potential (see Fig.~\ref{qp_pot}). As the density increases, the spectral function approaches zero at $\om=\mu$, due to the more degenerate behavior of the system. At $2\rho_0$, the low and high energy tails are associated to a stronger presence of correlations which populate states in energy regions far from the quasi-particle peak. These tails cause the peak to narrow, especially in the 2NF+3NF calculation, due to the fact that the spectral function must fulfill a sum rule at all densities \citep{Rio2007PhD}. In the 2NF case, the peak is broader but overall lower in value, which is why a region of low and high energy tails is still observed. This quasi-particle behavior for the spectral function could have been expected observing the minimum at zero momentum present in the imaginary part of the self-energy for $\om<\mu$, Fig.~\ref{imself}. The imaginary part of the self-energy enters both the numerator and the denominator of the expression for the $A(p,\om)$ \citep{Rio2007PhD}. The deeper the value for Im$\Sigma(p,\om)$, the smaller the value of the peak for the spectral function. It is interesting to note that, as the density increases, in the 2NF calculation the quasi-particle peak broadens and lowers in value. Conversely, if the contracted 3B force is included, the peak narrows even more as the density increases, in accordance to what is observed for the minimum of the imaginary part of the self-energy in Fig.~\ref{imself}. We must point out that the negative values which appear at 2$\rho_0$ when including 3NFs should be attributed to numerical errors when performing the calculation.

For momentum $p=2p_\mathrm F$, panel \subref{sf_2kF} of Fig.~\ref{spec_fun}, the behavior of the spectral function is similar to the zero momentum one but reflected in the particle region, $\om>\mu$. In this case the peak shifts to more repulsive energies the higher the density of the system, conversely to what happens in the zero momentum behavior. We note that at higher SP momentum, the minimum of the spectral function at $\om=\mu$ is much better resolved with respect to the zero momentum case, especially at density 2$\rho_0$. This is a consequence of the flattening of the cusp observed in the imaginary part of the self-energy (see Fig.~\subref*{imself2kF}). It is at this density that the largest effect due to the 2B density-dependent force appears, causing the quasi-particle peak to shift of $\sim50$ MeV to lower energies. Naively, one might expect an attractive effect on the saturation curve when including the 3NF. Note, however, that the energy dependence is always plotted with respect to the chemical potential, $\mu$. The variation in the latter is larger than the quasi-particle shift and results in an overall repulsive effect of the 3NF as the density increases, as observed in the quasi-particle potential at high densities and momenta (see Fig.~\ref{qp_pot}).

\begin{figure}[t!]
\begin{center}
\subfloat[]{\label{mom_Nogga}\includegraphics[width=0.88\textwidth]{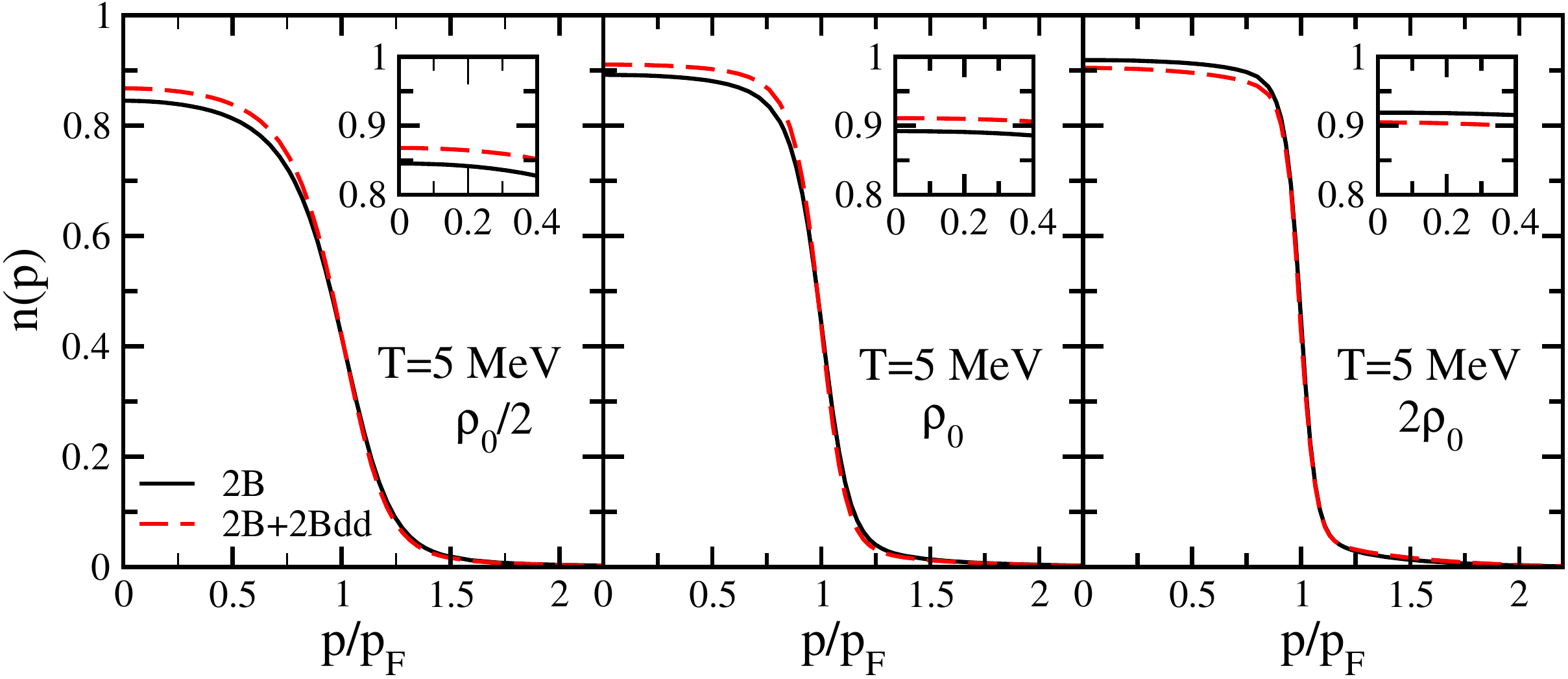}}
\hfill    \vskip .5cm
\subfloat[]{\label{mom_Nogga_log}\includegraphics[width=0.9\textwidth]{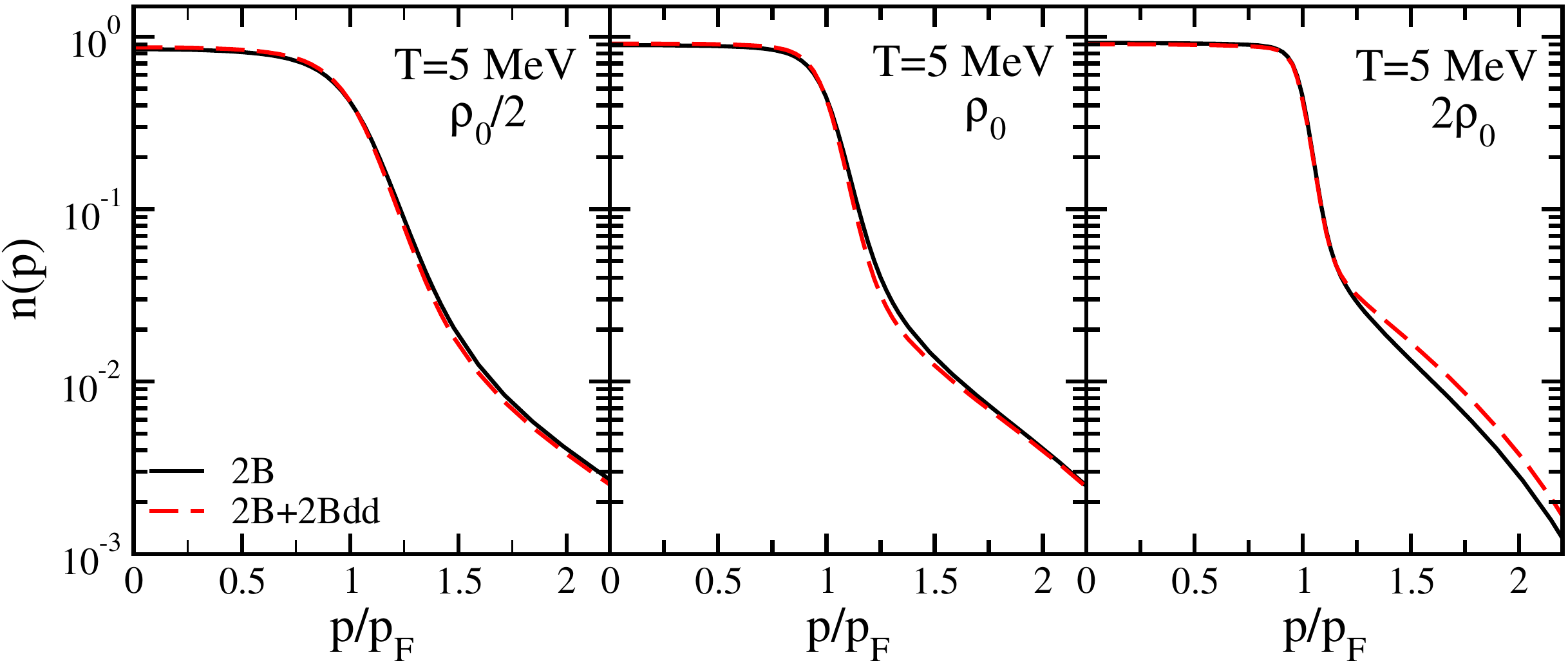}}
\caption{Momentum distribution in SNM using  2NF only at N3LO, black-solid lines, and including the density-dependent 2NF in the correlated average version with full regulator, red-dashed lines. Calculations are performed at $\rho$ = 0.08 - 0.16 - 0.32 fm$^{-3}$ in panels going from left to right. Insets in panel \protect\subref{mom_Nogga}, first row, focus on the depletion below the Fermi surface. Panel \protect\subref{mom_Nogga_log}, second row, presents the same results of panel \protect\subref{mom_Nogga} in logarithmic scale.}
\label{momdis_Nogga}
\end{center}
\end{figure}

At the Fermi momentum, panel \subref{sf_kF} of Fig.~\ref{spec_fun}, the spectral function presents a strongly peaked behavior located at energies $\om=\mu$, proving indeed that at low temperatures the quasi-particle energy must equal the chemical potential. In contrast to the behavior observed for the other momenta, at the Fermi momentum the peak narrows and has larger values as the density increases. This is due to the stronger degeneracy of the system for higher densities, which leads the spectral function to approach the zero temperature delta-peaked function at the chemical potential. As previously stated, for a non-interacting system, the spectral function results in a delta peaked at $\vep_{qp}(p_\textrm F)=\mu$, which is the behavior we observe in this more degenerate case. The narrowing of the peak drives high energy tails which grow with density. These tails, present for all momenta, are actually expected given the high energy behavior observed in Fig.~(\ref{imself_hightail}) for the imaginary part of the self-energy.

From a global perspective, we can affirm that the inclusion of the 2NFdd force doesn't induce strong modifications in the spectral function except for shifts in the QP peak. Nevertheless, a narrowing of the peaks is observed for all momenta as the density increases. This causes modifications in the off-shell behavior of the spectral function. This leads us to conclude that the inclusion of 3BFs must be considered not only at the QP level but in the entire off-shell dependency of the spectral function. As expected from the analysis of the imaginary and real part of the self-energy, modifications due to the different averaging procedures are negligible at saturation and sub-saturation densities, and are more visible for high densities. The effect of the different procedures will be more evident in the following analysis of the momentum distribution function.

Given the knowledge of the spectral function, via Eq.~(\ref{mom_dist}) we have direct access to the SP momentum distribution function. In Fig.~\subref*{mom_Nogga}, we show $n(p)$ in SNM at  $T=5$ MeV for three densities, $\rho$ = 0.08, 0.16, 0.32 fm$^{-3}$. We compare results obtained with and without the density-dependent 2NF, the former one calculated in the correlated average with the use of the full regulator function. The effect of the 3NF in the momentum distribution is relatively small for all densities. The logarithmic scale, in Fig.~\subref*{mom_Nogga_log}, gives us the possibility to appreciate, mainly at $2\rho_0$, a noticeable difference in $n(p)$  for momenta higher than the Fermi momentum. This enhancement of the distribution function at high momenta is a consequence of the added correlation by 3BFs which induces a larger high-momentum population.

\begin{figure}[t!]
\begin{center}
\subfloat[]{\label{mom_comp}\includegraphics[width=0.88\textwidth]{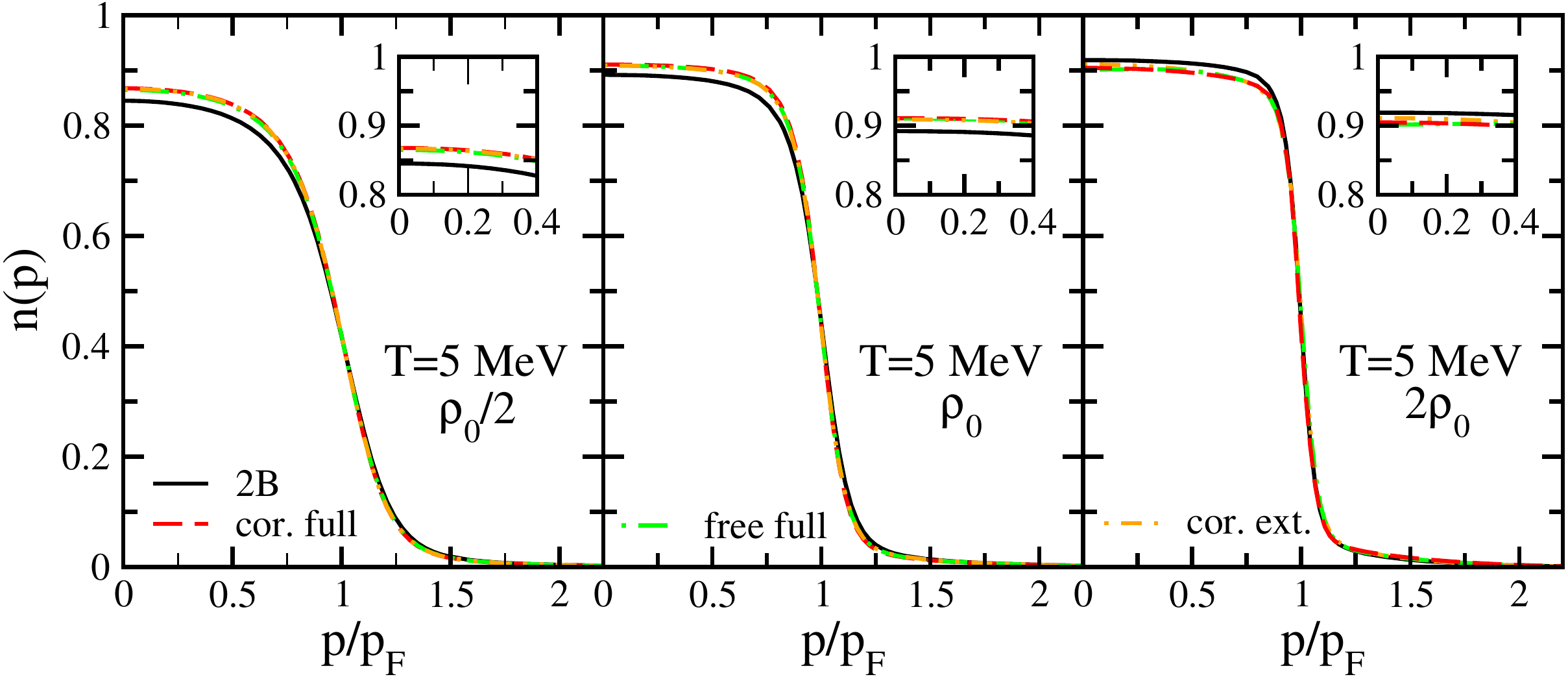}}
\hfill    \vskip .5cm
\subfloat[]{\label{mom_comp_log}\includegraphics[width=0.9\textwidth]{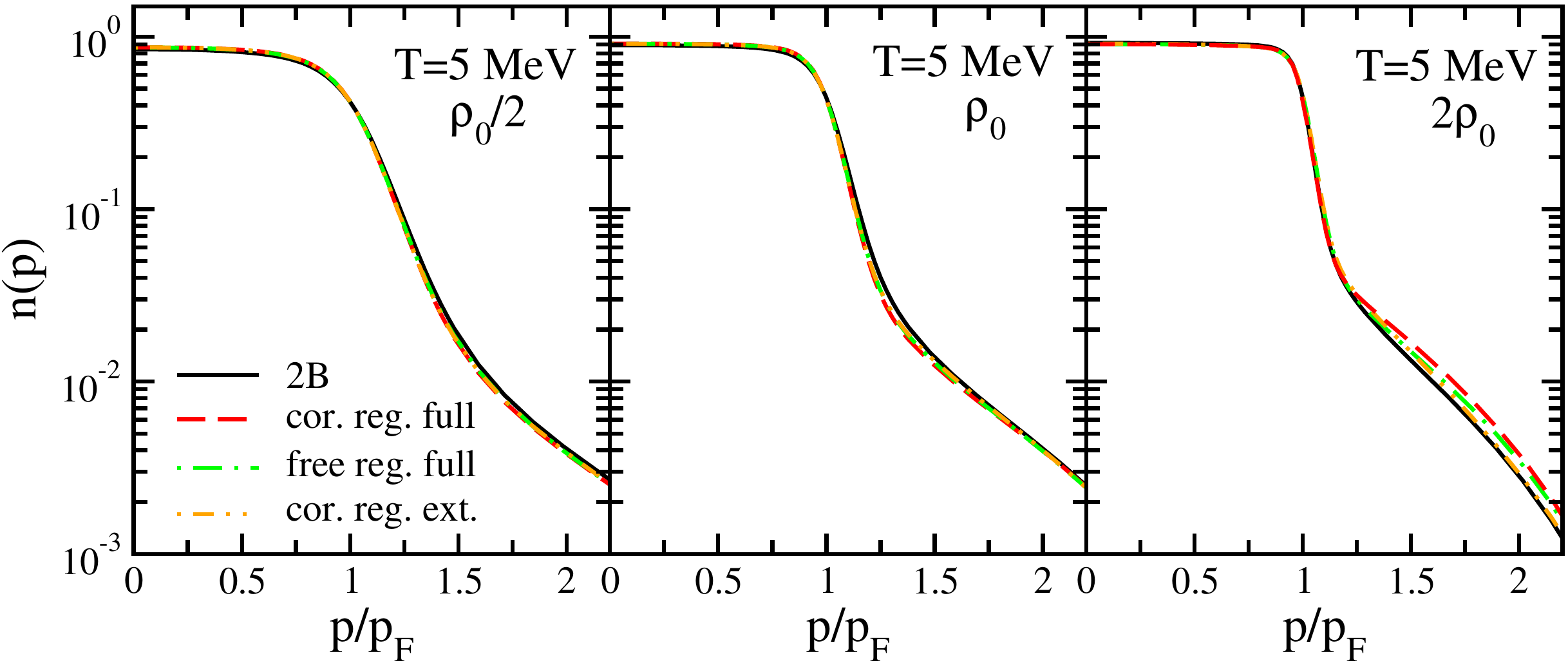}}
\caption{Momentum distribution in SNM in equal conditions as in Fig.~\ref{momdis_Nogga} but considering all averaging procedures for the 2NFdd force. Red-dashed, green-dot-dashed and orange-double-dot-dashed curves represent the inclusion of the 2NFdd obtained, respectively, in the correlated and free version with full regulator, and in the correlated version with the external one.}
\label{momdis_comp}
\end{center}
\end{figure}

In fact, we observe that 3NFs induce a noticeable density dependence in the depletion of the momentum distribution. This is shown in the insets of Fig.~\subref*{mom_Nogga}, which focus on the low-momentum region. The somewhat soft chiral 2NF induces a relatively small depletion, of order $10 \%$, as compared to traditional 2NF, which typically have $~15-20 \%$. For these traditional two-body potentials, the density dependence of the depletion is generally very soft and dominated by tensor correlations \citep{Rio2009Apr}. 3NF modify this behavior, possibly due to the additional tensor structures associated to the density-dependent 2NF explained in Sec. \ref{section:dd_n2lo}. At sub-saturation densities, the 3NF decreases the depletion. At density $\rho=0.08$ fm$^{-3}$, the zero momentum value for $n(p)$ goes from $n(0)_\mathrm{2NF+2NFdd}=0.868$ to $n(0)_\mathrm{2NF}=0.845$, driving a decrease in the kinetic energy value from $T_\mathrm{2NF}=24.35$ to $T_\mathrm{2NF+2NFdd}=23.53$. The difference is still small at $\rho$=0.16 fm$^{-3}$, within a few percent. Also in this case, we observe a higher depletion when considering only 2NF, i.e. $n(0)_\mathrm{2NF}=0.892$ versus $n(0)_\mathrm{2NF+2NFdd}=0.911$.  This causes once more a somewhat higher kinetic energy for the system without 3NF, that is, $T_\mathrm{2NF}=32.88$ MeV to be compared with $T_\mathrm{2NF+2NFdd}=32.02$ MeV. At higher densities, $\rho$=0.32 fm$^{-3}$, the 3NF induces a slightly larger depletion,  $n(0)_\mathrm{2NF+2NFdd}=0.905$, compared with $n(0)_\mathrm{2NF}=0.919$. Correspondingly the kinetic energy in the system with 3NF, $T_\mathrm{2NF+2NFdd}=47.98$ MeV, is a little bit higher than that with 2NF only, $T_\mathrm{2NF}=45.49$ MeV. This behavior leads us to conclude that the higher the density, the stronger the promotion of SP states to high momentum due to 3B forces. This could have been expected given the enhanced high energy tails at $2\rho_0$ observed in the imaginary part of the self-energy as also in the spectral function (see Fig.~\ref{imself_hightail} and right column of Fig.~\ref{spec_fun}). Nevertheless, this behavior is reversed at saturation and sub-saturation densities, where the inclusion of 3NFs lowers the effect of correlations in $n(p)$, inducing a smaller depletion.

We plot in Fig.~\subref*{mom_comp} the variations of the momentum distribution function due to the different averaging procedures in the construction of the density-dependent force. Concentrating on the low momentum region,  we observe negligible changes due to the use of the different density-dependent forces. The depletion varies with a percentage of less than 1\%. The small shifts in the depletion are seen better in the logarithmic scale presented in Fig.~\subref*{mom_comp_log}. In this scale, we see that the highest modifications are observed for high densities. If we look at density $2\rho_0$, a spread in the different curves is observed at high momentum. As already underlined, this is a consequence of the fact that the higher the density the stronger the effect induced from the different averaging procedures. We observe that the largest high-momentum population is induced by the correlated average with full regulator, leading to a more correlated behavior and a higher kinetic energy. Nonetheless, as we will see in the next section, the effect of the correlated average with full regulator provides less repulsive energy results at high densities with respect  to the other constructions of the averaged force. This once more validates the fact that the more regulated the momenta below $p_\textrm F$, the less repulsive the potential. Concluding, the effect of the different averaging procedures is mostly visible on quasi-particle properties, as observed in Fig.~\ref{qp_pot}, and will reflect mainly in bulk properties.

At this point, let us stress that in spite of the cutoff in both the 2NF and 3NF considered here, one finds a substantial population of high-momentum components. Similarly,  the spectral functions display qualitatively important tails at high energies. Traditional microscopic 2NF would yield even larger high-momentum components and fragmentation. Our calculations indicate the importance of considering such effects in many-body calculations even with relatively soft interactions. In fact, our approach is not affected by a cutoff in the many-body approximation, as applied on the contrary in the construction of the chiral interaction. Due to this construction, the high-momentum region should be free of population and we shouldn't observe states beyond the cutoff applied. In particular, let us stress that also when using soft interactions, the low-momentum SP properties are affected by correlations. 

%Studies towards a thorough analysis of high momentum states in the SP momentum distribution have been recently presented \citep{Rios2013}, where calculations are performed in the same approximation presented in this thesis but with the only use of 2B forces. Using different NN potentials, the authors show that the high momentum states induced by the short-range correlations have a universal trend, provided the momentum distribution is normalized to one. Furthermore it is observed that the distribution of these states is independent of the isospin asymmetry of the system.

%%%%%%%%%%%% Nuclear Matter %%%%%%%%%%%%%%%%%%

\section{Nuclear matter}
\label{section:nucl_matt}

We present in this section results for the total energy of symmetric nuclear matter obtained using the extended SCGF formalism described in Chap.~\ref{chapter:3BF_formalism} which includes consistently 3B forces. As explained in Chap.~\ref{chapter:eff_2b_int}, we include 3NF at N2LO in the chiral expansion in a density-dependent form obtained performing an average over the third particle. Averaging procedures vary according to correlated or free propagator used in the internal momentum integral, and on the two different regulator functions included, as explained after Eq.~(\ref{reg_fun}). In the correlated average we use the self-consistent SP propagator obtained through solution of the Dyson equation explained in Sec.~\ref{section:dyson_eq}. In the calculation, we include partial waves up to $J=4$ ($J=8$) in the dispersive (Hartree-Fock) contributions. The total energy is computed via the modified GMK sum rule defined in Eq~(\ref{gmk_3b}), where the 3B expectation value is evaluated as explained in the introduction to this chapter. When no 3B forces are included, the standard GMK sum rule of Eq~(\ref{gmk}) is applied.

\begin{figure}[t]
\begin{center}
\includegraphics[width=0.8\textwidth]{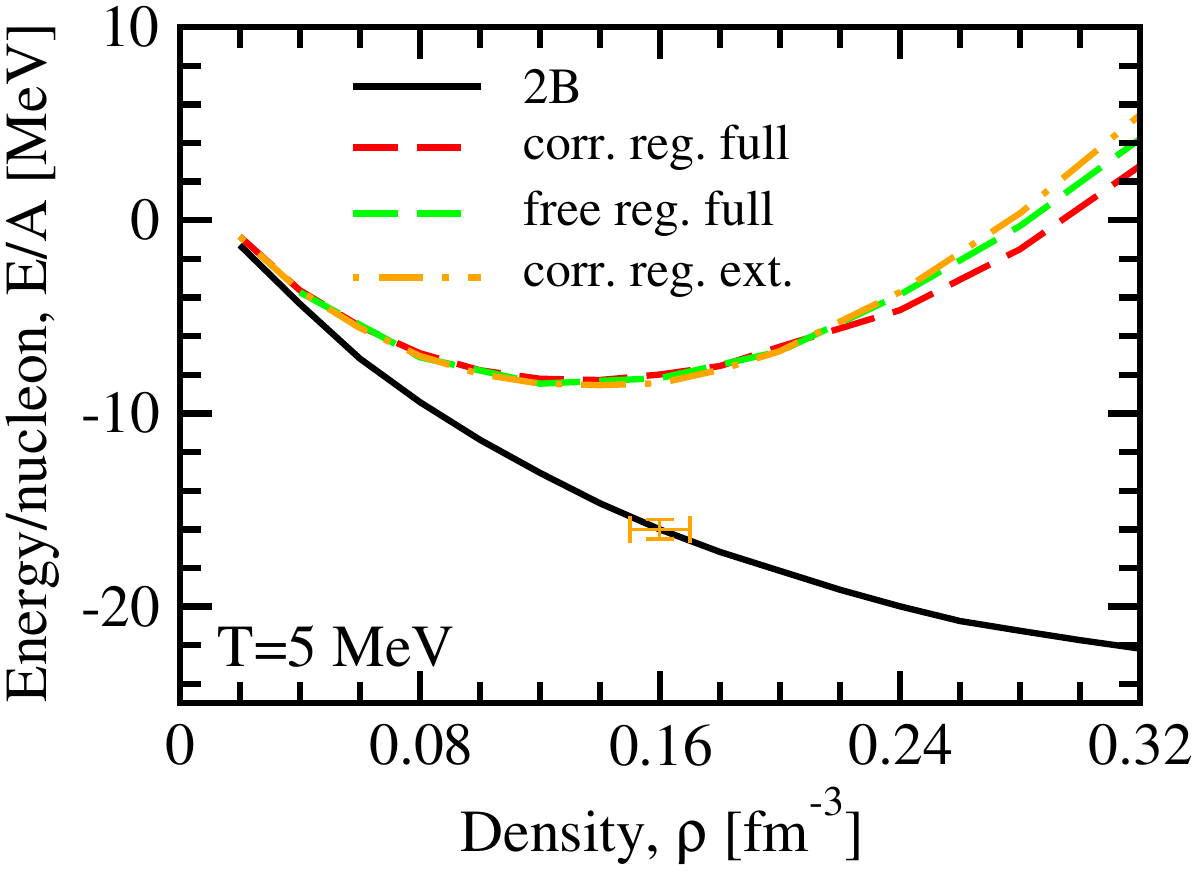}
\caption{Energy for SNM at $T=5$ MeV. The black-solid line depicts the NN N3LO calculation. Red-dashed, green-dot-dashed and orange-double-dot-dashed curves correspond to the inclusion of the 2NFdd, obtained, respectively, in the correlated and free version with full regulator, and in the correlated version with external regulator. The orange cross identifies the empirical saturation point.}
\label{snm_comp_Nogga}
\end{center}
\end{figure}

In Fig.~\ref{snm_comp_Nogga} we show the curves of the energy per nucleon obtained for symmetric nuclear matter at $T=5$ MeV. The effect of 3NF is striking. As expected from the partial wave analysis and from the study of the microscopic properties, the effect of 3B forces is to induce repulsion in the energy values. This repulsion increases with density and is the main cause of nuclear matter saturation. The 2B only calculation, obtained with NN N3LO of \citep{Ent2003}, saturates at high densities not visible in the range of the figure. To be more precise, saturation is observed at $\rho=0.42$ fm$^{-3}$ at an attractive energy value of $E\sim-23$ MeV. This high saturation density is cured by the inclusion of the density-dependent 2NF, which shifts the minima of the energy to densities close to the empirical value. In Fig.~\ref{snm_comp_Nogga} we use for the low energy constants $c_D$ and $c_E$ the same values used in Sec.~\ref{section:pw_analysis}, $c_D=-1.11$ and $c_E=-0.66$ \citep{Nog2006}. Note that in this calculation and in the following ones, where the N3LO force is used in the 2B sector, the LECs $c_1, c_3, c_4$ applied in the N2LO density-dependent force are taken in accordance to Ref.\citep{Ent2003}, i.e. $c_1=-0.81, c_3=-3.2, c_4=5.4$ all measured in GeV$^{-1}$. The cutoff applied on the 3NF is set at $\Lambda_\mathrm{3NF}=500$ MeV according to \citep{Nog2006}. We obtain a qualitatively good saturation density of $\rho \sim 0.14$  fm$^{-3}$ and a less attractive saturation energy of $\sim-9$ MeV with respect to the empirical value. This value increases of $\sim 1$ MeV in attraction if thermal effects are taken into account (see Fig.~\subref*{snmtdep}). At saturation density, the curves obtained performing the different averaging procedures follow what was expected from the partial wave analysis of Chap.~\ref{chapter:eff_2b_int}. A shift of maximum $\sim0.5$ MeV is observed for the saturation energies of the different curves. Specifically, at the saturation density $\rho=0.14$ fm$^{-3}$, matter is less bound when using the correlated average with the full regulator.  The free average with full regulator presents a slightly more bound value with respect to the previous one, and the correlated average with the external regulator provides the highest binding energy value. As already stressed, this shift is  less than 5\% going from the most attractive to the most repulsive result. A greater spreading in the energy curves is observed at high densities. This could have been expected looking at the high density behavior of the quasi-particle potential in Fig.~\ref{qp_pot}. Conversely to what is observed at saturation density, at double this value, the behavior for the curves appears reversed. The correlated average with full regulator leads now the lowest energy value. Nevertheless, we can say that in absolute energy values the effect of the different averages is the same, both at $\rho_0$ and at 2$\rho_0$. In fact, the correlated average with external regulator presents the highest value, both in attraction at $\rho_0$ as in repulsion at  $2\rho_0$. The free average with full regulator follows the former, and the smallest absolute values appear when using the full correlated average. This is a consequence of what we had argued in Sec.~\ref{section:pw_analysis}, where we stated that the higher the availability of momentum states below $p_\textrm F$, the stronger the effect of the 3NF, wether in attraction or in repulsion. Nevertheless we must comment on the fact that, except for a maximum energy shift of less than 4 MeV at high densities, the matching of results obtained with the different averaging procedure is remarkable.

\begin{figure}[t]
\begin{center}
\includegraphics[width=0.8\textwidth]{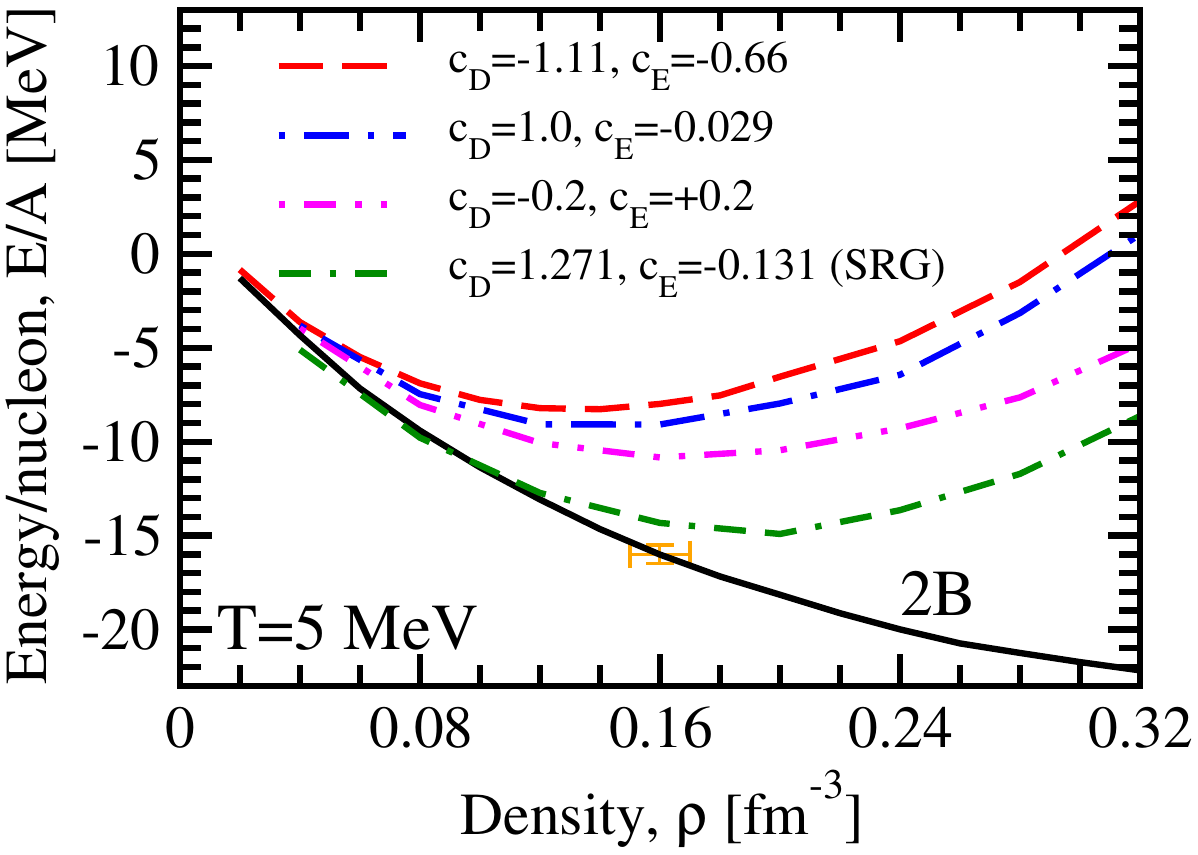}
\caption{Energy for SNM at $T=5$ MeV. The black-solid line depicts the NN N3LO calculation. Red-dashed, blue-dot-dashed, pink-double-dot-dashed and green-dot-double-dashed lines correspond to different couples of LECs, $c_D$ and $c_E$, as explained in the figure and in the text. The orange cross identifies the empirical saturation point.}
\label{snm_comp_cd_ce}
\end{center}
\end{figure}

Different approaches to obtain the two additional low-energy constants appearing in the 3N chiral force have been presented in the literature. These are usually based on fits in the few-body sector \citep{Nog2006,Nav2007Nov,Mar2013}. Consequently, we might expect quantitative changes to the saturation properties coming from different determinations of the LECs. We explore these differences in Fig.~\ref{snm_comp_cd_ce}, where we show the saturation curve at $T=5$ MeV obtained with four different combinations of $c_D$ and $c_E$. In the first three cases, the Entem-Machleidt N3LO potential \citep{Ent2003} has been complemented with the density-dependent force obtained for different couples of  LECs, derived in the correlated version with the full regulator, and with $\Lambda_\textrm{3NF}=500$ MeV. The choice of combinations is representative of the spread in LEC values associated to different fitting protocols. The first choice, $c_D=-1.11$ and $c_E=-0.66$, already used in Fig.~\ref{snm_comp_Nogga}, has been determined from the binding energies of $^3$H and $^4$He in exact few-body calculations \citep{Nog2006}. The second set, $c_D=1.0$ and $c_E=-0.029$, has been obtained with a local version of the 3NF, but fit to the $A=3$ system only \citep{Nav2007Nov}. The third set of LECs, with $c_D=-0.2$ and $c_E=+0.2$, has been used in Ref.~\citep{Kre2012} for SNM calculations on the basis of natural sizes. 

Curves corresponding to different values of $c_D$ and $c_E$ fall within a narrow band below saturation, which reaches a maximum spread of $\sim3$ MeV at $\rho_0$. In general, the modifications are mild even if the constants are changed considerably, indicating that 3NF contributions are small at low densities. Above $\rho \sim 0.16$ fm$^{-3}$, however, differences develop as density increases. These can be largely explained by the contact term and its contribution to the total energy, proportional to $c_E$. In the Hartree-Fock approximation, this reads:
\beq
\frac{E_{c_E}}{A} \simeq - \frac{3}{16} \frac{c_E}{f_\pi^4 \Lambda_\chi} \rho_f^2 = -5.5 \, c_E \left( \frac{\rho_f}{\rho_0} \right)^2 \text{ MeV.}
\enq
One  therefore expects positive values of $c_E$ to lead to more attractive contributions. Looking at Fig.~\ref{snm_comp_cd_ce}, we see that the curve obtained with the positive value of $c_E=+0.2$ is, as a matter of fact, the most attractive. The ones obtained with negative values of $c_E$ increase in repulsion as the absolute value of this constant rises. This is a direct consequence of the effect observed on $S$ partial waves (see Fig.~\ref{pot_S}). There, we had observed that the contact 3NF derived term could provide attraction or repulsion to the 2NF-only partial wave according to the sign in the LEC $c_E$.

In the last case presented in Fig.~\ref{snm_comp_cd_ce}, the green double-dash-dotted line, a similarity renormalization group (SRG) transformation has been applied to the NN N3LO potential. This transformation renormalizes the 2NF, suppressing off-diagonal matrix elements and giving rise to a universal low-momentum interaction. This is usually performed to make the 2NF more tractable in the many-body framework \citep{Bog2007,Bog2010}. The SRG evolution, however, introduces induced many-body forces and these need to be consistently taken into account. A full SRG evolution of 2NF and 3NF has been achieved only recently in momentum space, which is the relevant basis for nuclear matter calculations \citep{Heb2012,Heb2013Mar,Wen2013}. Here we follow a previous simpler strategy, whereby the SRG evolution is performed on the 2NF alone \citep{Heb2011} and the LECs of the 3NF are refitted for each scale at which the transformation is performed. In the specific case of Fig.~\ref{snm_comp_cd_ce} we choose the case where the renormalization scale on the 2NF is equal to the one applied on the 3NF, i.e. $\Lambda_\textrm{SRG}=\Lambda_\textrm{3NF}=2.0$ fm$^{-1}$. LECs $c_D=1.271$ and $c_E=-0.131$ are then chosen according to \citep{Heb2011}. The low-momentum potential calculation presents the most attractive result for the binding energy of symmetric nuclear matter, approaching the empirical saturation energy value around $\sim -15$ MeV, at a higher density of $\rho=0.20$ fm$^{-3}$ with respect to the empirical one. Overall, the different energy curves in Fig.~\ref{snm_comp_cd_ce} obtained with the four different couples of LECs reach a spread of $\sim 6$ MeV at empirical saturation density $\rho_0$; this spread doubles its value at $2\rho_0$. Conversely to what was previously discussed in relation to the energy contribution given by the contact term, even though the $c_E$ is negative in the SRG-evolved case, the energy is more attractive with respect to all other curves. This is due to the decimated high momenta in the renormalized-2NF, which is consequently less repulsive.
%%% this is not true, I cannot say this because the hamiltonian has changed. in the pnm case yes I can say it!!!
%We must underline that the spread in the energy curves between the SRG-evolved NN force case and the non evolved ones, leads us to think that nuclear matter is not perturbative in the chiral nuclear force. 
As tested in Ref.~\citep{Car2013Oct}, the perturbative regime of nuclear matter is reproduced when using evolved interactions, at least up to saturation density.

\begin{figure}[t]
\begin{center}
\includegraphics[width=0.8\textwidth]{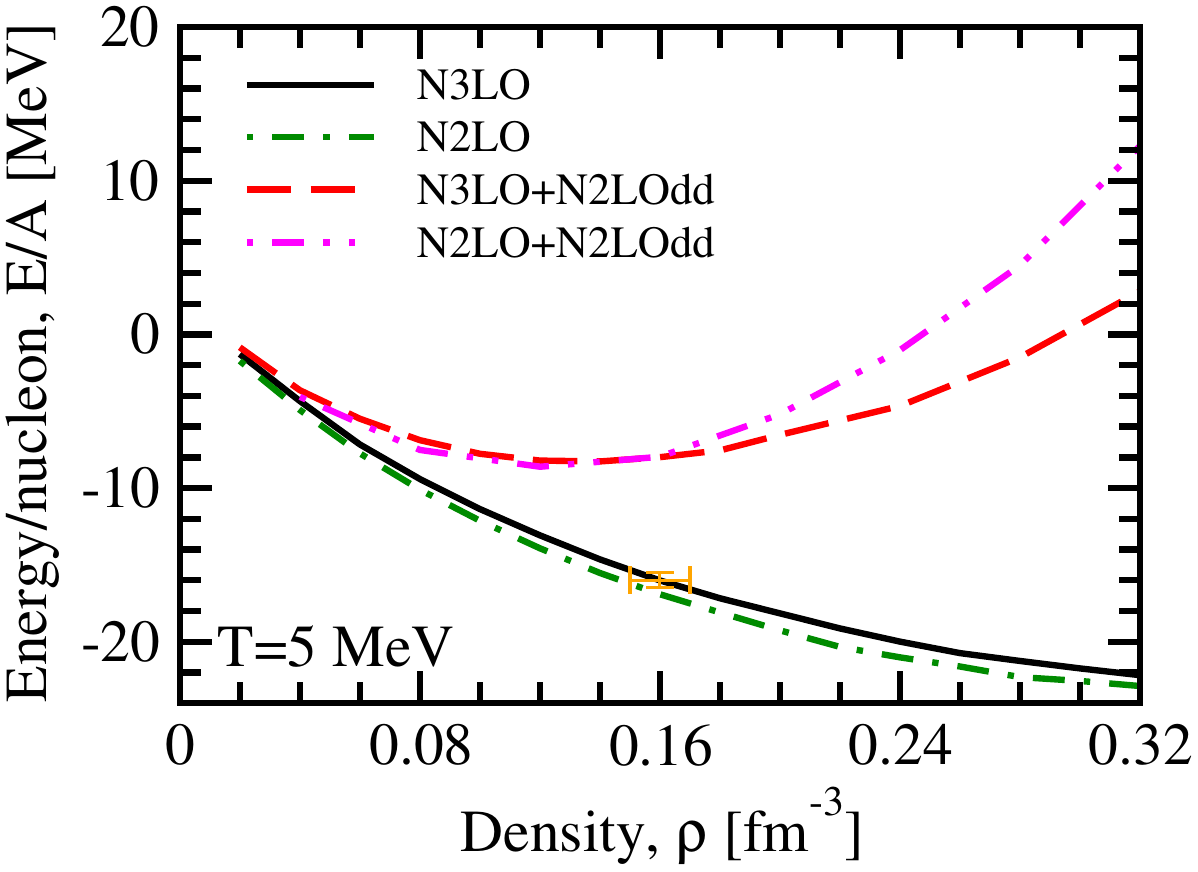}
\caption{Energy for SNM at $T=5$ MeV. The black-solid line depicts the NN N3LO calculation. The green-dot-dashed shows the NN N2LO optimized calculation. The red-dashed line corresponds to calculation performed with NN N3LO plus the N2LOdd obtained in the correlated version with full regulator, with LECs $c_D=-1.11, c_E=-0.66$. The pink-doubledot-dashed refers to calculation performed with the same N2LOdd but with  $c_D=-0.39, c_E=-0.389$, and $c_1,c_3,c_4$ according to \citep{Eks2013}.}
\label{snm_n3lo_n2lo}
\end{center}
\end{figure}

We must stress, at this point of the discussion, that in all calculations presented up to now, the order at which $\chi$EFT is implemented in the interaction is not equivalent in the 2B and 3B sector. A consistent measure of the convergence of the chiral expansion would only be provided if the energy results for the system where compared order by order with a consistent Hamiltonian. In other words, NLO, N2LO and N3LO results should yield smaller and smaller error bands. At N3LO, one would in principle also have to include four nucleon forces, as recently done for neutron matter in Ref.~\citep{Tew2013,Kru2013}, where a full N3LO calculation has been performed. 

For the purpose of consistency in the $\chi$EFT expansion, we present in Fig.~\ref{snm_n3lo_n2lo} results obtained for SNM with the use of an optimized 2NF at N2LO which was recently presented in Ref.~\citep{Eks2013}. We compare the energy per particle at $T=5$ MeV using the N3LO 2B Entem-Machleidt potential \citep{Ent2003} and the 2B N2LO of Ref.~\citep{Eks2013}. Results with and without the correlated averaged 3NF with full regulator are showed. For the N2LO 3NF which complements the optimized NN N2LO, we use the values of $c_D$ and $c_E$ quoted in Ref.~\citep{Hag2013}, i.e. $c_D=-0.39$ and $c_E=-0.389$. In both calculations, the cutoff on the 3NF contracted force is set to $\Lambda_\mathrm{3NF}=500$ MeV. In the full-N2LO calculation, LECs coming from the 2B sector, namely $c_1, c_3, c_4$, to be used in the N2LO 3NF are set to values reported in \citep{Eks2013}. Looking at the 2NF-only results, the curves obtained are very similar, with N2LO consistently more attractive than N3LO in the whole density range. As already observed in Fig.~\ref{snm_comp_Nogga}, the inclusion of 3NF is crucial to obtain saturation at realistic densities and energies. The full-N2LO calculation yields slightly more attractive results at low densities, moving the saturation energy to a smaller density value of $\rho=0.12$ fm$^{-3}$, with respect to the N3LO one. But conversely, it is more repulsive above saturation. This indicates a larger incompressibility at high densities for results obtained with the optimized version of the N2LO 2NF \citep{Eks2013}. For the full-N2LO calculation, our results are in fair agreement with respect to those performed in the Coupled Cluster many-body method, recently presented by authors in Ref.~\citep{Hag2013}. 

Overall, the differences in between the two calculations, full-N2LO or 2B-N3LO+3B-N2LO, can be ascribed mostly to the distinct structure of the 2NF itself, and furthermore on the different values of the LECs. Even though the calculations can be both valid on their own, a direct comparison is hard to carry out. An analysis of the kind would be fair if, in going from the N2LO to the N3LO order in the chiral expansion, the Hamiltonian would have been obtained consistently, which is not the present case. Conclusions on the compared quality of the results respect to the consistency in the chiral expansion are then difficult to drive.

%%%%%%%%%%%%%%%% Neutron matter  %%%%%%%%%%%%%%%%%%

\section{Neutron matter and the symmetry energy} 
\label{section:neut_matt}

Using the extended SCGF to include 3B forces, we compute in this section the bulk properties of PNM. The effect that we observe, due to the inclusion of the density-dependent 2NF, is repulsion at all densities. Similarly to the symmetric matter case, repulsion added by the 3B force rises with increasing density. In the case of pure neutron matter, where the total energy of the system is always a positive value, the rising of the energy for all densities contributes to stiffen the EOS of PNM. This effect is convenient  in order to yield higher values for the mass of neutron stars. As a matter of fact, recent astrophysical observations, where masses as high as twice the solar mass have been observed for neutron stars \citep{Dem2010,Ant2013}, have ruled out a variety of microscopic descriptions which lead to softer EOSs.

The case of pure neutron matter is particularly interesting for the implementation of nuclear chiral interactions. In fact, up to fourth order in the $\chi\textrm{EFT}$ expansion, i.e. N3LO, all many-body forces among neutrons are predicted. In other words, no further coupling constants, except for those present in the 2B sector, i.e. $c_1$ and $c_3$, need to be adjusted. In App.~\ref{chapter:dens_dep_terms} we demonstrate the vanishing for 3NF terms at N2LO proportional to $c_4$, $c_D$ and $c_E$. A similar demonstration can be done for 3NF terms at N3LO \citep{Epe2009Oct}. Recently, studies for the pure neutron matter system have been computed at full N3LO \citep{Tew2013,Kru2013}.

In PNM, as for the symmetric nuclear matter case, we perform calculations complementing the N3LO by  \citep{Ent2003} with the density-dependent force computed in neutron matter as detailed in Sec.~\ref{section:dd_n2lo}. We will also show results obtained using the newly optimized N2LO by \citep{Eks2013}, complemented with the same density-dependent force of  Sec.~\ref{section:dd_n2lo}, therefore being consistent in the order of the chiral expansion. Unless if otherwise stated, cutoff on the 3NF contracted force is set to $\Lambda_\mathrm{3NF}=500$ MeV.

\begin{figure}[t]
\begin{center}
\includegraphics[width=0.8\textwidth]{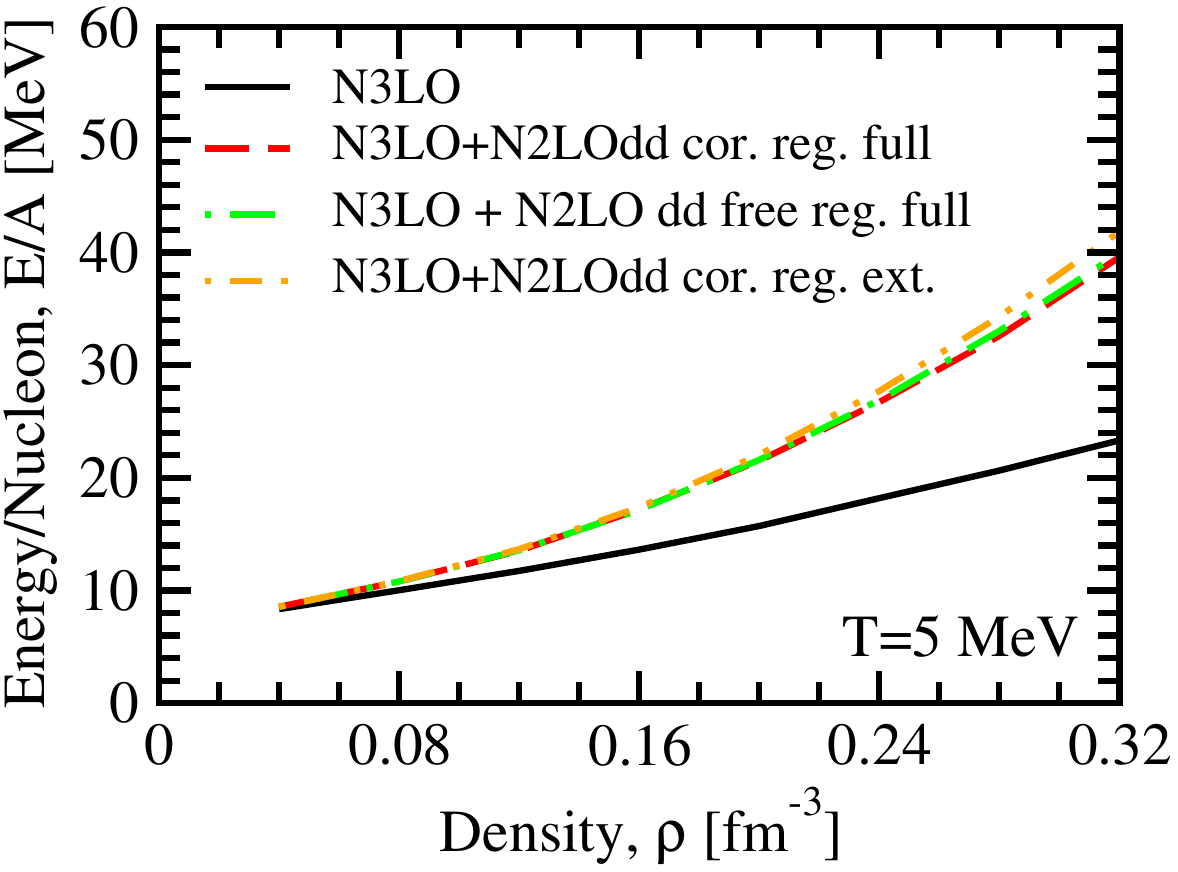}
\caption{Energy for PNM at $T=5$ MeV. The black-solid line depicts the NN N3LO calculation. Red-dashed, green-dot-dashed and orange-double-dot-dashed curves correspond to the inclusion of the 2NFdd, obtained, respectively, in the correlated and free version with full regulator, and in the correlated version with external regulator function.}
\label{pnm_comp}
\end{center}
\end{figure}

We present in Fig.~\ref{pnm_comp} the density dependence for the total energy per nucleon in pure neutron matter computed at finite temperature $T=5$ MeV. LECs values are taken in accordance to \citep{Ent2003}, $c_1=-0.81$ GeV$^{1}$ and $c_3=-3.2$ GeV$^{-1}$. In the 2B only case, calculated with the NN at N3LO by \citep{Ent2003}, values for the total energy of the system are around $\sim13$ MeV at empirical saturation density $\rho_0=0.16$ fm$^{-3}$. The energy gains a $\sim10$ MeV in repulsion at double saturation density $2\rho_0$. The inclusion of 3NFs contribute to enhance even more the energy values in the entire density range. This is a direct consequence of the modifications observed on the partial wave matrix elements in Fig.~\ref{pw_pnm}. Repulsion due to the density-dependent 2NF goes from less than $1$ MeV at density $\rho_0/2=0.08$ fm$^{-3}$, rising to $\sim4$ MeV at saturation density, and boosting up to $\sim15$ MeV at double saturation density.  We must outline that, thanks to the non-perturbative nature of our calculation, which performs an all-order ladder summation for the in-medium $T$-matrix interaction, we are not bound to low-density regions. Therefore, we can span with one same many-body technique a higher density range, whereas perturbative many-body calculations must perform some kind of extrapolation. 

Furthermore we present in Fig.~\ref{pnm_comp} the variation in the total energy curve due to the different averaging procedures in the construction of the density-dependent force. The difference is negligible for low densities, and becomes appreciable only around double saturation density. At $2\rho_0$, we observe an energy value of $\sim40$ MeV. Here, the three different averages reach a maximum spread of $\sim 1$ MeV. We note that the behavior at high densities is similar to the one observed in Fig.~\ref{snm_comp_Nogga} for SNM. Results obtained with the self-consistent correlated average with the use of the full regulator provide the less repulsive energies at these high densities. On the contrary, the most repulsive results are observed for the self-consistent correlated average but with the use of the external regulator function. Nevertheless, the overlap in between the different curves is striking, leading to an almost complete independence on the average procedure implemented. For an estimate of the $T=0$ energy results we can rely on the Sommerfeld expansion. Thermal effects slightly increase the zero temperature energies, in accordance to what was presented in Fig.~\subref*{pnmtdep}.

\begin{figure}[t]
\begin{center}
\includegraphics[width=0.8\textwidth]{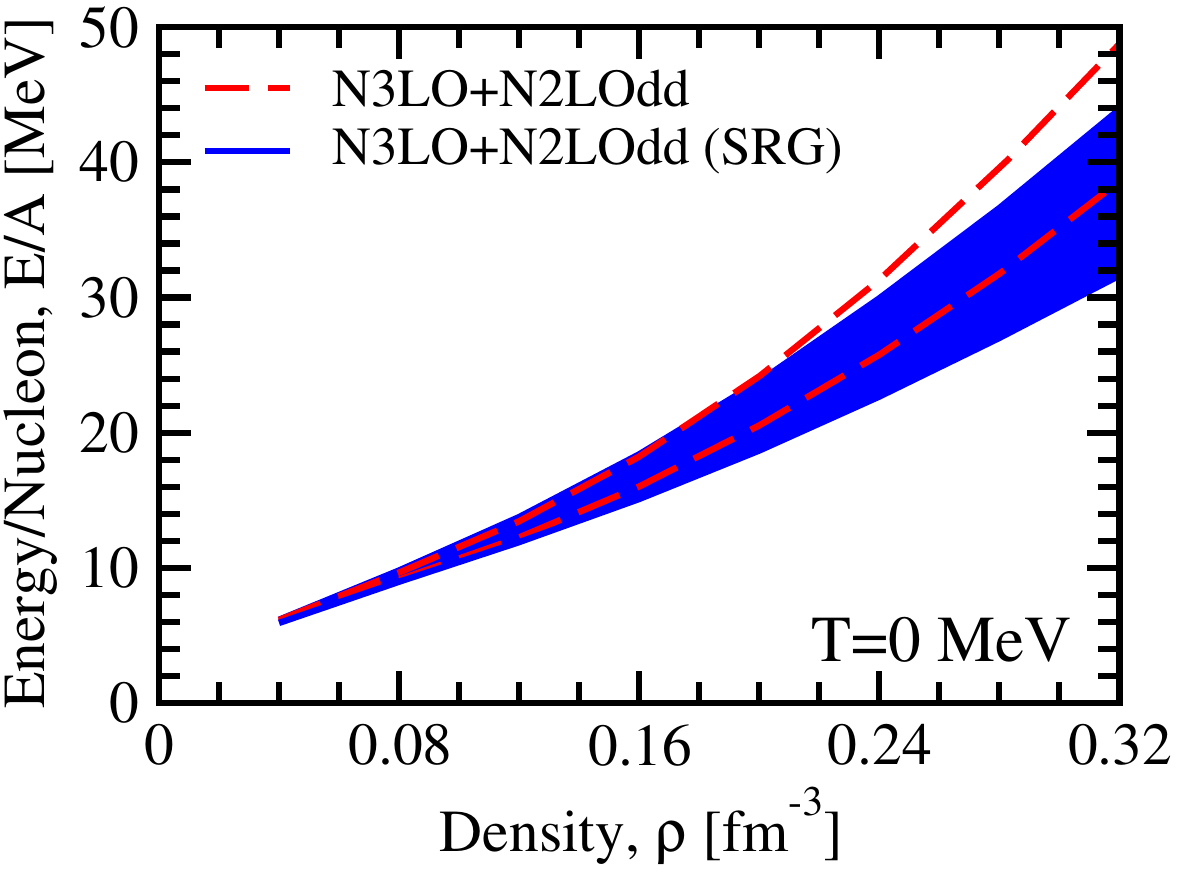}
\caption{Energy for PNM at $T=0$ MeV. Red-dashed lines depict the calculation for the NN N3LO plus the 2NFdd, obtained in the correlated version with full regulator. Blue-solid lines correspond to the calculation where the SRG evolution is applied on the NN N3LO, $\Lambda_\textrm{SRG}=2.0$ fm$^{-1}$. The band depends on the LECs values used in the calculation as explained in the text.}
\label{pnm_Kai_comp}
\end{center}
\end{figure}

Contrary to what was presented in the symmetric matter case, where theoretical uncertainties are dominated by the different couples of the additional LECs $c_D$ and $c_E$, in the case of PNM, variations in the calculations are dominated by uncertainties in the determination of the LECs $c_1$ and $c_3$ \citep{Heb2013Jul}. These LEC relate NN, $\pi$N and 3N interactions  in the chiral expansion and their determination from $\pi$N scattering is within uncertainties in agreement with the NN scattering extracted values. These lead to an error band for each constant equal to $c_1=-(0.7-1.4)$GeV$^{-1}$ and $c_3=-(3.2-5.7)$GeV$^{-1}$ \citep{Ent2003,Epe2005,Ren2003}. We show this error band in Fig.~\ref{pnm_Kai_comp}. Results are presented at zero temperature, following the extrapolation explained in the introduction to this chapter, and computing each curve in two different cases for the 2B force: the complete N3LO Entem-Machleidt force \citep{Ent2003} and the SRG-evolved NN N3LO force with a cutoff of $\Lambda=2.0$ fm$^{-1}$. In both cases the 2B force is complemented  with the density-dependent 2NF derived in neutron matter with the correlated average with full regulator function, with cutoff set to $\Lambda=2.5$fm$^{-1}$ in the former case, and $\Lambda=2.0$fm$^{-1}$ in the latter one. Results are depicted in a similar fashion to the ones presented in Ref.~\citep{Heb2013Jul} for an easier comparison. Upper curves correspond to the upper bound value for the LECs, i.e. $c_1=-1.4$GeV$^{-1}$ and $c_3=-5.7$GeV$^{-1}$ ; while lower curves correspond to the lower bound value, i.e. $c_1=-0.7$GeV$^{-1}$ and $c_3=-3.2$GeV$^{-1}$ . We observe that in both cases, of complete and evolved NN force, the spread in results due to theoretical uncertainties in the LECs ranges from less than 1 MeV at sub saturation densities, to $\sim4$ MeV at $\rho_0$, and doubles this value at $2\rho_0$, presenting a higher spread for the evolved case. Looking at Fig.~\ref{pnm_Kai_comp}, we observe that bands overlap perfectly up to almost $\rho=0.22$ fm$^{-3}$. In this sense, neutron matter can be fairly well described, up to these densities, with both the full and evolved 2B force, proving the perturbative behavior in this low-energy regime.

\begin{figure}[t]
\begin{center}
\includegraphics[width=0.8\textwidth]{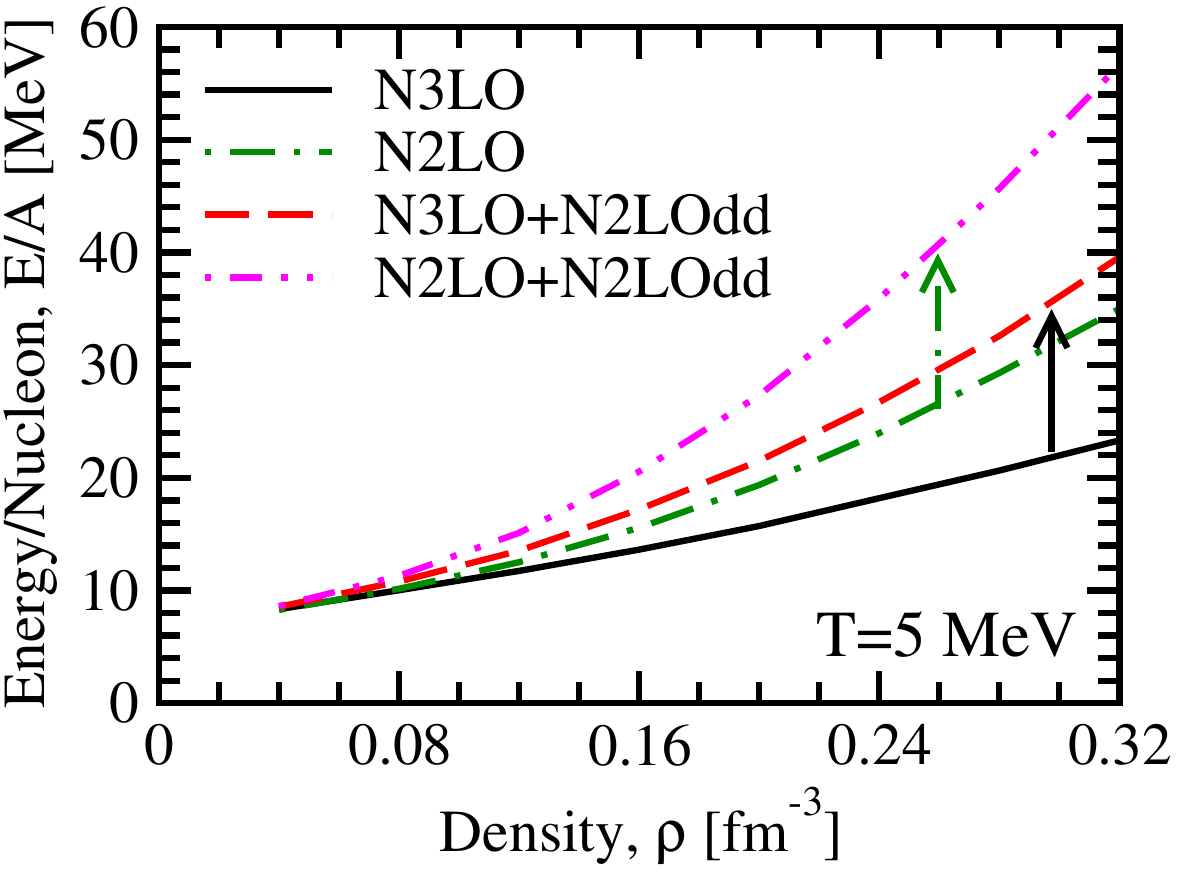}
\caption{Energy for PNM at $T=5$ MeV. The black-solid line depicts the NN N3LO calculation. The green-dot-dashed line corresponds to the NN N2LO optimized calculation. The red-dashed line corresponds to calculation performed with NN N3LO plus the N2LOdd obtained in the correlated version with full regulator. The pink-double-dot-dashed refers to calculation performed with the same N2LOdd but with $c_1,c_3$ according to \citep{Eks2013}, on top of the N2LO optimized potential. Arrows help identify the curves obtained respectively with the N2LO (green-dot-dashed) and with the N3LO 2NF (black-solid) in the 2B sector, and its modification when including 3BFs.}
\label{pnm_n3lo_n2lo}
\end{center}
\end{figure}

To be consistent in the order of the chiral expansion for the nuclear force, we perform, also for PNM, calculations at full N2LO using the newly optimized NN force from Ref.~\citep{Eks2013}. In the full N2LO calculation the LECs, $c_1, c_3$, used in the density-dependent force, are taken in accordance to \citep{Eks2013}. In the N3LO case, values are taken from \citep{Ent2003}. In Fig.~\ref{pnm_n3lo_n2lo}, we see that in both cases, of including or not the density-dependent 2NF, results obtained with the newly optimized N2LO force are more repulsive for all densities with respect to the N3LO ones. In the 2B case, this repulsion is around $\sim2$ MeV at $\rho_0=0.16$ fm$^{-3}$ and grows with the density up to $\sim10$ MeV at $2\rho_0$. In the 2B+3B calculation the repulsion between the N2LO and N3LO case is a bit stronger at all densities, with respect to values obtained with the 2B forces only. The repulsive effect approaches values of $\sim15$ MeV at double saturation density. As discussed for SNM, we ascribe this overall repulsion between the N2LO optimized and the N3LO calculations to the different construction of the 2NF force. It is interesting to note that the 2B N2LO potential provides higher energy values with respect to the 2B N3LO one. The curve obtained with the former potential approaches the one obtained with the 2B N3LO plus the density-dependent potential. If thermal effects are taken into account, curves obtained with the N2LO optimized force are in fair agreement with the Coupled Cluster calculations presented in Ref.~\citep{Hag2013}. Once again, it is difficult to evaluate the quality of our results in relation to the chiral expansion consistency, given the different structures of the N2LO and the N3LO forces. 

\begin{figure}[t]
\begin{center}
\includegraphics[width=0.8\textwidth]{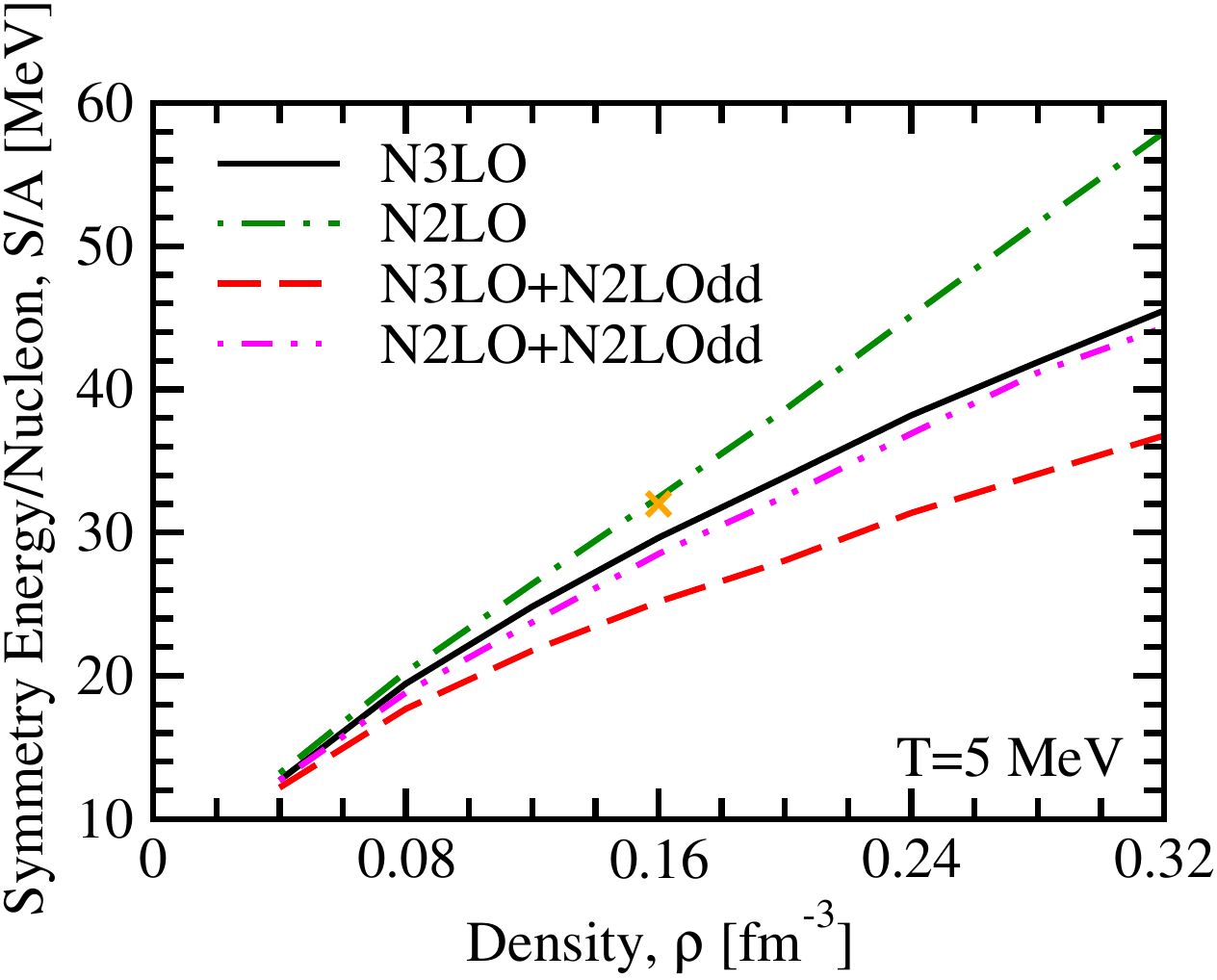}
\caption{Symmetry energy at $T=5$ MeV. The black-solid line depicts the NN N3LO calculation. The green-dot-dashed line depicts the NN N2LO optimized calculation. The red-dashed line corresponds to calculation performed with NN N3LO plus the N2LOdd obtained in the correlated version with the full regulator. The pink-double-dot-dashed line represents the calculation performed with the same N2LOdd on top of the N2LO optimized potential. The orange cross defines the accepted symmetry energy value at saturation density, 32 MeV.}
\label{sym_comp}
\end{center}
\end{figure}

We want to conclude this chapter with a study of the symmetry energy. As already outlined in the introduction to this chapter, a reliable estimation for the density dependence of the symmetry energy is crucial to understand and predict properties of isospin-rich nuclei and neutron stars. To compute the symmetry energy, one generally resorts to the parabolic formula, i.e. one assumes that the energy per particle (or any of its  components, kinetic and potential), has a quadratic dependence on asymmetry \citep{Vid2009}:
\begin{align}
	\frac{E}{A}(\rho,\alpha)= 	\frac{E}{A}(\rho,0) + \frac{S}{A}(\rho) \alpha^2 \,.
	\label{symmetry}
\end{align}
$\alpha=(N-Z)/(N+Z)$ defines the isospin asymmetry of the system, with $N$ and $Z$ neutrons and protons respectively. This immediately yields that the symmetry energy, $S(\rho)$, is given by the difference of PNM and SNM energies:
\begin{align}
	\frac{S}{A}(\rho) = \frac{E}{A}(\rho,1) - \frac{E}{A}(\rho,0) \, .
	\label{eq:diff}
\end{align}
The SCGF approach can be generalized to isospin asymmetric systems \citep{Fri2005}. The quadratic dependency of the energy with respect to the asymmetry parameter $\alpha$, Eq.~(\ref{symmetry}), has been validated using only 2NFs, however we don't expect 3NFs to alter this behavior \citep{Car2012}.

We plot in Fig.~\ref{sym_comp} the density dependence for the symmetry energy obtained using both the N3LO potential by \citep{Ent2003} and the optimized N2LO by \citep{Eks2013}. In both cases, we perform the calculation in the 2B-only case, and in the 2B+3B procedure. Curves are obtained from those presented in Fig.~\ref{snm_n3lo_n2lo} and Fig.~\ref{pnm_n3lo_n2lo} applying Eq.~(\ref{eq:diff}). It is interesting to note that when including the density-dependent force, either on top of the 2NF N2LO or N3LO potential, the values of the symmetry energy lower. This can be interpreted analyzing the different modifications which 3B forces induce on, respectively, the PNM and SNM system. In the latter case, the repulsion provided on energy values when including 3NFs is stronger with respect to the former case. This affects the subtraction in Eq.~(\ref{eq:diff}) providing smaller quantities for the symmetry energy when considering the density-dependent force in the calculation. To better understand this behavior, we present in Table~\ref{table} the results for the total energy per nucleon in PNM, SNM and the symmetry energy, at $T=5$ MeV and empirical saturation density $\rho_0$. We must point out that the effect of temperature, following the procedure presented in the introduction to this chapter, yields a decrease in the symmetry energy values at $T=5$ MeV of approximately $\sim 1$ MeV, with respect to the zero temperature ones. Furthermore, we present in Table~\ref{table} the value of the the slope of the symmetry energy at empirical saturation density. This quantity is measured as $L=3\rho_0\partial (S(\rho)/A)/\partial\rho|_{\rho=\rho_0}$, and its value is constrained to be in a range of $\sim50-80$ MeV \citep{Tsa2009}. 

\begin{table}[t!]
\begin{center}
\scalebox{1}{
	\begin{tabular}{c| c c c c}
		$\rho_0=0.16\textrm{fm}^{-3}$ & $E_\textrm{PNM}/A$ & $E_\textrm{SNM}/A$ & $S/A$ & $L$\\
		\hline 
		N3LO & 13.6 & -16.0 & 29.6 & 52.9 \\
		N2LOopt & 15.6 & -16.9 & 32.5 & 69.4 \\
	  	N3LO+N2LOdd & 17.2 & -7.99 & 25.2 & 40.8\\
		N2LOopt+N2LOdd & 20.5 & -7.96 & 28.5 & 53.9\\
	\end{tabular}
}
\end{center}
	\caption{First three columns on the right-side show respectively the total energy per nucleon in pure neutron matter, symmetric nuclear matter and the symmetry energy. Last column on the right-side shows the value of the slope of the symmetry energy, $L$. All quantities are measured in MeV at empirical saturation density $\rho=0.16$ fm$^{\textrm{-3}}$ and $T=5$ MeV for different NN interactions as explained in the first column on the left-side.}
	\label{table}
\end{table}

As expected from Fig.~\ref{sym_comp}, the result which best approaches the currently accepted $\sim 32$ MeV symmetry energy value at $\rho_0$  \citep{Tsa2009}  is obtained from the 2B N2LO calculation. Taking thermal effects into account, this lies slightly above of the accepted quantity. Subsequently, the symmetry energy obtained with the 2B N3LO potential is also a good estimate of $S/A$ at saturation density, settling a bit below $32$ MeV. The slope $L$ obtained with the N2LO force is higher with respect to the result given by the 2B N3LO. This is a consequence of the more repulsive PNM energy  together with a more attractive SNM one for the N2LO case, which enhances the value of the slope of the symmetry energy.  Results for $S/A$ obtained including the density-dependent forces lie farther below the accepted value, which is a consequence of what we argued before. The less attractive energy obtained for SNM together with a much more repulsive energy in PNM when including 3NFs, provides a smaller value for the symmetry energy. This effect can be traced back to the choice in the selected couple of LECs,  $c_D$ and $c_E$. An alternative choice in this selection could yield a more bound SNM system, with energies approaching the desired empirical saturation energy. Consequently, this would improve results for the symmetry energy. As a consequence, for lower results of the symmetry energy, lower values of the slope $L$ are observed when including 3NFs. We once more point out that, these smaller values for $L$ at saturation density, could be improved with a different choice of low-energy constants for the description of the SNM energy curve.

\clearpage{\pagestyle{empty}\cleardoublepage}

%%%%%%%%%%%%%%%%% Conclusions %%%%%%%%%%%%%%%%%%%%%

\chapter{Summary and Conclusions}
\label{chapter:sum_concl}

The idea, on which the development of this thesis is based, has been the consistent introduction of three-body forces in the formalism of the SCGF theory. This work can be considered as the first attempt to expand the theoretical formulation of the Green's functions approach for nuclear systems to include on a same footing 2B and 3B forces. A systematic approach to the inclusion of 3B interactions in the SCGF theory was a necessary step. We have extended a formalism that appeared long ago, at the end of the 60's \citep{Mart1959}. The effort put in performing this task has been pushed by a double ambition. On the one hand, it is a well-recognized fact that nuclear many-body calculations require the inclusion of 3NFs to provide reliable nuclear structure results. On the other, the inclusion of many-body forces, beyond the 2NF, is mandatory when dealing with calculations implemented with chiral forces. In conclusion, the wish to provide consistent calculations with the use of chiral interactions has been our main motivation along the development of this thesis. 
\\

The introductory chapter of this thesis has been devoted to a general overview of the many-body problem for infinite nuclear systems. Different many-body formalisms to deal with this problem have been briefly revised. All the methods presented are \emph{ab initio}, in the sense that the problem is treated from first principles, i.e. the Hamiltonian has to be defined and the Schr\"odinger equation has to be solved. The Hamiltonian is built up of microscopic 2B potentials which are constructed to fit experimental scattering data. In this sense, these potentials contain a repulsive short-range and tensor intermediate and long range part of the strong interaction. Most of the methods discussed here are nonperturbative, meaning that the correlations arising from the repulsive behavior of the potential are taken into account and incorporated either into the interaction or into the wave function. We also considered perturbative approaches. Conversely to the previous, the perturbative methods claim that calculations can be performed with evolved nuclear potentials as well. In the former approach, the short-range core strongly characterizes the NN potential. In the latter, by means of regularization techniques, universal low-momentum softer potentials can be defined. While on one side, the correlated behavior has always been considered as the necessary condition for nuclear matter saturation, on the other, many-body forces are unavoidable elements which cannot be neglected. 
%An infinite number of potentials are equally capable to reproduce the low-energy physics of nuclear systems, given that the choice is made efficient and convenient under the conditions of interest. 
Furthermore, a revision of the most important formulations for the microscopic potentials has been presented, including Argonne $v18$ \citep{Wir1995} and the CdBonn \citep{Mac2001} potentials. At the end of the introductory chapter, we have especially focused our attention on the chiral inspired approach to nuclear forces. Reviewing the basic ideas on which $\chi$EFT is based, we then outlined the state-of-the-art of chiral forces. 
\\

The second chapter has been devoted to the description of the Green's functions many-body approach, starting from the definition of the Hamiltonian with the inclusion of a 1B, 2B and 3B contributions. For a specific SP energy and momentum, the GF can describe the propagation of a hole or a particle which interacts with the many-body system. The information enclosed in the SP propagator can be accessed by the spectral function. The spectral function defines the probability of encountering a specific hole-state or particle-state in the many-body system, and its width in energy defines the lifetime of this state. This method goes beyond a mean-field approach, in that the spread in energy of the spectral function dictates that a certain range of energies is available for a given SP momentum. The hole spectral function is directly related to experimental data, as it is connected to the value of the removed strength in ($e,e'p$) experiments of knock-out reactions. We have then focused on the reformulation of the Galitskii-Migdal-Koltun sum rule \citep{Gal1958,Kol1974}. If the system interacts only via up to 2BFs, the sum rule provides the energy of the many-body ground state by means of the sole knowledge of the SP spectral function. When the 3B force is included in the Hamiltonian, the sum rule needs further information to be evaluated. There are two different ways to reformulate the sum rule:
\begin{itemize}
\item in one case, the sum rule implies the computation of the 2BF expectation value in the many-body ground state. This approach is preferable if the fully correlated 2B propagator is computed in the many-body approximation, by means, for example, of the $T$-matrix.
\item in the other case, the sum rule requires the knowledge of the 3BF expectation value in the many-body ground state. In nuclear physics, this latter quantity is expected to provide a smaller contribution to the total energy with respect to the 2BF expectation value. We would then rather use this version of the sum rule, especially if approximations are performed in the evaluation of the expectation value.
\end{itemize}
The second version of the sum rule is the one which has been implemented in the calculations presented in the thesis. Here, the expectation value of the 3BF needed to evaluate the GMK sum rule has been only computed at the HF level. Computing the 2BF expectation value at the HF level would have induced higher errors in the total energy value due to the bigger contribution given by the 2BF operator. In the last section of Chapt.~\ref{chapter:formalism}, we have revised the self-consistent solution of the SP propagator by means of Dyson's equation. We have highlighted the fact that the solution of this equation, by means of the irreducible self-energy, requires only the inclusion of 1PI connected diagrams. Furthermore, we have dedicated particular emphasis on the concept of skeleton diagrams which, if used in the definition of the irreducible self-energy, sum up big groups of nested diagrams. Finally, we stressed that the use of an irreducible self-energy formed of connected skeleton 1PI diagrams requires a truly self-consistent iterative solution of Dyson's equation. 
\\

In Chap.~\ref{chapter:3BF_formalism}, the full expansion of the self-energy including 1B, 2B and 3B contributions has been presented. A further characteristic in the diagrams to be included in the irreducible self-energy has been considered:
\begin{itemize}
\item the concept of \emph{interaction-irreducible diagram} has been introduced, following the definition of articulation vertices \citep{Bla1986}. An articulation vertex is such if it contains a portion which links a group of propagating lines that can be isolated and disconnected from the vertex itself. We have defined a diagram as an interaction-irreducible one if no articulation vertices can be encountered.
\end{itemize}
The concept of interaction-irreducible diagrams has then led us to define the effective interactions to be used in the construction of the irreducible self-energy:
\begin{itemize}
\item an effective 1B interaction which is the sum of three contributions: the original 1B interaction; a 1B average over the 2B interaction; and a 2B average over the 3B force. The 1B and 2B averages are performed using fully dressed propagators.
\item an effective 2B interaction which is the sum of the original 2B interaction plus a correlated 1B average over the 3B force.
\end{itemize}
This approach has been compared to the usual normal ordering procedure to define effective interactions when beyond 2B forces are included in the Hamiltonian. We considered our approach as a generalization of the usual normal ordering, in that the ordering of the Hamiltonian is done with respect to the correlated many-body wave function, instead of the Slater determinant used in the latter approach. Subsequently, we emphasized the fact that, in order to correctly evaluate the perturbative expansion of the self-energy by means of the effective interaction Hamiltonian, all interaction reducible diagrams must be disregarded. In fact, these are already taken into account by means of the effective interactions. \\
The complete perturbative expansion of the self-energy has been performed all the way up to third order, proving that the definition of the effective 1B and 2B interactions substantially reduces the  number of diagrams to be considered. This procedure substantially facilitates the enumeration of all the contributions which build the self-energy expansion. In addition, we solved the equation of motion for the propagator. This solution induces the so called hierarchy of EOMs for the GFs. In other words  the solution for the EOM of the $N$-body propagator, requires the knowledge of a different $N$-body propagator. In the case of including 3BFs, the EOM technique defines a hierarchy of equations that link each $N$-body GF to the ($N$ -1)-, ($N$+1)- and ($N$+2)-body GFs. A truncation for this hierarchy is necessary in order to solve the system of equations. The solution for the EOM of both the 1B and 2B propagator was formally presented without truncations, i.e. with exact expressions. Evaluating the solution of the EOM for the 1B and 2B propagators we encountered:
\begin{itemize}
\item from the EOM of the 1B propagator we have obtained an all-order definition of the irreducible self-energy in terms of 1PI interacting vertex functions. The 1B effective interaction defines directly the energy-independent  part of the self-energy, which includes the HF approximation for the SP self-energy. Next, the vertex interacting functions, the $\Gamma^{4-\textrm{pt}}$ and $\Gamma^{6-\textrm{pt}}$, define the energy-dependent part of the self-energy which is connected, respectively, to the 2B effective and 3B interaction.  
\item from the EOM of the 2B propagator, we have obtained a complete definition of the $\Gamma^{4-\textrm{pt}}$ vertex function. This includes terms containing both the 2B effective and the 3B interaction. Special attention was paid in considering all the exchange contributions when solving the EOM for the 2B propagator.
\end{itemize}
We then demonstrated the correspondence in between the EOM method and the up-to-third-order self-energy expansion performed in the first part of the chapter. This has proved, at least up to third order, the validity of both approaches to the self-energy. Subsequently, truncations on the $\Gamma^{4-\textrm{pt}}$ interacting vertex function were finally discussed, leading to specific many-body approximations:
\begin{itemize}
\item we have analyzed the ladder, ring and parquet approximations to the solution of the many-body problem, paying special attention to the ladder approximation used in this thesis. We saw how the inclusion of the 3B force enters these approximations only by means of the effective 2B interaction. Consequently, this leaves the aspect of the approximation formally unaltered, with respect to the one in the sole 2B case. 
\end{itemize}
The pedagogical approach followed in this chapter aimed at clarifing the use of effective interactions in nonperturbative approaches. The inclusion of 3BFs performed not from first principles,  i.e. from the definition of the Hamiltonian, has led in the past to an incorrect treatment of the three-body contributions in the many-body approximations.
\\

The fourth chapter has been dedicated to the calculation of the density-dependent 2BFs from chiral 3NFs at N2LO. Chap.~\ref{chapter:eff_2b_int} has been divided into two parts: the first part presenting the formal evaluation of the density-dependent force, and the second part dedicated to the partial-wave analysis of the potential matrix elements. In the evaluation of the 2B density-dependent force:
\begin{itemize}
\item we have performed the average over the third particle of the three 3B terms appearing at N2LO in the chiral expansion. This average has been computed with the use of the self-consistent correlated momentum distribution function obtained at each step in the iterative procedure of the SCGF method. In the average, a regulator function in terms of Jacobi coordinates has been considered. Expressions for the six potential terms deriving from the contraction of the 3NFs have been presented in the specific case of diagonal momenta. The average has been performed in both cases of SNM and PNM.
\end{itemize}
In the second part of the chapter, dedicated to the analysis of the partial waves of the potential matrix elements:
\begin{itemize}
\item we have studied the effect in SNM of each of the six density-dependent terms obtained at saturation density on top of the bare 2B N3LO force. The effects on $S$, $D$, $S-D$ mixing and $P$ waves have been studied in detail. This analysis has led us to the conclusion that the effect provided by the  $2\pi$ exchange contracted term which includes in-medium effects, i.e. $\tilde V_\textrm{TPE-3}^\textrm{3NF}$, and the contact contracted term, i.e. $\tilde V_\textrm{cont}^\textrm{3NF}$, which includes vertex corrections to the contact 2B term, introduce the strong repulsion observed in $S$ partial waves. This repulsive effect, which increases with the density of the system, is what provides a mechanism for nuclear matter saturation. 
\item the overall effect of the density-dependent force has been studied for the same partial waves, considering both a free and a correlated average for the contraction of the 3B force. The comparison between the two cases has led to small differences. Higher absolute values have been observed when using the free propagator, mainly in the $D$ and $S-D$ mixing partial waves, of around $\sim0.04$ fm in enhancement. This effect has been ascribed to the different depletion of the momentum distributions used in the averaging procedure, considering that the less the depletion, the higher the absolute values of the partial waves.
\item furthermore, the effect on partial waves given by the use of different regulator functions in the averaging procedure has been analyzed. In one case, the complete function in Jacobi coordinates has been used. In the other, an external regulator which doesn't affect the internal integrated momentum has been chosen. The effect on partial waves using an external regulator function has been mostly observed on $D$ and $S-D$ mixing partial waves, obtaining an effect similar to the one provided by the use of a free propagator in the averaging procedure, i.e. higher absolute values when considering a less depleted momentum distribution function. Some of the integrated functions which build the density-dependent force have also been analyzed considering the different averaging procedures.
\end{itemize} 
From the analysis performed on the partial waves in SNM, we have concluded that the repulsion which one observes on the total energy is mostly provided by the $S$ waves. Furthermore, from the analysis of the different averaging procedures, we have concluded that the small variations are not mainly related to the presence of high-momentum states, but mostly on how momenta below $p_\textrm F$ are regulated. If these momenta are less depleted, the absolute value of the contracted 3NF rises. In fact, at saturation density, the strongest effect in absolute values is provided by the correlated average non regulated; followed by the free average with internal regulation, and the smallest absolute values are provided by the fully regulated correlated average.\\
The effects on partial waves in the case of PNM have also been studied, evaluating the density-dependent contribution at saturation density:
\begin{itemize}
\item we have observed an overall repulsive effect provided by the density-dependent force on the $S$ and $P$ partial waves accessible in neutron matter. Also in this case we have analyzed the effects provided by the different averaging procedures. Similar conclusions as for the SNM case have been driven.
\end{itemize}
In addition, at the end of Chap.~\ref{chapter:eff_2b_int} we have presented the extrapolation followed for the diagonal potential matrix elements to non-diagonal elements in relative momentum. A qualitative validation for this extrapolation has been discussed, comparing the off-diagonal $S$ partial wave of the density-dependent $1\pi$ exchange 2B term, i.e $\tilde V_\textrm{TPE-2}^\textrm{3NF}$, to the one of the bare $1\pi$ exchange 2B term.
\\

In the final chapter, we have presented results for both microscopic and macroscopic properties of symmetric nuclear and pure neutron matter. In the introduction to Chap.~\ref{chapter:results}, we have described the effects caused by the finite temperature calculations. By means of a simple extrapolation formula, the effects of temperature have been evaluated at all densities considered, providing us with an estimation of the zero temperature results. Some details on the numerical inclusion of 3BFs in the SCGF formalism have also been introduced. \\
We started with the analysis of the microscopic properties of the SNM system:
\begin{itemize}
\item the imaginary part of the SP self-energy has been analyzed for three typical momenta at various densities, i.e. $p=0$, $p_\textrm F$, and $2p_\textrm F$ for $\rho_0/2$, $\rho_0$ and $2\rho_0$. We have observed an overall small effect when considering the 3B force in the calculation. Main differences were noticed at zero momenta in the position of the minima which appears at energies close and below $\mu$. Besides, at high densities, the density-dependent 2B force lowered the values in the minima which is observed beyond the chemical potential. At these high densities, especially at low momenta, the imaginary part of the self-energy approached zero at the chemical potential value, proving the stronger degeneracy of the system in these conditions.
\item we performed a study on the quasiparticle spectrum, focusing on the SP potential which is derived in this approach from the real part of the self-energy. We observed a visible repulsive effect provided by the density-dependent 2NF for all momenta. This modification increases with the density. Effects provided on the self-energy by the different averaging procedures were mainly visible only at $2\rho_0$. 
\item we then focused our analysis on the spectral function. Also on this microscopic quantity, the effect of 3B forces has proved to be small. The spectral function has been analyzed for three typical momenta at various densities, i.e. $p=0$, $p_\textrm F$, and $2p_\textrm F$ for $\rho_0/2$, $\rho_0$ and $2\rho_0$. At zero momenta, the overall change due to the effective 2BF has been that of narrowing the peak of the spectral function, and consequently inducing higher energy tails to preserve the energy weighed sum rule for the spectral functions. For increasing momenta, we found a narrowing of the quasi-particle peak of the spectral function caused by the 3B forces. The strongest effect was observed at high density and high momenta, causing also a shift in the quasi-particle peak position. The effect caused by the use of different averaging procedures was damped even more in the spectral function, with respect to the imaginary part of the self-energy, and very little discrepancies in the curves were observed. 
\item finally we analyzed the SP momentum distribution. The main effect has been tested on the depletion of the $n(p)$ at low momenta. At low densities, 3NF induce a smaller depletion, being in some sense a source of weaker correlations, while at high densities, the modification is reversed, providing a larger population of high-momentum states. Modifications due to the different averaging procedures were once again barely visible, providing small variations in the high-momentum tails of the $n(p)$ especially at $2\rho_0$. As before, we can ascribe this modification to the different regulation applied on momenta below $p_\textrm F$ which, at high densities has a more visible repercussion. 
\end{itemize}
All in all, the small consequences, observed on the microscopic properties, has lead us to think that the 3NFs effect can be well accounted for in the quasi-particle behavior of the system. However, modification on the high energy tails in the spectral function and changes in the population of high-momentum components in the momentum distribution, is a sign that changes provided by the 3B forces on the full off-shell dependency must be considered. Furthermore, the modifications provided by the different averages are mostly visible at high densities. \\
After analyzing the microscopic properties in the case of symmetric nuclear matter, we have focused our attention on the total energy of the many-body ground state in both cases of SNM and PNM. In the case of SNM:
\begin{itemize}
\item we have studied the effect of the inclusion of 3NFs in the calculation of the total energy. Repulsion with respect to the 2B calculation has been observed for all densities. An increasing repulsion with density was tested, providing the saturation of the energy around the empirical saturation density of nuclear matter. A $\sim 9$ MeV binding energy has been obtained, a bit smaller with respect to the empirical one.
\item computing the energy with the use of the different averages, in the construction of the density-dependent force, provided a striking agreement in between the different procedures. A stronger effect mostly at high densities has been observed, which follows directly the conclusions driven from the analysis of the microscopic properties. 
\item additionally, we have studied the effect of using different LECs in the intermediate and short-range 3NF derived terms. The use of different couples of $c_D$ and $c_E$ constants has led to a band governed by the value of the $c_E$. The less negative or more positive the value of the $c_E$, the more attractive the curve for all densities.  We have also compared our results with those obtained applying SRG regularization on the 2B potential. Results for the energy of SNM appear more attractive for all densities with respect to the non regularized calculations. 
\item for the sake of consistency in the chiral expansion, we have implemented the full N2LO chiral force in both the 2B and 3B sector. We have relied on a newly optimized version of the N2LO together with the density-dependent force derived in this thesis. While in the 2B only calculation, the N2LO force yields slightly more attractive results with respect  to the N3LO ones for all densities, in the 2B+3B case, the full N2LO calculation provides results which are a bit more attractive at low densities, but visibly more repulsive at higher ones. A comparison in terms of consistency in the chiral expansion is hampered by the different Hamiltonians which characterize the two cases.
\end{itemize}
Similarly to what has been studied in the SNM case, we have also investigated the properties in PNM:
\begin{itemize}
\item we have analyzed the effect of the 3BF on the total energy of the system. The effect of the inclusion of the density dependence force has been to provide repulsion at all densities. As in the SNM case, the changes due to the different averaging procedures have been mostly observed in the high density region. 
\item moreover, we have studied the error band in the energy curve of PNM given by the theoretical uncertainties in the  LECs, the $c_1$ and $c_3$ constants. Comparing with the 2B-SRG evolved calculations, we have proved the perturbative behavior of pure neutron matter in the $\chi$EFT expansion.
\item to be once again consistent in the chiral expansion, we have performed calculations at full N2LO. In this case, the sole 2B-N2LO calculation provides higher energy values with respect to the 2B-N3LO results, on the contrary to what was obtained in SNM. The 2B-N2LO calculated values approach those obtained with the 2B N3LO plus the density-dependent N2LO force. When considering the full 2N+3N force at N2LO in the calculation, results appeared to be more repulsive, especially at high densities. As in the SNM case, we ascribe these changes to the different construction of the 2B force, and to the different LECs used. 
\end{itemize}
As a conclusion to the chapter, we have presented calculations for the symmetry energy and its slope in density dependece, focusing especially on its values at saturation density. We have provided calculations with both the N2LO and N3LO in the 2B sector complemented with the density-dependent N2LO force. We observed how, in both cases, the inclusion of the density-dependent force lowers the values of the symmetry energies for all densities. This causes a slightly larger discrepancy with respect to the accepted value of $32$ MeV at saturation density. The 2B only calculations gave higher values for the symmetry energy in the entire density range studied, where the 2B-N2LO provided the best value at saturation density in comparison to the accepted one. Different choices in the LECs $c_D$ and $c_E$ in the SNM curve could help to improve on the 2B+3B results for the symmetry energy.
\\

We conclude that the inclusion of the 3BFs in the many-body calculation has proved to be fundamental to obtain the saturation mechanism in SNM at acceptable energy/density values and to provide a stiffer equation of state for PNM. This once again proves the necessity to consider many-body forces to render the theoretical calculations consistent with empirical and experimental results.
\\

This study can be considered as a first step to the analysis, from a nonperturbative point of view, of the low-temperature properties of SNM and PNM with the inclusion of 3BFs in the SCGF approach. Several next steps for improvement can be devised. \\
Higher-order terms in the 3B force expectation value, beyond the HF level, could be included in the correction to the GMK sum rule \citep{Cip2013}. If the dressed $G^{II}$ were computed, the alternative formulation of the sum rule could be used, computing the complete expectation value of the 2B force \citep{Car2013Nov}. In addition, the full calculation of the 1B effective term (see Eq.~(\ref{ueff})), considering the contraction of the 3BF with a dressed $G^{II}$, could also be implemented. \\
As far as the averaging procedure, calculations with matrix elements considering off-diagonal relative momenta can be seen as a step to be taken. This could be a way to test the validity of the extrapolation made on diagonal matrix elements to obtain the off-diagonal ones \citep{JWHol2010}.\\
From the point of view of the chiral expansion, a full N3LO calculation could be implemented \citep{Kru2013}. This would imply the definition of new density-dependent terms from the 3NFs appearing at N3LO in the chiral expansion. Nonetheless, a treatment for the 4NFs appearing at N3LO must then be sorted out. Accessibility to other formulations of the chiral potentials for different orders in $\chi$EFT (NLO, N2LO, N3LO, etc.)  with consistent Hamiltonians at each order, could provide the possibility to truly test the convergence of results in the chiral expansion \citep{Epe2002Dec1,Epe2005,Epe2009Oct}. \\
The high temperature behavior, $T\sim10-20$ MeV, could also be explored, analyzing how thermodynamical properties of both SMN and PNM are changed due to the inclusion of 3NFs \citep{Rio2009Feb,Som2009}. Above all, studies on $\beta$-stable nuclear matter could really help constrain the finite temperature EOS including both 2B and 3B forces \citep{Heb2013Jul}.\\

Furthermore, we would like to underline once again that, while the main motivation of our study has been nuclear systems, the extended SCGF formalism can be applied to other many-body systems, of either atomic or molecular nature. In this sense, the effort in deriving the full extension to include 3BFs from first principles, has been also to advance the formalism of the SCGF method to solve the quantum many-body problem.

\clearpage{\pagestyle{empty}\cleardoublepage}

\appendix
%%%%%%%%%%%%%%%% Appendix A: Feynman Rules %%%%%%%%%%%%

\chapter{Feynman rules}
\label{chapter:feynman_rules}

We present in this appendix the Feynman rules associated with the diagrams arising
in the perturbative expansion of Eq.~(\ref{gpert}). The rules are given both in time and energy formulation, and some specific examples will be considered at the end. We pay 
particular attention to non-trivial symmetry factors arising in diagrams that include many-body 
interactions.
We work with antisymmetrized matrix elements, but for practical purposes represent 
them by extended lines. 

We provide the Feynman diagram rules for a given $p$-body propagator, such as Eqs.~(\ref{g4pt}) 
and~(\ref{g6pt}). These arise from a trivial generalization of the perturbative 
expansion of the 1B propagator in Eq.~(\ref{gpert}). 
At  $k$-th order in perturbation theory, any contribution from the time-ordered product in 
Eq.~(\ref{gpert}), or its generalization, is represented by a diagram with $2p$ external 
lines and $k$ interaction lines (from here on called vertices), 
all connected by means of oriented fermion lines. 
These fermion lines arise from contractions between annihilation and creation operators,
$$
\,\,\underbracket[0.6pt][0.2em]{\!\!a^I_{\de}(t)a\!\!}\,{}^{I \, \dag}_{\gamma}(t') \equiv \lan \Phi_0^N | \T \left[ a^I_{\de}(t)a^{I \, \dag}_{\gamma}(t')\right] |\Phi_0^N\ran=i\hbar \,G^{(0)}_{\de\gamma}(t-t') .
$$ 
Applying the Wick theorem to any such arbitrary diagram, results in the following Feynman rules.
\begin{description}
\item[Rule 1] Draw all, topologically distinct and connected diagrams with $k$ vertices, and $p$ incoming and $p$ outgoing external lines, using directed arrows. For interaction vertices the external lines are not present.
\item[Rule 2] Each oriented fermion line represents a Wick contraction, leading to the unperturbed propagator  
$i\hbar G_{\al\be}^{(0)}(t-t')$ [or $i\hbar G_{\al\be}^{(0)}(\omega_i)$]. 
In time formulation, the $t$ and $t'$ label the times of the vertices at the end and at the beginning of the line. 
In energy formulation, $\omega_i$ denotes the energy carried by the propagator. 
\item[Rule 3] Each fermion line starting from and ending at the \emph{same} vertex is an 
equal-time propagator,  $-i\hbar G_{\al\be}^{(0)}(0^-)=\rho_{\al\be}^{(0)}$.
\item[Rule 4] For each 1B, 2B or 3B vertex, write down a factor $\frac{i}{\hbar} U_{\al \be}$, \, $-\frac{i}{\hbar} V_{\al\ga,\be\de}$  or  $-\frac{i}{\hbar} W_{\al\ga\xi,\be\de\ta}$, respectively. For effective interactions, the factors are $-\frac{i}{\hbar} \widetilde{U}_{\al \be}$, \, $-\frac{i}{\hbar} \widetilde{V}_{\al\ga,\be\de}$.
\end{description}
When propagator renormalization is considered, only skeleton diagrams are used in the 
expansion. In that case, the previous rules apply with the substitution 
$i \hbar G_{\alpha\beta}^{(0)} \to 
i \hbar G_{\alpha\beta}$.
Furthermore, note that  Rule 3 applies to diagrams embedded 
in the one-body effective interaction 
(see Fig.~\ref{ueffective}) and therefore they should not be considered explicitly in
an interaction-irreducible expansion. 
In calculating $\tilde U$, however, 
one should use the correlated $\rho_{\alpha\beta}$ instead of the unperturbed one. 
\begin{description}
\item[Rule 5] Include a factor $(-1)^{L}$ where $L$ is the number of closed fermion loops. This sign comes from the odd permutation of  operators needed to create a loop
and does not include loops of a single propagator, already accounted for by Rule 3.
\item[Rule 6] For a diagram representing a $2p$-point GF, add a factor $(-i/\hbar)$, whereas for a $2p$-point interaction vertex without external lines (such as $\Sigma^\star$ and $\Gamma^{2p-pt}$) add a factor $i\hbar$.
\end{description}
The next two rules require a distinction between the time and the energy representation. 
In the time representation:
\begin{description}
\item[Rule 7] Assign a time to each interaction vertex. All the fermion lines connected to the same vertex $i$ share the same time,~$t_i$. 
\item[Rule 8] Sum over all the internal quantum numbers and integrate over all internal times from $-\infty$ to $+\infty$. 
\end{description}
Alternatively, in energy representation:
\begin{description}
\item[Rule 7']  Label each fermion line with an energy $\omega_i$, 
under the \emph{constraint} that the total incoming energy equals the total outgoing energy at 
each interaction vertex, \hbox{$\sum_i\omega_i^{in}=\sum_i\omega_i^{out}$}.
\item[Rule 8'] Sum over all the internal quantum numbers and integrate over each independent internal energy, with an extra factor $\frac{1}{2\pi}$, i.e. $\int^{+\infty}_{-\infty} \frac{d\omega_i}{2\pi}$.
\end{description}

\begin{figure}[t]
\begin{center}
 \subfloat[]{\label{diagr_29}\includegraphics[width=.28\textwidth]{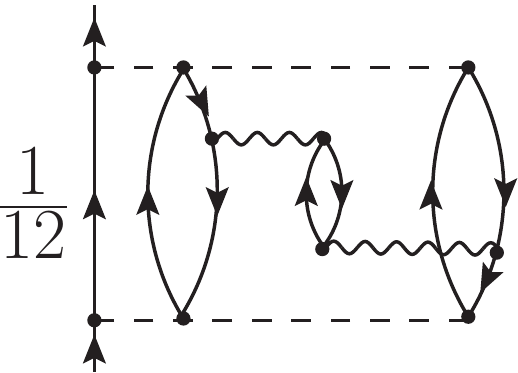}}
 \hspace{1cm}
  \subfloat[]{\label{diagr_27}\includegraphics[width=.28\textwidth]{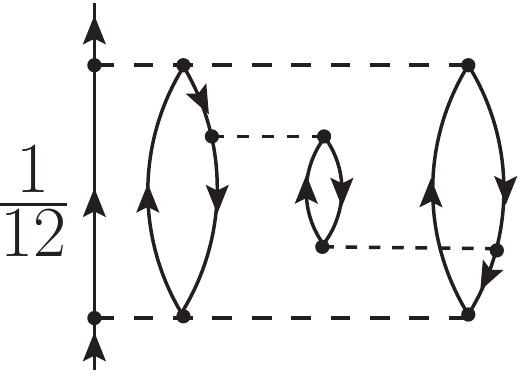}}
    \vfill
  \subfloat[]{\label{diagr_26}\includegraphics[width=.28\textwidth]{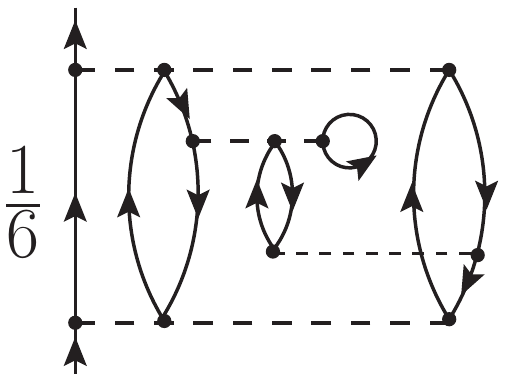}}
   \hspace{0.8cm}
  \subfloat[]{\label{diagr_28}\includegraphics[width=.32\textwidth]{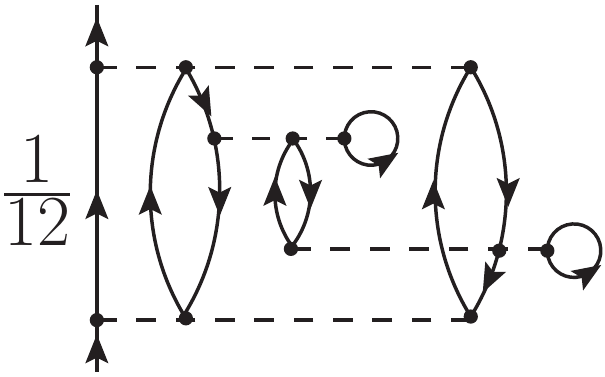}}
\caption{Examples of diagrams containing symmetric and interacting lines, with explicit symmetry 
factors. Diagrams (b) to (d) are obtained by expanding the effective
interaction of diagram (a) according to Eq (\ref{veff}). 
Swapping the 3B and 2B internal vertices in (c) gives a distinct, but topologically equivalent, contribution.}
\label{rule9-2a}
\end{center}
\end{figure}

Each diagram is then multiplied by a combinatorial factor S that  originates from the number of 
equivalent Wick contractions that lead to it. This symmetry factor 
represents the order of the symmetry group for one specific diagram or, in other words, 
the order of the permutation group of both open and closed lines, once the vertices are fixed. 
Its structure, assuming only 2BFs and 3BFs, is the following :
\beq
S=\frac{1}{k!}\frac{1} {[(2!)^2]^{q} [(3!)^2]^{k-q} }\binom{k}{q} \; C
= \prod_i S_i \; .
\label{diagsymfac}
\enq
Here, $k$ represents the order of expansion. 
$q$ ($k-q$) denotes the number of 2B (3B) vertices in the diagram.
The binomial factor counts the number of terms in the expansion $(V+W)^k$ 
that have $q$ factors of $V$ and $k-q$ factors of $W$.
Finally, $C$ is  the overall number of \emph{distinct} contractions and reflects 
the symmetries of the diagram. Stating general rules to find $C$ is not simple. 
For example, explicit simple rules valid for the well-known $\lambda \phi^4$ scalar  theory are still 
an object of debate \citep{Hue2012}. 
An explicit calculation for $C$ has to be carried out diagram by diagram. Eq.~(\ref{diagsymfac}) can normally be factorized in a product of factors $S_i$,
each due to a particular symmetry of the diagram. In the following, we list a series of specific examples which is,
by all  means, not exhaustive.
\begin{description}
 \item[Rule 9]  For each group of $n$ symmetric lines, or symmetric groups-of-lines as defined below, multiply by a symmetry factor $S_i$=$\frac{1}{n!}$. The overall symmetry factor of the diagram will be $S=\prod_i S_i$.
Possible cases include:
  \end{description}
\begin{enumerate}[(i)]
 \item {\em Equivalent lines}.  
 $n$ equally-oriented fermion lines are said to be equivalent if they start from the same initial vertex and end on the same final vertex.
 \item {\em Symmetric and interacting lines}.  
 $n$ equally-oriented fermion lines that start from the same initial vertex and end on the same final 
 vertex, but are linked via an interaction vertex to one or more closed fermion line blocks. 
 The factor arises as long as the diagram is {\em invariant} under the permutation of the two blocks.
 \item {\em Equivalent groups of lines}. 
 These are blocks of interacting lines (e.g. series of bubbles) that are equal to each other: 
           they all start from the same initial vertex and end on the same final vertex.
 \end{enumerate} 

 Rule 9-i  is the most well-known case and applies, for instance, to the two second-order diagrams 
 of Fig.~\ref{2ord}. Diagram \subref*{2ord_2B} has 2 upward-going equivalent lines and requires a symmetry factor $S_e$=$\frac1{2!}$. In contrast,�diagram \subref*{2ord_3B} has 3 upward-going equivalent lines and 2 downward-going equivalent lines, that give a factor $S_e$=$\frac1{2! \, 3!}$=$\frac1{12}$.
 
\begin{figure}[t]
\begin{center}
  \subfloat[]{\label{diagr_33}\includegraphics[width=.3\textwidth]{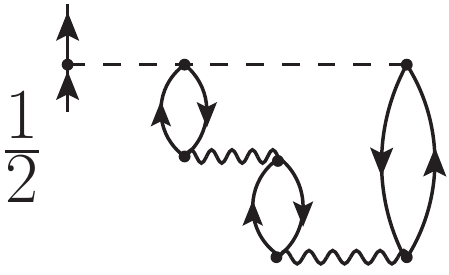}}
  \hspace{1cm}
  \subfloat[]{\label{diagr_31}\includegraphics[width=.3\textwidth]{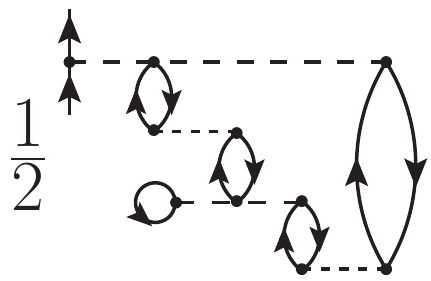}}
    \newline
  \subfloat[]{\label{diagr_12}\includegraphics[width=.3\textwidth]{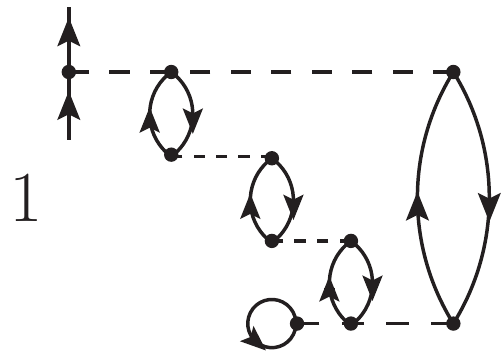}}
\caption{Examples of diagrams entering the static part of the self-energy. Applying rule 9-ii, 
diagrams (a) and (b) take a factor $S_{si}=\frac{1}{2}$ from the symmetry between the 
two bubbles attached to the upper three body vertex. 
The symmetry is broken in diagram (c), where the overall factor is  $S_{si}=1$  }
\label{rule9-2b}
\end{center}
\end{figure}

Figs.~\ref{rule9-2a} and~\ref{rule9-2b} give specific examples of the application of rule 9-ii. 
Diagram~\subref*{diagr_29} has 3 upward-going equivalent, non-interacting lines, which yield a
symmetry factor $S_e$=$\frac1{3!}$ due to rule 9-i. 
However, there are also two downward-going symmetric and equivalent lines, that 
interact through the exchange of a bubble and thus give rise to a factor $S_{si}$=$\frac1{2!}$. 
The total factor is therefore $S$=$S_{e} \times S_{si}$=$\frac1{12}$.
Let us now expand the two 2B effective interactions that are connected to the intermediate
bubble according to Eq.~(\ref{veff}).
Diagram~\subref*{diagr_29}  is now seen to contain three contributions, diagrams~\subref*{diagr_27} to \subref*{diagr_28}, with the symmetry factors shown in the figure.
Note that drawing the contracted 3B vertex above or below the bubble 
in \subref*{diagr_26} leads to two topologically equivalent diagrams that must only be drawn once, 
i.e. diagram \subref*{diagr_26}. However, since the diagram is no longer symmetric under the 
exchange of the 
two downward-going equivalent lines, rule 9-ii does not apply anymore and the $S_{si}$
factor is no longer needed. 

\begin{figure}
\begin{center}
 \includegraphics[width=1.\textwidth]{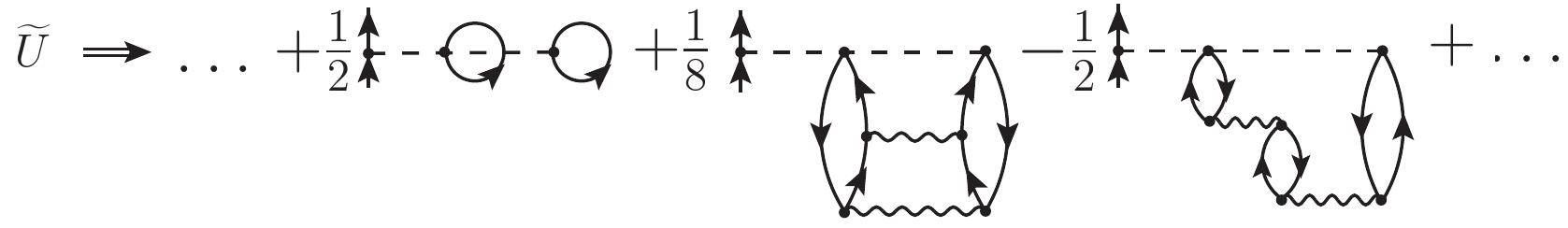}
\caption{Diagrams entering the effective one-body interaction, Eq.~(\ref{ueff}), obtained by 
substituting the right-hand side of Fig.~\ref{g4pt_2nd} into Eq.~(\ref{g4ptgamma}). The two bubble terms 
correctly reproduce the symmetric factor inferred by applying rules 9-i and 9-ii. }
\label{rule9-2u}
\end{center}
\end{figure}

A similar situation occurs when the two interacting fermion lines start and end on the same vertex, as in Fig.~\ref{rule9-2b}. Consider the left-most and right-most external fermion bubbles. 
In all three diagrams, they are
connected to each other by a 3B interaction vertex above and by a series of interactions and
medium polarizations below. 
The intermediate bubble interactions in diagrams \subref*{diagr_33} and \subref*{diagr_31} are symmetric under 
exchange. There are therefore two sets of symmetric interacting lines (the two up-going and two 
down-going fermion lines) and hence both diagrams take a factor $S_{si}=1/2$. 
In contrast, the two external loops in \subref*{diagr_12} are not symmetric under exchange due to the
lower 3B vertex. Rule 9-ii does not apply anymore and  $S_{si}=1$. 
If all the vertices between the external loops where equal (e.g. effective 2B terms $\tilde{V}$),
a factor $S_{si}$=$1/2$ would still apply.

The case of Fig.~\ref{rule9-2b} is of particular importance because these diagrams directly contribute to 
the energy-independent 1B effective interaction. 
In the EOM approach, these contributions arise from 
the first three terms on the right-hand side of Fig.~\ref{g4pt_2nd}. Note that the ladder diagram 
has a symmetry factor $S_{e}$=$1/2$ and that the exchange contribution in the bubble term has to be 
considered. Using these diagrams to the define the 2B propagator in Eq.~(\ref{g4ptgamma}) and 
inserting these in the last term of Eq.~(\ref{ueff}), one finds the contributions to $\tilde{U}$ shown 
in Fig.~\ref{rule9-2u}. The two bubble terms have summed up to form diagram \subref*{diagr_33}, each 
of them contributing a factor $1/4$ from Eq.~(\ref{ueff}). Consequently, the approach leads to 
the correct overall $S_{si}$=$1/2$ symmetry factor.
In our approach, there is no need to explicitly
compute these diagrams, since they are automatically included by Eq.~(\ref{ueff}).

\begin{figure}[t]
 \centering
  \subfloat[]{\label{diagr_32}\includegraphics[width=.28\textwidth]{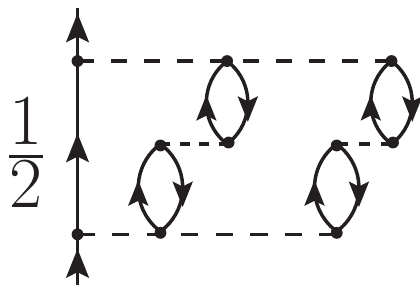}}
  \hspace{0.6cm}
  \subfloat[]{\label{diagr_30}\includegraphics[width=.25\textwidth]{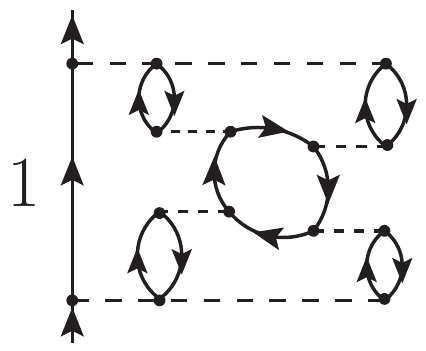}}
 \caption{ Examples of a diagram where equivalent group of lines are present and one where rule 9-iii does not apply. Swapping the two chains of bubbles in (a), one finds an identical diagram. This is precisely the case of rule 9-iii, which brings in a factor $S_{egl}$=$\frac{1}{2}$. Performing the same exchange in diagram (b) generates a graph where the direction of the internal loop is reversed. No symmetry rule applies here and $S_{egl}$=$1$}
 \label{rule9-3}
 \end{figure}

Finally, rule 9-iii applies to the diagram in Fig.~\subref*{diagr_32}. In this case, the two chains of bubble 
diagrams are equal and start and end at the same 3BF vertices. Hence, they are equivalent groups of 
lines and the diagram takes a factor $S_{egl}=\frac{1}{2}$. 
Diagram~\subref*{diagr_30} is different because the exchange of all the bubbles 
generates a diagram in which the direction of the internal fermion loop is reversed. 
Therefore no symmetry rule applies and the symmetry factor is just $S_{egl}=1$.
This is, however, topologically equivalent to the initial diagram and hence must be 
counted only once.

As an example of the application of the above Feynman rules, we give here the formulae for some of the diagrams in Fig.~\ref{3ord}. Let us start with the contribution depicted in diagram Fig.~\subref*{3ord_232B}. There are two sets of upward-going equivalent lines, which contribute to a symmetry factor $S_e=\frac{1}{2^2}$. Considering the overall factor of Eq.~(\ref{diagsymfac}) and the presence of one closed fermion loop, one finds:

\beqn
\nn &&
\Sigma^{(5c)}_{\al \be}(\om)=
- \frac{(i \hbar)^{4} }{4}
\int\frac{\d\om_1}{2\pi} \cdots \int\frac{\d\om_4}{2\pi}
\sum_{\substack{ \ga\de\nu \mu\ep\lm \\ \xi\eta\ta \sig\tau\chi}} 
\\\nn && \qquad
\times\widetilde{V}_{\al\ga,\de\nu}\gz_{\de\mu}(\om_1)\gz_{\nu\ep}(\om_2) 
W_{\mu\ep\lm,\xi\eta\ta}\gz_{\xi\sig}(\om_3)\gz_{\eta\tau}(\om_4) 
\\ && \qquad
\times\gz_{\ta\ga}(\om_1+\om_2-\om)
\widetilde{V}_{\sig\tau,\be\chi}
\gz_{\chi\lm}(\om_3+\om_4-\om)  \, .
\enqn

Diagrams \subref*{3ord_233B_1} and \subref*{3ord_233B_2} differ only for the orientation of a loop. Hence, there are
two pairs of equivalent lines in the first case and one pair and one triplet of equivalent lines in the second, which is reflected in their
different symmetry factors:

\beqn
\nn &&
\Sigma^{(5h)}_{\al \be}(\om)=\frac{(i \hbar)^{5} }{4}
\int\frac{\d\om_1}{2\pi} \cdots \int\frac{\d\om_5}{2\pi}
\sum_{\substack{ \ga\de\ep \\ \xi\ta\sig \mu\nu\lm \\ \eta\tau\phi \chi\zeta}} 
\\\nn && \quad
\times\widetilde{V}_{\al\ga,\de\ep}  \gz_{\de\xi}(\om_1)  \gz_{\nu\ga}(\om_2) 
W_{\xi\ta\sig, \mu\nu\lm}  \gz_{\mu\eta}(\om_3)\gz_{\chi\ta}(\om_4)\gz_{\ep\tau}(\om-\om_1+\om_2)
\\  && \quad
\times \, W_{\eta\tau\phi,\beta\chi\zeta } 
\, \gz_{\lm\phi}(\om_5) \, \gz_{\zeta \sig} (\om_2+\om_3+\om_5-\om_1-\om_4) 
\, ,\,\,\,
\enqn
\beqn
\nn&&
\Sigma^{(5i)}_{\al \be}(\om)=\frac{(i \hbar)^{5} }{12}
\int\frac{\d\om_1}{2\pi} \cdots \int\frac{\d\om_5}{2\pi}
\sum_{\substack{ \ga\de\ep \\ \xi\ta\sig \mu\nu\lm \\ \eta\tau\phi \chi\zeta}} 
\\\nn && \quad
\times\widetilde{V}_{\al\ga,\de\ep}  \gz_{\de\xi}(\om_1)  \gz_{\ep\ta}(\om_2) 
W_{\xi\ta\sig, \mu\nu\lm}  \gz_{\mu\eta}(\om_3)\gz_{\nu\tau}(\om_4)\gz_{\chi\ga}(\om_1+\om_2-\om) 
\\  && \quad
\times\, W_{\eta\tau\phi,\beta\chi\zeta } 
\, \gz_{\lm\phi}(\om_5) \, \gz_{\zeta \sig} (\om_3+\om_4+\om_5-\om_1-\om_2) 
\, .
\enqn

\clearpage{\pagestyle{empty}\cleardoublepage}

%%%%%%%%%%%% Appendix B: interaction irreducible diagrams %%%%%%%%%%%%%%

\chapter{Interaction-irreducible diagrams}
\label{chapter:int_irr_diag}

Interaction-irreducible diagrams can be used to distinguish between two 
different many-body effects in
the SCGF approach. On the one hand, effective interactions 
sum all the instantaneous contributions associated with ``averaging out" subgroups
of particles that lead to interaction-reducible diagrams. This has the advantage of
reducing drastically the number of diagrams at each order in the perturbative expansion.
It also gives rise to well-defined in-medium interactions. 
On the other hand, the remaining diagrams will now include higher-order terms summed
via the effective interaction itself. 

In this appendix, we prove that the perturbative expansion can be recast 
into a set containing only interaction-irreducible diagrams at any given order, as long as
properly defined effective interactions are used. 
The argument we propose has been often used to demonstrate how disconnected 
diagrams cancel out in the perturbative expansion. 
We now apply it to a slightly different case that requires extra care. 
We focus on the case of a diagram that includes only 2B and 3BFs. 
The extension to the general case of many-body forces should be straightforward.

Eq.~(\ref{gpert}) gives the perturbative expansion of the 1B GF in terms of the Hamiltonian, $H_1(t)$, in the interaction picture. The $k$-th order term of the perturbative expansion reads:
\beqn
\label{vterm}
 \nn &&
G_{\al\be}^{(k-{\rm th})}(t-t')=\left(\frac{-\ii}{\hbar}\right)^{k+1}\frac{1}{k!}\underbracket{\idotsint \textrm{d} t_{k}}_{k \,\,\textrm{terms}}
\\ && \qquad\qquad\quad
\times\lan \Phi_0^N | \T \left[ a^I_{\al}(t) a_{\be}^{I \dg} (t')H_1(t_1)\cdots H_1(t_k)\right] |\Phi_0^N\ran_\text{conn} \, .\,\,\,\,\,
\enqn
Only connected contributions are allowed and we take the interaction picture
 external creation and destruction operators to the left for convenience. Let us assume, without loss of generality, 
that the diagram is composed of $q$ 2B and $(k-q)$ 3B interaction operators. 
This gives rise to $\binom{k}{q}$ identical contributions when expanding
 \hbox{$H_1(t)=U(t)+V(t)+W(t)$} in the time-ordered product, as discussed right after Eq.~(\ref{diagsymfac}) in App.~\ref{chapter:feynman_rules}.
 
Let us denote with $O(t)$ a generic operator, representing either a 2B, 
$V(t)$, or a 3B, $W(t)$, potential. 
Suppose now that there is a sub-set of $m$ operators that are arbitrary connected to
each other, but that share the external links with a \emph{unique} operator, $O(t_n)$, outside 
the subset. In other words, $O(t_{n})$ is the only way to enter and exit the subset of $m$-
operators $\{O(t_{n+1}), \cdots, O(t_{n+m})\}$  as drawn below:
\begin{equation}
\label{red_cut}
O\!\!\!\overbracket{\,\,(t_1)\cdots O\!\!\!\underbracket{\,\,(t_{n-1})\,\cdot\,\,\, }}\!\!O\!\!\!\Bigg{\vert}\phantom{a}\!\!\!\overbracket{\underbracket{ \overbracket{\,(t_n)\cdot
\,\Big\{\,O \!\!}\,\,(t_{n+1})\cdots\!\!\!}\,\,\, O\!\!}\,\,(t_{n+m})\,\,\Big\}\,.
\end{equation}
$O(t_n)$ is also necessarily connected to the other interactions and, hence, this is an 
\emph{articulation vertex}. 
In general, there can be an arbitrary number of articulation vertices, such as  $O(t_n)$, at any 
given order. Each one of these vertices would isolate a particular subset of operators. The following arguments can be applied to each subset separately. 

For simplicity, let us restrict the argument to the simplest case of one articulation vertex only. Suppose that, among $m$ terms, there are $a$ 2B and $b$ 3B interactions, with $a+b=m$.
The number of time-ordered products $V(t)$ and $W(t)$ in Eq.~(\ref{vterm}) that is consistent with the above decomposition is
\begin{equation}
\binom{k}{q}\binom{q}{a}\binom{k-q}{b}=\binom{k}{m}\binom{m}{a}\binom{n}{q-a}\,,
\end{equation}
where $m+n=k$. 

Let us consider the case in which $O(t_n)$ is a 3B operator, with matrix elements 
$W_{\mu\ga\de,\ta\sig\xi}$ connected with four legs to the internal subset of $m$ vertices and 
with two legs to the rest of the diagram. We can factorize the
amplitude in Eq.~(\ref{vterm}) by adding an intermediate identity operator as follows:
\begin{eqnarray}
\nn
&&\frac{1}{n!}\binom{n}{q-a}\underbracket{\idotsint \textrm{d} t_{n}}_{n \,\,\textrm{terms}} 
\frac{1}{(3!)^2} \binom{3}{1}^2  
\\\nn && \times
\lan\Phi_0^N | \T \left[ a^I_{\al}(t) a^{I \dag}_{\be}(t')O(t_1) ~ \cdots ~ O(t_{n-1})\, a^{I \dag}_{\mu}(t_n^+)a^I_{\ta}(t_n)\right] |\Phi_0^N \ran \,\,W_{\mu\ga\de,\ta\sig\xi} \, \,\,  \\\nn
&& \quad \quad \times \quad
  \frac{1}{(m)!}\binom{m}{a}\underbracket{\idotsint  \textrm{d} t_{k}}_{m \,\textrm{terms} }   
 \\\nn &&
\times  \lan\Phi_0^N | \T \left[a^{I\dag}_{\ga}(t_n^+)a^{I\dag}_{\de}(t_n^+) a^I_{\xi}(t_n)a^I_{\sig}(t_n)O(t_{n+1})\cdots O(t_{k})\right]  |\Phi_0^N \ran  \de_{k,n+m}\,. 
\\
\label{redu}
\end{eqnarray} 
Note that the factorization of the time ordered product, by inserting a $|\Phi_0^N \ran \lan \Phi_0^N |$, is possible because the Wick theorem normal-orders these products with respect to the reference state, $| \Phi_0^N \ran$. In other words, 
both Eqs. (\ref{vterm}) and (\ref{redu}) lead to exactly the same results after all Wick contractions have been carried out.

All possible orders in which 
a general $O(t)$ enters Eq.~(\ref{redu}) are equivalent and are accounted for by 
the binomial factors.
The factor $\binom{3}{1}$ accounts for all the possible ways, eventually decided by contractions, in which the six creation/annihilation operators in $W(t_n)$ can be separated in the two factors [see also Eq.~(\ref{cd_factors}) below].
We also include an additional factor $\binom{3}{1}$ coming from all the possible ways to choose one creation/annihilation operator among the three possible pairs. 
The correct time ordering for creation and annihilation 
operators associated with $W(t_n)$ is preserved using 
$a^{\dag}(t^+_n)$. 

With this decomposition, we can identify the second line of Eq.~(\ref{redu}) as an $m$-th order contribution (with $a$ 2B and $(m-a)$ 3B operators) to the perturbative expansion of
$G^{4-{\rm pt}}_{\sig\xi,\ga\de}(t_n , t_n ; t_n^+,  t_n^+) = G^{II}_{\sig\xi,\ga\de}(t_n-t_n^+)$. 
Collecting all possible contributions of form (\ref{red_cut}) and (\ref{redu}) 
in which the first $n$ operators are unchanged, the $k$-th order interaction-reducible 
contribution to $G$ becomes:
\begin{eqnarray}
&&\nn
G_{\al\be}^{(k-{\rm th})}(t-t') \to \left(\frac{-\ii}{\hbar}\right)^{n+1}\frac{1}{n!}\binom{n}{q-a} \idotsint \textrm{d} t_{n} 
\\ \nn && \qquad\quad
 \times\lan\Phi_0^N | \T \left[ a^I_{\al}(t) a^{I\dag}_{\be}(t') O(t_1)\cdots O(t_{n-1})a^{I\dag}_{\mu}(t_n^+) a^I_{\ta}(t_n)\right] |\Phi_0^N \ran_{\textrm{int-irr}}
\\\nn &&  \qquad\qquad\quad
\quad \quad \times \underbracket{W_{\mu\ga\de,\ta\sig\xi} \,\frac{\ii\, \hbar}{(2!)^2}\, G^{{II}\, (m-{\rm th},\textrm{a})}_{\sig\xi,\ga\de}\,(t_n-t_n^+)}_{\mbox{$U^{\textrm{eff}}_{\mu\ta}$}} \; ,
\\
\label{redu2}
\end{eqnarray} 
where $G^{{II}\, (m-{\rm th},\textrm{a})}$ sums all the diagrams at $m$-th order with 
$a$ two-body operators. Note that the last term no longer depends on time and can be seen as an energy-independent correction to the 1B potential. We can automatically take into 
account these interaction-reducible terms by reformulating the initial hamiltonian to include 
the effective 1B vertex:
\begin{equation}
\widetilde{U}_{\mu\ta} \to U_{\mu\ta}+ W_{\mu\ga\de,\ta\sig\xi}\,\frac{\ii\hbar}{(2!)^2}\,\underbracket{G^{II}_{\sig\xi,\ga\de}\,(t- t^+)}_{\mbox{$-\frac{\ii}{\hbar}\,\rho^{2B}_{\sig\xi,\ga\de}$}}\,,
\end{equation}
where now we use an \emph{exact} $G^{II}$. The perturbative expansion obtained 
with this effective interaction should only contain interaction-irreducible diagrams  to avoid double 
counting.
 
Note that in Eq.~(\ref{redu2}) we automatically obtain the correct symmetry factor $1/(2!)^2$ associated with the contraction of $W$ with the two pairs of incoming and outgoing lines 
of $G^{II}$.
In the general case, a $c$-body vertex can be reduced to a $d$-body one (with $d<c$) by 
using a $(c-d)$-body GF.  The overall combinatorial factor in that case will be:
\begin{equation}
\frac{1}{(c!)^2}\left(\frac{c!}{d!(c-d)!}\right)^2=\underbrace{\frac{1}{(d!)^2}}_{\textrm{new vertex}}\,\,\underbrace{\frac{1}{((c-d)!)^2}}_{c-d\, \textrm{equal lines}} \, .
\label{cd_factors}
\end{equation} 
This yields both the correct combinatorial factors entering the new effective $d$-body vertex
and the symmetry factor associated with the contraction with the ($c-d$)-body GF.  
The above arguments can be  generalized to any starting $n$-body Hamiltonian.  
Applying these derivation to all possible cases for a 3B Hamiltonian leads to the effective interactions discussed in Eqs.~(\ref{ueff}) and~(\ref{veff}).

\clearpage{\pagestyle{empty}\cleardoublepage}

%%%%%%%%%%%% Appendix C: Density dependent 2NF terms %%%%%%%%%%%%%

\chapter{Complete expressions for the density-dependent force}
\label{chapter:dens_dep_terms}

In this appendix we would like to derive the complete expressions for the density-dependent terms obtained from contraction of the 3NFs at N2LO in the chiral expansion. In Sec.~\ref{section:dd_n2lo} we presented the expressions for the six in-medium contributions $\tilde V^\mathrm{3NF}$ only in the specific case of diagonal momentum, i.e. equal relative incoming and outgoing momentum $k=|{\bf k}|=|{\bf k'}|$. When considering off-diagonal momenta, the most general expression for a two-nucleon potential reads \citep{Erk1971}:
\beqn
\nn 
V({\bf k},{\bf k'})&=&V^s_c+\bd\tau_1\cdot\bd\tau_2V^v_c
\\\nn 	&&
+[V^s_\sigma+\bd\tau_1\cdot\bd\tau_2V^v_\sigma]\bd\sigma_1\cdot\bd\sigma_2
\\\nn &&
+[V^s_{\sigma q}+\bd\tau_1\cdot\bd\tau_2V^v_{\sigma q}]\bd\sigma_1\cdot{\bf q}\bd\sigma_2\cdot{\bf q}
\\\nn &&
+[V^s_{\sigma K}+\bd\tau_1\cdot\bd\tau_2V^v_{\sigma K}]\bd\sigma_1\cdot{\bf K}\bd\sigma_2\cdot{\bf K}
\\\nn &&
+[V^s_{SL}+\bd\tau_1\cdot\bd\tau_2V^v_{SL}]i(\bd\sigma_1+\bd\sigma_2)\cdot({\bf q}\times{\bf K})
\\ &&
+[V^s_{\sigma L}+\bd\tau_1\cdot\bd\tau_2V^v_{\sigma L}]\bd\sigma_1\cdot({\bf q}\times{\bf K})\bd\sigma_2\cdot({\bf q}\times{\bf K})\,.
\label{off-shell_vnn}
\enqn
In the previous expression we have introduced the total relative momentum ${\bf K}=({\bf k}+{\bf k'})/2$. Comparing with the on-shell expression written in Eq.~(\ref{on-shell_vnn}), we see that a further tensor structure has appeared, i.e. $\bd\sigma_1\cdot{\bf K}\bd\sigma_2\cdot{\bf K}$. Given the presence of this additional operatorial form, the partial wave analysis which we have followed from Ref.~\citep{JWHol2010} is not valid anymore, because in this approach this structure is disregarded. It can be shown that, in the diagonal momentum limit, this structure vanishes \citep{Erk1971}. 

For each of the three 3NF contributions, Eqs.~(\ref{tpe})-(\ref{ope})-(\ref{cont}), we need to evaluate Eq.~(\ref{dd3bf_new}). In doing so, we need to define the three different transferred momenta for each specific case of particles interchange defined in Eq.~(\ref{dd3bf_new}). We define the center of mass momentum ${\bf P}$ as \citep{Heb2010Jul}:
\beq
\label{ptot}
{\bf P}={\bf p}_1+{\bf p}_2= {\bf p'}_1+{\bf p'}_2\,,
\enq
and the relative incoming and outgoing momenta:
\beqn
{\bf k}&=&\frac{{\bf p}_1-{\bf p}_2}{2}\,,
\\\nn
{\bf k'}&=&\frac{{\bf p'}_1-{\bf p'}_2}{2}\,.
\label{k_in_out}
\enqn
We can then express the incoming and outgoing SP momenta of particle 1 and 2 as functions of ${\bf P}, {\bf k}, {\bf k'}$: 
\beqn
{\bf p}_1&=&\frac{{\bf P}}{2}+{\bf k}\,,
\\\nn
{\bf p'}_1&=&\frac{{\bf P}}{2}+{\bf k'}\,,
\\\nn
{\bf p}_2&=&\frac{{\bf P}}{2}-{\bf k}\,,
\\\nn
{\bf p'}_2&=&\frac{{\bf P}}{2}-{\bf k'}\,.
\enqn
Depending on which of the three exchange terms we are considering in Eq.~(\ref{dd3bf_new}),the transferred momenta ${\bf q}_i$, with $i=1,2,3$  will change. For the direct term we can rewrite
\beqn
\nn
{\bf q}_1&=&{\bf p'}_1-{\bf p}_1={\bf k'}-{\bf k}\,,
\\\label{transf_1}
{\bf q}_2&=&{\bf p'}_2-{\bf p}_2={\bf k}-{\bf k'}\,,
\\\nn
{\bf q}_3&=&{\bf p'}_3-{\bf p}_3=0\,.
\enqn
For the exchange term $P_{13}$ we have:
\beqn
\nn
{\bf q}_1&=&{\bf p'}_1-{\bf p}_3=\frac{{\bf P}}{2}+{\bf k'}-{\bf p}_3\,,
\\\label{transf_2}
{\bf q}_2&=&{\bf p'}_2-{\bf p}_2={\bf k}-{\bf k'}\,,
\\\nn
{\bf q}_3&=&{\bf p'}_3-{\bf p}_1={\bf p}_3-\frac{{\bf P}}{2}-{\bf k}\,.
\enqn
In the last case, the one with the exchange term $P_{23}$, we have;
\beqn
\nn
{\bf q}_1&=&{\bf p'}_1-{\bf p}_1={\bf k'}-{\bf k}\,,
\\\label{transf_3}
{\bf q}_2&=&{\bf p'}_2-{\bf p}_3=\frac{{\bf P}}{2}-{\bf k'}-{\bf p}_3\,,
\\\nn
{\bf q}_3&=&{\bf p'}_3-{\bf p}_2={\bf p}_3-\frac{{\bf P}}{2}+{\bf k}\,.
\enqn
In the approximation of zero center of mass momenta, i.e. ${\bf P}=0$, the previous expressions simplify. 

In the following expressions we define ${\bf q}_1=-{\bf q}_2={\bf q}$. For convenience we also define the following multiplying factors, which include the LECs:
\beq
\tilde c_1=\frac{c_1g_A^2M_\pi^2}{2F_\pi^4}\,, \quad 
\tilde c_3=\frac{c_3g_A^2}{4F_\pi^4}\,, \quad
\tilde c_4=\frac{c_4g_A^2}{8F_\pi^4}\,, \quad 
\tilde c_D=\frac{c_D g_A}{8F_\pi^4\Lambda_\chi}\,, \quad
\tilde c_E=\frac{c_E}{2F_\pi^4\Lambda_\chi}\,.
\label{LEC_new}
\enq

\subsubsection{Contraction of the 3NF two-pion-exchange term}
Starting with the 3NF TPE contribution, Eq.~(\ref{tpe}), we need to calculate the integral (see Eq.~(\ref{dd3bf_new})):
\beqn
\nn
&&\langle {\bf 1' 2'}|\tilde V^\mathrm{3NF}_\mathrm{OPE}
|{\bf 1 2}\rangle_A =
\mathrm{Tr}_{\sigma_3}\mathrm{Tr}_{\tau_3}
\int \frac{{\mathrm d}{\bf p}_3}{(2\pi)^3}n({\bf p}_3) f(k,k',p_3)
\\ \nn&& 
\langle {\bf 1' 2' 3'}|
\sum_{i\neq j\neq k} 
\frac{(\bd\sigma_i\cdot{\bf q}_i)(\bd\sigma_j\cdot{\bf q}_j)}{({\bf q}_i^2 + M_\pi^2)
({\bf q}_j^2 + M_\pi^2)}
F_{ijk}^{\alpha\beta}\tau_i^{\alpha}\tau_j^{\beta}
(1-P_{13}-P_{23})
|{\bf 1 2 3}\rangle_{A_{12}}\,, 
\\\label{dd3bf_tpe}
\enqn
with the tensor $F_{ijk}^{\alpha\beta}$ given in Eq.(\ref{tpe_tensor}). If we develop the sum over the three particle indices inside Eq.~(\ref{dd3bf_tpe}), we obtain (note that an implicit sum in indices $\alpha,\beta$ is considered) :
\beqn
\nn
&&\sum_{i\neq j\neq k} 
\frac{(\bd\sigma_i\cdot{\bf q}_i)(\bd\sigma_j\cdot{\bf q}_j)}{({\bf q}_i^2 + M_\pi^2)
({\bf q}_j^2 + M_\pi^2)}
F_{ijk}^{\alpha\beta}\tau_i^{\alpha}\tau_j^{\beta}
\\\nn &=&
\frac{(\bd\sigma_1\cdot{\bf q}_1)(\bd\sigma_2\cdot{\bf q}_2)}{({\bf q}_1^2 + M_\pi^2)
({\bf q}_2^2 + M_\pi^2)} [F_{123}^{\alpha\beta}\tau_1^{\alpha}\tau_2^{\beta}
+F_{213}^{\alpha\beta}\tau_2^{\alpha}\tau_1^{\beta}]
\\\nn && +
\frac{(\bd\sigma_1\cdot{\bf q}_1)(\bd\sigma_3\cdot{\bf q}_3)}{({\bf q}_1^2 + M_\pi^2)
({\bf q}_3^2 + M_\pi^2)} [F_{132}^{\alpha\beta}\tau_1^{\alpha}\tau_3^{\beta}
+F_{312}^{\alpha\beta}\tau_3^{\alpha}\tau_1^{\beta}]
\\\nn && +
\frac{(\bd\sigma_2\cdot{\bf q}_2)(\bd\sigma_3\cdot{\bf q}_3)}{({\bf q}_2^2 + M_\pi^2)
({\bf q}_3^2 + M_\pi^2)} [F_{231}^{\alpha\beta}\tau_2^{\alpha}\tau_3^{\beta}
+F_{321}^{\alpha\beta}\tau_3^{\alpha}\tau_2^{\beta}]
\\\nn &=&
2\left\{\frac{(\bd\sigma_1\cdot{\bf q}_1)(\bd\sigma_2\cdot{\bf q}_2)}{({\bf q}_1^2 + M_\pi^2)
({\bf q}_2^2 + M_\pi^2)} [\bd\tau_1\cdot\bd\tau_2(-\tilde c_1+\tilde c_3\,{\bf q}_1\cdot{\bf q}_2)\right.
\\\nn &&\quad\quad\quad\quad\quad\quad\quad\quad \left.
+\tilde c_4 (\bd\tau_1\times\bd\tau_2)\cdot\bd\tau_3 \, \bd\sigma_3\cdot({\bf q}_1\times{\bf q}_2)]\right.
\\\nn && \left.+
\frac{(\bd\sigma_1\cdot{\bf q}_1)(\bd\sigma_3\cdot{\bf q}_3)}{({\bf q}_1^2 + M_\pi^2)
({\bf q}_3^2 + M_\pi^2)} [\bd\tau_1\cdot\bd\tau_3(-\tilde c_1+\tilde c_3\,{\bf q}_1\cdot{\bf q}_3)\right.
\\\nn &&\quad\quad\quad\quad\quad\quad\quad\quad \left.
+\tilde c_4 (\bd\tau_1\times\bd\tau_3)\cdot\bd\tau_2 \, \bd\sigma_2\cdot({\bf q}_1\times{\bf q}_3)]\right.
\\\nn && \left.+
\frac{(\bd\sigma_2\cdot{\bf q}_2)(\bd\sigma_3\cdot{\bf q}_3)}{({\bf q}_2^2 + M_\pi^2)
({\bf q}_3^2 + M_\pi^2)} [\bd\tau_2\cdot\bd\tau_3(-\tilde c_1+\tilde c_3\,{\bf q}_2\cdot{\bf q}_3)\right.
\\ &&\quad\quad\quad\quad\quad\quad\quad\quad \left.
+\tilde c_4 (\bd\tau_2\times\bd\tau_3)\cdot\bd\tau_1 \, \bd\sigma_1\cdot({\bf q}_2\times{\bf q}_3)]\right\}\,.
\label{tpe_expansion}
\enqn
We then need to perform the trace over the spin/isospin of the third particle in expression Eq.~(\ref{tpe_expansion}), for each of the three exchanging terms, e.g. 1, $P_{13}$ and $P_{23}$. 

In the case of SNM, we have for the direct term, i.e. 1:
\beqn
\nn
&&
\mathrm{Tr}_{\sigma_3}\mathrm{Tr}_{\tau_3}\langle {\bf 1' 2' 3'}|\cdots
|{\bf 1 2 3}\rangle_{A_{12}}=
\\ && \qquad\quad
4\frac{(\bd\sigma_1\cdot{\bf q}_1)(\bd\sigma_2\cdot{\bf q}_2)}{({\bf q}_1^2 + M_\pi^2)
({\bf q}_2^2 + M_\pi^2)} [\bd\tau_1\cdot\bd\tau_2(-\tilde c_1+\tilde c_3\,{\bf q}_1\cdot{\bf q}_2)]\,,
\label{tpe_st_1}
\enqn
where, in the previous and in the following two equations, the dots on the left-hand side correspond to the term written in the last equality of Eq.~(\ref{tpe_expansion}), disregarding the multiplying factor 2. \\
In the case of PNM, Eq.~(\ref{tpe_st_1}) acquires a multiplying factor 1/2, as detailed in (\ref{pnm_tpe_1}). \\
In the case of SNM, we have for the exchange term $P_{13}$:
\beqn
\nn
&&
\mathrm{Tr}_{\sigma_3}\mathrm{Tr}_{\tau_3}\langle {\bf 1' 2' 3'}|\cdots
P_{13}|{\bf 1 2 3}\rangle_{A_{12}}=
\\\nn && \qquad
\frac{(\bd\sigma_2\cdot{\bf q}_2)}{({\bf q}_1^2 + M_\pi^2)
({\bf q}_2^2 + M_\pi^2)} \bd\tau_1\cdot\bd\tau_2 
\\\nn && \qquad\qquad
[\bd\sigma_1\cdot{\bf q}_1
(-\tilde c_1+(\tilde c_3+2\tilde c_4)
{\bf q}_1\cdot{\bf q}_2)-2\tilde c_4\bd\sigma_1\cdot{\bf q}_2q_1^2]
\\\nn && \qquad+ 
\frac{({\bf q}_1\cdot{\bf q}_3)+i\bd\sigma_1\cdot({\bf q}_1\times{\bf q}_3)}{({\bf q}_1^2 + M_\pi^2)
({\bf q}_3^2 + M_\pi^2)}
\\\nn && \qquad\qquad
[3(-\tilde c_1+\tilde c_3\,{\bf q}_1\cdot{\bf q}_3)+2i\tilde c_4\bd\tau_1\cdot\bd\tau_2
{\bd\sigma}_2\cdot({\bf q}_1\times{\bf q}_3)]
\\\nn && \qquad + 
\frac{(\bd\sigma_2\cdot{\bf q}_2)}{({\bf q}_2^2 + M_\pi^2)({\bf q}_3^2 + M_\pi^2)}
\bd\tau_1\cdot\bd\tau_2
\\\nn && \qquad\qquad
[\bd\sigma_1\cdot{\bf q}_3(-\tilde c_1+(\tilde c_3+2\tilde c_4){\bf q}_2\cdot{\bf q}_3)
-2\tilde c_4\bd\sigma_1\cdot{\bf q}_2q_3^2]\,;\quad
\\
\label{tpe_st_2}
\enqn
In the case of PNM, terms proportional to $c_4$ are zero because the structure $(\bd\tau_1\times\bd\tau_2)\cdot\bd\tau_3$, appearing in Eq.~(\ref{tpe_expansion}), leads to a vanishing term when calculated in a system of solely neutrons.  Contributions proportional to $c_1$ and $c_3$ change according to: the first and third line in Eq.~(\ref{tpe_st_2}) acquire a multiplying factor 1/4, and furthermore appear in an isoscalar form, i.e. no $\bd\tau_1\cdot\bd\tau_2$ multiplying, with an added -1/2 factor (see (\ref{pnm_tpe_2}) for explanation); the second line in Eq.~(\ref{tpe_st_2}) acquires a multiplying factor 1/3 (see (\ref{pnm_tpe_3}) for explanation). \\
In the case of SNM, we have for the exchange term $P_{23}$:
\beqn
\nn
&&
\mathrm{Tr}_{\sigma_3}\mathrm{Tr}_{\tau_3}\langle {\bf 1' 2' 3'}|\cdots
P_{23}|{\bf 1 2 3}\rangle_{A_{12}}
\\\nn &=&
\frac{(\bd\sigma_1\cdot{\bf q}_1)}{({\bf q}_1^2 + M_\pi^2)
({\bf q}_2^2 + M_\pi^2)} \bd\tau_1\cdot\bd\tau_2 
\\\nn && \qquad\qquad\qquad
[\bd\sigma_2\cdot{\bf q}_2(-\tilde c_1+(\tilde c_3+2\tilde c_4)\,{\bf q}_1\cdot{\bf q}_2)-
2\tilde c_4\bd\sigma_2\cdot{\bf q}_1\,q_2^2)]
\\\nn && + 
\frac{(\bd\sigma_1\cdot{\bf q}_1)}{({\bf q}_1^2 + M_\pi^2)({\bf q}_3^2 + M_\pi^2)}
\bd\tau_1\cdot\bd\tau_2
\\\nn && \qquad\qquad\qquad
[\bd\sigma_2\cdot{\bf q}_3(-\tilde c_1+(\tilde c_3+2\tilde c_4){\bf q}_1\cdot{\bf q}_3)
-2\tilde c_4\bd\sigma_2\cdot{\bf q}_1\,q_3^2]
\\\nn && + 
\frac{({\bf q}_2\cdot{\bf q}_3)+i\bd\sigma_2\cdot({\bf q}_2\times{\bf q}_3)}{({\bf q}_2^2 + M_\pi^2)
({\bf q}_3^2 + M_\pi^2)}
\\\nn && \qquad\qquad\qquad
[3(-\tilde c_1+\tilde c_3\,{\bf q}_2\cdot{\bf q}_3)+2i\tilde c_4\bd\tau_1\cdot\bd\tau_2
{\bd\sigma}_1\cdot({\bf q}_2\times{\bf q}_3)]\,.
\\
\label{tpe_st_3}
\enqn
In the case of PNM, the results follow the ones obtained for the contributions in Eq.~(\ref{tpe_st_2}). Terms proportional to $c_4$ in Eq.~(\ref{tpe_st_3}) are zero for the same reason as in Eq.~(\ref{tpe_st_2}).  Contributions proportional to $c_1$ and $c_3$ change according to: the first and second line in Eq.~(\ref{tpe_st_2}) acquire a multiplying factor 1/4, and furthermore appear in an isoscalar form, i.e. no $\bd\tau_1\cdot\bd\tau_2$ multiplying, with an added -1/2 factor (see (\ref{pnm_tpe_2}) for explanation); the third line in Eq.~(\ref{tpe_st_2}) acquires a multiplying factor 1/3 (see (\ref{pnm_tpe_3}) for explanation).

If we then perform the momentum integral over the third particle, substituting in Eq.~(\ref{tpe_st_1})-(\ref{tpe_st_2})-(\ref{tpe_st_3}) the expressions for the transferred momenta given in Eqs.~(\ref{transf_1})-(\ref{transf_2})-(\ref{transf_3}), we can rearrange the complete result in three different contributions. For SNM we have, writing explicitly the prefactors given in (\ref{LEC_new}):
\beq
\tilde V_\mathrm{TPE-1}^\mathrm{3NF}=\frac{g_A\,\rho_f}{2 F_\pi^4}
\frac{(\bd\sigma_1\cdot{\bf q})(\bd\sigma_2\cdot{\bf q})}{[q^2 + M_\pi^2]^2}
\bd\tau_1\cdot\bd\tau_2[2 c_1M_\pi^2+ c_3\,q^2]\,;
\label{tpe_dd_1_full}
\enq
\beqn
\tilde V_\mathrm{TPE-2}^\mathrm{3NF}&=& 
\frac{g_A^2}{16\pi^2F_\pi^4}\frac{\bd\tau_1\cdot\bd\tau_2}{{\bf q}^2 + M_\pi^2} 
\left\{
\frac{2}{[k'^2-k^2]}
\left[
\bd\sigma_1\cdot({\bf q}\times{\bf K})\bd\sigma_2\cdot({\bf q}\times{\bf K})\right.\right.
\\\nn &&\left.\left.
-\bd\sigma_1\cdot\bd\sigma_2[q^2K^2-({\bf q}\cdot{\bf K})^2]+\bd\sigma_1\cdot{\bf q}\bd\sigma_2\cdot{\bf q} k^2 + \bd\sigma_1\cdot{\bf K}\bd\sigma_2\cdot{\bf K} q^2
\right]\right.
\\\nn && \left. 
\left[
4c_1M_\pi^2\left[\Gamma_1(k)+\Gamma_0(k)-\Gamma_1(k')-\Gamma_0(k')\right]
\right.\right.
\\\nn && \left.\left.
-(c_3+c_4)[k'^2-k^2]\left[A(k)+\frac{\Gamma_2(k)}{k^2}+A(k')+\frac{\Gamma_2(k')}{k'^2}\right]\right.\right.
\\\nn && \left.\left.
-(c_3+c_4)q^2\left[A(k)+\frac{\Gamma_2(k)}{k^2}-A(k')-\frac{\Gamma_2(k')}{k'^2}\right]\right]\right.
\\\nn && \left.
+(\bd\sigma_1\cdot{\bf q})(\bd\sigma_2\cdot{\bf q})
\left[
4c_1M_\pi^2\left[\Gamma_1(k)+\Gamma_0(k)+\Gamma_1(k')+\Gamma_0(k')\right]\right.\right.
\\\nn && \left.\left.
+(c_3+c_4)[k'^2-k^2]\left[A(k)+\frac{\Gamma_2(k)}{k^2}-A(k')-\frac{\Gamma_2(k')}{k'^2}\right]\right.\right.
\\\nn && \left. \left.
-(c_3+c_4)q^2\left[A(k)+\frac{\Gamma_2(k)}{k^2}+A(k')+\frac{\Gamma_2(k')}{k'^2}\right]\right.\right.
\\ && \left. \left.
+4c_4 ({\cal I}(k)+{\cal I}(k'))\right]\right\}\,;
\label{tpe_dd_2_full}
\enqn
\beqn
\nn &&
V_\mathrm{TPE-3}^\mathrm{3NF}=\frac{g_A^2}{16\pi^2F_\pi^4}\Big\{
\\\nn && \qquad
-12c_1M_\pi^2\big[\Gamma_0(k)+\Gamma_0(k')-G_0(k,k')(2M_\pi^2+q^2)\big]
\\\nn &&   \qquad
-c_3\big[8k_F^3-6(2M_\pi^2+q^2)[\Gamma_0(k)+\Gamma_0(k')]
\\\nn && \qquad\qquad
-6k'\Gamma_1(k')(k'-kz)-6k\Gamma_1(k)(k-k'z)
\\\nn && \qquad\qquad
+3(2M_\pi^2+q^2)^2G_0(k,k')\big] 
%\\\nn &&  \qquad
%+ 4c_4 \bd\tau_1\cdot\bd\tau_2(\bd\sigma_1\cdot\bd\sigma_2\, q^2-\bd\sigma_1\cdot{\bf q}\bd\sigma_2\cdot{\bf q})G_2(k,k')
\\\nn &&  \qquad
-(3c_3+c_4\bd\tau_1\cdot\bd\tau_2)\,i(\bd\sigma_1+\bd\sigma_2)\cdot({\bf q}\times{\bf K})
\big[\Gamma_0(k)+\Gamma_0(k')
\\\nn &&  \qquad\qquad
+\Gamma_1(k)+\Gamma_1(k')-(2M_\pi^2+q^2)(G_0(k,k')+2G_1(k,k'))\big]
\\\nn &&  \qquad -12c_1M_\pi^2\,  i(\bd\sigma_1+\bd\sigma_2)\cdot({\bf q}\times{\bf K})
\big[G_0(k,k')+2G_1(k,k')\big]
\\\nn &&  \qquad
+4c_4\bd\tau_1\cdot\bd\tau_2\bd\sigma_1\cdot({\bf q}\times{\bf K})\bd\sigma_2\cdot({\bf q}\times{\bf K})
\\ && \qquad\qquad
\big[G_0(k,k')+4G_1(k,k')+4G_3(k,k')+G_2(k,k')\big]\Big\}\,.
\label{tpe_dd_3_full}
\enqn
In the case of PNM, contributions given in Eqs.~(\ref{tpe_dd_1_full})-(\ref{tpe_dd_2_full})-(\ref{tpe_dd_3_full}) change according to explanation given in (\ref{pnm_tpe_1}), (\ref{pnm_tpe_2}) and (\ref{pnm_tpe_3}) respectively. It can be demonstrated that the terms given in Eqs.~(\ref{tpe_dd_1_full})-(\ref{tpe_dd_2_full})-(\ref{tpe_dd_3_full}) are equivalent to those presented in Eq.~(\ref{tpe_dd_1})-(\ref{tpe_dd_2})-(\ref{tpe_dd_3}) when considering diagonal elements, i.e $k=k'$.

In the previous expression we have introduced the function:
\beq
\label{aintegral}
A(k)=[\Gamma_0(k)+2\Gamma_1(k)+\Gamma_3(k)]\,,
\enq
The functions $\Gamma_0(k), \Gamma_1(k), \Gamma_2(k), \Gamma_3(k), {\cal I}(k)$, which are integrals over a single pion propagator, have already been introduced in Sec.~\ref{section:dd_n2lo}. The function $G_0(k,k')$, which is an integral over the product of two different pion propagators, now reads:
\beq
\frac{G_{0,\star,\star\star}}{(2\pi)^2}(k,k')=\int \frac{{\mathrm d}{\bf k}_3}{(2\pi)^3}n({\bf p}_3)
\frac{\{p_3^0,p_3^2,p_3^4\}}{[[{\bf k'}+{\bf p}_3]^2+M_\pi^2][[{\bf p}_3+{\bf k}]^2+M_\pi^2]}\,.\qquad\,\,\,\,\,
\label{G_0_full} 
\enq
The rest of the functions, $G_1(k,k'), G_2(k,k'), G_3(k,k')$, and the auxiliary one $G_{1\star}(k,k')$, change according to:
\beqn
&&\nn G_1(k,k')=\frac{1}{4(1+z)}
\Big[\frac{\Gamma_0(k)}{k'^2}+\frac{\Gamma_0(k')}{k^2}
\\ && 
-\left(\frac{k^2+M_\pi^2}{k^2}+\frac{k'^2+M_\pi^2}{k'^2}\right)G_0(k,q)
-\frac{G_\star(k,k')}{k^2}-\frac{G_\star(k,k')}{k'^2}\Big]\,,\qquad\qquad
\label{G_1_full}
\enqn
\beqn
\nn
\label{G_1star_full}
&&G_{1\star}(k,k')=\frac{1}{4(1+z)}
\Big[\frac{3\Gamma_2(k)+k^2\Gamma_3(k)}{k'^2}+\frac{3\Gamma_2(k')+k'^2\Gamma_3(k')}{k^2}
\\ && 
-\left(\frac{k^2+M_\pi^2}{k^2}+\frac{k'^2+M_\pi^2}{k'^2}\right)G_\star(k,q)
-\frac{G_{\star\star}(k,k')}{k^2}-\frac{G_{\star\star}(k,k')}{k'^2}\Big]\,,\qquad\,\,\,
\enqn
\beq
\label{G_2_full}
G_2(k,k')=\frac{1}{2}(2M_\pi^2+k^2+k'^2)G_1(k,k')+G_\star(k,k')+G_{1\star}(k,k')\,,
\enq
\beqn
\nn
\label{G_3}
G_3(k,k')&=&\frac{1}{4(1+z)}
\Big[\frac{\Gamma_1(k)}{2k'^2}+\frac{\Gamma_1(k')}{2k^2}
\\\nn&& 
-2\left(\frac{k^2+M_\pi^2}{k^2}+\frac{k'^2+M_\pi^2}{k'^2}\right)G_1(k,k')
\\&&
-2\frac{G_{1\star}(k,k')}{k^2}-2\frac{G_{1\star}(k,k')}{k'^2}
-\frac{G_\star(k,k')}{k^2}-\frac{G_\star(k,k')}{k'^2}\Big]\,.\qquad\,\,\,\,\,
\enqn
where $z=\textrm{cos}\theta_{kk'}$.

\subsubsection{Contraction of the 3NF one-pion-exchange term}
Let us now perform the integration in Eq.~(\ref{dd3bf_new}) for the 3NF OPE contribution Eq.~(\ref{ope}):
\beqn
\nn &&
\langle {\bf 1' 2'}|\tilde V^\mathrm{3NF}_\mathrm{OPE}
|{\bf 1 2}\rangle_A =
\mathrm{Tr}_{\sigma_3}\mathrm{Tr}_{\tau_3}
\int \frac{{\mathrm d}{\bf p}_3}{(2\pi)^3}n({\bf p}_3)f(k,k',p_3)
\\\nn && \qquad
 \langle {\bf 1' 2' 3'}|-\sum_{i\neq j\neq k} \tilde c_D
\frac{\bd\sigma_j\cdot{\bf q}_j}{{\bf q}_j^2 + M_\pi^2}(\bd\tau_i\cdot\bd\tau_j)
(\bd\sigma_i\cdot{\bf q}_j)
\\ && \qquad\qquad\qquad\qquad\qquad\qquad\qquad\times
(1-P_{13}-P_{23})
|{\bf 1 2 3}\rangle_{A_{12}}\,.\qquad\quad
\label{dd3bf_ope}
\enqn
If we develop the sum inside Eq.~(\ref{dd3bf_ope}) we get:
\beqn
&&\nn
\sum_{i\neq j\neq k} 
\frac{\bd\sigma_j\cdot{\bf q}_j}{{\bf q}_j^2 + M_\pi^2}(\bd\tau_i\cdot\bd\tau_j)
(\bd\sigma_i\cdot{\bf q}_j) =
\\\nn && 
\left\{\frac{\bd\sigma_1\cdot{\bf q}_1}{{\bf q}_1^2 + M_\pi^2}
[(\bd\tau_2\cdot\bd\tau_1)(\bd\sigma_2\cdot{\bf q}_1) + (\bd\tau_3\cdot\bd\tau_1)(\bd\sigma_3\cdot{\bf q}_1)]
\right.
\\\nn 
&&+\left. \frac{\bd\sigma_2\cdot{\bf q}_2}{{\bf q}_2^2 + M_\pi^2}
[(\bd\tau_1\cdot\bd\tau_2)(\bd\sigma_1\cdot{\bf q}_2) + (\bd\tau_3\cdot\bd\tau_2)(\bd\sigma_3\cdot{\bf q}_2)]
\right.
\\ && + \left. 
 \frac{\bd\sigma_3\cdot{\bf q}_3}{{\bf q}_3^2 + M_\pi^2}
[(\bd\tau_1\cdot\bd\tau_3)(\bd\sigma_1\cdot{\bf q}_3) + (\bd\tau_2\cdot\bd\tau_3)(\bd\sigma_2\cdot{\bf q}_3)] 
\right\}\,.
\label{ope_expansion}
\enqn
We then need to perform the trace over spin/isospin of the third particle in Eq.~(\ref{ope_expansion}) for each of the three interchanging cases, e.g. 1, $P_{13}$ and $P_{23}$. \\
In the case of SNM we obtain for the direct term, i.e. 1:
\beqn
\nn
&& \mathrm{Tr}_{\sigma_3}\mathrm{Tr}_{\tau_3}\langle {\bf 1' 2' 3'}|\cdots
|{\bf 1 2 3}\rangle_{A_{12}}
\\ && = 4
\left\{\frac{\bd\sigma_1\cdot{\bf q}_1}{{\bf q}_1^2 + M_\pi^2}(\bd\tau_2\cdot\bd\tau_1)(\bd\sigma_2\cdot{\bf q}_1)+
\frac{\bd\sigma_2\cdot{\bf q}_2}{{\bf q}_2^2 + M_\pi^2}
(\bd\tau_1\cdot\bd\tau_2)(\bd\sigma_1\cdot{\bf q}_2)\right\}\,,\qquad\,\,\,\,\,\,
\label{ope_st_1}
\enqn
where, in the previous and in the two following equations, the dots written on the left-hand side correspond to term written in Eq.~(\ref{ope_expansion}). 
In the case of SNM we obtain for the exchange term $P_{13}$:
\beqn
\nn
&& \mathrm{Tr}_{\sigma_3}\mathrm{Tr}_{\tau_3}\langle {\bf 1' 2' 3'}|\cdots
P_{13}|{\bf 1 2 3}\rangle_A 
\\\nn && =\frac{\bd\sigma_1\cdot{\bf q}_1}{{\bf q}_1^2 + M_\pi^2} [(\bd\tau_2\cdot\bd\tau_1)(\bd\sigma_2\cdot{\bf q}_1)
+3\, (\bd\sigma_1\cdot{\bf q}_1)]
%\\\nn && 
+ 2 \frac{\bd\sigma_2\cdot{\bf q}_2}{{\bf q}_2^2 + M_\pi^2}
(\bd\tau_1\cdot\bd\tau_2)(\bd\sigma_1\cdot{\bf q}_2)
\\ && 
\quad\quad+ \frac{1}{{\bf q}_3^2 + M_\pi^2}
[3\,{\bf q}_3^2 +(\bd\tau_1\cdot\bd\tau_2)(\bd\sigma_1\cdot{\bf q}_3)(\bd\sigma_2\cdot{\bf q}_3)]\,;
\label{ope_st_2}
\enqn
In the case of SNM we obtain for the exchange term $P_{23}$:
\beqn
\nn
&& \mathrm{Tr}_{\sigma_3}\mathrm{Tr}_{\tau_3}\langle {\bf 1' 2' 3'}|\cdots
P_{23}|{\bf 1 2 3}\rangle_A 
\\\nn && = 
2 \frac{\bd\sigma_1\cdot{\bf q}_1}{{\bf q}_1^2 + M_\pi^2}
(\bd\tau_2\cdot\bd\tau_1)(\bd\sigma_2\cdot{\bf q}_1)
%\\\nn && 
+\frac{\bd\sigma_2\cdot{\bf q}_2}{{\bf q}_2^2 + M_\pi^2} [(\bd\tau_1\cdot\bd\tau_2)(\bd\sigma_1\cdot{\bf q}_2)
+3\, (\bd\sigma_2\cdot{\bf q}_2)]
\\ && \quad\quad+ \frac{1}{{\bf q}_3^2 + M_\pi^2}
[3\,{\bf q}_3^2 +(\bd\tau_1\cdot\bd\tau_2)(\bd\sigma_1\cdot{\bf q}_3)(\bd\sigma_2\cdot{\bf q}_3)]\,;
\label{ope_st_3}
\enqn
In the case of PNM, performing the trace over spin for expression given in Eq.~(\ref{ope_expansion}), considering all exchanges, e.g. 1, $P_{13}$ and  $P_{23}$, leads to a vanishing quantity, as was already explained at the end of Sec.~\ref{section:dd_n2lo}. \\
If we then perform the momentum integration, substituting in Eq.~(\ref{ope_st_1})-(\ref{ope_st_2})-(\ref{ope_st_3}) the expressions for the transferred momenta given in Eqs.~(\ref{transf_1})-(\ref{transf_2})-(\ref{transf_3}), we can rearrange the complete result into two contributions:
\beq
\tilde V_\mathrm{OPE-1}^\mathrm{3NF}=-\frac{c_D\,g_A\,\rho_f}{8\,F_\pi^4\,\Lambda_\chi}
\frac{(\bd\sigma_1\cdot{\bf q})(\bd\sigma_2\cdot{\bf q})}{{\bf q}^2 + M_\pi^2}
(\bd\tau_1\cdot\bd\tau_2)\,,
\label{ope_dd_1_full}
\enq
\beqn
\nn
&&\tilde V_\mathrm{OPE-2}^\mathrm{3NF}=\frac{c_Dg_A}{16\pi^2F_\pi^4\Lambda_\chi}
\left\{
\left[\big[\bd\sigma_1\cdot{\bf q}\bd\sigma_2\cdot{\bf q}+\bd\sigma_1\cdot{\bf K}\bd\sigma_2\cdot{\bf K}\big][A(k)+A(k')] \right.\right.
\\\nn && \quad \left.\left.
+\frac{[A(k)-A(k')]}{[k'^2-k^2]}\Big[%\right.\right.
%\\\nn && \qquad \left.\left.
\sigma_1\cdot{\bf K}\bd\sigma_2\cdot{\bf K}q^2+\bd\sigma_1\cdot{\bf q}\bd\sigma_2\cdot{\bf q}\left(\frac{k^2+k'^2}{2}-\frac{q^2}{4}\right)\right.\right.
\\\nn && \quad\left.\left.
+\bd\sigma_1\cdot\bd\sigma_2(q^2K^2-({\bf q}{\bf K})^2)+\bd\sigma_1\cdot({\bf q}\times{\bf K})\bd\sigma_2\cdot({\bf q}\times{\bf K}) \right.\right.
\\ && \quad\left.\left.
+(\Gamma_2(k)+\Gamma_2(k'))\bd\sigma_1\cdot\bd\sigma_2\Big](\bd\tau_1\cdot\bd\tau_2)\right]
+3({\cal I}(k)+{\cal I}(k'))\right\}\,.
\label{ope_dd_2_full}
\enqn
In the previous expressions we have used the explicit prefactor given in (\ref{LEC_new}) for the quantity $\tilde c_D$. Contributions given in Eqs.~(\ref{ope_dd_1_full})-(\ref{ope_dd_2_full}) are equivalent to those presented in Eq.~(\ref{ope_dd_1})-(\ref{ope_dd_2}) when considering the case of diagonal momentum elements, i.e. $k=k'$.

\subsubsection{Contraction of the 3NF contact term}
For the contact term Eq.~(\ref{cont}), we need to evaluate the integral:
\beqn
\nn
&&\langle {\bf 1' 2'}|\tilde V^\mathrm{3NF}_\mathrm{cont}
|{\bf 1 2}\rangle_A =
\mathrm{Tr}_{\sigma_3}\mathrm{Tr}_{\tau_3}
\int \frac{{\mathrm d}{\bf p}_3}{(2\pi)^3}n({\bf p}_3) f(k,k',p_3)
\\ && \quad\quad\quad\quad
 \langle {\bf 1' 2' 3'}|\sum_{j\neq k} \tilde{c}_E
\bd\tau_j \cdot \bd\tau_k (1-P_{13}-P_{23})|{\bf 1 2 3}\rangle_{A_{12}}\,.
\label{dd3bf_cont}
\enqn
Expanding the sum inside Eq.~(\ref{dd3bf_cont}) we get:
\beq
\sum_{j\neq k} \bd\tau_j \cdot \bd\tau_k =2\sum_{cycle}\bd\tau_j \cdot \bd\tau_k 
= 2 (\bd\tau_1\cdot\bd\tau_2+\bd\tau_2\cdot\bd\tau_3+\bd\tau_3\cdot\bd\tau_1)\,.
\label{cont_expansion}
\enq
In the case of SNM, the trace over spin/isospin over the third particle of Eq.~(\ref{cont_expansion}) leads to:
\beq
\mathrm{Tr}_{\sigma_3}\mathrm{Tr}_{\tau_3} \langle {\bf 1' 2' 3'}|\cdots|{\bf 1 2 3}\rangle_{A_{12}} = 4 \bd\tau_1\cdot\bd\tau_2\,;
\enq
\beq
\mathrm{Tr}_{\sigma_3}\mathrm{Tr}_{\tau_3} \langle {\bf 1' 2' 3'}|\cdots
P_{13}|{\bf 1 2 3}\rangle_{A_{12}} = 2 \bd\tau_1\cdot\bd\tau_2 +3\,;
\enq
\beq
\mathrm{Tr}_{\sigma_3}\mathrm{Tr}_{\tau_3} \langle {\bf 1' 2' 3'}|\cdots P_{23}|{\bf 1 2 3}\rangle_{A_{12}} = 2 \bd\tau_1\cdot\bd\tau_2+3\,.
\enq
The dots in the previous expressions correspond to the term given in Eq. (\ref{cont_expansion}) disregarding the multiplying factor 2. \\
In the case of PNM, evaluating the trace over spin of Eq.~(\ref{cont_expansion}), considering all exchanges, leads to a vanishing term. 

The integral over momentum in Eq.~(\ref{dd3bf_cont}) is trivial. Writing explicitly the prefactor $\tilde c_E$ given in Eq.~(\ref{LEC_new}), we have for the contact 3NF contracted term:
\beq
\label{cont_dd_full}
\tilde V_\mathrm{cont}^\mathrm{3NF}=-\frac{3 c_E\rho_f}{2 F_\pi^4\Lambda_\chi}\,.
\enq

\section{Numerical implementation}
Let us give some details on the numerical implementation for the calculation of the density-dependent force. We start with the definition of the mesh necessary to calculate the integral over the internal momenta ${\bf p}_3$ (see Eq.~(\ref{dd3bf_new})). We have two different cases:
\begin{itemize}
\item calculation with dressed distribution function $n({\bf p}_3)$. In this case we need to cover momenta up to a certain high value in which it is sure that the $n({\bf p}_3)$ has reached zero. We define three Gauss-Legendre meshes respectively from $0$ to $p_\textrm F/3$, from $p_\textrm F/3$ to $p_\textrm F+p_\textrm F/3$, from $p_\textrm F+p_\textrm F/3$ to $3p_\textrm F$. Meshes are chosen to cover accurately the behavior of the distribution function from momenta lower to those higher than $p_\textrm F$. Finally, high momenta after $p_\textrm F$ are represented through a tangential mesh. We have 50  points in the Gauss-Legendre meshes, and 500 in the tangential one. 
\item calculation with the undressed distribution function, i.e. a step function up to $p_\textrm F$. In this case we use a simple gauss mesh from 0 to $p_\textrm F$ of 50 points
\end{itemize}
For the external relative momenta, i.e. $k$ and $k'$, and the relative angle between them, i.e. $z$=cos$\theta$, we define gaussian meshes. For relative momenta we have $N_k=100$, for values from 0 to 1100 MeV. At $k\sim1000$ MeV it is certain that the potential is zero. After the complete evaluation of the density-dependent force $\tilde V^\textrm {3NF}$, we perform a tangential map of these values up to relative momenta $k\sim10^6$ MeV. A high-momentum mapping of the potential is necessary for the correct integration of the Lippmann-Schwinger equation to obtain the $T$-matrix \citep{Rio2007PhD}.

In the case of the average using the self-consistent momentum distribution function coming form the previous iterative step, we need to obtain this via Eq.~(\ref{mom_dist}). At each iterative step, the values of the imaginary and real part of the self-energy are stored, for different points in the energy and momentum space. The number of points in the energy mesh is $N_\om\sim6000$ for energies ranging from $\om=[-2200:10000]$ MeV. For the momentum mesh we have $N_k=70$, for SP momenta going from 0 to 3000 MeV. We then interpolate through a spline the values of the imaginary and real part of the self-energy to a fine energy mesh of $N_{\om,\textrm{spline}}=30000$. These values are used to define the spectral function (see Eqs.~(\ref{Sp_self})-(\ref{Sh_self})) necessary to evaluate Eq.~(\ref{mom_dist}). This last step is performed via a trapezoidal integral in the energy range. We then perform a linear interpolation of the obtained values of $n({\bf p})$ to the mesh of ${\bf p}_3$ defined for the integration of the quantities in the density-dependent force. Extrapolated values are set to zero.
\\

Let us give some hints on the general numerical implementation of the SCGF method. A schematic representation of each iterative step has been shown at the beginning of Chap.~\ref{chapter:results} in Fig.~\ref{num_impl}. Approximately $\sim25$ minutes are necessary to perform a complete iteration, as the one depicted in Fig.~\ref{num_impl}. Depending on the starting point of the iteration procedure, $\sim10$ iterations are needed to obtain converged results. These could be less for low densities and high temperatures, and if the starting point is a converged iteration for another density, temperature state of the system. The number of iterations for convergence increase for lower temperatures and high densities. The fundamental quantities that one has to calculate are the imaginary part of the 2B GF, the $T$-matrix and the self-energy. 

The first order approximation of the 2B propagator corresponds to the independent propagation of two fully dressed particles. This includes two terms, a direct and an exchange one, as depicted diagrammatically in Fig.~\ref{g4point}. This is built from an imaginary and a real part \citep{Rio2007PhD}. Once these two parts are obtained separately, an angle average is necessary to calculate the full $G^{II,f}$. This average is necessary to circumvent the coupling of partial waves with different total angular momentum $J$ which appear in $G^{II,f}$. The average is performed over the angle formed by the center of mass momentum and the relative momentum of the two nucleons. This strategy facilitates the solution of the Lippmann-Schwinger equation to evaluate the $T$-matrix. 

The Lippmann-Schwinger equation to be solved is then a one dimensional integral equation for each allowed combination of $J,\,S,\,T$, and at most two coupled values of $L$, due to the tensor component of the NN interaction. By means of a discretization procedure, the equation for the $T$-matrix is converted into a complex matrix equation which can be solved via standard numerical techniques \citep{Rio2007PhD}. A matrix inversion has to be performed to solve this equation. This can be quite demanding if the dimension of the matrix is large. It is then important to sample in a correct manner the number of integration mesh points without loosing physical information. This is achieved by sampling conveniently the region where $G^{II,f}$ is maximum in the relative momentum, and the high relative momentum region where, due to correlations,  $G^{II,f}$ might not be negligible. This task is the most time-consuming in the SCGF program \citep{Rio2007PhD}. 

A nice property of the SCGF method concerns the fact that there is no need to worry about poles in the inversion of this matrix. In fact in this approach, the integrand remains finite for all energies and momenta. There is however a pole in the bosonic function which is present in the $T$-matrix, that is canceled by a node of the same $T$-matrix (see \citep{Rio2007PhD} for details). We have already mentioned that at low temperatures, the appearance of bound states signals the onset of the pairing instability. This would directly appear as a pole in the matrix which has to be inverted to solve the Lippmann-Schwinger equation. However, this should be seen only below a critical temperature which is around $T_c\sim5$ MeV. Our calculations are at this border line. Especially in the case of symmetric nuclear matter, convergence at this temperature and for increasing density starts to be slow and difficult to achieve. This is due to the neutron-proton pairing in the coupled $^3S_1-^3D_1$ channel. In pure neutron matter, where this channel is not available, convergence is good all the way up to high densities, even for low temperatures. 

The remaining step in the SCGF method is the calculation of the self-energy from the $T$-matrix. The first quantity to be obtained is the imaginary part of the self-energy, Im$\Sigma^\star$. A momenta and energy integrals have to be performed, taking special care for the pole in energy of the Bose function in the $T$-matrix. Even though this pole is canceled by a node, specific numerical arrangements have to be taken in the definition of meshes. The real part of the self-energy is then obtained by means of a dispersion relation from the imaginary part, while the HF self-energy is calculated directly from the potential. To evaluate this quantity, the calculation of the momentum distribution is needed, which is obtained as described above. According to the mesh points in which the spectral function is needed, the interpolation is done on the imaginary and real part of the self-energy, and not directly on the spectral function. This is done in order to avoid incorrect samplings of the structure of the spectral functions which could induce numerical inaccuracies. We must point out that the energy mesh for the evaluation of the spectral function must be accurate enough to reproduce not only the quasi-particle peak region, but furthermore the low and high-energy tails which characterize the spectral function.

\clearpage{\pagestyle{empty}\cleardoublepage}

\backmatter % book mode only

\bibliographystyle{jmb}
\bibliography{biblio}
\clearpage{\pagestyle{empty}\cleardoublepage}

%\newpage
%\null\thispagestyle{empty}
%\clearpage{\pagestyle{empty}\cleardoublepage}

\end{document}